%% file: draft.tex
\documentclass[aos,preprint]{imsart}
\usepackage{silence}
\input{preamble}

\begin{document}

\begin{bibunit}[imsart-nameyear]

\begin{frontmatter}
\title{Testing Conditional Independence via the Spectral Generalized Covariance Measure: Beyond Euclidean Data}
\runtitle{Testing CI via SGCM} 

\begin{aug}
\author[A]{\fnms{Ryunosuke}~\snm{Miyazaki}}
\and
\author[A]{\fnms{Yoshimasa}~\snm{Uematsu}\ead[label=e2]{yoshimasa.uematsu@r.hit-u.ac.jp}}

\address[A]{Department of Social Data Science, Hitotsubashi University\printead[presep={,\ }]{e2}}
\end{aug}

\begin{abstract}
We propose a conditional independence (CI) test based on a new measure, the \emph{spectral generalized covariance measure} (SGCM). The SGCM is constructed by expressing the squared norm of the conditional cross-covariance operator in spectral coordinates and approximating it in finite dimensions using data-dependent bases obtained from empirical covariance operators. This avoids direct estimation of conditional mean embeddings and reduces nuisance estimation to a finite collection of scalar-valued regressions. On the theoretical side, under a doubly robust product-bias condition, we establish uniform bootstrap validity and uniform asymptotic size control, and derive nontrivial uniform power and uniform consistency over classes of projected separated alternatives. The analysis also clarifies the role of spectral truncation: stronger truncation relaxes nuisance-estimation requirements, whereas weaker truncation retains more of the projected signal. To support applications beyond Euclidean data, we develop characteristic-kernel constructions on general Polish spaces via a pullback principle and non-constant completely monotone transforms of continuous negative-type semimetrics, with closure under finite tensor products. These constructions cover examples such as distribution-valued data, curves in metric spaces, and manifold-valued observations. Simulations show near-nominal size in the main settings and competitive power across a range of challenging scenarios.
\end{abstract}

\begin{keyword}
\kwd{Characteristic kernel}
\kwd{Conditional cross-covariance operator}
\kwd{Double robustness}
\kwd{Random objects}
\kwd{Spectral decomposition}
\end{keyword}

\end{frontmatter}

\input{Body.tex}


\begin{funding}
This work was supported by JSPS KAKENHI Grant Numbers JP25K00625 and JP24K02904, and by Nomura Foundation Grant Number N25-4-F30-021. 
\end{funding}

\section*{Supplementary material}
\label{sec:supplementary-material}
The Supplementary Material includes technical details, extended proofs, additional simulations, and further information regarding characteristic kernels on general Polish spaces.

\putbib[ref]

\end{bibunit}

\clearpage
\begin{bibunit}[imsart-nameyear]

\appendix
\setcounter{section}{0}
\setcounter{subsection}{0}
\renewcommand{\thesection}{\Alph{section}}
\renewcommand{\thesubsection}{\thesection.\arabic{subsection}}
\setcounter{footnote}{0}
\setcounter{figure}{0}
\setcounter{table}{0}
\renewcommand{\theequation}{\thesection.\arabic{equation}}
\renewcommand{\thefigure}{\thesection.\arabic{figure}}
\renewcommand{\thetable}{\thesection.\arabic{table}}
\setcounter{equation}{0}

\makeatletter
\@addtoreset{theorem}{section}
\@addtoreset{equation}{section}
\@addtoreset{figure}{section}
\@addtoreset{table}{section}
\setcounter{theorem}{0}
\let\c@lemma\c@theorem
\let\c@proposition\c@theorem
\let\c@corollary\c@theorem
\let\c@definition\c@theorem
\let\c@assumption\c@theorem
\let\c@remark\c@theorem
\renewcommand{\thetheorem}{\thesection.\arabic{theorem}}
\let\thelemma\thetheorem
\let\theproposition\thetheorem
\let\thecorollary\thetheorem
\let\thedefinition\thetheorem
\let\theassumption\thetheorem
\let\theremark\thetheorem
\makeatother

\begin{center}
	{\Large Supplementary Material for} \\[7mm]
	\textbf{\Large Testing Conditional Independence via the Spectral Generalized Covariance Measure: Beyond Euclidean Data} \\[10mm]
	\textsc{\large Ryunosuke Miyazaki} {\large and} \textsc{\large Yoshimasa Uematsu} \\[5mm]
	\textit{\large Department of Social Data Science, Hitotsubashi University} \\[1mm]
\end{center}

\noindent\textbf{Abstract.}
This supplementary material provides technical details, extended proofs for the theoretical results, additional simulation scenarios, and further discussion on the characteristic kernels presented in the main manuscript.

\begingroup
\section*{Contents}
\setcounter{tocdepth}{2}
\csname @starttoc\endcsname{stoc}
\bigskip

\section*{Notation and conventions}

\setcounter{section}{0}
\setcounter{subsection}{0}

\suppTOCon

\input{Appendix.tex}

\suppTOCoff
\endgroup

\putbib[ref]

\end{bibunit}

\end{document}

%% file: Body.tex
\section{Introduction}\label{sec:intro}

Conditional independence (CI) is a fundamental notion in statistics and machine learning, with central roles in causal inference, graphical modelling and variable selection \citep{Dawid1979,Pearl2009,Candesetal2018}. In many modern applications, however, the variables of interest are not naturally represented as Euclidean vectors, but as random objects taking values in general metric
spaces, such as trajectories, distributions and compositional observations \citep{PetersenandMuller2019,DubeyandMuller2020,BritoandDias2022,StefanucciandMazzuco2022}. Motivated by such settings, we study nonparametric testing of
\[
H_0:X\indep Y\mid Z
\qquad\text{versus}\qquad
H_1:X\notindep Y\mid Z,
\]
based on i.i.d.\ observations from a distribution on a product of Polish spaces.

The main difficulty is to construct a CI test that remains statistically meaningful while being robust to high dimensionality and applicable to complex data structures in such general spaces. On the one hand, regression-based procedures are attractive in the presence of high-dimensional nuisance covariates, but often target surrogate notions of conditional association rather than CI itself \citep{ShahandPeters2020,Lundborgetal2022,Scheideggeretal2022}. On the other hand, kernel- and metric-based approaches can in principle capture more general alternatives, but direct estimation of conditional distributions or conditional mean embeddings in reproducing kernel Hilbert spaces (RKHSs) is often difficult and can complicate both size control and implementation \citep{Zhang2011,Wangetal2015a,Pogodinetal2025,Zhangetal2025,TangandLi2026}. For non-Euclidean random objects, this issue is further exacerbated by the fundamental need to construct suitable characteristic kernels on the underlying spaces.

We propose the spectral generalized covariance measure (SGCM), a new CI test based on a spectral representation of the conditional cross-covariance operator (CCCO). The key idea is to retain an exact operator-valued target for CI while bypassing the explicit estimation of conditional mean embeddings in RKHSs. Instead, we expand the target in data-dependent spectral coordinates and reduce nuisance estimation to a finite collection of scalar conditional mean regressions for spectral scores. This yields a regression-based implementation that can directly leverage modern machine-learning methods, while the non-pivotal null distribution is calibrated by a multiplier bootstrap.

Our main theoretical contribution is a uniform asymptotic theory for the proposed test. Under suitable spectral separation, regression-bias and non-degeneracy conditions, we establish uniform weak convergence under the null and uniform bootstrap validity, which together imply uniform asymptotic size control. On the alternative side, we derive nontrivial uniform power and uniform consistency over classes of projected separated alternatives. These results make explicit the role of spectral truncation: after truncation, the relevant signal is the projected CCCO, whereas the nuisance regressions need only satisfy a doubly robust product-bias
condition.

To support applications beyond Euclidean data, we also study kernel design on general Polish spaces. While the asymptotic theory above is formulated for the CCCO induced by a given choice of kernels, its interpretation as a CI test requires that the chosen kernels characterize CI. We therefore develop general constructions that guarantee this identification property. In particular, we establish a pullback principle for characteristic kernels and show that non-constant completely monotone transforms of continuous negative-type semimetrics yield bounded characteristic kernels, with closure under finite tensor products. These results provide a systematic route to valid kernel construction for CI testing on a broad class of non-Euclidean domains.

Simulations show competitive size--power performance, with near-nominal size in the main settings and reasonable robustness in more challenging designs.

\section{Definitions, background and the CI measure}
\label{sec:definition_and_Our_Measure}

\subsection{Notation and probability setup}\label{subsec:notation}

Let \(\cX,\cY,\cZ\) be Polish spaces and set \(\cS\coloneqq \cX\times\cY\times\cZ\). Write \(\cB(\cS)\) for the Borel \(\sigma\)-algebra on \(\cS\) and \(\cP(\cS)\) for the set of Borel probability measures on \(\cS\); we omit the argument \(\cS\) when it is clear from context. For each \(P\in\cP(\cS)\), let \(\{(X_i,Y_i,Z_i)\}_{i\ge 1}\) be an i.i.d.\ sequence with common law \(P\). For asymptotic arguments, we work on the canonical product space \((\Omega,\cF)\coloneqq (\cS^{\bbN},\cB(\cS^{\bbN}))\) equipped with \(\pr_P\coloneqq P^{\otimes\bbN}\), under which \((X_i,Y_i,Z_i)\) are the coordinate maps. For a family \(\cP'\subset\cP(\cS)\), we write \(V_n=o_{\cP'}(1)\) if \(\sup_{P\in\cP'}\pr_P(|V_n|>\varepsilon)\to 0\) for every \(\varepsilon>0\), and \(V_n=O_{\cP'}(1)\) if \(\lim_{M\to\infty}\limsup_{n\to\infty}\sup_{P\in\cP'}\pr_P(|V_n|>M)=0\).

Throughout, \(k_{\cX},k_{\cY},k_{\cZ}\) are bounded kernels on \(\cX,\cY,\cZ\), with RKHSs \(\cH_{\cX},\cH_{\cY},\cH_{\cZ}\), respectively. For Hilbert spaces \(\cH\) and \(\cH'\), write \(\cH\otimes\cH'\) for their Hilbert tensor product. Unless stated otherwise, \(\|\cdot\|\) and \(\langle\cdot,\cdot\rangle\) denote the norm and inner product of the ambient Hilbert space, while \(\|\cdot\|_{\mathrm{HS}}\), \(\|\cdot\|_{\mathrm{tr}}\), and \(\|\cdot\|_{\op}\) denote the Hilbert--Schmidt, trace, and operator norms. We assume that the feature maps \(x\mapsto k_{\cX}(x,\cdot)\), \(y\mapsto k_{\cY}(y,\cdot)\), and \(z\mapsto k_{\cZ}(z,\cdot)\) are Borel measurable. Under these conditions, the RKHSs are separable and the Bochner integrals below are well defined; see \citet[Theorem~2.4]{OwhadiandScovel2017}.

For \(P_X\in\cP(\cX)\), the kernel mean embedding is \(\int_{\cX} k(x,\cdot)\,P_X(\diff x)\). Given a regular conditional distribution \(P_{X\mid Z=z}\), the conditional mean embedding (CME) is \(\mu_{X\mid Z}(z)\coloneqq \int_{\cX} k(x,\cdot)\,P_{X\mid Z=z}(\diff x)\); we use the analogous notation \(\mu_{Y\mid Z}\) for \(Y\). A kernel \(k\) is called characteristic if the map \(P_X\mapsto \int_{\cX} k(x,\cdot)\,P_X(\diff x)\) is injective. It is called \(L^2\)-universal if its RKHS \(\cH_k\) is dense in \(L^2(P)\) for every \(P\in\cP(\cX)\). It is called integrally strictly positive definite (ISPD) if \(\iint k(x,x')\,\eta(\diff x)\eta(\diff x')>0\) for every nonzero finite signed Borel measure \(\eta\).

\subsection{Conditional cross-covariance operator}
\label{subsec:CCCO}

Our population target is the conditional cross-covariance operator (CCCO). While closely related objects appear in \citet{Fukumizuetal2004} and \citet{Pogodinetal2025}, we use a CME-based representation that is convenient for both the CI characterization and construction of the test statistic. 
Using residual feature maps $\epsilon_{X\mid Z} \coloneqq k_{\cX}(X,\cdot)-\mu_{X\mid Z}(Z)$ and $\epsilon_{Y\mid Z} \coloneqq k_{\cY}(Y,\cdot)-\mu_{Y\mid Z}(Z)$, we define $\Psi\in \cH_{XYZ}\coloneqq \cH_{\cX}\otimes\cH_{\cY}\otimes\cH_{\cZ}$ by
\begin{equation}\label{eq:def_Psi}
    \Psi\coloneqq \epsilon_{X\mid Z}\otimes \epsilon_{Y\mid Z}\otimes k_{\cZ}(Z,\cdot),
\end{equation}
whose expectation yields the CCCO:
\begin{equation}\label{ccco}
    \Sigma_{XYZ\mid Z}\coloneqq \E_P[\Psi].
\end{equation}

It is immediate that $X \indep Y \mid Z$ implies $\Sigma_{XYZ\mid Z}=0$. To obtain the converse, we impose the following condition on the kernels.

\begin{assumption}\label{ass:kernel_characteristic}
At least one of the following holds: (i) \(k_{\cX}\otimes k_{\cZ}\) and \(k_{\cY}\) are characteristic; (ii) \(k_{\cY}\otimes k_{\cZ}\) and \(k_{\cX}\) are characteristic; (iii) \(k_{\cX}\) and \(k_{\cY}\) are characteristic, and \(k_{\cZ}\) is \(L^2\)-universal.
\end{assumption}

\begin{proposition}\label{prop:CI_V}
Under Assumption~\ref{ass:kernel_characteristic}, for any Borel probability measure \(P\in\cP(\cS)\), \(X\indep Y\mid Z\) if and only if \(\Sigma_{XYZ\mid Z}=0\).
\end{proposition}

Assumption~\ref{ass:kernel_characteristic} is weaker than the \(L^2\)-universality imposed by \citet{Pogodinetal2025}. On general Polish spaces, tensor products of characteristic exponential-type kernels remain characteristic; see Corollary~\ref{cor:tensor_char_cmon} in Section \ref{subsec:char_theory}. On Euclidean spaces, more general criteria for characteristic product kernels are available \citep{SzaboandSriperumbudur2018}. Alternatively, one may replace tensor-product characteristicity by \(L^2\)-universality.

\begin{remark}[Relation to the classical CCCO and MMD]
Up to a canonical isomorphism, $\Sigma_{XYZ\mid Z}$ coincides with the classical CCCO of \citet{Fukumizuetal2004} and corresponds to the MMD between the joint law and its conditionally independent coupling. Proofs are provided in the Supplementary Material (Section~\ref{subsec:CCCO_MMD}).
\end{remark}

\section{Methodology: spectral generalized covariance measure}\label{sec:Methodology}

\subsection{Spectral representation and test statistic}

Building on the CME-based representation of the CCCO, we rewrite \(\left\|\Sigma_{XYZ\mid Z}\right\|^2\) in spectral coordinates and construct a finite-dimensional empirical analogue.

Let $\Psi_i$, $\epsilon_{X\mid Z}^{(i)}$, and $\epsilon_{Y\mid Z}^{(i)}$ be i.i.d.\ copies of the random elements defined in \eqref{eq:def_Psi}. A direct plug-in implementation would estimate $\mu_{X\mid Z}$ and $\mu_{Y\mid Z}$ as RKHS-valued regressions, plug them into $\Psi$, and compute the squared norm of the resulting empirical average. However, this approach requires RKHS-valued nonparametric regression and introduces regularization bias directly into the test statistic. We instead work with a spectral representation, which reduces the infinite-dimensional nuisance estimation problem to a finite collection of scalar conditional mean regressions.

Specifically, by the separability of the RKHSs, we may decompose $\left\|\Sigma_{XYZ\mid Z}\right\|^2$ using Parseval's identity. For arbitrary orthonormal bases $\{e_p\}_{p\geq 1} \subset \cH_{\cX}$ and $\{f_q\}_{q\geq 1} \subset \cH_{\cY}$, define the coordinatewise residuals by
\begin{align*}
\zeta_{X\mid Z,p}^{(i)} &\coloneqq \bigl\langle \epsilon_{X\mid Z}^{(i)}, e_p \bigr\rangle
= e_p(X_i) - \E\left[e_p(X_i)\mid Z_i\right], \\
\zeta_{Y\mid Z,q}^{(i)} &\coloneqq \bigl\langle \epsilon_{Y\mid Z}^{(i)}, f_q \bigr\rangle
= f_q(Y_i) - \E\left[f_q(Y_i)\mid Z_i\right].
\end{align*}
Then we obtain the desired spectral representation:
\begin{align}
\left\|\Sigma_{XYZ\mid Z}\right\|^2
&= \left\langle \E[\Psi_1], \E[\Psi_2]\right\rangle
 = \E\left[\left\langle \Psi_1,\Psi_2\right\rangle\right] \notag\\
&= \E\left[
\bigl\langle \epsilon_{X\mid Z}^{(1)}, \epsilon_{X\mid Z}^{(2)} \bigl\rangle
\bigl\langle \epsilon_{Y\mid Z}^{(1)}, \epsilon_{Y\mid Z}^{(2)} \bigr\rangle
k_{\cZ}(Z_1,Z_2)
\right] \notag\\
&= \E\left[
\sum_{p,q=1}^{\infty}
\zeta_{X\mid Z,p}^{(1)} \zeta_{X\mid Z,p}^{(2)}
\zeta_{Y\mid Z,q}^{(1)} \zeta_{Y\mid Z,q}^{(2)}
k_{\cZ}(Z_1,Z_2)
\right].
\label{eq:ccco-parseval}
\end{align}

Let $\cD_n\coloneqq\{(X_i,Y_i,Z_i)\}_{i=1}^n$ be an i.i.d.\ sample from $P$. For the theoretical development, we partition the index set $\{1,\dots,n\}$ into two disjoint subsets $I_1$ and $I_2$ of sizes $n_1$ and $n_2$, respectively. 

First, using $I_2$, we construct data-dependent basis functions, $\{\hat e_p\}_{p=1}^{n_2}$ and $\{\hat f_q\}_{q=1}^{n_2}$. Specifically, we employ the eigenfunctions of the empirical (uncentered) covariance operators
\[
\hat C_{XX}\coloneqq \frac{1}{n_2}\sum_{\ell\in I_2} k_{\cX}(X_{\ell},\cdot)\otimes k_{\cX}(X_{\ell},\cdot),\quad
\hat C_{YY}\coloneqq \frac{1}{n_2}\sum_{\ell\in I_2} k_{\cY}(Y_{\ell},\cdot)\otimes k_{\cY}(Y_{\ell},\cdot).
\]
By the spectral theorem for compact self-adjoint operators, we obtain orthonormal eigensystems \(\{(\hat\lambda_p,\hat e_p)\}_{p=1}^{n_2}\) and \(\{(\hat\mu_q,\hat f_q)\}_{q=1}^{n_2}\) such that
\[
\hat C_{XX}=\sum_{p=1}^{n_2}\hat\lambda_p\,\hat e_p\otimes \hat e_p,
\qquad
\hat C_{YY}=\sum_{q=1}^{n_2}\hat\mu_q\,\hat f_q\otimes \hat f_q,
\]
with eigenvalues ordered decreasingly and \(\|\hat e_p\|_{\cH_{\cX}}=\|\hat f_q\|_{\cH_{\cY}}=1\). To obtain a finite approximation of \eqref{eq:ccco-parseval}, we project onto the top $L_X$ and $L_Y$ eigenfunctions, $\{\hat e_p\}_{p=1}^{L_X}$ and $\{\hat f_q\}_{q=1}^{L_Y}$. Unlike data-independent expansions, such as random Fourier features \citep{RahimiandRecht2007}, this spectral truncation yields a data-adaptive low-rank approximation. It reduces the effective dimension while adapting to the empirical covariance structure, without requiring a Euclidean parametrization of the underlying Polish spaces. For the explicit construction of these eigenfunctions from the kernel evaluations of the samples, see Section \ref{sec:const_eigen_function} in Supplementary Material.

Second, we construct the test statistic on $I_1$ using the basis functions derived from $I_2$. For each $p\in[L_X]$ and $q\in[L_Y]$, we fit one-dimensional nonparametric regressions of $\{\hat e_p(X_i)\}_{i\in I_1}$ and $\{\hat f_q(Y_i)\}_{i\in I_1}$ on $\{Z_i\}_{i\in I_1}$. Because the response variables in these regression tasks are scalar-valued, standard machine learning methods, such as random forests, gradient boosting machines, or deep neural networks, can be used to estimate the conditional expectations $\E[\hat e_p(X)\mid \hat e_p, Z=\cdot]$ and $\E[\hat f_q(Y)\mid \hat f_q, Z=\cdot]$. Denoting the resulting regression estimators by $\hat{\E}[\hat e_p(X)\mid \hat e_p, Z=\cdot]$ and $\hat{\E}[\hat f_q(Y)\mid \hat f_q, Z=\cdot]$, we obtain the residual scores for $i\in I_1$: 
\begin{equation}
\hat\zeta^{(i)}_{X\mid Z,p} \coloneqq \hat e_p(X_i) - \hat{\E}[\hat e_p(X)\mid \hat e_p, Z=Z_i],
\qquad
\hat\zeta^{(i)}_{Y\mid Z,q} \coloneqq \hat f_q(Y_i) - \hat{\E}[\hat f_q(Y)\mid \hat f_q, Z=Z_i].
\end{equation}
Thus, nuisance estimation is reduced to scalar regressions, rather than direct RKHS-valued CME estimation \citep{Lietal2022b}.

Finally, with a positive definite kernel $k_{\cZ}$, we define the SGCM statistic as
\begin{equation}\label{eq:test_statistic}
n_1\hat T_{n_1} \coloneqq \frac{1}{n_1}\sum_{i,j \in I_1}\sum_{p=1}^{L_X}\sum_{q=1}^{L_Y}
\hat\zeta^{(i)}_{X \mid Z , p}\,\hat\zeta^{(i)}_{Y \mid Z , q}\,
\hat\zeta^{(j)}_{X \mid Z , p}\,\hat\zeta^{(j)}_{Y \mid Z , q}\,
k_{\cZ}(Z_i,Z_j).
\end{equation}
This is a kernel-weighted quadratic form in the residual scores, aggregated over observation pairs through \(k_{\cZ}(Z_i,Z_j)\).

In practice, the truncation levels $L_X$ and $L_Y$ are chosen adaptively to retain a pre-specified fraction of variance explained (FVE; \citealp{PetersenandMuller2016}). Specifically, $L_X$ and $L_Y$ are set as the smallest integers satisfying $\sum_{p=1}^{L_X}\hat{\lambda}_p \ge \tau \sum_{p=1}^{n_2}\hat{\lambda}_p$ and $\sum_{q=1}^{L_Y}\hat{\mu}_q \ge \tau \sum_{q=1}^{n_2}\hat{\mu}_q$ for a given threshold $\tau \in (0,1)$. This choice governs a statistical trade-off: smaller truncation dimensions reduce the complexity of downstream nuisance regressions and can improve size control, whereas larger dimensions retain more of the cross-covariance signal and can improve power. In Section 4, we denote these levels as \((L_{X,n}, L_{Y,n})\) to indicate their sample-size dependence and formalize this size--power trade-off. Sensitivity of the empirical results to the FVE cutoff is examined in Section~\ref{sec:supp_fve} of the Supplementary Material.

\subsection{Testing procedure}\label{subsec:proc}

Under \(H_0\), the null distribution of \(n_1\hat T_{n_1}\) is non-pivotal, as shown in Section~\ref{sec:theory}. We therefore calibrate the test by a multiplier bootstrap, which avoids refitting the nuisance regressions used to construct \(n_1\hat T_{n_1}\).

Let \(\{W_i^\ast\}_{i\in I_1}\) be i.i.d.\ multipliers, independent of the data, with \(\E[W_i^\ast]=0\) and \(\E[(W_i^\ast)^2]=1\) (e.g., standard normal, Rademacher, or Mammen's two-point multipliers \citep{Mammen1993}). Define the bootstrap analogue of \(n_1\hat T_{n_1}\) by
\begin{equation}\label{eq:bootstrap_statistic}
n_1 \hat T^{\ast}_{n_1}
\coloneqq
\frac{1}{n_1}\sum_{i,j \in I_1}\sum_{p=1}^{L_X}\sum_{q=1}^{L_Y}
\hat\zeta^{(i)}_{X \mid Z , p}\,\hat\zeta^{(i)}_{Y \mid Z , q}\,
\hat\zeta^{(j)}_{X \mid Z , p}\,\hat\zeta^{(j)}_{Y \mid Z , q}\,
k_{\cZ}(Z_i,Z_j)\,W_i^\ast W_j^\ast.
\end{equation}

For a prescribed nominal level \(\alpha\in(0,1)\), the bootstrap testing procedure is as follows.
\begin{steplist}
\item Compute \(n_1 \hat T_{n_1}\) as defined in \eqref{eq:test_statistic}.
\item For \(b=1,\dotsc,B\), generate an i.i.d.\ copy
\(\{W_i^{\ast(b)}\}_{i\in I_1}\) of the multipliers and compute
\(n_1 \hat T^{\ast(b)}_{n_1}\) by replacing \(W_i^\ast\) with \(W_i^{\ast(b)}\) in \eqref{eq:bootstrap_statistic}.
\item Let \(Q_{n,1-\alpha}^{\ast(B)}\) be the empirical \((1-\alpha)\)-quantile of
\(\{n_1 \hat T^{\ast(b)}_{n_1}\}_{b=1}^{B}\).
\item Reject \(H_0:X\indep Y\mid Z\) if $n_1 \hat T_{n_1} > Q_{n,1-\alpha}^{\ast(B)}$.
\end{steplist}

\section{Theoretical properties}\label{sec:theory}
\subsection{Setting}
We establish a uniform asymptotic theory for the proposed test over a mildly regular subclass $\tilde{\cP}\subset\cP$. Such a restriction is necessary: over the unrestricted model class $\cP$, any CI test with uniform validity can have only trivial power \citep{ShahandPeters2020}. By working on $\tilde{\cP}$, we obtain uniform size control together with nontrivial uniform power. In particular, these uniform guarantees imply the usual pointwise results as a special case. 

Throughout this section, the kernels are treated as given, and the asymptotic theory is developed for the resulting CCCO and its truncated projections. In particular, no characteristicity (Assumption~\ref{ass:kernel_characteristic}) is needed for the size and power results below. Conditions under which these operators characterize CI were given in Section~\ref{sec:definition_and_Our_Measure}, while concrete kernel constructions ensuring those conditions are developed in Section~\ref{sec:kernels}.

Let $\cP_0 \coloneqq \{P \in \cP : X \indep Y \mid Z\}$ and $\cP_1 \coloneqq \cP \setminus \cP_0$ denote the null and alternative classes, respectively. Let $\tilde{\cP}_0 \coloneqq \tilde{\cP}\cap\cP_0$ and $\tilde{\cP}_1 \coloneqq \tilde{\cP}\cap\cP_1$ denote the corresponding subclasses of $\tilde{\cP}$. For each $P\in\tilde{\cP}$, let
\[
C_{XX}\coloneqq \E_P\left[k_{\cX}(X,\cdot)\otimes k_{\cX}(X,\cdot)\right],
\qquad
C_{YY}\coloneqq \E_P\left[k_{\cY}(Y,\cdot)\otimes k_{\cY}(Y,\cdot)\right]
\]
be the uncentred covariance operators. Let \(\{\lambda_j(P)\}_{j\ge 1}\) and \(\{\mu_j(P)\}_{j\ge 1}\) denote the eigenvalues of \(C_{XX}\) and \(C_{YY}\), respectively. For $m\in\bbN$, let $\Pi_{X,m}$ and $\Pi_{Y,m}$ denote the orthogonal projections onto the first $m$ eigendirections of $C_{XX}$ and $C_{YY}$, respectively. Also let $\Pi_{X,\infty}$ and $\Pi_{Y,\infty}$ denote the orthogonal projections onto $\overline{\range(C_{XX})}$ and $\overline{\range(C_{YY})}$, respectively.

Henceforth, $n_1+n_2=n$ and all asymptotics are as $n\to\infty$; when bootstrap quantiles are used, we also assume $B=B_n\to\infty$. Let $\{L_{X,n}\}_{n\ge 1}$ and $\{L_{Y,n}\}_{n\ge 1}$ be nondecreasing sequences of positive integers representing the truncation levels. Formulating the theory in terms of these deterministic sequences allows us to cover both bounded and diverging truncation regimes. We write $L_\infty^X\coloneqq \lim_{n\to\infty}L_{X,n}\in\bbN\cup\{\infty\}$ and $L_\infty^Y\coloneqq \lim_{n\to\infty}L_{Y,n}\in\bbN\cup\{\infty\}$ for their limits. Using these truncation levels, we project the \(X\)- and \(Y\)-components of \(\Psi\) onto the corresponding leading eigenspaces, while leaving the \(\cH_{\cZ}\)-component unchanged. Specifically, define
\[
\Pi_n\coloneqq \Pi_{X,L_{X,n}}\otimes \Pi_{Y,L_{Y,n}}\otimes I,
\qquad
\Pi_\infty\coloneqq \Pi_{X,L_\infty^X}\otimes \Pi_{Y,L_\infty^Y}\otimes I,
\]
where \(I\) denotes the identity operator on \(\cH_{\cZ}\). Let $\hat{\Pi}_n$ be its empirical counterpart based on $\hat{C}_{XX}$ and $\hat{C}_{YY}$. Thus, \(\Pi_n\Psi\) is the population analogue of the truncated representation used in the test statistic, and \(\Pi_\infty\Psi\) is its limiting version.

\subsection{Uniform regularity conditions}

We impose the following four regularity conditions uniformly over $\tilde{\cP}$. First, given the truncation levels \(L_{X,n}\) and \(L_{Y,n}\), define the uniform eigengaps as
\[
g_{L_{X,n}}^X
\coloneqq
\inf_{P\in\tilde{\cP}}
\left\{\lambda_{L_{X,n}}(P)-\lambda_{L_{X,n}+1}(P)\right\},
\qquad
g_{L_{Y,n}}^Y
\coloneqq
\inf_{P\in\tilde{\cP}}
\left\{\mu_{L_{Y,n}}(P)-\mu_{L_{Y,n}+1}(P)\right\}.
\]

\begin{assumption}\label{ass:simplicity_gap_unif}
Assume that \(g_{L_{X,n}}^X\sqrt{n_2}\to \infty\) and
\(g_{L_{Y,n}}^Y\sqrt{n_2}\to \infty\).
\end{assumption}

Next, for \(p\in[L_{X,n}]\) and \(q\in[L_{Y,n}]\), define
$m_{X,p}(z)\coloneqq \E_P\left[\hat e_p(X)\mid Z=z,\hat e_p\right]$ and $m_{Y,q}(z)\coloneqq \E_P\left[\hat f_q(Y)\mid Z=z,\hat f_q\right]$. Let \(\hat m_{X,p}\) and \(\hat m_{Y,q}\) denote their estimators. Set
\[
B_{X,n_1}^2
\coloneqq
\frac{1}{n_1}\sum_{i\in I_1}\sum_{p=1}^{L_{X,n}}
\left\{\hat m_{X,p}(Z_i)-m_{X,p}(Z_i)\right\}^2,
\enspace
B_{Y,n_1}^2
\coloneqq
\frac{1}{n_1}\sum_{i\in I_1}\sum_{q=1}^{L_{Y,n}}
\left\{\hat m_{Y,q}(Z_i)-m_{Y,q}(Z_i)\right\}^2.
\]

\begin{assumption}\label{ass:doubly_robust_uniform}
Assume that \(B_{X,n_1}=o_{\tilde{\cP}}(1)\),
\(B_{Y,n_1}=o_{\tilde{\cP}}(1)\), and
\(B_{X,n_1}B_{Y,n_1}=o_{\tilde{\cP}}(n_1^{-1/2})\).
\end{assumption}

\begin{assumption}\label{ass:fixed-basis-uniform_tail}
Assume that there exist orthonormal systems
\(\left\{\tilde e_p\right\}_{p=1}^{L_\infty^X}\subset \cH_{\cX}\) and
\(\left\{\tilde f_q\right\}_{q=1}^{L_\infty^Y}\subset \cH_{\cY}\), and an
orthonormal basis \(\left\{\tilde g_r\right\}_{r\geq 1}\subset \cH_{\cZ}\),
such that
\[
\lim_{K\to\infty}\sup_{P\in\tilde{\cP}}
\left\{
\sum_{p=K+1}^{L_\infty^{X}}\E_P\left[\tilde e_p(X)^2\right]
+
\sum_{q=K+1}^{L_\infty^Y}\E_P\left[\tilde f_q(Y)^2\right]
+
\sum_{r=K+1}^{\infty}\E_P\left[\tilde g_r(Z)^2\right]
\right\}
=0.
\]
\end{assumption}

Finally, we define the population second-moment operator of the limiting projected signal by
\[
C_P^{(\infty)}\coloneqq \E_P\left[\left(\Pi_\infty\Psi\right)\otimes\left(\Pi_\infty\Psi\right)\right].
\]
If \(L_\infty^X=L_\infty^Y=\infty\), then, by definition,
\(\Pi_\infty\) projects onto the full relevant \(X\)- and \(Y\)-subspaces, and thus
\(\Pi_\infty\Psi=\Psi\) almost surely. Consequently, \(C_P^{(\infty)}=\E_P\left[\Psi\otimes\Psi\right]\) follows.

\begin{assumption}\label{ass:non_degenerate_unif}
Assume that \(\inf_{P\in\tilde{\cP}}\bigl\|C_P^{(\infty)}\bigr\|_{\mathrm{op}}>0\).
\end{assumption}

Assumption~\ref{ass:simplicity_gap_unif} ensures eigenspace stability, while Assumption~\ref{ass:doubly_robust_uniform} reduces the bias analysis to the product-rate condition \(B_{X,n_1} B_{Y,n_1}=o_{\tilde{\cP}}(n_1^{-1/2})\). Relative to \citet{Pogodinetal2025}, this is weaker in that we do not require each nuisance family to be controlled separately on the square-root scale; it suffices that their interaction term be \(o_{\tilde{\cP}}(n_1^{-1/2})\). Under CME smoothness conditions with \(\beta_X,\beta_Y\ge 1\), and a uniform \(Z\)-side trace bound \(\mathrm{tr}\left(C_{ZZ}(C_{ZZ}+\lambda I)^{-1}\right)\le Q\lambda^{-p}\) for some \(p\in(0,1]\), one obtains \(B_{X,n_1}=O_{\tilde{\cP}}(n_1^{-\rho_X})\) and \(B_{Y,n_1}=O_{\tilde{\cP}}(n_1^{-\rho_Y})\), where \(\rho_X=\beta_X/\{2(\beta_X+p)\}\) and \(\rho_Y=\beta_Y/\{2(\beta_Y+p)\}\). Hence Assumption~\ref{ass:doubly_robust_uniform} holds whenever \(\rho_X+\rho_Y>1/2\); see Section~\ref{sec:uniform-krr-cme} of the Supplementary Material for detailed sufficient conditions. Assumption~\ref{ass:fixed-basis-uniform_tail} provides the uniform tightness needed for the Gaussian limit, and Assumption~\ref{ass:non_degenerate_unif} rules out uniform degeneracy of the limiting null law; see Sections~\ref{sec:spectral_gap_growing_gap} and \ref{sec:suff_cond_uniform_tightness}.

\subsection{Uniform bootstrap validity and asymptotic size}
\label{subsec:uniform_bootstrap_and_size}

For the test based on \eqref{eq:test_statistic}, define the test function by
\begin{equation}
\phi_n^{(B_n)} \coloneqq \mathds{1}\left\{n_1\hat T_{n_1}>Q_{n,1-\alpha}^{\ast(B_n)}\right\},
\end{equation}
where $Q_{n,1-\alpha}^{\ast(B_n)}$ is the critical value defined in Section~\ref{subsec:proc}. Moreover, for each \(P\in\tilde{\cP}\), define
\begin{equation}
F_P(t)
\coloneqq
\pr\left(\sum_{\ell=1}^{\infty}\xi_{\ell,P}G_\ell^2\le t\right),
\qquad t\in\bbR,
\end{equation}
where \(\{\xi_{\ell,P}\}_{\ell\ge 1}\) are the positive eigenvalues of \(C_P^{(\infty)}\), counted with multiplicity, and \(\{G_\ell\}_{\ell\ge 1}\) are independent standard normal variables. Here and below, \(\pr^{\ast}(\,\cdot\mid \cD_n)\) denotes conditional probability with respect to the bootstrap multipliers $\{W_i^*\}_{i \geq 1}$, given the data $\cD_n$.

The next theorem shows that \(F_P\) provides the relevant asymptotic benchmark under the null, and that the bootstrap consistently approximates it uniformly over \(\tilde{\cP}\). The theorem below continues to hold if the nuisance regressions are estimated by fixed-\(K\) fold cross-fitting; in that case, for \(i\in I_k\), \(\hat m_{X,p}(Z_i)\) and \(\hat m_{Y,q}(Z_i)\) are understood as the out-of-fold predictions \(\hat m_{X,p}^{(-k)}(Z_i)\) and \(\hat m_{Y,q}^{(-k)}(Z_i)\). A detailed discussion of this extension is given in Remark~\ref{rem:extension_to_k_fold} in Supplementary Material.

\begin{theorem}[Uniform asymptotic theory and size control]
\label{thm:uniform_size_control}
Suppose that \(\tilde{\cP}\subset\cP\) satisfies Assumptions~\ref{ass:simplicity_gap_unif}--\ref{ass:non_degenerate_unif}. Then the following statements hold:
\begin{enumerate}
\item
\emph{Uniform weak convergence under the null:}
\[
\lim_{n\to\infty}\sup_{P\in\tilde{\cP}_0}\sup_{t\in\bbR}
\left|
\pr_P\left(n_1\hat T_{n_1}\le t\right)-F_P(t)
\right|
= 0.
\]

\item
\emph{Uniform bootstrap consistency:} For every \(\varepsilon>0\),
\[
\lim_{n\to\infty}\sup_{P\in\tilde{\cP}}
\pr_P
\left(
\sup_{t\in\bbR}
\left|
\pr^{\ast}\left(n_1\hat T_{n_1}^{\ast}\le t\mid \cD_n\right)-F_P(t)
\right|
>\varepsilon
\right)
= 0.
\]

\item
\emph{Uniform asymptotic size control:}
\[
\lim_{n\to\infty}\sup_{P\in\tilde{\cP}_0}
\left|
\pr_P\left(\phi_n^{(B_n)}=1\right)-\alpha
\right|
= 0.
\]
\end{enumerate}
\end{theorem}

The bootstrap consistently approximates the limiting law, and the resulting test has asymptotic level \(\alpha\) uniformly over \(\tilde{\cP}_0\). In particular, slower growth of the truncation levels \(L_n^{X}\) and \(L_{Y,n}\) relaxes both the spectral-separation requirement in Assumption~\ref{ass:simplicity_gap_unif} and the regression requirement in Assumption~\ref{ass:doubly_robust_uniform}, at the cost of reduced sensitivity to higher-order components; this trade-off enters the power analysis below.

\subsection{Uniform power under projected separated alternatives}
\label{subsec:uniform_power}

We next turn to alternatives. Under truncation, detectability is governed by the projected operator \(\Pi_n\Sigma_{XYZ\mid Z}\). For \(c>0\), define
\begin{equation}
\label{eq:projected_alt_class_uniform}
\tilde{\cP}_{1,n}(c)
\coloneqq
\left\{
P\in\tilde{\cP}_1:
\left\|\Pi_n\Sigma_{XYZ\mid Z}\right\|\ge \frac{c}{\sqrt{n_1}}
\right\}.
\end{equation}
That is, \(\tilde{\cP}_{1,n}(c)\) consists of alternatives whose projected signal remains separated from zero at the \(n_1^{-1/2}\) scale after truncation.

The next condition is used only for the power analysis. It requires that empirical projection does not substantially reduce the projected signal uniformly over \(\tilde{\cP}_{1,n}(c)\).

\begin{assumption}
\label{ass:uniform_power_projected_signal_preservation_Pi_n}
For every fixed \(c>0\), assume that
\[
\sqrt{n_1}
\left(
\left\|\Pi_n\Sigma_{XYZ\mid Z}\right\|
-
\left\|\hat\Pi_n\Sigma_{XYZ\mid Z}\right\|
\right)_+
=
O_{\tilde{\cP}_{1,n}(c)}(1).
\]
\end{assumption}

If \(L_\infty^X,L_\infty^Y<\infty\), then the eigengaps are eventually bounded away from zero, so Assumption~\ref{ass:uniform_power_projected_signal_preservation_Pi_n} holds whenever \(n_1/n_2=O(1)\). A sufficient condition covering the diverging-truncation regime is given in Supplementary Material (Section~\ref{subsec:sufficient_projected_signal_preservation}).

\begin{theorem}[Uniform power under projected separated alternatives]
\label{thm:uniform_power_monotone}
Suppose that \(\tilde{\cP}\subset\cP\) satisfies Assumptions~\ref{ass:simplicity_gap_unif}--\ref{ass:uniform_power_projected_signal_preservation_Pi_n}, and that all conditional expectations in \(\hat T_{n_1}\) and \(\hat T_{n_1}^{\ast}\) are estimated from an auxiliary sample independent of the data used to construct the test and bootstrap statistics. Then the following statements hold:
\begin{enumerate}
\item
\emph{Nontrivial uniform power.}
For any \(\beta\in(\alpha,1)\), there exists \(c>0\) such that
\[
\liminf_{n\to\infty}\inf_{P\in\tilde{\cP}_{1,n}(c)}
\pr_P\left(\phi_n^{(B_n)}=1\right)\ge \beta.
\]

\item
\emph{Uniform consistency.}
For any sequence \(c_n\to\infty\),
\[
\lim_{n\to\infty}\inf_{P\in\tilde{\cP}_{1,n}(c_n)}
\pr_P\left(\phi_n^{(B_n)}=1\right)=1.
\]
\end{enumerate}
\end{theorem}

Thus, detectability under truncation is governed by \(\|\Pi_n\Sigma_{XYZ\mid Z}\|\), rather than by the full operator \(\|\Sigma_{XYZ\mid Z}\|\). The auxiliary-sample condition in Theorem~\ref{thm:uniform_power_monotone} is imposed only for the power analysis: unlike the size result, it avoids self-fitting effects and yields a cleaner separation between the projected signal and the remainder. Accordingly, Theorem~\ref{thm:uniform_power_monotone} yields nontrivial uniform power once the projected signal is separated from zero at the \(n_1^{-1/2}\) scale by a sufficiently large constant, and uniform consistency whenever it dominates that scale. More broadly, Section~\ref{sec:theory} is formulated in terms of the projected CCCO signal induced by the chosen kernels and does not itself require the kernels to be characteristic. In Supplementary Material, Section \ref{sec:suffic_cond_PSS} gives sufficient conditions under which the projected signal strength condition in \eqref{eq:projected_alt_class_uniform} is satisfied. The same viewpoint also yields consistency under fixed alternatives and along broader sequences of local alternatives; we defer these extensions to Supplementary Material (Section~\ref{subsec:local_power_diverging}). To interpret these asymptotic guarantees as guarantees for conditional independence testing, it remains to ensure that conditional dependence is reflected in a nonzero CCCO signal; the next section addresses this issue through kernel choice.

\section{Kernel designs for random objects}\label{sec:kernels}
\subsection{A pullback principle for characteristic kernels}\label{subsec:char_theory}

The asymptotic guarantees in Section~\ref{sec:theory} are established for the CCCO induced by a given choice of kernels. By Proposition~\ref{prop:CI_V}, these guarantees become guarantees for CI testing once Assumption~\ref{ass:kernel_characteristic} holds. The purpose of this section is therefore to provide concrete kernel constructions under which that assumption is satisfied on general Polish spaces.

We develop a general framework for constructing characteristic kernels on Polish spaces. Although such kernels are well understood on separable Hilbert spaces \citep{WynneandDuncan2022} and on metric spaces of strong negative type \citep{Ziegeletal2024}, the latter condition can be restrictive in non-Euclidean applications; for example, it fails on important spaces such as $L^1$ \citep{Lyons2013}. Our approach avoids this restriction by combining a pullback principle with geometric or learned representations. 

Let \(\cM\) and \(\cN\) be Polish spaces and let \(\phi:\cM\to\cN\) be Borel measurable. For a kernel \(k_{\cN}\) on \(\cN\), define its pullback kernel \(\phi^\ast k_{\cN}\) on \(\cM\) by
\[
(\phi^\ast k_{\cN})(x,y)\coloneqq k_{\cN}(\phi(x),\phi(y)),
\quad x,y\in\cM.
\]
Also define the class of Borel probability measures on \(\cM\) for which the pullback-kernel mean embedding is well defined:
\begin{equation}
\cP_{\phi^\ast k_{\cN}}(\cM)
\coloneqq
\left\{
P\in\cP(\cM):
\int_{\cM}\sqrt{(\phi^\ast k_{\cN})(x,x)}\,P(\mathrm{d}x)<\infty
\right\}. \label{cond:integrable}
\end{equation}
Denote by \(\mu^{\phi^\ast k_{\cN}}(P)\) and \(\phi_\sharp P\) the kernel mean embedding of \(P\) with respect to \(\phi^\ast k_{\cN}\) and the pushforward of \(P\) under \(\phi\), respectively. The next theorem shows that equality of the pullback-kernel mean embeddings is equivalent to equality of the pushforward measures under \(\phi\).

\begin{theorem}
\label{thm:pullback}
Suppose that \(k_{\cN}\) is characteristic on \(\cN\), or more weakly, on the image \(\phi(\cM)\). Then, for any \(P,Q\in\cP_{\phi^\ast k_{\cN}}(\cM)\),
\[
\mu^{\phi^\ast k_{\cN}}(P)=\mu^{\phi^\ast k_{\cN}}(Q)
\quad\text{if and only if}\quad
\phi_{\sharp}P=\phi_{\sharp}Q.
\]
In particular, if \(\phi\) is injective, then \(\phi^\ast k_{\cN}\) is characteristic on \(\cM\) with respect to \(\cP_{\phi^\ast k_{\cN}}(\cM)\).
\end{theorem}

\begin{remark}\label{rem:pullback-borel-iso-auto}
The equivalence in Theorem~\ref{thm:pullback} holds for general measurable spaces. In that general setting, the final statement should be formulated under the assumption that \(\phi:\cM\to\phi(\cM)\) is an isomorphism of measurable spaces. If \(\phi\) is merely injective and measurable, the induced \(\sigma\)-algebra on \(\cM\) may be strictly coarser, failing to separate all probability measures. In the present Polish/Borel setting, this subtlety is resolved because an injective Borel measurable map into a Polish space induces a Borel isomorphism onto its image. Moreover, if \(k_{\cN}\) is bounded, then the integrability condition in \eqref{cond:integrable} is automatic, so \(\cP_{\phi^\ast k_{\cN}}(\cM)=\cP(\cM)\).
\end{remark}

\begin{remark}\label{rem:pullback-ispd-spd}
If \(\phi:\cM\to\phi(\cM)\) is an isomorphism of measurable spaces, then the pushforward argument used in the proof of Theorem~\ref{thm:pullback} shows that pullback preserves stronger nondegeneracy properties of the restricted kernel \(k_{\cN}|_{\phi(\cM)\times\phi(\cM)}\). Specifically, \(\phi^\ast k_{\cN}\) is ISPD on \(\cM\) with respect to finite signed Borel measures (Section \ref{subsec:notation}) if and only if \(k_{\cN}|_{\phi(\cM)\times\phi(\cM)}\) satisfies the analogous property on \(\phi(\cM)\). Likewise, \(\phi^\ast k_{\cN}\) is strictly positive definite on \(\cM\) if and only if \(k_{\cN}|_{\phi(\cM)\times\phi(\cM)}\) is strictly positive definite on \(\phi(\cM)\). Consequently, if \(k_{\cN}|_{\phi(\cM)\times\phi(\cM)}\) is ISPD, then \(\phi^\ast k_{\cN}\) is characteristic.
\end{remark}


Theorem~\ref{thm:pullback} provides a general device for modular kernel design and, in particular, links our framework to representation-based constructions. When the original domain \(\cM\) lacks a convenient linear structure, one may work through a measurable representation \(\phi:\cM\to\cN\) into a structured space \(\cN\), such as a hyperbolic space \citep{NickelandKiela2017} or an \(L^1\) space \citep{DezaandLaurent1997}. The theorem remains applicable even when \(\phi\) is not injective: in that case, the pullback kernel identifies probability measures on \(\cM\) exactly through their pushforward laws under \(\phi\). Thus, the resulting CI procedure operates on the distributional information retained by the chosen representation.

We now specialize this principle to the canonical setting of negative-type semimetrics. Recall that a \emph{semimetric} \(\rho\) on \(\cM\) is a nonnegative, symmetric function such that \(\rho(x,x)=0\) for all \(x\in\cM\), and \(\rho(x,y)=0\) implies \(x=y\), but the triangle inequality need not hold. It is said to be of \emph{negative type} if $\sum_{i,j=1}^n \alpha_i\alpha_j \rho(x_i,x_j)\le 0$ for all \(n\ge 2\), \(x_i\in\cM\), and coefficients \(\alpha_i\) satisfying \(\sum_{i=1}^n \alpha_i=0\). A standard result states that any negative-type semimetric admits an isometric embedding into a Hilbert space \citep{Sejdinovicetal2013}. Equivalently, fixing an arbitrary base point \(x_0\in\cM\), the distance-induced kernel $k(x,y)\coloneqq \{\rho(x,x_0)+\rho(y,x_0)-\rho(x,y)\}/2$ is positive definite, and its RKHS feature map \(\phi:\cM\to\cH_k\), \(\phi(x)\coloneqq k(\cdot,x)\), satisfies $\rho(x,y)=\|\phi(x)-\phi(y)\|_{\cH_k}^2$. 

We also consider \emph{completely monotone} functions \(f\in C^\infty(0,\infty)\), which satisfy \((-1)^n f^{(n)}(t)\ge 0\) for all \(n\ge 0\) and \(t>0\). The next proposition shows that composing a negative-type semimetric with such a function yields a characteristic kernel.

\begin{proposition}\label{prop:cmonotone_nonconstant_characteristic}
Let \(\cM\) be a Polish space equipped with a continuous semimetric \(\rho\) of negative type. Let $f:[0,\infty)\to\bbR$ be a completely monotone function on $(0,\infty)$ that is not identically constant and satisfies $f(0)\coloneqq \lim_{t\downarrow 0} f(t)<\infty$. Then
\[
k_f(u,v)\coloneqq f(\rho(u,v)),
\quad u,v\in\cM,
\]
is bounded, continuous, and positive definite. Moreover, \(k_f\) is characteristic. In fact, it is ISPD, and hence also strictly positive definite.
\end{proposition}

Standard choices for $f$ include the exponential $f(t)=e^{-\gamma t}$ and the rational quadratic $f(t)=(1+ct)^{-\alpha}$ ($\alpha,c,\gamma>0$); see \citet{AlzerandBerg2006} for further examples.

We next extend this construction to tensor product kernels on product spaces. This is particularly useful for combining heterogeneous objects and, in our setting, provides a direct route to Assumption~\ref{ass:kernel_characteristic}. 

\begin{corollary}\label{cor:tensor_char_cmon}
Let \(\cN = \prod_{\ell=1}^L \cN_{\ell}\) be a product of Polish spaces. For each \(\ell\in\{1,\dots,L\}\), let \(\rho_{\ell}\) be a continuous semimetric of negative type on \(\cN_{\ell}\), and let \(f_\ell\) satisfy the conditions of Proposition~\ref{prop:cmonotone_nonconstant_characteristic}. Then the tensor product kernel
\[
k_{\otimes,L}(u,v)\coloneqq\prod_{\ell=1}^L f_{\ell}(\rho_{\ell}(u_{\ell},v_{\ell})), \quad u,v \in \cN,
\]
is bounded, continuous, and positive definite. Moreover, \(k_{\otimes,L}\) is characteristic. In fact, it is ISPD, and hence also strictly positive definite.
\end{corollary}

\subsection{Illustrative Spaces}\label{subsec:illustrative_example}

We illustrate the framework on several non-Euclidean domains. For examples based on negative-type semimetrics, it suffices to simply verify this negative-type property, since Proposition~\ref{prop:cmonotone_nonconstant_characteristic} ensures that composing them with any non-constant completely monotone function yields a characteristic kernel.

\begin{example}[Spheres, simplices, and probability densities]
On the unit sphere \(\bbS(\cH)\) of a separable Hilbert space \(\cH\) (including \(\bbR^d\)), the geodesic distance \(d_{\bbS}(x,y)\coloneqq \arccos(\langle x,y\rangle_{\cH})\) has the property that \(d_{\bbS}^{\,q}\) is of negative type for \(0<q\le 1\) \citep[see][and the Supplementary Material (Proposition~\ref{prop:sphere_nd})]{Istas2012}. This geometry transfers naturally to compositional data: the square-root map \(f\mapsto \sqrt{f}\) embeds the simplex \(\Delta^{d-1}\), or the space of probability densities \(\mathfrak{D}\coloneqq\{f\ge 0:\int_{\cX} f\,\diff \mu=1\}\) over a measure space, injectively into the positive orthant of a sphere. The induced Fisher--Rao distance \(d_{\mathrm{FR}}(f,g)\coloneqq \arccos(\int_{\cX} \sqrt{fg}\,\diff\mu)\) therefore inherits this negative-type property, and hence characteristic kernels can be constructed for compositional and density-valued data.
\end{example}

\begin{example}[Time-varying random objects]
To model evolving random objects, such as trajectories on a sphere or time-varying networks, consider the space \(L^p(\cT,\cM)\) of measurable functions from a time domain \(\cT\) into a metric space \((\cM,d_{\cM})\), equipped with
\[
D_p(f,g)\coloneqq \left\{\int_{\cT} d_{\cM}(f(t),g(t))^p \nu(\diff t)\right\}^{1/p}.
\]
By Supplementary Material (Section~\ref{negative_definite_Hilbert_sphere}), if \(\cM\) is Polish and \(d_{\cM}^{\,q}\) is of negative type, then \(L^p(\cT,\cM)\) is Polish and \(D_p^{\,q}\) is of negative type for \(0<q\le p\). 
\end{example}

\begin{example}[Manifold-valued data]
Let \((\cM,g)\) be a Riemannian manifold. On a Hadamard manifold (e.g., the space of symmetric positive definite matrices equipped with the affine-invariant geodesic distance), the logarithm map \(\Log_o:\cM\to T_o\cM\) is globally defined and injective for every \(o\in\cM\). While this map is generally not distance-preserving, it provides a global representation into the Hilbert space \(T_o\cM\). Therefore, Theorem~\ref{thm:pullback} shows that any characteristic kernel \(k_{T_o\cM}\) on \(T_o\cM\) induces a characteristic kernel on \(\cM\) through
\[
k_{\cM}(x,y)\coloneqq k_{T_o\cM}(\Log_o(x),\Log_o(y)).
\]
More generally, on arbitrary, possibly infinite-dimensional manifolds, a global characteristic kernel can be constructed by patching local pullback kernels via a continuous partition of unity subordinate to a countable atlas. We refer to the Supplementary Material (Section~\ref{negative_definite_Hilbert_sphere}) for details.
\end{example}



\section{Simulation study}\label{sec:simulation}

\subsection{Setting}\label{subsec:simu_setting}
We study the finite-sample behavior of the SGCM test across several data-generating processes (DGPs). We report empirical size and power at level \(\alpha=0.05\) based on \(1{,}000\) Monte Carlo replications, using a multiplier bootstrap with \(B=2{,}000\) i.i.d.\ Gaussian multipliers. An empirical comparison with Mammen and Rademacher multipliers is reported in Table~\ref{tab:sgcm_multipliers}; for the corresponding theoretical comparison, based on the role of the multiplier fourth moment in the sufficient separation constant, see Section~\ref{sec:supp_multiplier_theory} of the Supplementary Material. Conditional expectations are estimated by five-fold cross-fitting with XGBoost \citep{ChenandGuestrin2016}, and we split the sample into \(I_2\) (20\%) for eigenfunction estimation and \(I_1\) (80\%) for test statistic evaluation. Sensitivity to the FVE cutoff and to the sample split ratio is examined in Sections~\ref{sec:supp_fve} and \ref{sec:supp_split_ratio} of the Supplementary Material.

For \(\cS\in\{\cX,\cY,\cZ\}\), we use the distance-induced exponential kernel \(k_{\cS}(s,s')=\exp(-\gamma_{\cS} d_{\cS}(s,s'))\) with the median heuristic \(\gamma_{\cS}=\left(\operatorname{med}_{i<j} d_{\cS}(s_i,s_j)\right)^{-1}\). We consider Euclidean data with the Euclidean distance (Section~\ref{subsec:euclid}), and distributional data with the 1-- and 2--Wasserstein, Fisher--Rao (FR), and Hellinger distances (Section~\ref{subsec:distribution}).

\subsection{Euclidean data}\label{subsec:euclid}

We benchmark our CI test against representative methods drawn from regression-, kernel-, metric-, and classification-based families:

\begin{description}[
  font=\normalfont\textsc,
  leftmargin=!,      
  labelwidth=!,
  widest=MMDCI,      
  labelsep=0.8em,
  itemsep=0.2ex,
  topsep=0.3ex,
  parsep=0pt,
  align=left
]
  \item[GCM]   \emph{Regression-based}: Generalized Covariance Measure \citep{ShahandPeters2020}.
  \item[WGCM]  \emph{Regression-based}: Weighted Generalized Covariance Measure \citep{Scheideggeretal2022}.
  \item[CDCOV] \emph{Metric-based}: Conditional Distance Covariance test \citep{Wangetal2015a}.
  \item[KCI]   \emph{Kernel-based}: Kernel Conditional Independence test \citep{Zhang2011}.
  \item[MMDCI] \emph{Kernel-based}: Doubly robust MMD-based conditional independence test with conditional generators \citep{Zhangetal2025}.
  \item[CCIT]  \emph{Others}: Classifier Conditional Independence Test \citep{Senetal2017}.
  \item[SGCM]  \emph{Proposed}: our Spectral Generalized Covariance Measure.
\end{description}
Here, all the regression-based methods (GCM, WGCM) employ XGBoost as in our implementation. 

For each scenario \texttt{(a2)} and \texttt{(a4)}, we consider three designs for the alternative hypotheses. The first design (DGP1-1) is a simple linear perturbation. The following two (DGP1-2 and DGP1-3) are designed to probe regression-based tests in more detail, as they involve an unobserved latent variable $U$, inducing a conditional dependence between $X$ and $Y$ given $Z$.
\begin{description}[
  leftmargin=!,
  labelwidth=!,
  widest=DGP 1-3.,
  labelsep=0.6em,
  itemsep=0.2ex,
  topsep=0.3ex,
  parsep=0pt,
  align=left
]
  \item[DGP 1-1.] \(Y \leftarrow Y + 0.2\,X\).
  \item[DGP 1-2.] \(U\sim N(0,1)\), with transformations \(X \leftarrow X+0.3(U^2-1)\) and \(Y \leftarrow Y+0.3\,U\).
  \item[DGP 1-3.] Same transformations as in DGP 1-2, but with \(U \leftarrow f_1(Z)+\varepsilon_{U}\) where \(\varepsilon_{U}\sim N(0,1)\).
\end{description}
We consider $n\in\{100,200,300,400\}$. The key distinction is that GCM is consistent only when $\E[\cov(X,Y\mid Z)]\neq 0$, whereas WGCM requires $\cov(X,Y\mid Z)\neq 0$ with positive probability. Accordingly, both can be weak in DGP1-2, while in DGP1-3 this limitation is specific to GCM. For further details, see Section~\ref{sec:reg_based_power} of the Supplementary Material.

\runinhead{Results for low-dimensional settings}
Table~\ref{tab:side_by_side_low_dim} summarizes the size and power results for the low-dimensional settings. In terms of size, SGCM is close to the nominal level across scenarios and sample sizes. GCM and WGCM behave similarly, while KCI is unstable, with substantial size inflation in \((a4)\) and also at smaller sample sizes in \((a2)\). CDCOV, MMDCI, and CCIT exhibit severe size distortions throughout and are therefore not size-valid in these designs.

Restricting to size-valid procedures, regression baselines are strongest under the additive alternative (DGP1-1). Under DGP1-2, SGCM achieves nontrivial power increasing with \(n\), whereas GCM and WGCM remain close to the nominal level. Under DGP1-3, SGCM reaches high power and dominates WGCM, while GCM remains essentially powerless. Overall, SGCM provides reliable size control and robust power across the more challenging alternatives.

\begin{table*}[htbp]
  \centering
  \caption{Empirical size (\(\alpha=0.05\)) and power summary. Left: DGP \texttt{(a2)}. Right: DGP \texttt{(a4)}.}
  \label{tab:side_by_side_low_dim}
  \renewcommand{\arraystretch}{0.9}

  \begin{minipage}[t]{0.49\textwidth}
    \centering
    \footnotesize
    \textbf{DGP \texttt{(a2)}}\\[0.6ex]
    \resizebox{\linewidth}{!}{%
    \begin{tabular}{@{}llcccc@{}}
      \toprule
      \multicolumn{2}{c}{} & \multicolumn{4}{c}{Sample Size} \\
      \cmidrule(lr){3-6}
      DGP & Method & 100 & 200 & 300 & 400 \\
      \midrule
      \multirow{7}{*}{\shortstack[l]{Size}}
      & \textsc{gcm}   & 0.088 & 0.070 & 0.064 & 0.065 \\
      & \textsc{wgcm}  & 0.131 & 0.086 & 0.078 & 0.078 \\
      & \textsc{cdcov} & 1.000 & 1.000 & 1.000 & 1.000 \\
      & \textsc{kci}   & 0.176 & 0.297 & 0.070 & 0.066 \\
      & \textsc{mmdci} & 1.000 & 1.000 & 1.000 & 1.000 \\
      & \textsc{ccit}  & 0.447 & 0.423 & 0.439 & 0.450 \\
      & \textsc{sgcm}  & 0.059 & 0.053 & 0.069 & 0.073 \\
      \midrule
      \multirow{7}{*}{\shortstack[l]{Power \\ (DGP1-1)}}
      & \textsc{gcm}   & 0.574 & 0.841 & 0.953 & 0.976 \\
      & \textsc{wgcm}  & 0.538 & 0.678 & 0.774 & 0.850 \\
      & \textsc{cdcov} & ---   & ---   & ---   & ---   \\
      & \textsc{kci}   & 0.785 & 0.972 & 0.810 & 0.907 \\
      & \textsc{mmdci} & ---   & ---   & ---   & ---   \\
      & \textsc{ccit}  & ---   & ---   & ---   & ---   \\
      & \textsc{sgcm}  & 0.242 & 0.534 & 0.723 & 0.828 \\
      \midrule
      \multirow{7}{*}{\shortstack[l]{Power \\ (DGP1-2)}}
      & \textsc{gcm}   & 0.076 & 0.064 & 0.065 & 0.072 \\
      & \textsc{wgcm}  & 0.110 & 0.081 & 0.081 & 0.086 \\
      & \textsc{cdcov} & ---   & ---   & ---   & ---   \\
      & \textsc{kci}   & 0.190 & 0.480 & 0.709 & 0.859 \\
      & \textsc{mmdci} & ---   & ---   & ---   & ---   \\
      & \textsc{ccit}  & ---   & ---   & ---   & ---   \\
      & \textsc{sgcm}  & 0.155 & 0.335 & 0.486 & 0.655 \\
      \midrule
      \multirow{7}{*}{\shortstack[l]{Power \\ (DGP1-3)}}
      & \textsc{gcm}   & 0.072 & 0.057 & 0.058 & 0.064 \\
      & \textsc{wgcm}  & 0.159 & 0.279 & 0.434 & 0.553 \\
      & \textsc{cdcov} & ---   & ---   & ---   & ---   \\
      & \textsc{kci}   & 0.596 & 0.944 & 0.980 & 0.999 \\
      & \textsc{mmdci} & ---   & ---   & ---   & ---   \\
      & \textsc{ccit}  & ---   & ---   & ---   & ---   \\
      & \textsc{sgcm}  & 0.326 & 0.694 & 0.921 & 0.979 \\
      \bottomrule
    \end{tabular}%
    }
  \end{minipage}
  \hfill
  \begin{minipage}[t]{0.49\textwidth}
    \centering
    \footnotesize
    \textbf{DGP \texttt{(a4)}}\\[0.6ex]
    \resizebox{\linewidth}{!}{%
    \begin{tabular}{@{}llcccc@{}}
      \toprule
      \multicolumn{2}{c}{} & \multicolumn{4}{c}{Sample Size} \\
      \cmidrule(lr){3-6}
      DGP & Method & 100 & 200 & 300 & 400 \\
      \midrule
      \multirow{7}{*}{\shortstack[l]{Size}}
      & \textsc{gcm}   & 0.138 & 0.090 & 0.096 & 0.086 \\
      & \textsc{wgcm}  & 0.208 & 0.118 & 0.102 & 0.091 \\
      & \textsc{cdcov} & 1.000 & 1.000 & 1.000 & 1.000 \\
      & \textsc{kci}   & 0.985 & 0.999 & 0.981 & 0.997 \\
      & \textsc{mmdci} & 1.000 & 1.000 & 1.000 & 1.000 \\
      & \textsc{ccit}  & 0.465 & 0.464 & 0.412 & 0.441 \\
      & \textsc{sgcm}  & 0.073 & 0.085 & 0.095 & 0.071 \\
      \midrule
      \multirow{7}{*}{\shortstack[l]{Power \\ (DGP1-1)}}
      & \textsc{gcm}   & 0.669 & 0.863 & 0.968 & 0.984 \\
      & \textsc{wgcm}  & 0.649 & 0.743 & 0.822 & 0.884 \\
      & \textsc{cdcov} & ---   & ---   & ---   & ---   \\
      & \textsc{kci}   & ---   & ---   & ---   & ---   \\
      & \textsc{mmdci} & ---   & ---   & ---   & ---   \\
      & \textsc{ccit}  & ---   & ---   & ---   & ---   \\
      & \textsc{sgcm}  & 0.339 & 0.616 & 0.777 & 0.873 \\
      \midrule
      \multirow{7}{*}{\shortstack[l]{Power \\ (DGP1-2)}}
      & \textsc{gcm}   & 0.088 & 0.075 & 0.075 & 0.079 \\
      & \textsc{wgcm}  & 0.154 & 0.100 & 0.097 & 0.094 \\
      & \textsc{cdcov} & ---   & ---   & ---   & ---   \\
      & \textsc{kci}   & ---   & ---   & ---   & ---   \\
      & \textsc{mmdci} & ---   & ---   & ---   & ---   \\
      & \textsc{ccit}  & ---   & ---   & ---   & ---   \\
      & \textsc{sgcm}  & 0.153 & 0.316 & 0.475 & 0.624 \\
      \midrule
      \multirow{7}{*}{\shortstack[l]{Power \\ (DGP1-3)}}
      & \textsc{gcm}   & 0.100 & 0.079 & 0.075 & 0.066 \\
      & \textsc{wgcm}  & 0.212 & 0.285 & 0.456 & 0.566 \\
      & \textsc{cdcov} & ---   & ---   & ---   & ---   \\
      & \textsc{kci}   & ---   & ---   & ---   & ---   \\
      & \textsc{mmdci} & ---   & ---   & ---   & ---   \\
      & \textsc{ccit}  & ---   & ---   & ---   & ---   \\
      & \textsc{sgcm}  & 0.255 & 0.544 & 0.805 & 0.907 \\
      \bottomrule
    \end{tabular}%
    }
  \end{minipage}
\end{table*}

\runinhead{High-dimensional DGPs}
We follow the post-nonlinear noise model of \citet{Shietal2021}. For \(d\in\{10,20,30,40,50\}\), draw \(Z\sim N(0,I_d)\) and \(\varepsilon_X,\varepsilon_Y\overset{\text{i.i.d.}}{\sim} N(0,1)\), and set
\[
X=\sin\left(\beta_X^{\top}Z+0.5\,\varepsilon_X\right),\qquad
Y=\cos\left(\beta_Y^{\top}Z+bX+0.5\,\varepsilon_Y\right),
\]
where \(\beta_X,\beta_Y\in\bbR^d\) have i.i.d.\ coordinates from \(\mathrm{Unif}([-1,1])\) and are normalized by \(\beta_X\leftarrow \beta_X/\left\|\beta_X\right\|_1\), \(\beta_Y\leftarrow \beta_Y/\left\|\beta_Y\right\|_1\). We set \(b=0\) under \(H_0\) and \(b=0.5\) under \(H_1\).

\begin{figure}
    \centering
    \includegraphics[width=1\linewidth]{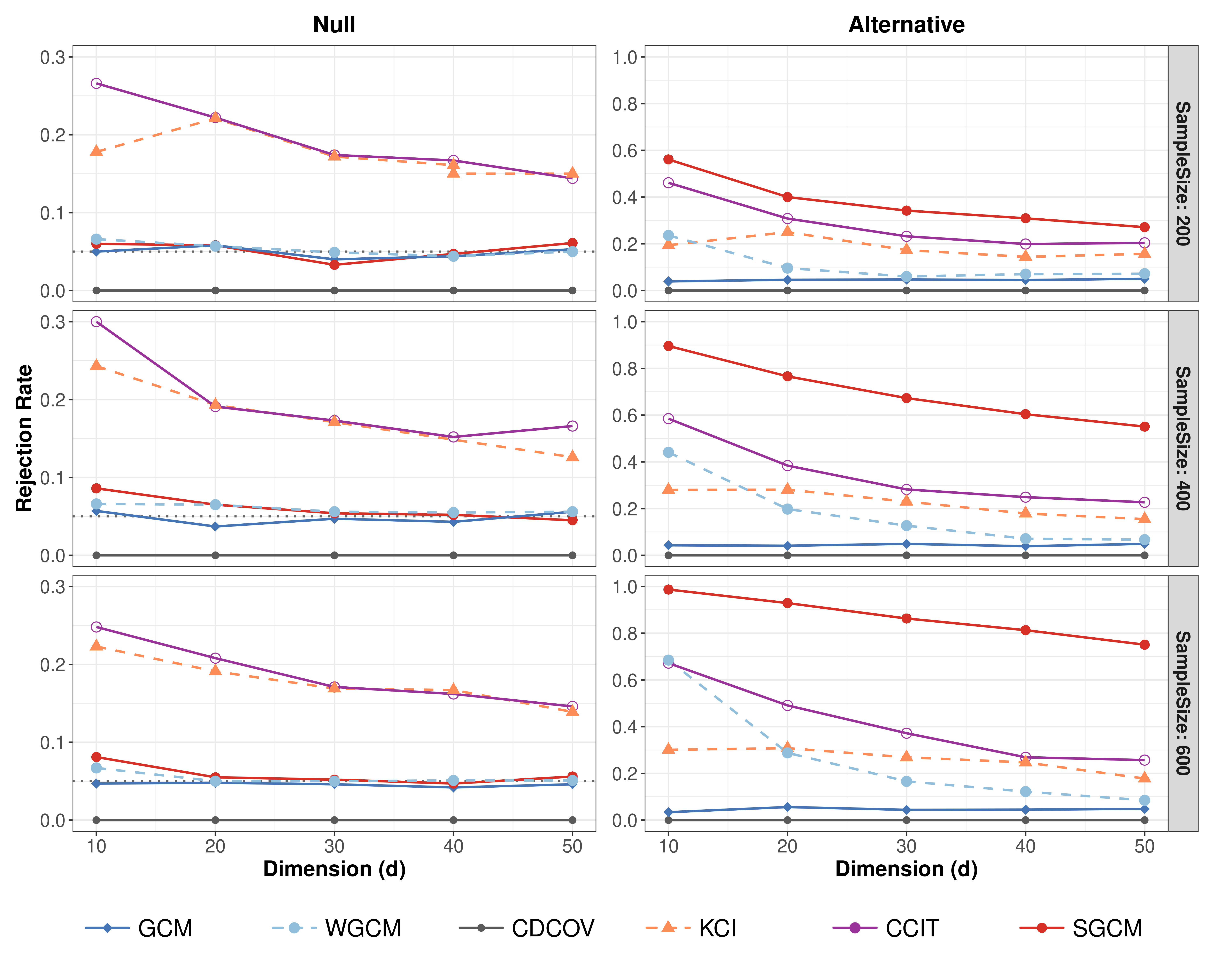}
    \caption{Empirical rejection rates by dimension for the high-dimensional post-nonlinear noise model. Left column: null; right column: alternative. Rows correspond to sample sizes \(n\in\{200,400,600\}\) (top to bottom).}
    \label{fig:result_dgp_b}
\end{figure}

\runinhead{Results for high-dimensional settings}
Figure~\ref{fig:result_dgp_b} reports rejection rates under \(H_0\) (left) and \(H_1\) (right) as functions of \(d\) for \(n\in\{200,400,600\}\); MMDCI is omitted due to computational cost. Under \(H_0\), SGCM, GCM, and WGCM stay close to the nominal level uniformly over \((d,n)\), whereas KCI and CCIT inflate size and CDCOV is conservative. Among size-valid procedures, SGCM attains the highest power across \((d,n)\), with power decreasing in \(d\) and increasing in \(n\); WGCM is second-best and GCM is weakest.

\subsection{Distributional data}\label{subsec:distribution}
We consider the case where \(X\) and \(Y\) are probability distributions observed through point clouds. We compare the 1-- and 2--Wasserstein, Fisher--Rao (FR), and Hellinger distances. For univariate distributions, \(d_{\cW_1}\) and \(d_{\cW_2}\) reduce to \(L^1/L^2\) distances between empirical quantile functions. For FR and Hellinger, we estimate densities by kernel density estimation and compute the corresponding distances from the inner product of square-root densities.

\runinhead{DGPs}
Each observation consists of a pair \((X_i,Y_i)\) of univariate empirical distributions together with a Euclidean covariate \(Z_i\in\bbR ^{10}\). For every \(i\), both \(X_i\) and \(Y_i\) are observed as point clouds of common size \(m=100\). We report results for sample sizes \(n\in\left\{200,400,600\right\}\). Generate covariates and noise via
\begin{equation*}
  Z_i\stackrel{\mathrm{i.i.d.}}{\sim} N(0,I_{10}),
  \qquad
  \varepsilon_{X,i},\varepsilon_{Y,i}\stackrel{\mathrm{i.i.d.}}{\sim} N(0,1),
\end{equation*}
mutually independent. Draw \(\beta_X,\beta_Y\in\bbR ^{10}\) with coordinates i.i.d.\ from \(\mathrm{Unif}([-1,1])\), then set \(\beta_X\leftarrow \beta_X/\left\|\beta_X\right\|_1\) and \(\beta_Y\leftarrow \beta_Y/\left\|\beta_Y\right\|_1\). A signal parameter
\(
  c\in\left\{0,\;0.05,\;0.10,\;0.15,\;0.20\right\}
\)
controls conditional dependence by sharing \(\varepsilon_{X,i}\) across \(X\) and \(Y\) in the alternatives below: \(c=0\) corresponds to \(H_0: X\indep Y \mid Z\), whereas \(c>0\) induces dependence of controlled strength. All remaining SGCM settings follow Section~\ref{subsec:euclid}.
\begin{description}[
  leftmargin=!,
  labelwidth=!,
  widest=DGP 2-2.,
  labelsep=0.6em,
  itemsep=0.6ex,
  topsep=0.3ex,
  parsep=0pt,
  align=left
]
\item[DGP 2-1.]
Define the latent means
\begin{equation*}
  \mu_{X,i}=\beta_X^{\top}Z_i+0.5\,\varepsilon_{X,i},
  \qquad
  \mu_{Y,i}(c)=\beta_Y^{\top}Z_i+c\,\varepsilon_{X,i}+0.5\,\varepsilon_{Y,i}.
\end{equation*}
Conditional on \((Z_i,\varepsilon_{X,i},\varepsilon_{Y,i})\), sample
\begin{equation*}
  X_i=\{X_{i\ell}\}_{\ell=1}^{m},
  \quad
  X_{i\ell}\stackrel{\mathrm{i.i.d.}}{\sim}  N(\mu_{X,i},1),
  \qquad
  Y_i=\{Y_{i\ell}\}_{\ell=1}^{m},
  \quad
  Y_{i\ell}\stackrel{\mathrm{i.i.d.}}{\sim}  N(\mu_{Y,i}(c),1).
\end{equation*}

\item[DGP 2-2.]
Let
\begin{equation*}
  s_{X,i}=\beta_X^{\top}Z_i+0.5\,\varepsilon_{X,i},
  \qquad
  s_{Y,i}(c)=\beta_Y^{\top}Z_i+c\,\varepsilon_{X,i}+0.5\,\varepsilon_{Y,i},
\end{equation*}
and define variances \(\sigma_{X,i}^2=\exp(s_{X,i})\) and \(\sigma_{Y,i}^2(c)=\exp(s_{Y,i}(c))\). Conditional on these,
\begin{equation*}
  X_{i\ell}\stackrel{\mathrm{i.i.d.}}{\sim}  N(0,\sigma_{X,i}^2),
  \qquad
  Y_{i\ell}\stackrel{\mathrm{i.i.d.}}{\sim}  N(0,\sigma_{Y,i}^2(c)),
  \qquad \ell=1,\dots,m.
\end{equation*}
\end{description}

\runinhead{Results for Distributional Data}
Table~\ref{tab:results_distributional} summarizes size (\(c=0\)) and power (\(c>0\)). Across metrics and sample sizes, the empirical size at \(c=0\) is close to the nominal level. In the mean-varying design (DGP 2-1), all four metrics yield nearly identical power. In the variance-varying design (DGP 2-2), \(d_{\cW_1}\) and \(d_{\cW_2}\) remain closely matched, whereas FR and Hellinger are less sensitive to variance changes and exhibit lower power as \(c\) and \(n\) increase.

\begin{table*}[htbp]
\centering
\caption{Empirical size ($\alpha=0.05$ at $c=0$) and power ($c>0$) under the Gaussian bootstrap. Left: DGP 2-1. Right: DGP 2-2.}
\label{tab:results_distributional}

\begin{minipage}[t]{0.49\textwidth}
\centering
\footnotesize
\textbf{Mean-varying (DGP 2-1)}\\[0.6ex]
\resizebox{\linewidth}{!}{%
\begin{tabular}{@{}llccccc@{}}
\toprule
& & \multicolumn{5}{c}{Signal Strength \((c)\)} \\
\cmidrule(l){3-7}
\(n\) & Metric & 0.00 & 0.05 & 0.10 & 0.15 & 0.20 \\
\midrule
\multirow{4}{*}{200}
& 1--Wasserstein & 0.051 & 0.066 & 0.105 & 0.193 & 0.329 \\
& 2--Wasserstein & 0.044 & 0.052 & 0.100 & 0.186 & 0.302 \\
& Fisher--Rao    & 0.049 & 0.060 & 0.096 & 0.182 & 0.315 \\
& Hellinger      & 0.056 & 0.068 & 0.103 & 0.175 & 0.304 \\
\midrule
\multirow{4}{*}{400}
& 1--Wasserstein & 0.046 & 0.080 & 0.192 & 0.414 & 0.682 \\
& 2--Wasserstein & 0.049 & 0.074 & 0.199 & 0.422 & 0.680 \\
& Fisher--Rao    & 0.050 & 0.073 & 0.178 & 0.394 & 0.659 \\
& Hellinger      & 0.053 & 0.076 & 0.177 & 0.392 & 0.656 \\
\midrule
\multirow{4}{*}{600}
& 1--Wasserstein & 0.049 & 0.113 & 0.301 & 0.635 & 0.899 \\
& 2--Wasserstein & 0.056 & 0.102 & 0.304 & 0.644 & 0.894 \\
& Fisher--Rao    & 0.055 & 0.100 & 0.277 & 0.614 & 0.883 \\
& Hellinger      & 0.056 & 0.099 & 0.278 & 0.597 & 0.872 \\
\bottomrule
\end{tabular}%
}
\end{minipage}
\hfill
\begin{minipage}[t]{0.49\textwidth}
\centering
\footnotesize
\textbf{Variance-varying (DGP 2-2)}\\[0.6ex]
\resizebox{\linewidth}{!}{%
\begin{tabular}{@{}llccccc@{}}
\toprule
& & \multicolumn{5}{c}{Signal Strength \((c)\)} \\
\cmidrule(l){3-7}
\(n\) & Metric & 0.00 & 0.05 & 0.10 & 0.15 & 0.20 \\
\midrule
\multirow{4}{*}{200}
& 1--Wasserstein & 0.063 & 0.085 & 0.161 & 0.277 & 0.438 \\
& 2--Wasserstein & 0.065 & 0.079 & 0.156 & 0.289 & 0.443 \\
& Fisher--Rao    & 0.059 & 0.057 & 0.090 & 0.161 & 0.281 \\
& Hellinger      & 0.046 & 0.060 & 0.094 & 0.166 & 0.278 \\
\midrule
\multirow{4}{*}{400}
& 1--Wasserstein & 0.045 & 0.096 & 0.252 & 0.538 & 0.783 \\
& 2--Wasserstein & 0.054 & 0.108 & 0.275 & 0.546 & 0.797 \\
& Fisher--Rao    & 0.046 & 0.064 & 0.147 & 0.338 & 0.592 \\
& Hellinger      & 0.044 & 0.057 & 0.150 & 0.346 & 0.597 \\
\midrule
\multirow{4}{*}{600}
& 1--Wasserstein & 0.051 & 0.125 & 0.413 & 0.768 & 0.952 \\
& 2--Wasserstein & 0.053 & 0.140 & 0.433 & 0.779 & 0.959 \\
& Fisher--Rao    & 0.054 & 0.099 & 0.252 & 0.533 & 0.829 \\
& Hellinger      & 0.048 & 0.092 & 0.252 & 0.515 & 0.826 \\
\bottomrule
\end{tabular}%
}
\end{minipage}

\end{table*}

\section{Discussion}

We proposed a conditional independence test based on the spectral generalized covariance measure (SGCM) for complex and non-Euclidean data. The construction avoids direct RKHS-valued nonparametric regression for conditional mean embeddings by reducing the problem to coordinatewise real-valued regressions. Under a doubly robust product-bias condition, we established uniform bootstrap validity, uniform asymptotic size control, nontrivial uniform power, and uniform consistency for the truncated procedure used in practice, with the power analysis formulated over classes of projected separated alternatives.

A central implication of the theory is that, once truncation is introduced, power is governed by the signal retained in the working representation. In our analysis, this appears through the projected norm \(\|\Pi_n \Sigma_{XYZ \mid Z}\|\), which makes explicit the trade-off between statistical sensitivity and dimensional reduction. The guarantees in Section~\ref{sec:theory} therefore concern the truncated, kernel-induced target, rather than the full untruncated operator. Practical performance correspondingly depends on what part of the conditional cross-covariance structure is retained after projection.

Kernel choice is likewise not merely a computational detail. It determines how the variables are represented in RKHSs and hence what notion of conditional dependence is probed by the procedure. Section~\ref{sec:kernels} provides general constructions that ensure valid kernels in broad non-Euclidean settings, but selecting kernels that are both interpretable and effective for a given application remains an open problem. Developing principled approaches to kernel choice is therefore an important direction for future work.

In this paper, we use principal-component truncation with an FVE rule. This choice is simple and stable in our experiments, but it is not the only reasonable one. Alternatives such as coefficient thresholding, shrinkage or penalization, low-rank kernel approximations, or other data-adaptive basis constructions may be preferable in some settings. A general data-driven rule that targets power while preserving valid size and remains computationally scalable is not yet available. Developing such rules is also an important direction for future work.

%% file: Appendix.tex
Throughout this appendix, assume that the kernels are bounded, and write
\[
K_X^2 \coloneqq \sup_{x\in\cX} k_{\cX}(x,x),\qquad
K_Y^2 \coloneqq \sup_{y\in\cY} k_{\cY}(y,y),\qquad
K_Z^2 \coloneqq \sup_{z\in\cZ} k_{\cZ}(z,z).
\]
We write \(P_X,P_Y,P_Z\) for the marginals of \(X,Y,Z\), and \(P_{XZ},P_{YZ}\) for the joint laws of \((X,Z)\) and \((Y,Z)\). For \(1\le p\le\infty\), let \(L_X^p=L^p(\cX,P_X)\), and define \(L_Y^p,L_Z^p,L_{XZ}^p,L_{YZ}^p\) analogously. Also write \(S_2(E,F)\) for the Hilbert--Schmidt operators from \(E\) to \(F\), \(\phi_{\cX}(x)=k_{\cX}(x,\cdot)\), \(\phi_{\cY}(y)=k_{\cY}(y,\cdot)\), and \(\phi_{\cZ}(z)=k_{\cZ}(z,\cdot)\), and let \(\pr^\ast,\E^\ast,\var^\ast\) denote probability, expectation, and variance conditional on \(\cD_n\). 

Let \(B\) be a separable Banach space, and define
\[
\mathrm{BL}_1(B)
\coloneqq
\left\{
f:B\to\bbR
:\,
\|f\|_{\infty}\le 1,\ \mathrm{Lip}(f)\le 1
\right\},
\]
where
\[
\|f\|_{\infty}\coloneqq \sup_{x\in B}|f(x)|,
\qquad
\mathrm{Lip}(f)\coloneqq \sup_{x\neq y}\frac{|f(x)-f(y)|}{\|x-y\|_B}.
\]
For \(\mu,\nu\in\cP(B)\), define the bounded--Lipschitz metric by
\[
d_{\mathrm{BL},B}(\mu,\nu)
\coloneqq
\sup_{f\in \mathrm{BL}_1(B)}
\left|
\int_B f\,\mathrm{d}\mu-\int_B f\,\mathrm{d}\nu
\right|.
\]
For \(\mu_n,\mu \in \cP(\cS)\), where \(\cS\) is Polish, we write \(\mu_n \to_w \mu\) for weak convergence; equivalently, \(d_{\mathrm{BL},\cS}(\mu_n,\mu)\to 0\).

For any real number $x$, its positive part is denoted by $(x)_+ \coloneqq \max(0,x)$. The complement of a set $A$ is denoted by $A^c$. We denote by $N(\mu, \Sigma)$ and $\cN(\mu, \Sigma)$ the Gaussian distributions on a Euclidean space and a Hilbert space, respectively.

\section{Proofs and related results for the conditional cross-covariance operator}\label{Appendix:NCCCO}
\subsection{Proof of the main result in Section~2}
We write \(\ddot X\coloneqq(X,Z)\) and $\ddot\cX= \cX\times\cZ$. By classical results on product RKHSs, the RKHS \(\cH_{\ddot{\cX}}\) of the product kernel
\(k_{\ddot{\cX}}((x,z),(x',z'))=k_{\cX}(x,x')k_{\cZ}(z,z')\)
is canonically isometric to the Hilbert tensor product \(\cH_{\cX}\otimes\cH_{\cZ}\); see, e.g., \citet[Section~4.2]{SteinwartandChristmann2008}, \citet[Section~5.5]{PaulsenandRaghupathi2016}.

\begin{proof}[of Proposition~2.1]
Since the kernels are bounded, all Bochner expectations and conditional Bochner expectations below are well defined.

By Lemma~\ref{lem:CME_decompose} (identifying \(\cH_{\ddot{\cX}}\cong\cH_{\cX}\otimes\cH_{\cZ}\)),
\begin{equation*}
\E\left[k_{\ddot{\cX}}((X,Z),\cdot) \mid Z\right]
 = 
\E\left[k_{\cX}(X, \cdot) \mid Z\right]\otimes k_{\cZ}(Z,\cdot)
\quad\text{in }\cH_{\cX}\otimes\cH_{\cZ}\ \text{a.s.}
\end{equation*}
Consequently,
\begin{equation}\label{eq:centered-tensor}
k_{\ddot{\cX}}((X,Z),\cdot)-\mu_{\ddot X\mid Z}(Z)
 = 
\left(k_{\cX}(X, \cdot)-\mu_{X\mid Z}(Z)\right)\otimes k_{\cZ}(Z,\cdot)
\quad\text{a.s.}
\end{equation}

Set
\[
\Sigma_{\ddot X Y\mid Z}
\coloneqq
\E\Bigl[
\bigl(k_{\ddot{\cX}}(\ddot X,\cdot)-\mu_{\ddot X\mid Z}(Z)\bigr)\otimes
\bigl(k_{\cY}(Y,\cdot)-\mu_{Y\mid Z}(Z)\bigr)
\Bigr]
\in \cH_{\ddot{\cX}}\otimes\cH_{\cY}.
\]
By \eqref{eq:centered-tensor}, this corresponds to
\[
\widetilde\Sigma_{XZY\mid Z}
=
\E\Bigl[
\bigl(k_{\cX}(X,\cdot)-\mu_{X\mid Z}(Z)\bigr)\otimes
k_{\cZ}(Z,\cdot)\otimes
\bigl(k_{\cY}(Y,\cdot)-\mu_{Y\mid Z}(Z)\bigr)
\Bigr]
\in \cH_{\cX}\otimes\cH_{\cZ}\otimes\cH_{\cY}.
\]
If \(U_{(23)}(u\otimes v\otimes w)=u\otimes w\otimes v\), then
\begin{equation}\label{eq:perm-equivalence}
\Sigma_{XYZ\mid Z}=0
\iff
\widetilde\Sigma_{XZY\mid Z}=0
\iff
\Sigma_{\ddot X Y\mid Z}=0.
\end{equation}

Assuming \(X\indep Y\mid Z\), fix arbitrary \(h_{\cX}\in\cH_{\cX}\), \(h_{\cY}\in\cH_{\cY}\), \(h_{\cZ}\in\cH_{\cZ}\). Since
\(c(Z)\coloneqq \left\langle \phi_{\cZ}(Z),h_{\cZ}\right\rangle\)
is \(\sigma(Z)\)-measurable and bounded, Proposition~2.6.31 in \cite{Hytonenetal2016} and $X \indep Y \mid Z$ give
\begin{align*}
&\E\left[
\left\langle \phi_{\cX}(X)-\mu_{X\mid Z}(Z),h_{\cX}\right\rangle
\left\langle \phi_{\cY}(Y)-\mu_{Y\mid Z}(Z),h_{\cY}\right\rangle
\,c(Z)
\right]\\ 
&\quad = \E\left[\E[\left\langle \phi_{\cX}(X)-\mu_{X\mid Z}(Z),h_{\cX}\right\rangle \mid Z]\E[\left\langle \phi_{\cY}(Y)-\mu_{Y\mid Z}(Z),h_{\cY}\right\rangle \mid Z] \,c(Z) \right]=0
\end{align*}
Hence \(\left\langle \Sigma_{XYZ\mid Z},\,h_{\cX}\otimes h_{\cY}\otimes h_{\cZ}\right\rangle=0\) for all elementary tensors. By density of the algebraic tensor product in \(\cH_{\cX}\otimes\cH_{\cY}\otimes\cH_{\cZ}\), it follows that \(\Sigma_{XYZ\mid Z}=0\).

By \eqref{eq:perm-equivalence}, we have \(\Sigma_{\ddot X Y\mid Z}=0\).
Under Assumption~1 (i), both \(k_{\ddot{\cX}}=k_{\cX}\otimes k_{\cZ}\) and \(k_{\cY}\) are characteristic. Because each kernel is bounded and characteristic by Assumption~1 (i), the direct sum \(\cH_{\ddot\cX}+\bbR\) is dense in \(L^2(P_{XZ})\) for any probability measure \(P_{XZ}\) on \((\ddot{\cX},\cB(\ddot\cX))\), and similarly \(\cH_{\cY}+\bbR\) is dense in \(L^2(P_Y)\) for any probability measure $P_Y$ on $(\cY, \cB(\cY))$ \cite[Proposition~5]{Fukumizuetal2009}.

Define
\[
B(f,g)\coloneqq \E\left[(f-\E[f\mid Z])(g-\E[g\mid Z])\right],
\qquad
f\in L_{XZ}^2,\ g\in L_Y^2.
\]
Since conditional expectation is an \(L^2\)-contraction,
\[
|B(f,g)|
\le
\|f-\E[f\mid Z]\|_{L^2}\,\|g-\E[g\mid Z]\|_{L^2}
\le
4\|f\|_{L_{XZ}^2}\|g\|_{L_Y^2},
\]
so \(B\) is continuous. Moreover, for \(f_H\in\cH_{\ddot\cX}\) and \(g_H\in\cH_{\cY}\),
\[
\langle \Sigma_{\ddot X Y\mid Z},f_H\otimes g_H\rangle = B(f_H,g_H),
\]
hence \(\Sigma_{\ddot X Y\mid Z}=0\) implies \(B=0\) on \(\cH_{\ddot\cX}\times\cH_{\cY}\), and therefore on \((\cH_{\ddot\cX}+\bbR)\times(\cH_{\cY}+\bbR)\). By denseness of these spaces in \(L_{XZ}^2\) and \(L_Y^2\), it follows that \(B(f,g)=0\) for all \(f\in L_{XZ}^2\) and \(g\in L_Y^2\). Thus, Corollary~\ref{cor:daudin_XZ} implies \(X\indep Y\mid Z\).

The case of Assumption~1~(ii) is identical by symmetry, exchanging the roles of \(X\) and \(Y\).

Now assume Assumption~1~(iii). For \(f\in L_X^2\) and \(g\in L_Y^2\), write
\[
\tilde f\coloneqq f(X)-\E[f(X)\mid Z],
\qquad
\tilde g\coloneqq g(Y)-\E[g(Y)\mid Z],
\qquad
M_{f,g}(Z)\coloneqq \E[\tilde f\,\tilde g\mid Z].
\]
From \(\Sigma_{XYZ\mid Z}=0\), we have
\[
\E[\tilde f_H\,\tilde g_H\,h_H(Z)]=0
\qquad
\forall\,f_H\in\cH_{\cX}+\bbR,\ g_H\in\cH_{\cY}+\bbR,\ h_H\in\cH_{\cZ}.
\]
Since bounded kernels imply \(f_H(X),g_H(Y)\in L^\infty\), it follows that \(M_{f_H,g_H}\in L_Z^2\), and hence
\[
\langle M_{f_H,g_H},h_H\rangle_{L^2(P_Z)}=0
\qquad
\forall\,h_H\in\cH_{\cZ}.
\]
By density of \(\cH_{\cZ}\) in \(L^2(P_Z)\), this gives \(M_{f_H,g_H}=0\) in \(L^2(P_Z)\).

Next choose \(f_n\in\cH_{\cX}+\bbR\) and \(g_n\in\cH_{\cY}+\bbR\) with \(f_n\to f\) in \(L_X^2\) and \(g_n\to g\) in \(L_Y^2\). Writing \(\tilde f_n=f_n(X)-\E[f_n(X)\mid Z]\) and \(\tilde g_n=g_n(Y)-\E[g_n(Y)\mid Z]\), the \(L^2\)-contraction property gives \(\tilde f_n\to \tilde f\) and \(\tilde g_n\to \tilde g\) in \(L^2\). Hence
\[
\|M_{f_n,g_n}-M_{f,g}\|_{L_Z^1}
\le
\E|\tilde f_n\tilde g_n-\tilde f\tilde g|
\to 0,
\]
so \(M_{f,g}=0\) a.s. Since \(f\) and \(g\) were arbitrary, Proposition~\ref{prop:daudin_CI_char} yields \(X\indep Y\mid Z\).
\end{proof}

\begin{lemma}\label{lem:CME_decompose}
Assume Assumption~\ref{ass:kernel_characteristic}. Define the product kernel on \(\cX\times\cZ\) by
\begin{equation*}
k_{\ddot\cX}((x,z),(x',z'))\coloneqq k_{\cX}(x,x')\,k_{\cZ}(z,z'),
\end{equation*}
and denote its RKHS by \(\cH_{\ddot\cX}\).
Then, as \(\cH_{\ddot\cX}\)-valued random variables,
\begin{equation*}
\E\left[k_{\ddot\cX}\bigl((X,Z),\cdot\bigr)\mid Z\right]
 = 
\E\left[k_{\cX}(X, \cdot)\mid Z\right]\otimes k_{\cZ}(Z,\cdot),
\end{equation*}
where we identify \(\cH_{\ddot\cX}\) isometrically with the Hilbert tensor product \(\cH_{\cX}\otimes\cH_{\cZ}\) and use the elementary tensor on the right-hand side.
\end{lemma}
\begin{proof}
Let \(U:\cH_{\cX}\otimes\cH_{\cZ}\to\cH_{\ddot{\cX}}\) be the canonical isometry determined by
\[
U\bigl(\phi_{\cX}(x)\otimes\phi_{\cZ}(z)\bigr)=k_{\ddot{\cX}}((x,z),\cdot).
\]
Since
\[
\langle \phi_{\cX}(x)\otimes\phi_{\cZ}(z),\phi_{\cX}(x')\otimes\phi_{\cZ}(z')\rangle
=
k_{\cX}(x,x')k_{\cZ}(z,z')
=
\langle k_{\ddot{\cX}}((x,z),\cdot),k_{\ddot{\cX}}((x',z'),\cdot)\rangle,
\]
we may identify \(\cH_{\ddot\cX}\cong \cH_{\cX}\otimes\cH_{\cZ}\).

Let \(\cD\) be the algebraic span of simple tensors \(h_{\cX}\otimes h_{\cZ}\), which is dense in \(\cH_{\cX}\otimes\cH_{\cZ}\). For \(h_{\cX}\otimes h_{\cZ}\in\cD\),
\begin{align*}
\left\langle \E[\phi_{\cX}(X)\otimes\phi_{\cZ}(Z)\mid Z],\,h_{\cX}\otimes h_{\cZ}\right\rangle
&=
\E\!\left[\langle \phi_{\cX}(X),h_{\cX}\rangle \langle \phi_{\cZ}(Z),h_{\cZ}\rangle \mid Z\right] \\
&=
\langle \phi_{\cZ}(Z),h_{\cZ}\rangle\,\E[\langle \phi_{\cX}(X),h_{\cX}\rangle\mid Z] \\
&=
\left\langle \E[\phi_{\cX}(X)\mid Z]\otimes\phi_{\cZ}(Z),\,h_{\cX}\otimes h_{\cZ}\right\rangle .
\end{align*}
Hence the two \(\cH_{\cX}\otimes\cH_{\cZ}\)-valued random variables agree on a dense set of test vectors, and therefore agree almost surely. Applying \(U\) gives the claim.
\end{proof}

\subsection{Relation to the classical CCCO and the MMD}\label{subsec:CCCO_MMD}

We record two equivalent representations of the CCCO used in Remark~1: the identification with the classical CCCO of \cite{Fukumizuetal2004}, and the MMD representation based on the conditionally independent coupling \(P^{CI}\).
\begin{proposition}\label{prop:equiv_cccov_YZ_concise}
Assume Assumption~\ref{ass:kernel_characteristic}. 
Let the product kernel on \(\ddot\cY =\cY\times\cZ\) be
\(k_{\ddot{\cY}}((y,z),(y',z'))\coloneqq
k_{\cY}(y,y')\,k_{\cZ}(z,z')\)
with RKHS \(\cH_{\ddot{\cY}}\), and let 
\(U_{YZ}\colon \cH_{\cY}\otimes\cH_{\cZ}\xrightarrow{\ \cong\ }\cH_{\ddot{\cY}}\)
be the canonical isometry.

Define the conditional cross-covariance operator
\(\Sigma_{X\ddot{Y} \mid Z}\in\cH_{\cX}\otimes\cH_{\ddot{\cY}}\) by
\begin{equation*}
\Sigma_{X\ddot{Y} \mid Z} =\Sigma_{X\ddot Y}-\Sigma_{XX}^{1/2}V_{XZ}V_{Z\ddot Y}\Sigma_{\ddot Y\ddot Y}^{1/2}.
\end{equation*}

Then the canonical product-kernel identification yields
\begin{equation}\label{eq:concise_YZ_equiv}
\Sigma_{X\ddot{Y} \mid Z}
= 
\left(I\otimes U_{YZ}\right)\,\Sigma_{XYZ\mid Z}
\in \cH_{\cX}\otimes\cH_{\ddot{\cY}}.
\end{equation}
In particular,
\begin{equation*}
    \|\Sigma_{X\ddot{Y} \mid Z}\|_{\cH_{\cX}\otimes \cH_{\ddot{\cY}}} = \|\Sigma_{XYZ \mid Z}\|_{\cH_{\cX} \otimes \cH_{\cY} \otimes \cH_{\cZ}}
\end{equation*}
\end{proposition}
\begin{proof}
Fix \(f_{\cX}\in\cH_{\cX}\) and write \(h_{\ddot Y}=U_{YZ}\left(h_{\cY}\otimes h_{\cZ}\right)\) with \(h_{\cY}\in\cH_{\cY}\), \(h_{\cZ}\in\cH_{\cZ}\).
By \citet[Proposition~5]{Fukumizuetal2004} applied to \(\ddot Y=(Y,Z)\),
\begin{equation*}
\left\langle f_{\cX}\otimes h_{\ddot Y},\,\Sigma_{X\ddot{Y} \mid Z}\right\rangle
= 
\E\left[\cov\left(\left\langle f_{\cX},k_{\cX}(X, \cdot)\right\rangle,\ 
\left\langle h_{\ddot Y},k_{\ddot{\cY}}((Y,Z),\cdot)\right\rangle\ \middle|\ Z\right)\right].
\end{equation*}
Using \(U_{YZ}\) and the product-kernel evaluation,
\begin{equation*}
\left\langle h_{\ddot Y},k_{\ddot{\cY}}((Y,Z),\cdot)\right\rangle
= 
\left\langle h_{\cY},k_{\cY}(Y, \cdot)\right\rangle\,
\left\langle h_{\cZ},k_{\cZ}(Z,\cdot)\right\rangle.
\end{equation*}
By \citet[Lemma~2]{ParkandMuandet2020},
\begin{equation*}
\left\langle f_{\cX}\otimes h_{\cY}\otimes h_{\cZ}, \Sigma_{XYZ\mid Z}\right\rangle
=
\E\left[\cov\left(\left\langle f_{\cX},k_{\cX}(X, \cdot)\right\rangle, 
\left\langle h_{\cY},k_{\cY}(Y, \cdot)\right\rangle\mid  Z\right) 
\left\langle h_{\cZ},k_{\cZ}(Z,\cdot)\right\rangle\right].
\end{equation*}
Combining the last two displays gives
\begin{equation*}
\left\langle f_{\cX}\otimes h_{\ddot Y},\,\Sigma_{X\ddot{Y} \mid Z}\right\rangle
= 
\left\langle f_{\cX}\otimes h_{\ddot Y},\ \left(I\otimes U_{YZ}\right)\Sigma_{XYZ\mid Z}\right\rangle.
\end{equation*}
Since elementary tensors are total, \eqref{eq:concise_YZ_equiv} follows.
\end{proof}

Since \(\cX\), \(\cY\), and \(\cZ\) are Polish spaces, regular conditional distributions \(P_{X\mid Z=z}\) and \(P_{Y\mid Z=z}\) exist. Fix such versions and define the conditionally independent coupling \(P^{\mathrm{CI}}\) by
\[
P^{\mathrm{CI}}(A\times B\times C)
\coloneqq
\int_C P_{X\mid Z=z}(A)\,P_{Y\mid Z=z}(B)\,P_Z(\diff z).
\]
Equivalently,
\[
P^{\mathrm{CI}}(\diff x,\diff y,\diff z)
=
P_{X\mid Z=z}(\diff x)\,P_{Y\mid Z=z}(\diff y)\,P_Z(\diff z).
\]
This is well defined up to \(P_Z\)-null modifications of the regular conditional distributions.
\begin{proposition}[MMD representation of the CCCO]
\label{prop:ccco_as_mmd}
Let
\[
K\bigl((x,y,z),(x',y',z')\bigr)
\coloneqq
k_{\cX}(x,x')\, k_{\cY}(y,y')\, k_{\cZ}(z,z')
\]
be the product kernel on \(\cX \times \cY \times \cZ\). Then
\begin{equation}
\label{eq:Sigma_as_mean_embedding_difference}
\Sigma_{XYZ \mid Z}
=
\mu_P^{(K)} - \mu_{P^{\mathrm{CI}}}^{(K)}
\in
\cH_{\cX} \otimes \cH_{\cY} \otimes \cH_{\cZ}.
\end{equation}
Consequently,
\begin{equation}
\label{eq:Sigma_as_MMD}
\left\| \Sigma_{XYZ \mid Z} \right\|
=
\operatorname{MMD}_K(P,P^{\mathrm{CI}}),
\qquad
\left\| \Sigma_{XYZ \mid Z} \right\|^2
=
\operatorname{MMD}_K^2(P,P^{\mathrm{CI}}).
\end{equation}
\end{proposition}

\begin{remark}
By Proposition~\ref{prop:ccco_as_mmd}, the CCCO signal is exactly the MMD between \(P\) and its conditionally independent coupling \(P^{\mathrm{CI}}\). While such a kernel discrepancy is not comparable with Wasserstein distances in full generality, Sinkhorn divergences interpolate between MMD and optimal transport, and, once suitable restrictions are imposed on the distribution class, bounds of the form \(W_p \lesssim \operatorname{MMD}^{\delta}\) become available. Thus, over such classes, a large transport discrepancy can be interpreted as a strong CCCO signal; see \cite{VayerandGribonval2023,Feydyetal2019}.
\end{remark}

\begin{proof}
Because the kernels are bounded, all Bochner integrals below are well defined. By the definition of \(P^{\mathrm{CI}}\) and Fubini's theorem,
\begin{align*}
\mu_{P^{\mathrm{CI}}}^{(K)}
&=
\iiint
k_{\cX}(x,\cdot)\otimes k_{\cY}(y,\cdot)\otimes k_{\cZ}(z,\cdot)\,
P^{\mathrm{CI}}(\diff x,\diff y,\diff z) \\
&=
\E\left[
\mu_{X\mid Z}(Z)\otimes \mu_{Y\mid Z}(Z)\otimes k_{\cZ}(Z,\cdot)
\right].
\end{align*}
Also,
\begin{align*}
\Sigma_{XYZ\mid Z}
&=
\E\left[
\left(k_{\cX}(X,\cdot)-\mu_{X\mid Z}(Z)\right)
\otimes
\left(k_{\cY}(Y,\cdot)-\mu_{Y\mid Z}(Z)\right)
\otimes
k_{\cZ}(Z,\cdot)
\right] \\
&=
\mu_P^{(K)}
-
\E\left[k_{\cX}(X,\cdot)\otimes \mu_{Y\mid Z}(Z)\otimes k_{\cZ}(Z,\cdot)\right] \\
&\quad
-
\E\left[\mu_{X\mid Z}(Z)\otimes k_{\cY}(Y,\cdot)\otimes k_{\cZ}(Z,\cdot)\right]
+
\mu_{P^{\mathrm{CI}}}^{(K)}.
\end{align*}
Both mixed terms equal \(\mu_{P^{\mathrm{CI}}}^{(K)}\) by conditioning on \(Z\). Hence
\[
\Sigma_{XYZ\mid Z}=\mu_P^{(K)}-\mu_{P^{\mathrm{CI}}}^{(K)},
\]
which is \eqref{eq:Sigma_as_mean_embedding_difference}. Taking norms yields \eqref{eq:Sigma_as_MMD}.
\end{proof}

\subsection{Daudin’s characterization of conditional independence on Polish spaces}
We use the usual factorization definition of \(X\indep Y\mid Z\) via regular conditional probabilities. For later use, we record a Polish-space version of Daudin's characterization \citep{Daudin1980} in terms of vanishing conditional covariances of centered \(L^2\)-functions.
\begin{proposition}\label{prop:daudin_CI_char}
Let \((\cX,\cY,\cZ)\) be Polish spaces, and let \(X\colon \Omega\to\cX\), \(Y\colon \Omega\to\cY\), \(Z\colon \Omega\to\cZ\) be random elements on a probability space \((\Omega,\cF,\pr)\).
Assume that for all \(f\in L^2_X\) and \(g\in L^2_Y\),
\begin{equation*}
\E\left[\left(f(X)-\E[f(X) \mid Z]\right)\left(g(Y)-\E[g(Y) \mid Z]\right)\mid Z\right]=0
\quad \text{a.s.-}P_Z.
\end{equation*}
Then \(X\indep Y\mid Z\).
\end{proposition}
\begin{proof}[of Proposition~\ref{prop:daudin_CI_char}]
    Taking indicators $f=\mathds 1_{A}$, $g=\mathds 1_{B}$ with $A\in\cB(\cX)$, $B\in\cB(\cY)$ (these belong to $L^2$ on a probability space), we obtain almost surely
\begin{equation*}
\pr(X\in A,\,Y\in B\mid Z)=\pr(X\in A\mid Z)\,\pr(Y\in B\mid Z).
\end{equation*}
Taking indicators establishes the factorization for any given pair of Borel sets $(A, B)$ almost surely. However, the definition of CI requires this identity to hold simultaneously for all $A \in \cB(\cX)$ and $B \in \cB(\cY)$ outside a single null set. To bridge this gap from a pointwise to a simultaneous almost-sure statement, we employ a standard measure-theoretic argument. Let $\cP_{\cX}\subset\cB(\cX)$ and $\cP_{\cY}\subset\cB(\cY)$ be countable $\pi$-systems that generate the respective Borel $\sigma$-algebras (e.g. finite intersections of balls with rational centers and radii). For each $A\in\cP_{\cX}, B\in\cP_{\cY}$, the above factorization holds almost surely; denote by $N_{A,B}$ the corresponding $P_Z$-null set where it may fail. Let $N\coloneqq\bigcup_{A,B}N_{A,B}$, still $P_Z$-null. Fix $z\in\cZ\setminus N$. The proof proceeds in two steps by applying the \(\pi\)-\(\lambda\) theorem twice.

First, fix an arbitrary set $A \in \cP_{\cX}$. We define a class of sets $\Lambda_A \subset \cB(\cY)$ as follows:
\begin{equation*}
\Lambda_{A} \coloneqq \{B \in \cB(\cY) : P_{XY\mid Z=z}(A\times B) = P_{X\mid Z=z}(A)P_{Y\mid Z=z}(B)\}.
\end{equation*}
It is a standard exercise to verify that $\Lambda_A$ is a $\lambda$-system. By hypothesis, the product identity holds for all sets in the \(\pi\)-system $\cP_{\cY}$, which implies that $\cP_{\cY} \subset \Lambda_A$. Since $\Lambda_A$ is a $\lambda$-system containing the $\pi$-system $\cP_{\cY}$ that generates $\cB(\cY)$, the $\pi$-$\lambda$ theorem \citep[Theorem~1.19]{Klenke2020} yields that $\sigma(\cP_{\cY}) \subset \Lambda_A$, and thus $\Lambda_A = \cB(\cY)$.

Second, having shown that the equality holds for any $A \in \cP_{\cX}$ and all $B \in \cB(\cY)$, we now extend it to all $A \in \cB(\cX)$. Define the class $\Lambda \subset \cB(\cX)$ by:
\begin{equation*}
\Lambda \coloneqq \{A \in \cB(\cX) : P_{XY\mid Z=z}(A\times B) = P_{X\mid Z=z}(A)P_{Y\mid Z=z}(B) \text{ for all } B \in \cB(\cY)\}.
\end{equation*}
Again, $\Lambda$ is a $\lambda$-system. From the conclusion of our first step, we know that $\cP_{\cX} \subset \Lambda$. Since $\Lambda$ is a $\lambda$-system containing the $\pi$-system $\cP_{\cX}$ that generates $\cB(\cX)$, we may once more apply the $\pi$-$\lambda$ theorem to conclude that $\Lambda = \cB(\cX)$.

Therefore, the identity $P_{XY\mid Z=z}(A\times B) = P_{X\mid Z=z}(A)P_{Y\mid Z=z}(B)$ holds for all $A \in \cB(\cX)$ and $B \in \cB(\cY)$. This is precisely the statement that the conditional measure $P_{XY\mid Z=z}$ is equal to the product of its marginals $P_{X\mid Z=z} \otimes P_{Y\mid Z=z}$. As this holds for all $z \in \cZ \setminus N$ where $P_Z(N)=0$, we have shown $X \indep Y \mid Z$.
\end{proof}

\begin{corollary}\label{cor:daudin_XZ}
Under the setting of Proposition~\ref{prop:daudin_CI_char}, suppose that for every \(f\in L^2_{XZ}\) and \(g\in L^2_Y\),
\begin{equation*}
\E\left[\left(f(X,Z)-\E[f(X,Z) \mid Z]\right)\left(g(Y)-\E[g(Y) \mid Z]\right)\mid Z\right]=0
\quad \text{a.s.-}P_Z.
\end{equation*}
Then \(X\indep Y\mid Z\).
\end{corollary}

\begin{proof}[of Corollary~\ref{cor:daudin_XZ}]
Work under the setting of the preceding proposition. Fix \(\tilde f\in L^2_X\), \(g\in L^2_Y\), and \(h\in L^{\infty}_Z\).
First, \(f(x,z)\coloneqq \tilde f(x)\,h(z)\) belongs to \(L^2_{XZ}\), since
\begin{equation*}
\E\left[\bigl|\tilde f(X)\,h(Z)\bigr|^2\right]
\le \left\|h\right\|_{\infty}^2\,\E\left[\tilde f(X)^2\right]
<\infty .
\end{equation*}
Moreover, by the tower property and the fact that \(h(Z)\) is \(\sigma(Z)\)-measurable and bounded,
\begin{equation*}
f(X,Z)-\E[f(X,Z) \mid Z]
=h(Z)\,\Bigl(\tilde f(X)-\E\bigl[\tilde f(X)\,\mid Z\bigr]\Bigr).
\end{equation*}
Applying the hypothesis with this \(f\) and the given \(g\) yields, almost surely,
\begin{equation*}
0
=\E\left[\left(f(X,Z)-\E[f(X,Z) \mid Z]\right)\left(g(Y)-\E[g(Y) \mid Z]\right)\mid Z\right]
=h(Z)\,\Phi(Z),
\end{equation*}
where
\begin{equation*}
\Phi(Z)\coloneqq
\E\left[\bigl(\tilde f(X)-\E\bigl[\tilde f(X)\mid Z\bigr]\bigr)\bigl(g(Y)-\E[g(Y) \mid Z]\bigr)\mid Z\right]
\end{equation*}
is \(\sigma(Z)\)-measurable and integrable.

Since the equality \(h(Z)\,\Phi(Z)=0\) holds for all \(h\in L^{\infty}_Z\), it holds in particular for all indicator functions \(h=\mathds 1_C\) with \(C\in\cB(\cZ)\). Hence \(\mathds 1_C(Z)\,\Phi(Z)=0\) a.s.\ for every Borel set \(C\), which implies \(\Phi(Z)=0\) a.s.-\(P_Z\) (take \(C=\{\Phi>0\}\) and \(C=\{\Phi<0\}\), both \(Z\)-measurable). Consequently, for every \(\tilde f\in L^2_X\) and \(g\in L^2_Y\),
\begin{equation*}
\E\left[\bigl(\tilde f(X)-\E\bigl[\tilde f(X)\mid Z\bigr]\bigr)\bigl(g(Y)-\E[g(Y) \mid Z]\bigr)\mid Z\right]=0
\quad\text{a.s.-}P_Z.
\end{equation*}
Thus the assumption of Proposition~\ref{prop:daudin_CI_char} is satisfied with test functions depending only on \(X\) (for \(\tilde f\)) and on \(Y\) (for \(g\)), and the proposition yields \(X\indep Y\mid Z\).
\end{proof}

\section{Explicit construction of the empirical eigenfunctions}\label{sec:const_eigen_function}
We record the standard Gram-matrix representation of the empirical eigenfunctions used in Section~3.1.
Since the arguments for \(X\) and \(Y\) are identical, we treat only the \(X\)-side. Since
\begin{equation*}
  \hat C_{XX}
  =
  \frac{1}{n_2}
  \sum_{i\in I_2}
  k_{\cX}(X_i,\cdot)\otimes k_{\cX}(X_i,\cdot),
\end{equation*}
any eigenfunction $\hat e_p$ with a nonzero eigenvalue $\hat\lambda_p$ lies in the span of $\{k_{\cX}(X_i,\cdot)\}_{i\in I_2}$. Writing $\hat e_p(\cdot) = \sum_{i\in I_2} \alpha_p[i]\,k_{\cX}(X_i,\cdot)$ for some $\alpha_p\in\bbR^{n_2}$, the eigenvalue equation $\hat C_{XX}\hat e_p = \hat\lambda_p \hat e_p$ reduces via the reproducing property to the finite-dimensional problem $K_X\alpha_p = n_2\hat\lambda_p\alpha_p$, where $(K_X)_{ij}=k_{\cX}(X_i,X_j)$.

Thus, for any eigenpair $(\kappa_p, u_p)$ of $K_X$ with $\kappa_p>0$ and $\|u_p\|_2=1$, setting $\alpha_p = u_p / \sqrt{\kappa_p}$ yields the normalized empirical eigenfunction
\begin{equation*}
  \hat e_p(\cdot)
  =
  \frac{1}{\sqrt{\kappa_p}}
  \sum_{i\in I_2} u_p[i]\;k_{\cX}(X_i,\cdot)
\end{equation*}
with eigenvalue $\hat\lambda_p = \kappa_p / n_2$. The construction for \(\hat f_q\) is identical after replacing \(X\) by \(Y\).
\section{Asymptotic properties of the SGCM test statistic}\label{sec:asymp_properties_SGCM}
In the sequel, $\hat{\Psi}_i^{(L_{X,n}, L_{Y,n})}$, is then given by
\begin{equation*}
\hat{\Psi}_i^{(L_{X,n}, L_{Y,n})} \coloneqq \hat u^{(i)}_X\otimes \hat u^{(i)}_Y\otimes \phi_{\cZ}(Z_i),
\end{equation*}
where
\begin{equation*}
\hat u^{(i)}_X\coloneqq \sum_{p=1}^{L_{X,n}}\hat\zeta^{(i)}_{X\mid Z,p}\,\hat e_p\in\cH_{\cX},
\qquad
\hat u^{(i)}_Y\coloneqq \sum_{q=1}^{L_{Y,n}}\hat\zeta^{(i)}_{Y\mid Z,q}\,\hat f_q\in\cH_{\cY}.
\end{equation*}

For $i \in I_1$, $p\in\{1,\dots,L_{X,n}\}$ and $q\in\{1,\dots,L_{Y,n}\}$ define
\begin{equation*}
\begin{aligned}
\zeta^{(i)}_{X\mid Z,p} &\coloneqq \hat e_p(X_i)-\E[\hat{e}_p(X_i) \mid \hat e_p, Z_i],
&
\zeta^{(i)}_{Y\mid Z,q} &\coloneqq \hat f_q(Y_i)-\E[\hat{f}_q(Y_i) \mid \hat f_q , Z_i].
\end{aligned}
\end{equation*}

Let the uncentered covariance operators \(C_{XX}\) and \(C_{YY}\) admit spectral decompositions
\[
C_{XX} e_p = \lambda_p e_p
\quad\text{in}\ \cH_{\cX},
\qquad
C_{YY} f_q = \mu_q f_q
\quad\text{in}\ \cH_{\cY},
\]
where \(\{e_p\}_{p\ge 1}\subset\cH_{\cX}\) and \(\{f_q\}_{q\ge 1}\subset\cH_{\cY}\) form orthonormal systems of eigenfunctions, and the corresponding eigenvalues are arranged in nonincreasing order.
We define the population spectral projectors as
\[
\Pi_{X,L_{X,n}} \coloneqq \sum_{p=1}^{L_{X,n}} e_p \otimes e_p,
\qquad
\Pi_{Y,L_{Y,n}} \coloneqq \sum_{q=1}^{L_{Y,n}} f_q \otimes f_q.
\]
Similarly, let \(\hat\Pi_{X,L_{X,n}}\) and \(\hat \Pi_{Y,L_{Y,n}}\) be the empirical orthogonal projectors on \(\cH_{\cX}\) and \(\cH_{\cY}\) constructed from the empirical eigenfunctions \(\hat e_p\) and \(\hat f_q\):
\[
\hat\Pi_{X,L_{X,n}} \coloneqq \sum_{p=1}^{L_{X,n}} \hat e_p \otimes \hat e_p,
\qquad
\hat \Pi_{Y,L_{Y,n}} \coloneqq \sum_{q=1}^{L_{Y,n}} \hat f_q \otimes \hat f_q.
\]

A fixed-\(K\) fold cross-fitted version can also be accommodated, but we do not adopt that formulation here. Instead, we present the theory in the basic notation introduced above, in which the conditional expectations entering the residual scores are estimated on the same sample used to construct the test statistic. The cross-fitted version is obtained by replacing these in-sample regression fits with their out-of-fold counterparts, without altering the main structure of the argument. See Remark~\ref{rem:extension_to_k_fold} for further details.

\subsection{Uniform asymptotic theory for size and power}
Throughout, for a distribution function \(F\) on \(\bbR\), write
\begin{equation}\label{eq:def_quantile_functional}
q_F(u)\coloneqq \inf\{t\in\bbR:F(t)\ge u\},
\qquad
u\in(0,1).
\end{equation}
\runinhead{Null limit, bootstrap approximation, and size control}
\begin{proof}[of Theorem~\ref{thm:uniform_size_control}]
The proof has three steps: the uniform null limit for \(n_1\hat T_{n_1}\), the uniform conditional bootstrap approximation, and the transfer to asymptotic size control.

\noindent\textbf{Part I: Uniform null limit law for \(n_1\hat T_{n_1}\).}

Let
\begin{equation}
S_{n,P}^{(\infty)}
\coloneqq
\frac{1}{\sqrt{n_1}}\sum_{i\in I_1}\Psi_{i,P}^{(\infty)},
\qquad
\Psi_{i,P}^{(\infty)}\coloneqq \Pi_\infty\Psi_i,
\qquad
F_P(t)\coloneqq \pr_P\left(\|G_P^{(\infty)}\|^2\le t\right).
\label{eq:def_Sn_and_FP_projected}
\end{equation}

By Proposition~\ref{prop:bias-negligible}, Lemma~\ref{lem:scaled-proj-replacement-uniform} and \ref{lem:coord-truncation-unified},
\begin{equation}\label{eq:size_step1_replacement}
n_1\hat T_{n_1}-\left\|S_{n,P}^{(\infty)}\right\|^2
=
o_{\tilde{\cP}_0}(1).
\end{equation}

Next, by Theorem~2 in \cite{Lundborgetal2022} (applied to \(\Psi_{i,P}^{(\infty)}\)) and Assumption~\ref{ass:fixed-basis-uniform_tail},
\begin{equation}\label{eq:BL_unif_CLT_projected}
\Delta_n
\coloneqq
\sup_{P\in\tilde{\cP}_0}
d_{\mathrm{BL}, \cH}\left(
\cL\left(S_{n,P}^{(\infty)}\right),
\cL\left(G_P^{(\infty)}\right)
\right)
\longrightarrow 0.
\end{equation}
By Assumption~\ref{ass:non_degenerate_unif},
\begin{equation}\label{eq:def_xi_lower_projected}
\underline{\xi}
\coloneqq
\inf_{P\in\tilde{\cP}_0}\left\|C_P^{(\infty)}\right\|_{\op}
>0.
\end{equation}

We now apply Lemma~\ref{lem:uniform_kolmogorov_distance_squared_norm} with
\[
X_{n,P}=S_{n,P}^{(\infty)},\qquad
G_P=G_P^{(\infty)},\qquad
\cP=\tilde{\cP}_0.
\]
Using \eqref{eq:BL_unif_CLT_projected} and \eqref{eq:def_xi_lower_projected}, we obtain
\begin{equation}\label{eq:squared_norm_cdf_unif_projected}
\sup_{P\in\tilde{\cP}_0}\sup_{t\in\bbR}
\left|
\pr_P\left(\left\|S_{n,P}^{(\infty)}\right\|^2\le t\right)-F_P(t)
\right|
\to 0.
\end{equation}

It remains to transfer the limit from \(\|S_{n,P}^{(\infty)}\|^2\) to \(n_1\hat T_{n_1}\). For \(\eta>0\), set
\[
B_n(\eta)
\coloneqq
\sup_{P\in\tilde{\cP}_0}
\pr_P\left(
\left|n_1\hat T_{n_1}-\|S_{n,P}^{(\infty)}\|^2\right|>\eta
\right),
\qquad
A_{n,P}(t)\coloneqq \pr_P\left(\|S_{n,P}^{(\infty)}\|^2\le t\right).
\]
Then \(B_n(\eta)\to 0\) for every fixed \(\eta>0\). The standard inclusion argument gives
\[
\left|
\pr_P(n_1\hat T_{n_1}\le t)-A_{n,P}(t)
\right|
\le
B_n(\eta)+A_{n,P}(t+\eta)-A_{n,P}(t-\eta).
\]
Adding and subtracting \(F_P\), and then taking suprema over \(P\) and \(t\), yields
\begin{equation}\label{eq:uniform_transfer_bound}
\sup_{P\in\tilde{\cP}_0}\sup_{t\in\bbR}
\left|
\pr_P(n_1\hat T_{n_1}\le t)-F_P(t)
\right|
\le
B_n(\eta)
+
3\sup_{P\in\tilde{\cP}_0}\sup_{u\in\bbR}
|A_{n,P}(u)-F_P(u)|
+
\omega(\eta),
\end{equation}
where
\[
\omega(\eta)
\coloneqq
\sup_{P\in\tilde{\cP}_0}\sup_{t\in\bbR}
\{F_P(t+\eta)-F_P(t-\eta)\}.
\]

By \eqref{eq:squared_norm_cdf_unif_projected},
\begin{equation}\label{eq:AnP_FP_unif_to_zero}
\sup_{P\in\tilde{\cP}_0}\sup_{u\in\bbR}
\left|A_{n,P}(u)-F_P(u)\right|
\to 0.
\end{equation}
By Corollary~\ref{cor:uniform_modulus_continuity_FP}, together with
\eqref{eq:def_xi_lower_projected},
\begin{equation}\label{eq:omega_eta_bound}
\omega(\eta)
\le
2\Phi\left(\sqrt{\frac{2\eta}{\underline{\xi}}}\right)-1
\longrightarrow 0
\qquad (\eta\downarrow 0).
\end{equation}

Hence, by \eqref{eq:uniform_transfer_bound} and \eqref{eq:AnP_FP_unif_to_zero}, for every fixed \(\eta>0\),
\begin{equation}\label{eq:limsup_eta_size}
\limsup_{n\to\infty}
\sup_{P\in\tilde{\cP}_0}\sup_{t\in\bbR}
\left|
\pr_P\left(n_1\hat T_{n_1}\le t\right)-F_P(t)
\right|
\le
\omega(\eta).
\end{equation}
Letting \(\eta\downarrow 0\) and using \eqref{eq:omega_eta_bound}, we conclude that
\begin{equation}\label{eq:uniform_null_limit_final}
\sup_{P\in\tilde{\cP}_0}\sup_{t\in\bbR}
\left|
\pr_P\left(n_1\hat T_{n_1}\le t\right)-F_P(t)
\right|
\to 0,
\end{equation}
which proves (1).

\noindent\textbf{Part II: Uniform conditional bootstrap approximation.}

For \(P\in\tilde{\cP}\), define
\[
S_{n,P}^{(\infty)\ast}
\coloneqq
\Pi_\infty\left(\frac{1}{\sqrt{n_1}}\sum_{i\in I_1}\Psi_i W_i^\ast\right),
\qquad
\hat F_{n,P}^{(\infty)\ast}(t)
\coloneqq
\pr_P^\ast\left(\left\|S_{n,P}^{(\infty)\ast}\right\|^2\le t \,\middle|\, \cD_n\right),
\qquad t\in\bbR.
\]
Also set
\[
X_{n,P}^{(\infty)\ast}
\coloneqq
S_{n,P}^{(\infty)\ast},
\qquad
\mu_{n,P}^{(\infty)\ast}
\coloneqq
\cL_P^\ast\left(X_{n,P}^{(\infty)\ast}\right),
\qquad
\nu_P
\coloneqq
\cN\left(0,C_P^{(\infty)}\right),
\]
and
\[
\hat C_{n,P}^{(\infty)}
\coloneqq
\frac{1}{n_1}\sum_{i\in I_1}
\left(\Pi_\infty\Psi_i\right)\otimes\left(\Pi_\infty\Psi_i\right).
\]

First, by Lemma~\ref{lem:multiplier_bootstrap_biases_negligible_uniform}(iii), for every
\(\eta>0\) and every \(\varepsilon>0\),
\begin{equation}
\label{eq:part2-bootstrap-negligibility-infty}
\sup_{P\in\tilde{\cP}}
\pr_P\left(
\pr_P^\ast\left(
\left|
n_1\hat T_{n_1}^\ast
-
\left\|S_{n,P}^{(\infty)\ast}\right\|^2
\right|>\eta
\,\middle|\,
\cD_n
\right)>\varepsilon
\right)\to 0.
\end{equation}
Hence it is enough to establish the desired conditional bootstrap cdf convergence for
\(\left\|S_{n,P}^{(\infty)\ast}\right\|^2\), and then transfer the conclusion back to
\(n_1\hat T_{n_1}^\ast\).

Next, apply Theorem~\ref{thm:verification-conditional-hilbert-BL} with
\[
\pi_{n,P}=\mu_{n,P}^{(\infty)\ast},
\qquad
S_{n,P}^\ast=S_{n,P}^{(\infty)\ast},
\qquad
\nu_{n,P}^\ast=\cN\left(0,\hat C_{n,P}^{(\infty)}\right),
\qquad
\widehat C_{n,P}=\hat C_{n,P}^{(\infty)},
\qquad
C_P=C_P^{(\infty)}.
\]
We verify its assumptions in turn. Assumptions~\ref{ass:bootstrap-verification-V1}--\ref{ass:bootstrap-verification-V3}
follow from the same multiplier-bootstrap verification argument as in Part~I, now applied directly to the oracle array
\(\{\Pi_\infty\Psi_i : i\in I_1\}\), together with the boundedness of the kernels and
Assumption~\ref{ass:fixed-basis-uniform_tail}. Assumption~\ref{ass:bootstrap-verification-V4} is exactly the uniform trace-norm covariance consistency
\begin{equation}
\label{eq:part2-trace-convergence-infty}
\sup_{P\in\tilde{\cP}}
\pr_P\left(
\left\|
\hat C_{n,P}^{(\infty)}-C_P^{(\infty)}
\right\|_{\mathrm{tr}}>\varepsilon
\right)\to 0
\qquad
\text{for every }\varepsilon>0,
\end{equation}
which is provided by Lemma~\ref{lem:uniform_trace_lln_projected_covariance}. Therefore, for every \(\eta>0\),
\begin{equation}
\label{eq:part2-BL-fixed-target-infty}
\sup_{P\in\tilde{\cP}}
\pr_P\left(
d_{\mathrm{BL},\cH}
\left(
\mu_{n,P}^{(\infty)\ast},
\nu_P
\right)>\eta
\right)\to 0.
\end{equation}

By Assumption~\ref{ass:non_degenerate_unif}, the boundedness of the kernels, and
Corollary~\ref{cor:uniform_anti_concentration_gaussian_hilbert_norm}, the Gaussian family
\[
\left\{\nu_P:P\in\tilde{\cP}\right\}
=
\left\{\cN\left(0,C_P^{(\infty)}\right):P\in\tilde{\cP}\right\}
\]
satisfies Assumption~\ref{ass:bootstrap-AC-transfer}. Applying
Theorem~\ref{thm:abstract-transfer-squared-norm-bootstrap} with
\[
X_{n,P}^\ast=S_{n,P}^{(\infty)\ast},
\qquad
G_P\sim \cN\left(0,C_P^{(\infty)}\right),
\]
and using \eqref{eq:part2-BL-fixed-target-infty}, we conclude that, for every \(\varepsilon>0\),
\begin{equation}
\label{eq:part2-infty-cdf-conv-to-FP}
\sup_{P\in\tilde{\cP}}
\pr_P\left(
\sup_{t\in\bbR}
\left|
\hat F_{n,P}^{(\infty)\ast}(t)-F_P(t)
\right|>\varepsilon
\right)\to 0.
\end{equation}

Finally, define
\[
A_{n,P}^\ast
\coloneqq
n_1\hat T_{n_1}^\ast,
\qquad
B_{n,P}^{(\infty)\ast}
\coloneqq
\left\|S_{n,P}^{(\infty)\ast}\right\|^2,\qquad
D_{n,P}^{(\infty)\ast}(\eta)
\coloneqq
\pr_P^\ast\left(
\left|
A_{n,P}^\ast-B_{n,P}^{(\infty)\ast}
\right|>\eta
\,\middle|\,
\cD_n
\right).
\]
For any \(t\in\bbR\) and \(\eta>0\), the elementary inclusions
\[
\left\{A_{n,P}^\ast\le t\right\}
\subset
\left\{B_{n,P}^{(\infty)\ast}\le t+\eta\right\}
\cup
\left\{\left|A_{n,P}^\ast-B_{n,P}^{(\infty)\ast}\right|>\eta\right\},
\]
and
\[
\left\{B_{n,P}^{(\infty)\ast}\le t-\eta\right\}
\subset
\left\{A_{n,P}^\ast\le t\right\}
\cup
\left\{\left|A_{n,P}^\ast-B_{n,P}^{(\infty)\ast}\right|>\eta\right\}
\]
yield
\[
\left|
\hat F_{n,P}^\ast(t)-\hat F_{n,P}^{(\infty)\ast}(t)
\right|
\le
D_{n,P}^{(\infty)\ast}(\eta)
+
\hat F_{n,P}^{(\infty)\ast}(t+\eta)
-
\hat F_{n,P}^{(\infty)\ast}(t-\eta),
\]
where \(\hat F_{n,P}^\ast(t)\coloneqq \pr_P^\ast(A_{n,P}^\ast\le t\mid\cD_n)\).
Adding and subtracting \(F_P\), we obtain
\begin{equation}
\label{eq:part2-transfer-master-infty}
\sup_{t\in\bbR}
\left|
\hat F_{n,P}^\ast(t)-F_P(t)
\right|
\le
D_{n,P}^{(\infty)\ast}(\eta)
+
3\sup_{u\in\bbR}
\left|
\hat F_{n,P}^{(\infty)\ast}(u)-F_P(u)
\right|
+
\omega_P(\eta),
\end{equation}
where
\[
\omega_P(\eta)
\coloneqq
\sup_{t\in\bbR}
\left\{
F_P(t+\eta)-F_P(t-\eta)
\right\}.
\]
Since
\[
\omega(\eta)
\coloneqq
\sup_{P\in\tilde{\cP}}\omega_P(\eta)\to 0
\qquad (\eta\downarrow 0)
\]
by Corollary~\ref{cor:uniform_modulus_continuity_FP}, we may choose \(\eta>0\) so that
\(\omega(\eta)<\varepsilon/3\). Then \eqref{eq:part2-transfer-master-infty} implies
\begin{equation}
\label{eq:part2-final-bound-infty}
\sup_{P\in\tilde{\cP}}
\pr_P\left(
\sup_{t\in\bbR}
\left|
\hat F_{n,P}^\ast(t)-F_P(t)
\right|>\varepsilon
\right)
\le
\sup_{P\in\tilde{\cP}}
\pr_P\left(
D_{n,P}^{(\infty)\ast}(\eta)>\frac{\varepsilon}{3}
\right)
+
\sup_{P\in\tilde{\cP}}
\pr_P\left(
\sup_{u\in\bbR}
\left|
\hat F_{n,P}^{(\infty)\ast}(u)-F_P(u)
\right|>\frac{\varepsilon}{9}
\right).
\end{equation}
The first term on the right-hand side converges to \(0\) by
\eqref{eq:part2-bootstrap-negligibility-infty}, and the second by
\eqref{eq:part2-infty-cdf-conv-to-FP}. Therefore,
\begin{equation}
\label{eq:part2-actual-cdf-conv-to-FP-infty}
\sup_{P\in\tilde{\cP}}
\pr_P\left(
\sup_{t\in\bbR}
\left|
\hat F_{n,P}^\ast(t)-F_P(t)
\right|>\varepsilon
\right)\to 0,
\end{equation}
which proves assertion~(2) of Theorem~\ref{thm:uniform_size_control}.

\noindent\textbf{Part III: Uniform asymptotic size.}

For \(P\in\tilde{\cP}_0\), define
\[
T_{n,P}\coloneqq n_1\hat T_{n_1},
\qquad
F_{n,P}(t)\coloneqq \pr_P\left(T_{n,P}\le t\right),
\qquad
\hat F_{n,P}^{\ast(B_n)}(t)
\coloneqq
\frac{1}{B_n}\sum_{b=1}^{B_n}
\mathds 1\left\{n_1\hat T_{n_1}^{\ast(b)}\le t\right\}.
\]
Consistent with the definition of the quantile functional in \eqref{eq:def_quantile_functional}, we abbreviate the $(1-\alpha)$-quantiles of $F_P$ and $\hat F_{n,P}^{\ast(B_n)}$ as follows:
\[
q_P \coloneqq q_{F_P}(1-\alpha),
\qquad
\hat q_{n,P}^{\ast(B_n)} \coloneqq q_{\hat F_{n,P}^{\ast(B_n)}}(1-\alpha).
\]

By definition of the test,
\[
\left\{\phi_n^{(B_n)}=1\right\}
=
\left\{T_{n,P}>\hat q_{n,P}^{\ast(B_n)}\right\}
\qquad \text{under }\pr_P.
\]

We apply Lemma~\ref{lem:finiteB_bootstrap_quantile_uniform_consistency} with
\(
u=1-\alpha\) and index family \(\tilde{\cP}=\tilde{\cP}_0\).
Its assumption (i) is exactly Part II of Theorem~\ref{thm:uniform_size_control}, and its assumption (ii) is Lemma~\ref{lem:uniform_quantile_regularity_gaussian_limit}. Hence, for every \(\delta>0\),
\begin{equation}
\label{eq:partIII_quantile_consistency}
r_n(\delta)
\coloneqq
\sup_{P\in\tilde{\cP}_0}
\pr_P\left(\left|\hat q_{n,P}^{\ast(B_n)}-q_P\right|>\delta\right)
\longrightarrow 0.
\end{equation}

By Corollary~\ref{cor:uniform_anti_concentration_gaussian_hilbert_norm}, there exists a deterministic modulus \(\omega(\delta)\downarrow 0\) as \(\delta\downarrow 0\) such that
\begin{equation}
\label{eq:partIII_uniform_modulus}
\sup_{P\in\tilde{\cP}_0}\sup_{t\in\bbR}
\left|F_P(t+\delta)-F_P(t)\right|
\le
\omega(\delta).
\end{equation}
In particular, each \(F_P\) is continuous. Since \(q_P=q_{F_P}(1-\alpha)\), it follows that
\(
F_P(q_P)=1-\alpha\) for all \(P\in\tilde{\cP}_0\).

Fix \(\delta>0\), and define the event
\[
E_{n,P}(\delta)\coloneqq \left\{\left|\hat q_{n,P}^{\ast(B_n)}-q_P\right|\le \delta\right\}.
\]
On \(E_{n,P}(\delta)\), we have \(q_P-\delta\le\hat q_{n,P}^{\ast(B_n)}\le q_P+\delta\), and therefore
\[
\left\{T_{n,P}>q_P+\delta\right\}
\subset
\left\{T_{n,P}>\hat q_{n,P}^{\ast(B_n)}\right\}
\subset
\left\{T_{n,P}>q_P-\delta\right\}.
\]
Hence,
\begin{equation}
\label{eq:partIII_sandwich_prob}
\pr_P\left(T_{n,P}>q_P+\delta\right)-\pr_P\left(E_{n,P}(\delta)^c\right)
\le
\pr_P\left(T_{n,P}>\hat q_{n,P}^{\ast(B_n)}\right)
\le
\pr_P\left(T_{n,P}>q_P-\delta\right)+\pr_P\left(E_{n,P}(\delta)^c\right).
\end{equation}

By Part I of Theorem~\ref{thm:uniform_size_control},
\begin{equation}
\label{eq:partIII_an_to_zero}
    a_n \coloneqq \sup_{P\in\tilde{\cP}_0}\sup_{t\in\bbR} \left|F_{n,P}(t)-F_P(t)\right| \longrightarrow 0.
\end{equation}

Using \eqref{eq:partIII_uniform_modulus}--\eqref{eq:partIII_an_to_zero}, we can bound the absolute difference directly. On the event $E_{n,P}(\delta)$, we have $q_P-\delta \le \hat q_{n,P}^{\ast(B_n)} \le q_P+\delta$, which implies
\begin{align*}
    \left| \pr_P\left(T_{n,P}>\hat q_{n,P}^{\ast(B_n)}\right) - \alpha \right|
    &\le
    \max_{\pm} \left| \pr_P\left(T_{n,P}>q_P \pm \delta\right) - \alpha \right| + \pr_P\left(E_{n,P}(\delta)^c\right) \\
    &\le
    \max_{\pm} \left| \left(1-F_P(q_P \pm \delta)\right) - \left(1-F_P(q_P)\right) \right| + a_n + \pr_P\left(E_{n,P}(\delta)^c\right) \\
    &=
    \max_{\pm} \left| F_P(q_P) - F_P(q_P \pm \delta) \right| + a_n + \pr_P\left(E_{n,P}(\delta)^c\right) \\
    &\le
    \omega(\delta) + a_n + \pr_P\left(E_{n,P}(\delta)^c\right).
\end{align*}

Therefore, for every \(P\in\tilde{\cP}_0\),
\[
\left|
\pr_P\left(T_{n,P}>\hat q_{n,P}^{\ast(B_n)}\right)-\alpha
\right|
\le
\omega(\delta)+a_n+\pr_P\left(E_{n,P}(\delta)^c\right).
\]
Taking \(\sup_{P\in\tilde{\cP}_0}\) and recalling \(\pr_P(E_{n,P}(\delta)^c)\le r_n(\delta)\),
\begin{equation}
\label{eq:partIII_master_bound}
\sup_{P\in\tilde{\cP}_0}
\left|
\pr_P\left(\phi_n^{(B_n)}=1\right)-\alpha
\right|
=
\sup_{P\in\tilde{\cP}_0}
\left|
\pr_P\left(T_{n,P}>\hat q_{n,P}^{\ast(B_n)}\right)-\alpha
\right|
\le
\omega(\delta)+a_n+r_n(\delta).
\end{equation}

\noindent
\textit{Step 4: Conclusion.}

Fix $\varepsilon>0$. By \eqref{eq:partIII_uniform_modulus}, we can choose $\delta>0$ such that $\omega(\delta)<\varepsilon/3$. Then, by \eqref{eq:partIII_an_to_zero} and \eqref{eq:partIII_quantile_consistency}, there exists $N$ such that $a_n<\varepsilon/3$ and $r_n(\delta)<\varepsilon/3$ for all $n\ge N$. Substituting these into \eqref{eq:partIII_master_bound} yields
\[
    \sup_{P\in\tilde{\cP}_0} \left| \pr_P\left(\phi_n^{(B_n)}=1\right)-\alpha \right| < \varepsilon \qquad (n\ge N).
\]
Since \(\varepsilon>0\) was arbitrary,
\[
\lim_{n\to\infty}\sup_{P\in\tilde{\cP}_0}
\left|
\pr_P\left(\phi_n^{(B_n)}=1\right)-\alpha
\right|
=0.
\]
This proves Part III.
\end{proof}
\runinhead{Uniform power under projected separated alternatives}
\begin{proof}[of~Theorem~\ref{thm:uniform_power_monotone}]
Fix \(P\in\tilde{\cP}\). By assumption and Corollary~\ref{cor:uniform_bias_control_alternative},
\[
n_1\hat T_{n_1}
=
\left\|
\hat\Pi_n
\left(
n_1^{-1/2}\sum_{i\in I_1}\Psi_i
\right)
\right\|^2
+
o_{\tilde{\cP}_{1,n}(c)}(1).
\]
Decomposing the sample mean into the centered term and \(\sqrt{n_1}\Sigma_{XYZ\mid Z}\), and applying the reverse triangle inequality, we get
\[
\sqrt{n_1\hat T_{n_1}}
\ge
\sqrt{n_1}\|\hat\Pi_n\Sigma_{XYZ\mid Z}\|
-
\left\|
\hat\Pi_n
\left(
n_1^{-1/2}\sum_{i\in I_1}(\Psi_i-\Sigma_{XYZ\mid Z})
\right)
\right\|.
\]
Define
\[
\Delta_{n,P}
\coloneqq
\sqrt{n_1}\left(\|\Pi_n\Sigma_{XYZ\mid Z}\|-\|\hat\Pi_n\Sigma_{XYZ\mid Z}\|\right)_+,
\qquad
R_{n,P}
\coloneqq
\left\|
\hat\Pi_n
\left(
n_1^{-1/2}\sum_{i\in I_1}(\Psi_i-\Sigma_{XYZ\mid Z})
\right)
\right\|.
\]
Since \(\sqrt{n_1}\|\hat\Pi_n\Sigma_{XYZ\mid Z}\|\ge \sqrt{n_1}\|\Pi_n\Sigma_{XYZ\mid Z}\|-\Delta_{n,P}\), on \(\tilde{\cP}_{1,n}(c)\) we have
\[
\sqrt{n_1\hat T_{n_1}}
\ge
c-\Delta_{n,P}-R_{n,P}.
\]
Hence
\[
\pr_P\left(\phi_n^{(B_n)}=0\right)
\le
\pr_P\left(
\Delta_{n,P}+R_{n,P}+\left(Q_{n,1-\alpha}^{\ast(B_n)}\right)^{1/2}\ge c
\right).
\]

We first record two uniform stochastic bounds. For \(R_{n,P}\), note that \(\hat\Pi_n\) is an orthogonal projection, hence \(\|\hat\Pi_n\|_{\op}\le 1\). Since the kernels are bounded, and thus \(\|\Psi\|^2\le 16K_X^2K_Y^2K_Z^2\) almost surely, uniformly over \(P\in\tilde{\cP}\). Using independence and centering,
\begin{equation}
\E_P\left[R_{n,P}^2\right]
\le
\E_P\left[
\left\|
n_1^{-1/2}\sum_{i\in I_1}(\Psi_i-\Sigma_{XYZ\mid Z})
\right\|^2
\right]
=
\E_P\left[\|\Psi-\Sigma_{XYZ\mid Z}\|^2\right]
\le
\E_P\left[\|\Psi\|^2\right]
\le
16K_X^2K_Y^2K_Z^2.
\label{eq:uniform_bound_R_n_P}
\end{equation}
Hence \(R_{n,P}=O_{\tilde{\cP}}(1)\).

Next, recall that \(q_P=q_{F_P}(1-\alpha)\). Since \(Q_{n,1-\alpha}^{\ast(B_n)}\) is precisely the finite-\(B_n\) bootstrap critical value, we may identify it with \(\hat q_{n,P}^{\ast(B_n)}\). Hence, by Part~II of Theorem~\ref{thm:uniform_size_control}, Lemma~\ref{lem:uniform_quantile_regularity_gaussian_limit}, and Lemma~\ref{lem:finiteB_bootstrap_quantile_uniform_consistency},
\[
\left|Q_{n,1-\alpha}^{\ast(B_n)}-q_P\right|
=
\left|\hat q_{n,P}^{\ast(B_n)}-q_P\right|
=
o_{\tilde{\cP}}(1).
\]
Moreover, since \(F_P\) is the law of \(\sum_{\ell\ge 1}\xi_{\ell,P}G_\ell^2\), we have
\(\E_P\left[\sum_{\ell\ge 1}\xi_{\ell,P}G_\ell^2\right]=\tr(C_P^{(\infty)})\le \E_P\|\Psi\|^2\le 16K_X^2K_Y^2K_Z^2\). Hence, by Markov's inequality,
\[
\sup_{P\in\tilde{\cP}} q_{P}\le \frac{16K_X^2K_Y^2K_Z^2}{\alpha},
\]
and therefore
\begin{equation}
Q_{n,1-\alpha}^{\ast(B_n)}=O_{\tilde{\cP}}(1).
\label{eq:uniform_bound_bootstrap_cv}
\end{equation}

We now prove the two claims.

\medskip\noindent
\textbf{Part I: Nontrivial uniform power.}
Fix \(\beta\in(\alpha,1)\) and let \(\eta\coloneqq 1-\beta\in(0,1-\alpha)\). By Assumption~\ref{ass:uniform_power_projected_signal_preservation_Pi_n}, \(\Delta_{n,P}=O_{\tilde{\cP}_{1,n}(c_0)}(1)\) for every fixed \(c_0>0\). By \eqref{eq:uniform_bound_R_n_P} and \eqref{eq:uniform_bound_bootstrap_cv}, both \(R_{n,P}\) and \(\left(Q_{n,1-\alpha}^{\ast(B_n)}\right)^{1/2}\) are uniformly bounded in probability; hence
\[
\Delta_{n,P}+R_{n,P}+\left(Q_{n,1-\alpha}^{\ast(B_n)}\right)^{1/2}
=
O_{\tilde{\cP}_{1,n}(c_0)}(1).
\]
Therefore there exists \(M>0\) such that
\[
\limsup_{n\to\infty}
\sup_{P\in\tilde{\cP}_{1,n}(c_0)}
\pr_P\left(
\Delta_{n,P}+R_{n,P}+\left(Q_{n,1-\alpha}^{\ast(B_n)}\right)^{1/2}>M
\right)
\le
\eta.
\]
Choose \(c>M\). Since \(\tilde{\cP}_{1,n}(c)\subset \tilde{\cP}_{1,n}(c_0)\), it follows that
\[
\limsup_{n\to\infty}
\sup_{P\in\tilde{\cP}_{1,n}(c)}
\pr_P\left(\phi_n^{(B_n)}=0\right)
\le
\eta,
\]
and thus
\[
\liminf_{n\to\infty}
\inf_{P\in\tilde{\cP}_{1,n}(c)}
\pr_P\left(\phi_n^{(B_n)}=1\right)
\ge
1-\eta
=
\beta.
\]

\medskip\noindent
\textbf{Part II: Uniform consistency.}
Let \(c_n\to\infty\), and fix \(\eta>0\). Again, Assumption~\ref{ass:uniform_power_projected_signal_preservation_Pi_n} gives \(\Delta_{n,P}=O_{\tilde{\cP}_{1,n}(1)}(1)\), while \eqref{eq:uniform_bound_R_n_P} and \eqref{eq:uniform_bound_bootstrap_cv} yield \(R_{n,P}=O_{\tilde{\cP}}(1)\) and \(\left(Q_{n,1-\alpha}^{\ast(B_n)}\right)^{1/2}=O_{\tilde{\cP}}(1)\). Hence
\[
\Delta_{n,P}+R_{n,P}+\left(Q_{n,1-\alpha}^{\ast(B_n)}\right)^{1/2}
=
O_{\tilde{\cP}_{1,n}(1)}(1).
\]
So there exists \(M>0\) such that
\[
\limsup_{n\to\infty}
\sup_{P\in\tilde{\cP}_{1,n}(1)}
\pr_P\left(
\Delta_{n,P}+R_{n,P}+\left(Q_{n,1-\alpha}^{\ast(B_n)}\right)^{1/2}>M
\right)
\le
\eta.
\]
Since \(c_n\to\infty\), for all sufficiently large \(n\) we have \(c_n>M\) and \(\tilde{\cP}_{1,n}(c_n)\subset \tilde{\cP}_{1,n}(1)\). Therefore, for all such \(n\),
\[
\sup_{P\in\tilde{\cP}_{1,n}(c_n)}
\pr_P\left(\phi_n^{(B_n)}=0\right)
\le
\sup_{P\in\tilde{\cP}_{1,n}(1)}
\pr_P\left(
\Delta_{n,P}+R_{n,P}+\left(Q_{n,1-\alpha}^{\ast(B_n)}\right)^{1/2}>M
\right).
\]
Taking \(\limsup_{n\to\infty}\) yields
\[
\limsup_{n\to\infty}
\sup_{P\in\tilde{\cP}_{1,n}(c_n)}
\pr_P\left(\phi_n^{(B_n)}=0\right)
\le
\eta.
\]
Since \(\eta>0\) was arbitrary, we conclude that
\[
\lim_{n\to\infty}
\sup_{P\in\tilde{\cP}_{1,n}(c_n)}
\pr_P\left(\phi_n^{(B_n)}=0\right)
=
0,
\]
equivalently,
\[
\lim_{n\to\infty}
\inf_{P\in\tilde{\cP}_{1,n}(c_n)}
\pr_P\left(\phi_n^{(B_n)}=1\right)
=
1.
\]
This completes the proof.
\end{proof}

\subsection{Approximation errors for the statistic and bootstrap}
\runinhead{Basic RKHS identities and boundedness}
\begin{lemma}\label{lem:ip-repr}
Fix an index $i$ and let $\hat e_p\in\cH_{\cX}$ be constructed from an auxiliary sample that is independent of $\left(X_i,Z_i\right)$.
Then
\begin{equation*}
\zeta^{(i)}_{X\mid Z,p} \coloneqq  \hat e_p(X_i)- \E\left[\hat e_p(X_i) \mid Z_i,\hat e_p\right]
= 
\left\langle \hat e_p,\ k_{\cX}(X_i,\cdot)-\mu_{X\mid Z}(Z_i)\right\rangle.
\end{equation*}
\end{lemma}

\begin{proof}
By the reproducing property, we have \(\hat e_p(X_i) = \langle \hat e_p, k_{\cX}(X_i,\cdot) \rangle\). Since \(\hat e_p\) is \(\sigma(Z_i, \hat e_p)\)-measurable and the kernel is bounded, standard properties of Bochner conditional expectations \citep[see, e.g.,][]{Hytonenetal2016} allow us to pass the expectation inside the inner product. Combined with the independence of the auxiliary sample from \((X_i, Z_i)\), this yields
\[
\E\left[\hat e_p(X_i) \mid Z_i, \hat e_p\right]
=
\left\langle \hat e_p, \E\left[k_{\cX}(X_i,\cdot) \mid Z_i, \hat e_p\right] \right\rangle
=
\left\langle \hat e_p, \E\left[k_{\cX}(X_i,\cdot) \mid Z_i\right] \right\rangle
=
\left\langle \hat e_p, \mu_{X\mid Z}(Z_i) \right\rangle.
\]
Subtracting this from the reproducing identity immediately yields the claim.
\end{proof}

\begin{lemma}\label{lem:bounded-M}
Assume the kernels are bounded. Then, almost surely,
\begin{equation*}
\sup_{i\in I_1}\ \sum_{p=1}^{L_{X,n}}\left(\zeta^{(i)}_{X\mid Z,p}\right)^2 \leq   4K_X^2,
\qquad
\sup_{i\in I_1}\ \sum_{q=1}^{L_{Y,n}}\left(\zeta^{(i)}_{Y\mid Z,q}\right)^2 \leq   4K_Y^2.
\end{equation*}
\end{lemma}

\begin{proof}
By Lemma~\ref{lem:ip-repr} and Bessel's inequality \citep[Corollary~4.10]{Conway2007}, we obtain
\begin{equation*}
\sum_{p=1}^{L_{X,n}}\left(\zeta^{(i)}_{X\mid Z,p}\right)^2
=
\sum_{p=1}^{L_{X,n}}\left\langle \hat e_p,\ \epsilon_{X\mid Z}^{(i)}\right\rangle^2
\leq 
\left\|\epsilon_{X\mid Z}^{(i)}\right\|^2
\leq 
2\left\|k_{\cX}(X_i,\cdot)\right\|^2
+2\left\|\mu_{X\mid Z}(Z_i)\right\|^2 
\le 4K_X^2 \quad \text{a.s.}
\end{equation*}
Taking the supremum over \(i \in I_1\) yields the result for \(X\). The bound for \(Y\) follows by symmetry.
\end{proof}

\begin{lemma}\label{lem:diag-to-uniform}
Assume that the kernel $k_{\cZ}$ is bounded. Then it holds that $\sup_{z,z'\in\cZ} |k_{\cZ}(z,z')| \le K_Z^2$.
\end{lemma}

\begin{proof}
For any \(z,z'\in\cZ\), the reproducing property and the Cauchy--Schwarz inequality imply
\[
\left|k_{\cZ}(z,z')\right|
=
\left|\left\langle k_{\cZ}(z,\cdot), k_{\cZ}(z',\cdot)\right\rangle\right|
\le
\left(k_{\cZ}(z,z)k_{\cZ}(z',z')\right)^{1/2}
\le
K_Z^2.
\]
Taking the supremum over \(z,z'\in\cZ\) completes the proof.
\end{proof}
\runinhead{Regression-induced approximation error} 
\begin{proposition}[Uniform negligibility of regression-induced bias under \(H_0\)]
\label{prop:bias-negligible}
Define
\begin{equation*}
n_1\,\tilde T_{n_1}^{(L_{X,n},L_{Y,n})} \coloneqq
\frac{1}{n_1}\sum_{i,j \in I_1}\sum_{p=1}^{L_{X,n}}\sum_{q=1}^{L_{Y,n}}
k_{\cZ}(Z_i, Z_j)\,
\zeta^{(i)}_{X\mid Z,p}\,\zeta^{(i)}_{Y\mid Z,q}\,
\zeta^{(j)}_{X\mid Z,p}\,\zeta^{(j)}_{Y\mid Z,q}.
\end{equation*}
Then,
\begin{equation*}
n_1\left(\hat T_{n_1}-\tilde T_{n_1}^{(L_{X,n},L_{Y,n})}\right)=o_{\tilde{\cP}_0}(1).
\end{equation*}

\end{proposition}
\begin{proof}
We suppress the dependence on \(P\in\tilde{\cP}_0\) in the notation; all stochastic orders below are uniform over \(\tilde{\cP}_0\).

Define
\begin{equation}
\label{eq:bias_negligible_Shat}
\hat S_{n_1}\coloneqq \frac{1}{\sqrt{n_1}}\sum_{i\in I_1}\hat{\Psi}_i^{(L_{X,n},L_{Y,n})}.
\end{equation}
Using \(\langle \hat e_p,\hat e_{p'}\rangle=\delta_{pp'}\), \(\langle \hat f_q,\hat f_{q'}\rangle=\delta_{qq'}\), and
\(\langle \phi_{\cZ}(Z_i),\phi_{\cZ}(Z_j)\rangle=k_{\cZ}(Z_i,Z_j)\), we have
\begin{equation*}
\left\|\hat S_{n_1}\right\|^2
=\frac{1}{n_1}\sum_{i,j \in I_1}
\left(\sum_{p=1}^{L_{X,n}}\hat\zeta^{(i)}_{X\mid Z,p}\hat\zeta^{(j)}_{X\mid Z,p}\right)
\left(\sum_{q=1}^{L_{Y,n}}\hat\zeta^{(i)}_{Y\mid Z,q}\hat\zeta^{(j)}_{Y\mid Z,q}\right)k_{\cZ}(Z_i, Z_j)
\,=\,n_1\hat T_{n_1}.
\end{equation*}
Similarly, for \(u^{(i)}_X \coloneqq \sum_{p=1}^{L_{X,n}} \zeta^{(i)}_{X\mid Z,p} \hat e_p\) and \(u^{(i)}_Y \coloneqq \sum_{q=1}^{L_{Y,n}} \zeta^{(i)}_{Y\mid Z,q} \hat f_q\), let
\begin{equation} \label{eq:bias_negligible_S}
S_{n_1} \coloneqq \frac{1}{\sqrt{n_1}}\sum_{i\in I_1} u^{(i)}_X \otimes u^{(i)}_Y \otimes \phi_{\cZ}(Z_i), \qquad \|S_{n_1}\|^2 = n_1 \tilde T_{n_1}^{(L_{X,n},L_{Y,n})}.
\end{equation}
Hence, by \eqref{eq:bias_negligible_Shat} and \eqref{eq:bias_negligible_S},
\begin{equation*}
n_1\left(\hat T_{n_1}-\tilde T_{n_1}^{(L_{X,n},L_{Y,n})}\right)
=\left\|\hat S_{n_1}\right\|^2-\left\|S_{n_1}\right\|^2
=\left\|\hat S_{n_1}-S_{n_1}\right\|^2+2\left\langle S_{n_1},\,\hat S_{n_1}-S_{n_1}\right\rangle.
\end{equation*}
Therefore it suffices to show
\begin{equation*}
\left\|\hat S_{n_1}-S_{n_1}\right\|^2=o_{\tilde{\cP}_0}(1)
\qquad\text{and}\qquad
\left|\left\langle S_{n_1},\hat S_{n_1}-S_{n_1}\right\rangle\right|=o_{\tilde{\cP}_0}(1).
\end{equation*}

Write
\[
\hat\zeta^{(i)}_{X\mid Z,p}=\zeta^{(i)}_{X\mid Z,p}-b^{(i)}_{X,p},
\qquad
\hat\zeta^{(i)}_{Y\mid Z,q}=\zeta^{(i)}_{Y\mid Z,q}-b^{(i)}_{Y,q}.
\]
As in the pointwise decomposition,
\[
\hat S_{n_1}-S_{n_1}=- B_{n_1}^{(1)}- B_{n_1}^{(2)}+B_{n_1}^{(3)},
\]
where
\begin{equation*}
B_{n_1}^{(1)}\coloneqq \frac{1}{\sqrt{n_1}}\sum_{i\in I_1}
\left(\sum_{p=1}^{L_{X,n}} b^{(i)}_{X,p}\,\hat e_p\right)\otimes
\left(\sum_{q=1}^{L_{Y,n}} \zeta^{(i)}_{Y\mid Z,q}\,\hat f_q\right)\otimes \phi_{\cZ}(Z_i),
\end{equation*}
\begin{equation*}
B_{n_1}^{(2)}\coloneqq \frac{1}{\sqrt{n_1}}\sum_{i\in I_1}
\left(\sum_{p=1}^{L_{X,n}} \zeta^{(i)}_{X\mid Z,p}\,\hat e_p\right)\otimes
\left(\sum_{q=1}^{L_{Y,n}} b^{(i)}_{Y,q}\,\hat f_q\right)\otimes \phi_{\cZ}(Z_i),
\end{equation*}
\begin{equation*}
B_{n_1}^{(3)}\coloneqq \frac{1}{\sqrt{n_1}}\sum_{i\in I_1}
\left(\sum_{p=1}^{L_{X,n}} b^{(i)}_{X,p}\,\hat e_p\right)\otimes
\left(\sum_{q=1}^{L_{Y,n}} b^{(i)}_{Y,q}\,\hat f_q\right)\otimes \phi_{\cZ}(Z_i).
\end{equation*}
Therefore,
\begin{equation*}
\left\|\hat S_{n_1}-S_{n_1}\right\|^2
\le 3\sum_{r=1}^{3}\left\|B_{n_1}^{(r)}\right\|^2,
\end{equation*}
so it suffices to prove
\[
\left\|B_{n_1}^{(r)}\right\|^2=o_{\tilde{\cP}_0}(1),\qquad r=1,2,3.
\]

For \(B_{n_1}^{(1)}\), orthonormality and
\(\langle \phi_{\cZ}(Z_i),\phi_{\cZ}(Z_j)\rangle=k_{\cZ}(Z_i, Z_j)\) give
\begin{equation*}
\left\|B_{n_1}^{(1)}\right\|^2
=\frac{1}{n_1}\sum_{i,j \in I_1}\sum_{p=1}^{L_{X,n}}\sum_{q=1}^{L_{Y,n}}
k_{\cZ}(Z_i, Z_j)\,b^{(i)}_{X,p}b^{(j)}_{X,p}\,\zeta^{(i)}_{Y\mid Z,q}\zeta^{(j)}_{Y\mid Z,q}.
\end{equation*}
Split this into diagonal and off-diagonal parts.

For the diagonal part (\(i=j\)), by Lemma~\ref{lem:diag-to-uniform}, Lemma~\ref{lem:bounded-M}, and Cauchy--Schwarz,
\begin{equation*}
\frac{1}{n_1}\sum_{i\in I_1}\sum_{p=1}^{L_{X,n}}\sum_{q=1}^{L_{Y,n}}
k_{\cZ}(Z_i,Z_i)\left(b^{(i)}_{X,p}\right)^2\left(\zeta^{(i)}_{Y\mid Z,q}\right)^2
\le
\frac{4K_Y^2K_Z^2}{n_1}\sum_{i\in I_1}\sum_{p=1}^{L_{X,n}}\left(b^{(i)}_{X,p}\right)^2.
\end{equation*}
By Assumption~\ref{ass:doubly_robust_uniform},
\[
\frac{1}{n_1}\sum_{i\in I_1}\sum_{p=1}^{L_{X,n}}\left(b^{(i)}_{X,p}\right)^2=o_{\tilde{\cP}_0}(1),
\]
hence the diagonal contribution is \(o_{\tilde{\cP}_0}(1)\).

For the off-diagonal part (\(i\neq j\)), let
\begin{equation*}
\cF_n\coloneqq \sigma(Z_i\colon i\in I_1)\vee\sigma((X_\ell,Y_\ell,Z_\ell)\colon \ell\in I_2).
\end{equation*}
Under \(H_0\) and conditional on \(\cF_n\), the evaluation observations remain independent across \(i\in I_1\), and the oracle coefficients are centered, so for \(i\neq j\),
\begin{equation}
\label{eq:offdiag_condmean_zero_Y}
\E_P\left[\zeta^{(i)}_{Y\mid Z,q}\zeta^{(j)}_{Y\mid Z,q}\mid \cF_n\right]=0,
\qquad i\neq j.
\end{equation}
Therefore the off-diagonal term has conditional mean zero given \(\cF_n\). Applying Lemma S6 in \cite{Lundborgetal2024PCM} yields
\(
\left\|B_{n_1}^{(1)}\right\|^2=o_{\tilde{\cP}_0}(1)\).

By symmetry, the same argument gives
\(
\left\|B_{n_1}^{(2)}\right\|^2=o_{\tilde{\cP}_0}(1)
\).

For \(B_{n_1}^{(3)}\), by orthonormality and Assumption~\ref{ass:doubly_robust_uniform},
\begin{equation*}
\left\|B_{n_1}^{(3)}\right\|^2
=\frac{1}{n_1}\sum_{i,j \in I_1}k_{\cZ}(Z_i, Z_j)
\left(\sum_{p=1}^{L_{X,n}} b^{(i)}_{X,p}b^{(j)}_{X,p}\right)
\left(\sum_{q=1}^{L_{Y,n}} b^{(i)}_{Y,q}b^{(j)}_{Y,q}\right) \le n_1 K_Z^2B_{X,n_1}^2B_{Y,n_1}^2=o_{\tilde{\cP}_0}(1).
\end{equation*}

Thus
\[
\left\|\hat S_{n_1}-S_{n_1}\right\|^2=o_{\tilde{\cP}_0}(1).
\]
Moreover, by Cauchy--Schwarz and the uniform null tightness \(\left\|S_{n_1}\right\|^2=O_{\tilde{\cP}_0}(1)\) (from Assumption~\ref{ass:doubly_robust_uniform} together with the uniform versions of Lemmas~S5--S6 in \cite{Lundborgetal2024PCM}),
\begin{equation*}
\left|\left\langle S_{n_1},\,\hat S_{n_1}-S_{n_1}\right\rangle\right|
\le \left\|S_{n_1}\right\|\left\|\hat S_{n_1}-S_{n_1}\right\|
=o_{\tilde{\cP}_0}(1).
\end{equation*}
Combining the displays proves
\[
n_1\left|\hat T_{n_1}-\tilde T_{n_1}^{(L_{X,n},L_{Y,n})}\right|=o_{\tilde{\cP}_0}(1).
\]
\end{proof}
\begin{remark}[Extension to fixed-$K$ fold cross-fitting]
\label{rem:extension_to_k_fold}
The above argument extends verbatim to the case where the regression estimators are constructed by fixed-$K$ fold cross-fitting. In particular, no substantive change in the proof is needed beyond replacing Assumption~\ref{ass:doubly_robust_uniform} by its cross-fitted analogue.

Let $I_1,\dots,I_K$ be a partition of the evaluation sample, and for each fold $k\in[K]$, let $\hat m_{X,p}^{(-k)}$ and $\hat m_{Y,q}^{(-k)}$ denote the estimators trained without using observations in $I_k$. Define
\[
(B_{X,n}^{\mathrm{cf}})^2
\coloneqq
\frac{1}{n_1}\sum_{k=1}^K\sum_{i\in I_k}\sum_{p=1}^{L_{X,n}}
\left\{\hat m_{X,p}^{(-k)}(Z_i)-m_{X,p}(Z_i)\right\}^2,\enspace
(B_{Y,n}^{\mathrm{cf}})^2
\coloneqq
\frac{1}{n_1}\sum_{k=1}^K\sum_{i\in I_k}\sum_{q=1}^{L_{Y,n}}
\left\{\hat m_{Y,q}^{(-k)}(Z_i)-m_{Y,q}(Z_i)\right\}^2.
\]
If \(B_{X,n}^{\mathrm{cf}}=o_{\tilde{\cP}}(1)\), \(B_{Y,n}^{\mathrm{cf}}=o_{\tilde{\cP}}(1)\), and \(B_{X,n}^{\mathrm{cf}} B_{Y,n}^{\mathrm{cf}}=o_{\tilde{\cP}}(n^{-1/2})\), then the same conclusion continues to hold under fixed-$K$ fold cross-fitting. Indeed, the proof uses only the conditional mean-zero property for the off-diagonal terms, namely \eqref{eq:offdiag_condmean_zero_Y} and its symmetric analogue with $X$ and $Y$ interchanged. The same property continues to hold under fixed-$K$ fold cross-fitting, since the out-of-fold prediction errors are measurable with respect to the corresponding training data, while under $H_0$ the oracle residuals remain conditionally centered given that training data. Hence the argument is unchanged.

Consequently, all subsequent null size results remain valid under fixed-$K$ fold cross-fitting after replacing Assumption~\ref{ass:doubly_robust_uniform} by the above cross-fitted version. In particular, no further modification is needed in the statements or proofs of the asymptotic size theory under the null.
\end{remark}
\begin{corollary}[Uniform sample-splitting bias control under alternatives]
\label{cor:uniform_bias_control_alternative}
Let \(\tilde S_{n,P}\) and \(\widehat S_{n,P}\) denote the oracle and plug-in empirical tensors on the evaluation fold \(I_1\), and define
\[
b_{n,P}\coloneqq \hat \Pi_n\left(\widehat S_{n,P}-\tilde S_{n,P}\right).
\]
Assume that the conditional expectation estimators used in the plug-in residual coordinates are trained on an auxiliary sample independent of \(I_1\cup I_2\). Uniformly over \(P\in\tilde{\cP}\),
\[
\left\|\widehat S_{n,P}-\tilde S_{n,P}\right\|=o_{\tilde{\cP}}(1),
\qquad
\left\|b_{n,P}\right\|=o_{\tilde{\cP}}(1).
\]
In particular, the regression-induced bias is asymptotically negligible uniformly over \(\tilde{\cP}\), without assuming \(H_0\).
\end{corollary}
\begin{proof}
The proof is the same decomposition as in Proposition~\ref{prop:bias-negligible}. The same norm inequality gives
\[
\left\|\widehat S_{n,P}-\tilde S_{n,P}\right\|^2
\le 3\sum_{r=1}^{3}\left\|B^{(r)}_{n,P}\right\|^2.
\]

For \(B^{(1)}_{n,P}\) and \(B^{(2)}_{n,P}\), split into diagonal and off-diagonal parts exactly as before. The diagonal parts are handled in the same way, using the standing boundedness assumptions together with
\[
B_{X,n_1}^2(L_{X,n})=o_{\tilde{\cP}}(1),
\qquad
B_{Y,n_1}^2(L_{Y,n})=o_{\tilde{\cP}}(1).
\]
For the off-diagonal parts, define
\[
\cF_n'
\coloneqq
\sigma\left(Z_i:i\in I_1\right)\vee
\sigma\left((X_\ell,Y_\ell,Z_\ell):\ell\in I_2\right)\vee
\sigma\left(\text{auxiliary training sample}\right).
\]
By construction, the regression errors \(b^{(i)}_{X\mid Z,p}\) and \(b^{(i)}_{Y\mid Z,q}\) are \(\cF_n'\)-measurable. Conditional on \(\cF_n'\), the evaluation observations remain independent across \(i\in I_1\), and the oracle coefficients are conditionally centered. Therefore the off-diagonal terms have conditional mean zero given \(\cF_n'\), and the same application of Lemma~S6 in \cite{Lundborgetal2024PCM} yields
\[
\left\|B^{(1)}_{n,P}\right\|^2=o_{\tilde{\cP}}(1),
\qquad
\left\|B^{(2)}_{n,P}\right\|^2=o_{\tilde{\cP}}(1).
\]

For \(B^{(3)}_{n,P}\), the same deterministic bound as in Proposition~\ref{prop:bias-negligible} gives
\[
\left\|B^{(3)}_{n,P}\right\|^2
\lesssim
n_1 B_{X,n_1}^2(L_{X,n})B_{Y,n_1}^2(L_{Y,n})
=o_{\tilde{\cP}}(1)
\]
by Assumption~\ref{ass:doubly_robust_uniform}. Hence
\[
\left\|\widehat S_{n,P}-\tilde S_{n,P}\right\|^2=o_{\tilde{\cP}}(1),
\qquad
\left\|\widehat S_{n,P}-\tilde S_{n,P}\right\|=o_{\tilde{\cP}}(1).
\]
Finally, since $\|\hat \Pi_n\|_{\mathrm{op}}\leq 1$, we have
\[
\left\|b_{n,P}\right\|
\le
\left\|\widehat S_{n,P}-\tilde S_{n,P}\right\|
=
o_{\tilde{\cP}}(1).
\]
\end{proof}

\runinhead{Empirical-to-population projector replacement}

\begin{lemma}[Uniform replacement of empirical by population truncation projectors]
\label{lem:scaled-proj-replacement-uniform}
Define
\begin{equation*}
\tilde T_{n_1}^{(L_{X,n},L_{Y,n})}
\coloneqq
\left\|
\left(\hat\Pi_{X,L_{X,n}}\otimes \hat \Pi_{Y,L_{Y,n}}\otimes I\right)\Sigma_{n_1}
\right\|^2,
\qquad
T_{n_1}^{(L_{X,n},L_{Y,n})}
\coloneqq
\left\|
\left(\Pi_{X,L_{X,n}}\otimes \Pi_{Y,L_{Y,n}}\otimes I\right)\Sigma_{n_1}
\right\|^2.
\end{equation*}
Assume \(X \indep Y \mid Z\) and the uniform eigengap conditions in Assumption~\ref{ass:simplicity_gap_unif}. Then
\begin{equation*}
n_1\left|\tilde T_{n_1}^{(L_{X,n},L_{Y,n})}-T_{n_1}^{(L_{X,n},L_{Y,n})}\right|
=
o_{\tilde{\cP}_0}(1).
\end{equation*}
\end{lemma}

\begin{proof}
Fix \(P\in\tilde{\cP}_0\). Define
\[
U\coloneqq \hat\Pi_n\Sigma_{n_1},
\qquad
V\coloneqq \Pi_n\Sigma_{n_1}.
\]
Then \(\tilde T_{n_1}^{(L_{X,n},L_{Y,n})}=\|U\|^2\) and \(T_{n_1}^{(L_{X,n},L_{Y,n})}=\|V\|^2\). By the polarization identity,
\begin{equation*}
n_1\left|\tilde T_{n_1}^{(L_{X,n},L_{Y,n})}-T_{n_1}^{(L_{X,n},L_{Y,n})}\right|
=
n_1\left|\|U\|^2-\|V\|^2\right| 
\le
n_1\left(\|U\|+\|V\|\right)\|U-V\|.
\end{equation*}
Since \(\|\hat\Pi_n\|_{\mathrm{op}}\le 1\) and \(\|\Pi_n\|_{\mathrm{op}}\le 1\),
\begin{equation}
\label{eq:proj_replacement_first_bound}
n_1\left|\tilde T_{n_1}^{(L_{X,n},L_{Y,n})}-T_{n_1}^{(L_{X,n},L_{Y,n})}\right|
\le
2n_1\|\Sigma_{n_1}\|\,\|(\hat\Pi_n-\Pi_n)\Sigma_{n_1}\|
\le
2n_1\|\Sigma_{n_1}\|^2\|\hat\Pi_n-\Pi_n\|_{\mathrm{op}}.
\end{equation}

We next control \(\|\hat\Pi_n-\Pi_n\|_{\mathrm{op}}\) uniformly.

Setting \(\Delta_{V,L_{V,n}} \coloneqq \hat\Pi_{V,L_{V,n}}-\Pi_{V,L_{V,n}}\) for \(V\in\{X,Y\}\), the expansion \(\hat\Pi_n-\Pi_n = \Delta_{X,L_{X,n}}\otimes \Pi_{Y,L_{Y,n}}\otimes I + \Pi_{X,L_{X,n}}\otimes \Delta_{Y,L_{Y,n}}\otimes I + \Delta_{X,L_{X,n}}\otimes \Delta_{Y,L_{Y,n}}\otimes I\) yields
\begin{equation} \label{eq:Pi_diff_tensor_bound}
\|\hat\Pi_n-\Pi_n\|_{\mathrm{op}} \le \|\Delta_{X,L_{X,n}}\|_{\mathrm{op}} + \|\Delta_{Y,L_{Y,n}}\|_{\mathrm{op}} + \|\Delta_{X,L_{X,n}}\|_{\mathrm{op}}\|\Delta_{Y,L_{Y,n}}\|_{\mathrm{op}}.
\end{equation}

We treat the \(X\)-part; the \(Y\)-part is identical. Let
\[
A_P\coloneqq C_{XX}(P),\qquad
\hat A_P\coloneqq \hat C_{XX}(P),\qquad
H_{X,P}\coloneqq \hat A_P-A_P.
\]
By Assumption~\ref{ass:simplicity_gap_unif}, \(g_{L_{X,n}}^X>0\) for all large \(n\), and \(g_{L_{X,n}}^X\sqrt{n_2}\to\infty\).

Define the midpoint \(r_{X,P} \coloneqq (\lambda_{L_{X,n}}(P)+\lambda_{L_{X,n}+1}(P))/2\) and the event \(\Omega_{X,n,P} \coloneqq \{\|H_{X,P}\|_{\mathrm{op}} < g_{L_{X,n}}^X/4\}\). Since \(\gamma_{X,P}^{(L_{X,n})}\ge g_{L_{X,n}}^X\), Weyl's inequality implies that on \(\Omega_{X,n,P}\), \(\hat C_{XX}(P)\) has exactly \(L_{X,n}\) eigenvalues above \(r_{X,P}\). Hence the empirical projector onto the leading \(L_{X,n}\) eigenspace equals the spectral projector of \(\hat C_{XX}(P)\) associated with \((r_{X,P},\infty)\). The Davis--Kahan \(\sin\Theta\) theorem \citep{DavisandKahan1970} then gives, on \(\Omega_{X,n,P}\),
\[
\|\Delta_{X,L_{X,n}}\|_{\mathrm{op}}
\le
\frac{2}{\gamma_{X,P}^{(L_{X,n})}}\|H_{X,P}\|_{\mathrm{op}}
\le
\frac{2}{g_{L_{X,n}}^X}\|\hat C_{XX}(P)-C_{XX}(P)\|_{\mathrm{op}}.
\]
Since \(\|\Delta_{X,L_{X,n}}\|_{\mathrm{op}}\le 2\) always, for any \(\varepsilon>0\),
\begin{align}
\sup_{P\in\tilde{\cP}_0}\pr_P\left(\|\Delta_{X,L_{X,n}}\|_{\mathrm{op}}>\varepsilon\right)
&\le
\sup_{P\in\tilde{\cP}_0}\pr_P\left(\Omega_{X,n,P}^c\right)
+
\sup_{P\in\tilde{\cP}_0}\pr_P\left(\frac{2}{g_{L_{X,n}}^X}\|\hat C_{XX}(P)-C_{XX}(P)\|_{\mathrm{op}}>\varepsilon\right) \nonumber\\
&\le
\sup_{P\in\tilde{\cP}_0}\pr_P\left(\|\hat C_{XX}(P)-C_{XX}(P)\|_{\mathrm{HS}}>\frac{g_{L_{X,n}}^X}{4}\right) \nonumber\\
&\quad+
\sup_{P\in\tilde{\cP}_0}\pr_P\left(\|\hat C_{XX}(P)-C_{XX}(P)\|_{\mathrm{HS}}>\frac{\varepsilon g_{L_{X,n}}^X}{2}\right).
\label{eq:DeltaX_prob_bound}
\end{align}
Here we used $\|\cdot\|_{\mathrm{op}}\le \|\cdot\|_{\mathrm{HS}}$. Under the boundedness of the kernels, Lemma \ref{lem:cov_op_uniformly} implies that there exists a constant $C<\infty$ independent of $\tilde{\cP}$ such that
\[
\sup_{P\in\tilde{\cP}_0}\E_P\left[\|\hat C_{XX}(P)-C_{XX}(P)\|_{\mathrm{HS}}\right]\le \frac{C}{\sqrt{n_2}}.
\]
Applying Markov's inequality to both terms on the right-hand side of \eqref{eq:DeltaX_prob_bound} gives
\[
\sup_{P\in\tilde{\cP}_0}\pr_P\left(\|\Delta_{X,L_{X,n}}\|_{\mathrm{op}}>\varepsilon\right)
\le
\frac{4C}{g_{L_{X,n}}^X\sqrt{n_2}}
+
\frac{2C}{\varepsilon g_{L_{X,n}}^X\sqrt{n_2}}
\longrightarrow 0.
\]
Thus, \(\|\Delta_{X,L_{X,n}}\|_{\mathrm{op}} = o_{\tilde{\cP}_0}(1)\) and \(\|\Delta_{Y,L_{Y,n}}\|_{\mathrm{op}} = o_{\tilde{\cP}_0}(1)\). Combining these with \eqref{eq:Pi_diff_tensor_bound} yields
\begin{equation} \label{eq:Pi_diff_uniform_small}
\|\hat\Pi_n-\Pi_n\|_{\mathrm{op}} = o_{\tilde{\cP}_0}(1).
\end{equation}

Finally, under \(X\indep Y\mid Z\), the summands defining \(\sqrt{n_1}\Sigma_{n_1}\) are uniformly square-integrable over \(P\in\tilde{\cP}_0\), hence
\begin{equation}
\label{eq:Sigma_uniform_tight}
\sqrt{n_1}\|\Sigma_{n_1}\|=O_{\tilde{\cP}_0}(1),
\qquad\text{equivalently}\qquad
n_1\|\Sigma_{n_1}\|^2=O_{\tilde{\cP}_0}(1).
\end{equation}
Substituting \eqref{eq:Pi_diff_uniform_small} and \eqref{eq:Sigma_uniform_tight} into \eqref{eq:proj_replacement_first_bound} gives
\[
n_1\left|\tilde T_{n_1}^{(L_{X,n},L_{Y,n})}-T_{n_1}^{(L_{X,n},L_{Y,n})}\right|
=
o_{\tilde{\cP}_0}(1),
\]
as claimed.
\end{proof}

\runinhead{Truncation replacement at the limiting projector}
\begin{lemma}[Uniform coordinate truncation replacement at the limiting projector]
\label{lem:coord-truncation-unified}
Define
\[
T_{n_1}^{(L_{X,n},L_{Y,n})} \coloneqq \|\Pi_n\Sigma_{n_1}\|^2,
\qquad
T_{n_1}^{(\infty)} \coloneqq \|\Pi_\infty\Sigma_{n_1}\|^2.
\]
Then
\[
n_1\left|T_{n_1}^{(L_{X,n},L_{Y,n})}-T_{n_1}^{(\infty)}\right|
=
o_{\tilde{\cP}_0}(1).
\]
\end{lemma}

\begin{proof}
Write
\[
S_{n,P}^{(L_{X,n},L_{Y,n})}
\coloneqq
\Pi_n\left(\frac{1}{\sqrt{n_1}}\sum_{i\in I_1}\Psi_i\right),
\qquad
S_{n,P}^{(\infty)}
\coloneqq
\Pi_\infty\left(\frac{1}{\sqrt{n_1}}\sum_{i\in I_1}\Psi_i\right).
\]
Then
\[
n_1 T_{n_1}^{(L_{X,n},L_{Y,n})}
=
\left\|S_{n,P}^{(L_{X,n},L_{Y,n})}\right\|^2,
\qquad
n_1 T_{n_1}^{(\infty)}
=
\left\|S_{n,P}^{(\infty)}\right\|^2.
\]
Hence
\[
n_1\left|T_{n_1}^{(L_{X,n},L_{Y,n})}-T_{n_1}^{(\infty)}\right|
=
\left|
\left\|S_{n,P}^{(L_{X,n},L_{Y,n})}\right\|^2
-
\left\|S_{n,P}^{(\infty)}\right\|^2
\right|
\le
\left(
\left\|S_{n,P}^{(L_{X,n},L_{Y,n})}\right\|
+
\left\|S_{n,P}^{(\infty)}\right\|
\right)
\left\|
S_{n,P}^{(L_{X,n},L_{Y,n})}-S_{n,P}^{(\infty)}
\right\|.
\]
It therefore suffices to show that
\[
\left\|S_{n,P}^{(L_{X,n},L_{Y,n})}\right\|
+
\left\|S_{n,P}^{(\infty)}\right\|
=
O_{\tilde{\cP}_0}(1)
\qquad\text{and}\qquad
\left\|
S_{n,P}^{(L_{X,n},L_{Y,n})}-S_{n,P}^{(\infty)}
\right\|
=
o_{\tilde{\cP}_0}(1).
\]

We first prove the uniform \(O_{\tilde{\cP}_0}(1)\) bound. Under \(X\indep Y\mid Z\), the summands \(\Psi_i\) are centered, so
\[
\E_P\left[
\left\|S_{n,P}^{(L_{X,n},L_{Y,n})}\right\|^2
\right]
=
\E_P\left[\left\|\Pi_n\Psi\right\|^2\right]
\le
\E_P\left[\left\|\Psi\right\|^2\right]
\le
16K_X^2K_Y^2K_Z^2.
\]
The same argument gives
\[
\E_P\left[
\left\|S_{n,P}^{(\infty)}\right\|^2
\right]
=
\E_P\left[\left\|\Pi_\infty\Psi\right\|^2\right]
\le
16K_X^2K_Y^2K_Z^2.
\]
Hence both norms are \(O_{\tilde{\cP}_0}(1)\) by Markov's inequality.

Next, write
\[
S_{n,P}^{(L_{X,n},L_{Y,n})}-S_{n,P}^{(\infty)}
=
\frac{1}{\sqrt{n_1}}\sum_{i\in I_1}(\Pi_n-\Pi_\infty)\Psi_i.
\]
Again using \(X\indep Y\mid Z\), the summands are centered and independent, so
\[
\E_P\left[
\left\|
S_{n,P}^{(L_{X,n},L_{Y,n})}-S_{n,P}^{(\infty)}
\right\|^2
\right]
=
\E_P\left[
\left\|(\Pi_n-\Pi_\infty)\Psi\right\|^2
\right].
\]

Now decompose
\[
\Pi_n-\Pi_\infty
=
(\Pi_{X,L_{X,n}}-\Pi_{X,L_{X,\infty}})\otimes \Pi_{Y,L_{Y,n}}\otimes I
+
\Pi_{X,L_{X,\infty}}\otimes(\Pi_{Y,L_{Y,n}}-\Pi_{Y,L_{Y,\infty}})\otimes I.
\]
Therefore, by \((a+b)^2\le 2a^2+2b^2\), contractivity of orthogonal projections, and kernel boundedness,
\[
\E_P\left[
\left\|(\Pi_n-\Pi_\infty)\Psi\right\|^2
\right]
\le
8K_Y^2K_Z^2\,
\E_P\left[
\left\|
(\Pi_{X,L_{X,n}}-\Pi_{X,L_{X,\infty}})\epsilon_{X\mid Z}
\right\|^2
\right]
\]
\[
\quad
+
8K_X^2K_Z^2\,
\E_P\left[
\left\|
(\Pi_{Y,L_{Y,n}}-\Pi_{Y,L_{Y,\infty}})\epsilon_{Y\mid Z}
\right\|^2
\right].
\]
By Lemma~\ref{lem:uniform-tail-eps-general-limit}, both expectations on the right-hand side converge to \(0\) uniformly over \(P\in\tilde{\cP}_0\). Hence
\[
\E_P\left[
\left\|
S_{n,P}^{(L_{X,n},L_{Y,n})}-S_{n,P}^{(\infty)}
\right\|^2
\right]
\longrightarrow 0
\]
uniformly over \(P\in\tilde{\cP}_0\), and another application of Markov's inequality yields
\[
\left\|
S_{n,P}^{(L_{X,n},L_{Y,n})}-S_{n,P}^{(\infty)}
\right\|
=
o_{\tilde{\cP}_0}(1).
\]

Combining the preceding bounds,
\[
n_1\left|T_{n_1}^{(L_{X,n},L_{Y,n})}-T_{n_1}^{(\infty)}\right|
=
o_{\tilde{\cP}_0}(1),
\]
as claimed.
\end{proof}

\begin{lemma}[Uniform residual-tail control for general truncation limits]
\label{lem:uniform-tail-eps-general-limit}
Under Assumption~\ref{ass:fixed-basis-uniform_tail}, we have
\[
\sup_{P\in\tilde{\cP}_0}
\E_P\left[
\left\|
\left(\Pi_{X,L_{X,\infty}}-\Pi_{X,L_{X,n}}\right)\epsilon_{X\mid Z}
\right\|^2
\right]
\longrightarrow 0.
\]
The same conclusion holds with \(X\), \(\Pi_{X,\cdot}\), \(L_{X,n}\), \(L_{X,\infty}\), and
\(\epsilon_{X\mid Z}\) replaced by \(Y\), \(\Pi_{Y,\cdot}\), \(L_{Y,n}\), \(L_{Y,\infty}\), and
\(\epsilon_{Y\mid Z}\), respectively.
\end{lemma}
\begin{proof}
We prove the \(X\)-statement; the \(Y\)-statement is identical.

If \(L_{\infty}^X<\infty\), then since \((L_{X,n})\) is nondecreasing and integer-valued, \(L_{X,n}=L_{\infty}^X\) for all sufficiently large \(n\). Thus
\[
\left(\Pi_{X,L_{\infty}^X}-\Pi_{X,L_{X,n}}\right)\epsilon_{X\mid Z}=0
\]
eventually, and the claim is immediate.

Now suppose \(L_{\infty}^X=\infty\), so \(\Pi_{X,L_{\infty}^X}=\Pi_X\). By \(\|a-b\|^2\le 2\|a\|^2+2\|b\|^2\), and conditional Jensen,
\begin{equation}
\label{eq:tail_epsX_factor4}
\E_P\left[\left\|(\Pi_X-\Pi_{X,L_{X,n}})\epsilon_{X\mid Z}\right\|^2\right]
\le
4\,\E_P\left[\left\|(\Pi_X-\Pi_{X,L_{X,n}})k_{\cX}(X,\cdot)\right\|^2\right].
\end{equation}

Since \(k_{\cX}(X,\cdot)\in \overline{\range(C_{XX})}\) almost surely, Parseval's identity yields
\[
\E_P\left[\left\|(\Pi_X-\Pi_{X,L_{X,n}})k_{\cX}(X,\cdot)\right\|^2\right]
=
\sum_{j>L_{X,n}}\lambda_j(P),
\]
where \(\{\lambda_j(P)\}_{j\ge 1}\) are the positive eigenvalues of \(C_{XX}\) under \(P\). Let \(\{\tilde e_p\}_{p=1}^{L_{\infty}^X}\) be the orthonormal system from Assumption~\ref{ass:fixed-basis-uniform_tail}. By Ky Fan's variational principle,
\[
\sum_{j>L_{X,n}}\lambda_j(P)
\le
\sum_{p>L_{X,n}}\E_P\left[\tilde e_p(X)^2\right].
\]
Taking the supremum over \(P\in\tilde{\cP}_0\) and using Assumption~\ref{ass:fixed-basis-uniform_tail},
\[
\sup_{P\in\tilde{\cP}_0}
\E_P\left[\left\|(\Pi_X-\Pi_{X,L_{X,n}})k_{\cX}(X,\cdot)\right\|^2\right]
\longrightarrow 0.
\]
The claim follows from \eqref{eq:tail_epsX_factor4}.
\end{proof}

\begin{corollary}[Uniform strong truncation replacement when \(L_{X,n},L_{Y,n}\to\infty\)]
\label{cor:coord-truncation-infinite}
Under the assumptions of Lemma~\ref{lem:coord-truncation-unified}, if \(L_{X,n}\to\infty\) and \(L_{Y,n}\to\infty\), then
\[
\Pi_\infty=\Pi_X\otimes \Pi_Y\otimes I
\quad\text{and}\quad
T_{n_1}^{(\infty)}=\|\Sigma_{n_1}\|^2.
\]
Hence, with \(\bar T_{n_1}\coloneqq \|\Sigma_{n_1}\|^2\),
\begin{equation*}
n_1\left|T_{n_1}^{(L_{X,n},L_{Y,n})}-\bar T_{n_1}\right|
=
o_{\tilde{\cP}_0}(1).
\end{equation*}
\end{corollary}

\begin{proof}
If \(L_{X,n}\to\infty\) and \(L_{Y,n}\to\infty\), then by convention \(\Pi_{X,L_{\infty}^X}=\Pi_X\) and \(\Pi_{Y,L_{\infty}^Y}=\Pi_Y\), so \(\Pi_\infty=\Pi_X\otimes\Pi_Y\otimes I\). The conclusion is exactly Lemma~\ref{lem:coord-truncation-unified}.
\end{proof}
\runinhead{Bootstrap analogue}
\begin{lemma}[Uniform negligibility of multiplier-bootstrap biases]
\label{lem:multiplier_bootstrap_biases_negligible_uniform}
Assume Assumptions~\ref{ass:simplicity_gap_unif},
\ref{ass:doubly_robust_uniform}, and \ref{ass:fixed-basis-uniform_tail} hold uniformly over
\(\tilde{\cP}\). 

Let \(\widehat S_{n,P}^\ast\) denote the actual multiplier-bootstrap process (so that
\(n_1\hat T_{n_1}^\ast=\|\widehat S_{n,P}^\ast\|^2\)), and define
\begin{equation*}
S_{n,P}^{(L_{X,n},L_{Y,n})\ast}
\coloneqq
\Pi_n\left(\frac{1}{\sqrt{n_1}}\sum_{i\in I_1}\Psi_i W_i^\ast\right),
\qquad
S_{n,P}^{(\infty)\ast}
\coloneqq
\Pi_\infty\left(\frac{1}{\sqrt{n_1}}\sum_{i\in I_1}\Psi_i W_i^\ast\right).
\end{equation*}

Then the following hold uniformly over \(P\in\tilde{\cP}\):

\begin{enumerate}
\item[(i)] For every \(\eta>0\) and \(\varepsilon>0\),
\begin{equation*}
\sup_{P\in\tilde{\cP}}
\pr_P\left(
\pr_P\left(
\left\|\widehat S_{n,P}^\ast-S_{n,P}^{(L_{X,n},L_{Y,n})\ast}\right\|>\eta
\,\middle|\,\cD_n
\right)>\varepsilon
\right)\to 0.
\end{equation*}

\item[(ii)] For every \(\eta>0\) and \(\varepsilon>0\),
\begin{equation*}
\sup_{P\in\tilde{\cP}}
\pr_P\left(
\pr_P\left(
\left\|S_{n,P}^{(L_{X,n},L_{Y,n})\ast}-S_{n,P}^{(\infty)\ast}\right\|>\eta
\,\middle|\,\cD_n
\right)>\varepsilon
\right)\to 0.
\end{equation*}

\item[(iii)] For every \(\eta>0\) and \(\varepsilon>0\),
\begin{equation*}
\sup_{P\in\tilde{\cP}}
\pr_P\left(
\pr_P\left(
\left|n_1\hat T_{n_1}^\ast-\left\|S_{n,P}^{(\infty)\ast}\right\|^2\right|>\eta
\,\middle|\,\cD_n
\right)>\varepsilon
\right)\to 0.
\end{equation*}
\end{enumerate}
\end{lemma}
\begin{proof}
We repeatedly use the conditional multiplier isometry:
for any \(\cD_n\)-measurable vectors \(a_i\in\cH\),
\begin{equation}
\label{eq:multiplier-isometry}
\E^\ast\left\|
\frac{1}{\sqrt{n_1}}\sum_{i\in I_1} a_i W_i^\ast
\right\|^2
=
\frac{1}{n_1}\sum_{i\in I_1}\|a_i\|^2.
\end{equation}
This follows from conditional independence and the moment conditions on \(W_i^\ast\).

We first evaluate the regression-induced bootstrap error. Let \(\widetilde S_{n,P}^\ast\) denote the multiplier bootstrap process built from the oracle residual coordinates but with the empirical truncation basis (equivalently, the same empirical projector as \(\widehat S_{n,P}^\ast\), but oracle residual coordinates in place of plug-in residual coordinates). Set
\begin{equation*}
\Delta_{n,P}^{\mathrm{reg},\ast}
\coloneqq
\widehat S_{n,P}^\ast-\widetilde S_{n,P}^\ast
=
\frac{1}{\sqrt{n_1}}\sum_{i\in I_1}\delta_{i,P}^{\mathrm{reg}}\,W_i^\ast,
\end{equation*}
where \(\delta_{i,P}^{\mathrm{reg}}\) is the pointwise plug-in minus oracle tensor on the evaluation fold. By the same decomposition used in the proof of Proposition~\ref{prop:bias-negligible} and Corollary~\ref{cor:uniform_bias_control_alternative}, and Assumption~\ref{ass:doubly_robust_uniform}, we obtain
\begin{equation} \label{eq:reg-pointwise-L2-small}
\frac{1}{n_1}\sum_{i\in I_1}\left\|\delta_{i,P}^{\mathrm{reg}}\right\|^2
\le
C\{ B_{X,n_1}^2+B_{Y,n_1}^2+n_1B_{X,n_1}^2B_{Y,n_1}^2 \}
=
o_{\tilde{\cP}}(1),
\end{equation}
where a constant \(C<\infty\) depends only on \((K_X,K_Y,K_Z)\). Using \eqref{eq:multiplier-isometry},
\begin{equation*}
\E^\ast\left[\left\|\Delta_{n,P}^{\mathrm{reg},\ast}\right\|^2\right]
=
\frac{1}{n_1}\sum_{i\in I_1}\left\|\delta_{i,P}^{\mathrm{reg}}\right\|^2
=
o_{\tilde{\cP}}(1).
\end{equation*}
Therefore, by conditional Markov's inequality, for every \(\eta>0\) and \(\varepsilon>0\),
\begin{equation}
\label{eq:reg-bootstrap-negl}
\sup_{P\in\tilde{\cP}}
\pr_P\left(
\pr^\ast\left(
\left\|\widehat S_{n,P}^\ast-\widetilde S_{n,P}^\ast\right\|>\eta
\right)>\varepsilon
\right)\to 0.
\end{equation}

Next, we address the empirical-vs-population projector bootstrap error. Define
\begin{equation*}
\Delta_{n,P}^{\mathrm{proj},\ast}
\coloneqq
\widetilde S_{n,P}^\ast-S_{n,P}^{(L_{X,n},L_{Y,n})\ast}
=
\frac{1}{\sqrt{n_1}}\sum_{i\in I_1}\left((\hat\Pi_n-\Pi_n)\Psi_i\right)W_i^\ast.
\end{equation*}
By \eqref{eq:multiplier-isometry},
\begin{equation*}
\E^\ast\left[\left\|\Delta_{n,P}^{\mathrm{proj},\ast}\right\|^2\right]
=
\frac{1}{n_1}\sum_{i\in I_1}\left\|(\hat\Pi_n-\Pi_n)\Psi_i\right\|^2
\le
\|\hat\Pi_n-\Pi_n\|_{\mathrm{op}}^2
\cdot
\frac{1}{n_1}\sum_{i\in I_1}\|\Psi_i\|^2.
\end{equation*}
Since \(\Psi_i=\epsilon_{X\mid Z}^{(i)}\otimes\epsilon_{Y\mid Z}^{(i)}\otimes\phi_{\cZ}(Z_i)\), and \(\|\Psi_i\|^2 \leq 16K_X^2K_Y^2K_Z^2\),
\begin{equation*}
\E^\ast\left[\left\|\Delta_{n,P}^{\mathrm{proj},\ast}\right\|^2\right]
\le
16K_X^2K_Y^2K_Z^2\,\|\hat\Pi_n-\Pi_n\|_{\mathrm{op}}^2.
\end{equation*}
The Davis--Kahan argument used in the proof of Lemma~\ref{lem:scaled-proj-replacement-uniform} yields
\(
\|\hat\Pi_n-\Pi_n\|_{\mathrm{op}}=o_{\tilde{\cP}}(1)\) under Assumption~\ref{ass:simplicity_gap_unif} and kernel boundedness. Therefore, \(\E^\ast[\|\Delta_{n,P}^{\mathrm{proj},\ast}\|^2] = o_{\tilde{\cP}}(1)\), and another conditional Markov bound gives, for every \(\eta,\varepsilon>0\),
\begin{equation}
\label{eq:proj-bootstrap-negl}
\sup_{P\in\tilde{\cP}}
\pr_P\left(
\pr_P^\ast\left(
\left\|\widetilde S_{n,P}^\ast-S_{n,P}^{(L_{X,n},L_{Y,n})\ast}\right\|>\eta \mid \cD_n
\right)>\varepsilon
\right)\to 0.
\end{equation}

Turning to the truncation-to-\(\Pi_\infty\) bootstrap error, we define
\begin{equation*}
\Delta_{n,P}^{\mathrm{tail},\ast}
\coloneqq
S_{n,P}^{(L_{X,n},L_{Y,n})\ast}-S_{n,P}^{(\infty)\ast}
=
\frac{1}{\sqrt{n_1}}\sum_{i\in I_1}\left((\Pi_n-\Pi_\infty)\Psi_i\right)W_i^\ast.
\end{equation*}
By \eqref{eq:multiplier-isometry} and the i.i.d.\ property of the evaluation fold, taking the expectation yields
\begin{equation*}
\E_P\left[\E^\ast\left[\left\|\Delta_{n,P}^{\mathrm{tail},\ast}\right\|^2\right]\right]
=
\E_P\left[\frac{1}{n_1}\sum_{i\in I_1}\left\|(\Pi_n-\Pi_\infty)\Psi_i\right\|^2\right]
=
\E_P\left[\left\|(\Pi_n-\Pi_\infty)\Psi\right\|^2\right].
\end{equation*}
We then repeat the single-summand \(V^{(1)}\), \(V^{(2)}\), \(V^{(3)}\) decomposition from the proof of Lemma~\ref{lem:coord-truncation-unified} (the one-observation second-moment bounds, which do not use \(X\indep Y\mid Z\)). Combining these with Lemma~\ref{lem:uniform-tail-eps-general-limit}, we obtain
\begin{equation*}
\sup_{P\in\tilde{\cP}}
\E_P\left[\left\|(\Pi_n-\Pi_\infty)\Psi\right\|^2\right]\to 0.
\end{equation*}
Hence \(\E^\ast[\|\Delta_{n,P}^{\mathrm{tail},\ast}\|^2] = o_{\tilde{\cP}}(1)\) uniformly, and conditional Markov's inequality yields, for every \(\eta,\varepsilon>0\),
\begin{equation}
\label{eq:tail-bootstrap-negl}
\sup_{P\in\tilde{\cP}}
\pr_P\left(
\pr_P^\ast\left(
\left\|S_{n,P}^{(L_{X,n},L_{Y,n})\ast}-S_{n,P}^{(\infty)\ast}\right\|>\eta
\mid \cD_n\right)>\varepsilon
\right)\to 0.
\end{equation}
This establishes statement (ii).

Consequently, we can bound the total finite-dimensional error to prove (i). By the triangle inequality,
\begin{equation*}
\widehat S_{n,P}^\ast-S_{n,P}^{(L_{X,n},L_{Y,n})\ast}
=
\left(\widehat S_{n,P}^\ast-\widetilde S_{n,P}^\ast\right)
+
\left(\widetilde S_{n,P}^\ast-S_{n,P}^{(L_{X,n},L_{Y,n})\ast}\right).
\end{equation*}
Thus, (i) immediately follows from \eqref{eq:reg-bootstrap-negl} and \eqref{eq:proj-bootstrap-negl} via the conditional union bound.

Finally, we transfer these bounds to the quadratic form to prove (iii). Setting \(D_{n,P}^\ast\coloneqq \widehat S_{n,P}^\ast-S_{n,P}^{(\infty)\ast}\), we observe that
\begin{equation*}
\left|n_1\hat T_{n_1}^\ast-\left\|S_{n,P}^{(\infty)\ast}\right\|^2\right|
=
\left|\left\|\widehat S_{n,P}^\ast\right\|^2-\left\|S_{n,P}^{(\infty)\ast}\right\|^2\right|
\le
2\left\|S_{n,P}^{(\infty)\ast}\right\|\,\left\|D_{n,P}^\ast\right\|
+\left\|D_{n,P}^\ast\right\|^2.
\end{equation*}
By conditional Markov's inequality and Cauchy--Schwarz, for every \(\eta>0\),
\begin{equation}
\pr^\ast\left(
\left|n_1\hat T_{n_1}^\ast-\left\|S_{n,P}^{(\infty)\ast}\right\|^2\right|>\eta
\right)
\le
\frac{2}{\eta}
\left(\E^\ast\left[\left\|S_{n,P}^{(\infty)\ast}\right\|^2\right]\right)^{1/2}
\left(\E^\ast\left[\left\|D_{n,P}^\ast\right\|^2\right]\right)^{1/2}+
\frac{1}{\eta}\E^\ast\left[\left\|D_{n,P}^\ast\right\|^2\right].
\end{equation}
Moreover, by \eqref{eq:multiplier-isometry},
\begin{equation*}
\E^\ast\left[\left\|S_{n,P}^{(\infty)\ast}\right\|^2\right]
=
\frac{1}{n_1}\sum_{i\in I_1}\left\|\Pi_\infty\Psi_i\right\|^2
\le
\frac{1}{n_1}\sum_{i\in I_1}\left\|\Psi_i\right\|^2
\le 16K_X^2K_Y^2K_Z^2
\quad\text{a.s.}
\end{equation*}
Hence, it remains only to show that \(\E^\ast[\|D_{n,P}^\ast\|^2]=o_{\tilde{\cP}}(1)\). This follows from the decomposition
\begin{equation*}
D_{n,P}^\ast
=
\left(\widehat S_{n,P}^\ast-S_{n,P}^{(L_{X,n},L_{Y,n})\ast}\right)
+
\left(S_{n,P}^{(L_{X,n},L_{Y,n})\ast}-S_{n,P}^{(\infty)\ast}\right),
\end{equation*}
the inequality \(\|u+v\|^2\le 2\|u\|^2+2\|v\|^2\), and the bounds proved in the preceding arguments. Therefore, (iii) follows by one final conditional Markov argument.
\end{proof}
\subsection{Covariance convergence and Gaussian comparison}
\runinhead{Empirical covariance convergence}

For \(P\in\tilde{\cP}\), define
\[
\Psi_i^{(\infty)} \coloneqq \Pi_\infty \Psi_i,
\qquad
C_{n,P}^{(\infty)}
\coloneqq
\frac{1}{n_1}\sum_{i\in I_1}
\Psi_i^{(\infty)}\otimes \Psi_i^{(\infty)},
\qquad
C_P^{(\infty)}
\coloneqq
\E_P\left[\Psi^{(\infty)}\otimes \Psi^{(\infty)}\right].
\]

\begin{lemma}[Uniform trace-norm LLN for the \(\Pi_\infty\)-projected oracle covariance]
\label{lem:uniform_trace_lln_projected_covariance}
Assume that Assumption~\ref{ass:fixed-basis-uniform_tail} holds uniformly over \(\tilde{\cP}\). Then, without assuming \(X \indep Y \mid Z\),
\[
\left\|C_{n,P}^{(\infty)}-C_P^{(\infty)}\right\|_{\mathrm{tr}}
=
o_{\tilde{\cP}}(1).
\]
Equivalently, for every \(\varepsilon>0\),
\[
\sup_{P\in\tilde{\cP}}
\pr_P\left(
\left\|C_{n,P}^{(\infty)}-C_P^{(\infty)}\right\|_{\mathrm{tr}}>\varepsilon
\right)\to 0.
\]
\end{lemma}

\begin{proof}
Fix \(P\in\tilde{\cP}\), and suppress the dependence on \(P\) in the notation. Let
\(\{u_j\}_{j\ge 1}\) be the fixed orthonormal basis of
\(\cH_{\cX}\otimes\cH_{\cY}\otimes\cH_{\cZ}\) obtained by enumerating the tensor-product basis underlying Assumption~\ref{ass:fixed-basis-uniform_tail}, and let
\[
Q_J \coloneqq \sum_{j=1}^J u_j\otimes u_j
\]
denote the orthogonal projection onto \(\mathrm{span}\{u_1,\dots,u_J\}\).
Write
\[
\Psi_{i,J}^{(\infty)} \coloneqq Q_J \Psi_i^{(\infty)},
\qquad
C_{n,P,J}^{(\infty)}
\coloneqq
\frac{1}{n_1}\sum_{i\in I_1}
\Psi_{i,J}^{(\infty)}\otimes \Psi_{i,J}^{(\infty)},
\qquad
C_{P,J}^{(\infty)}
\coloneqq
\E_P\left[\Psi_J^{(\infty)}\otimes \Psi_J^{(\infty)}\right].
\]
Then
\[
\left\|C_{n,P}^{(\infty)}-C_P^{(\infty)}\right\|_{\mathrm{tr}}
\le
\left\|C_{n,P}^{(\infty)}-C_{n,P,J}^{(\infty)}\right\|_{\mathrm{tr}}
+
\left\|C_{n,P,J}^{(\infty)}-C_{P,J}^{(\infty)}\right\|_{\mathrm{tr}}
+
\left\|C_{P,J}^{(\infty)}-C_P^{(\infty)}\right\|_{\mathrm{tr}}.
\]

We first control the two truncation terms. By the rank-one trace inequality
\[
\|u\otimes u-v\otimes v\|_{\mathrm{tr}}
\le
\|u+v\|\,\|u-v\|,
\]
together with Cauchy--Schwarz,
\[
\left\|C_{n,P}^{(\infty)}-C_{n,P,J}^{(\infty)}\right\|_{\mathrm{tr}}
\le
\left(
\frac{1}{n_1}\sum_{i\in I_1}\left\|\Psi_i^{(\infty)}+\Psi_{i,J}^{(\infty)}\right\|^2
\right)^{1/2}
\left(
\frac{1}{n_1}\sum_{i\in I_1}\left\|\Psi_i^{(\infty)}-\Psi_{i,J}^{(\infty)}\right\|^2
\right)^{1/2}.
\]
Since \(\Pi_\infty\) and \(Q_J\) are orthogonal projections, \(\|\Psi_i^{(\infty)}\|\le \|\Psi_i\|\) and \(\|\Psi_{i,J}^{(\infty)}\|\le \|\Psi_i^{(\infty)}\|\). By kernel boundedness,
\(\|\Psi_i\|^2\le 16K_X^2K_Y^2K_Z^2\) almost surely. Hence the first factor is bounded uniformly by a deterministic constant \(C<\infty\), and therefore
\[
\left\|C_{n,P}^{(\infty)}-C_{n,P,J}^{(\infty)}\right\|_{\mathrm{tr}}
\le
C\left(
\frac{1}{n_1}\sum_{i\in I_1}\left\|(I-Q_J)\Psi_i^{(\infty)}\right\|^2
\right)^{1/2}.
\]
Taking expectations and using Markov's inequality, it is enough to show that
\[
\sup_{P\in\tilde{\cP}}
\E_P\left[\left\|(I-Q_J)\Psi^{(\infty)}\right\|^2\right]\to 0
\qquad (J\to\infty).
\]
This is exactly the fixed-basis uniform tail control for \(\Psi^{(\infty)}\), which follows from Assumption~\ref{ass:fixed-basis-uniform_tail} by the same coefficient expansion used in the proof of Lemma~\ref{lem:uniform_quantile_regularity_gaussian_limit}. Consequently,
\[
\sup_{P\in\tilde{\cP}}
\pr_P\left(
\left\|C_{n,P}^{(\infty)}-C_{n,P,J}^{(\infty)}\right\|_{\mathrm{tr}}>\varepsilon
\right)
\longrightarrow 0
\qquad (J\to\infty)
\]
uniformly in \(n\). The same argument, with the expectation in place of the empirical average, gives
\[
\sup_{P\in\tilde{\cP}}
\left\|C_{P,J}^{(\infty)}-C_P^{(\infty)}\right\|_{\mathrm{tr}}
\longrightarrow 0
\qquad (J\to\infty).
\]

It remains to control the finite-dimensional term
\(\left\|C_{n,P,J}^{(\infty)}-C_{P,J}^{(\infty)}\right\|_{\mathrm{tr}}\) for fixed \(J\).
Write
\[
a_{i,r}\coloneqq \left\langle \Psi_i^{(\infty)},u_r\right\rangle,
\qquad
1\le r\le J.
\]
Then \(Q_J\Psi_i^{(\infty)}=\sum_{r=1}^J a_{i,r}u_r\), so \(C_{n,P,J}^{(\infty)}-C_{P,J}^{(\infty)}\) is represented on \(\mathrm{span}\{u_1,\dots,u_J\}\) by the \(J\times J\) matrix
\[
\Gamma_{n,P,J}
\coloneqq
\left(
\frac{1}{n_1}\sum_{i\in I_1} a_{i,r}a_{i,s}
-
\E_P[a_{1,r}a_{1,s}]
\right)_{1\le r,s\le J}.
\]
Since \(|a_{i,r}|\le \|\Psi_i^{(\infty)}\|\le \|\Psi_i\|\le 4K_XK_YK_Z\) almost surely, every entry of \(\Gamma_{n,P,J}\) is the empirical mean of a uniformly bounded scalar array. Thus, for each fixed pair \((r,s)\), Chebyshev's inequality gives
\[
\sup_{P\in\tilde{\cP}}
\pr_P\left(
\left|
\frac{1}{n_1}\sum_{i\in I_1} a_{i,r}a_{i,s}
-
\E_P[a_{1,r}a_{1,s}]
\right|>\eta
\right)
\le
\frac{C}{n_1\eta^2},
\]
where \(C<\infty\) depends only on the kernel bounds. Since there are only finitely many pairs \((r,s)\) with \(1\le r,s\le J\), a union bound yields
\[
\sup_{P\in\tilde{\cP}}
\pr_P\left(
\max_{1\le r,s\le J}
\left|
\frac{1}{n_1}\sum_{i\in I_1} a_{i,r}a_{i,s}
-
\E_P[a_{1,r}a_{1,s}]
\right|>\eta
\right)\to 0
\qquad (n\to\infty).
\]
Because \(J\) is fixed, all norms on \(J\times J\) matrices are equivalent, and hence
\[
\left\|C_{n,P,J}^{(\infty)}-C_{P,J}^{(\infty)}\right\|_{\mathrm{tr}}
=
o_{\tilde{\cP}}(1)
\qquad \text{for each fixed } J.
\]

Now fix \(\varepsilon>0\). Choose \(J\) so large that the two truncation terms are each smaller than \(\varepsilon/3\) uniformly over \(P\), and then let \(n\to\infty\) so that the finite-dimensional term is smaller than \(\varepsilon/3\) uniformly over \(P\). The preceding decomposition then yields
\[
\sup_{P\in\tilde{\cP}}
\pr_P\left(
\left\|C_{n,P}^{(\infty)}-C_P^{(\infty)}\right\|_{\mathrm{tr}}>\varepsilon
\right)\to 0.
\]
This proves the claim.
\end{proof}

\begin{lemma}\label{lem:cov_op_uniformly}
Uniformly over \(\cP\), we have
\[
\|\hat C_{XX}-C_{XX}\|_{\mathrm{HS}} = O_{\cP}(n_2^{-1/2}),
\qquad
\|\hat C_{XX}-C_{XX}\|_{\mathrm{HS}}^2 = O_{\cP}(n_2^{-1}).
\]
Analogous rates hold for \(\hat C_{YY}\) and \(C_{YY}\).
\end{lemma}

\begin{proof}
For \(i \in I_2\), define \(Z_i \coloneqq \phi_{\cX}(X_i)\otimes \phi_{\cX}(X_i)\), so that \(\hat C_{XX} = \frac{1}{n_2}\sum_{i\in I_2}Z_i\) and \(C_{XX} = \E_P[Z_i]\). By the reproducing property and the boundedness of the kernel, \(\|Z_i\|_{\mathrm{HS}} = k_{\cX}(X_i,X_i) \le K_X^2\) almost surely.

Since the observations in \(I_2\) are i.i.d.\ under any \(P \in \cP\), bounding the variance of the empirical mean by the second moment yields
\[
\E_P\left\|\hat C_{XX}-C_{XX}\right\|_{\mathrm{HS}}^2
\le
\frac{1}{n_2}\E_P\left\|Z_i\right\|_{\mathrm{HS}}^2
\le
\frac{K_X^4}{n_2}
\qquad \text{for all } P\in\cP.
\]
By Markov's inequality, for any \(M>0\),
\[
\sup_{P\in\cP}
\pr_P\left(
\left\|\hat C_{XX}-C_{XX}\right\|_{\mathrm{HS}} > \frac{M}{\sqrt{n_2}}
\right)
\le
\frac{K_X^4}{M^2}.
\]
Since the right-hand side vanishes as \(M \to \infty\) independently of \(n_2\) and \(P\), we conclude that
\(\|\hat C_{XX}-C_{XX}\|_{\mathrm{HS}} = O_{\cP}(n_2^{-1/2})\). The rate for the squared norm follows immediately. The proof for \(\hat C_{YY}\) and \(C_{YY}\) is identical.
\end{proof}

\runinhead{From bounded-Lipschitz convergence to squared norms}
\begin{lemma}[Uniform anti-concentration for Gaussian Hilbert squared norms]
\label{lem:uniform_anti_concentration_gaussian_hilbert_squared_norm}
For each \(P\in\cP\), let \(G_P\) be a centered Gaussian random element in a separable Hilbert space \(\cH\) with covariance operator \(C_P\). Let \(\lambda_{P,1}\) denote the largest eigenvalue of \(C_P\), and assume that
\[
\inf_{P\in\cP}\lambda_{P,1}\ge \underline{\lambda}>0.
\]
Then, for every \(\delta>0\),
\[
\sup_{P\in\cP}\sup_{t\in\bbR}
\pr_P\left(\left|\|G_P\|^2-t\right|\le \delta\right)
\le
2\Phi\left(\sqrt{\frac{2\delta}{\underline{\lambda}}}\right)-1.
\]
Consequently,
\[
\lim_{\delta\downarrow 0}
\sup_{P\in\cP}\sup_{t\in\bbR}
\pr_P\left(\left|\|G_P\|^2-t\right|\le \delta\right)
=0.
\]
\end{lemma}

\begin{proof}
Fix \(P\in\cP\). Since \(G_P\) is a centered Gaussian random element in a separable Hilbert space, its covariance operator \(C_P\) is self-adjoint, positive, and trace class. Hence there exist an orthonormal basis \(\{e_{P,j}\}_{j\ge 1}\) and eigenvalues \(\{\lambda_{P,j}\}_{j\ge 1}\) such that \(C_P e_{P,j}=\lambda_{P,j}e_{P,j}\), \(\lambda_{P,1}\ge \lambda_{P,2}\ge \cdots \ge 0\), and \(\sum_{j=1}^\infty \lambda_{P,j}<\infty\). On an extension of the probability space, let \(\{Z_j\}_{j\ge 1}\) be i.i.d.\ \(N(0,1)\), and define \(S_P^\sharp \coloneqq \sum_{j=1}^\infty \lambda_{P,j} Z_j^2\). Since the summands are nonnegative and \(\E_P[S_P^\sharp]=\sum_{j=1}^\infty \lambda_{P,j}<\infty\), the series is finite \(\pr_P\)-a.s. By the Karhunen--Lo\`eve expansion, \(S_P^\sharp \stackrel{d}{=} \|G_P\|^2\). Therefore it suffices to bound
\[
\sup_{t\in\bbR}\pr_P\left(\left|S_P^\sharp-t\right|\le \delta\right).
\]

Write \(S_P^\sharp=X_P+Y_P\), where \(X_P\coloneqq \lambda_{P,1}Z_1^2\) and \(Y_P\coloneqq \sum_{j=2}^\infty \lambda_{P,j}Z_j^2\). Then \(X_P\) and \(Y_P\) are independent. For \(h>0\), let \(Q_W(h)\coloneqq \sup_{a\in\bbR}\pr_P(W\in[a,a+h])\). If \(U\) and \(V\) are independent, then
\[
Q_{U+V}(h)\le Q_U(h),
\]
because for every \(a\in\bbR\),
\[
\pr_P(U+V\in[a,a+h])
=
\E_P\left[\pr_P\left(U\in[a-V,a+h-V]\,\middle|\,V\right)\right]
\le Q_U(h).
\]
Taking the supremum over \(a\) proves the claim. Applying this with \(U=X_P\), \(V=Y_P\), and \(h=2\delta\), we obtain
\[
\sup_{t\in\bbR}\pr_P\left(\left|S_P^\sharp-t\right|\le \delta\right)
=
Q_{S_P^\sharp}(2\delta)
\le
Q_{X_P}(2\delta).
\]

It remains to bound \(Q_{X_P}(2\delta)\). Let \(Z\sim N(0,1)\) and \(Y_\lambda\coloneqq \lambda Z^2\) for \(\lambda>0\). Then \(Y_\lambda\) has density \(f_\lambda(x)=\frac{1}{\sqrt{2\pi \lambda x}}e^{-x/(2\lambda)}\mathds{1}_{\{x>0\}}\), which is strictly decreasing on \((0,\infty)\) because \(\frac{d}{dx}\log f_\lambda(x)=-\frac{1}{2x}-\frac{1}{2\lambda}<0\). Hence among all intervals of length at most \(2\delta\), the maximal mass is attained by \([0,2\delta]\), so \(Q_{Y_\lambda}(2\delta)=\pr_P(Y_\lambda\in[0,2\delta])\). Therefore
\[
Q_{X_P}(2\delta)
=
\pr_P\left(Z_1^2\le \frac{2\delta}{\lambda_{P,1}}\right)
\le
\pr_P\left(Z_1^2\le \frac{2\delta}{\underline{\lambda}}\right)
=
2\Phi\left(\sqrt{\frac{2\delta}{\underline{\lambda}}}\right)-1.
\]

Combining the preceding bounds yields
\[
\sup_{t\in\bbR}\pr_P\left(\left|\|G_P\|^2-t\right|\le \delta\right)
=
\sup_{t\in\bbR}\pr_P\left(\left|S_P^\sharp-t\right|\le \delta\right)
\le
2\Phi\left(\sqrt{\frac{2\delta}{\underline{\lambda}}}\right)-1.
\]
Since the right-hand side is independent of \(P\), taking the supremum over \(P\in\cP\) gives
\[
\sup_{P\in\cP}\sup_{t\in\bbR}
\pr_P\left(\left|\|G_P\|^2-t\right|\le \delta\right)
\le
2\Phi\left(\sqrt{\frac{2\delta}{\underline{\lambda}}}\right)-1.
\]
Letting \(\delta\downarrow 0\), the continuity of \(\Phi\) and \(\Phi(0)=1/2\) imply
\[
\lim_{\delta\downarrow 0}
\sup_{P\in\cP}\sup_{t\in\bbR}
\pr_P\left(\left|\|G_P\|^2-t\right|\le \delta\right)
=0.
\]
This completes the proof.
\end{proof}

\begin{corollary}[Uniform modulus of continuity of the distribution functions]
\label{cor:uniform_modulus_continuity_FP}
Under the assumptions of Lemma~\ref{lem:uniform_anti_concentration_gaussian_hilbert_squared_norm}, for every \(\eta>0\), we have
\[
\sup_{P\in\cP}\sup_{t\in\bbR}
\left\{F_P(t+\eta)-F_P(t-\eta)\right\}
\le
2\Phi\left(\sqrt{\frac{2\eta}{\underline{\lambda}}}\right)-1.
\]
In particular,
\[
\lim_{\eta\downarrow 0}
\sup_{P\in\cP}\sup_{t\in\bbR}
\left\{F_P(t+\eta)-F_P(t-\eta)\right\}
=0.
\]
\end{corollary}

\begin{proof}
Fix \(P\in\cP\), \(t\in\bbR\), and \(\eta>0\). By definition of \(F_P\),
\[
F_P(t+\eta)-F_P(t-\eta)
=
\pr_P\left(t-\eta<S_P\le t+\eta\right)
\le
\pr_P\left(\left|S_P-t\right|\le \eta\right).
\]
Taking \(\sup_{t\in\bbR}\) and then \(\sup_{P\in\cP}\), and applying
Lemma~\ref{lem:uniform_anti_concentration_gaussian_hilbert_squared_norm} with \(\delta=\eta\), we obtain
\[
\sup_{P\in\cP}\sup_{t\in\bbR}
\left\{F_P(t+\eta)-F_P(t-\eta)\right\}
\le
2\Phi\left(\sqrt{\frac{2\eta}{\underline{\lambda}}}\right)-1.
\]
The final limit follows immediately from the continuity of \(\Phi\) at \(0\).
\end{proof}

\begin{corollary}[Uniform anti-concentration for Gaussian Hilbert norms]
\label{cor:uniform_anti_concentration_gaussian_hilbert_norm}
If \(\sup_{P\in\cP}\E_P[\|G_P\|^2] < \infty\) in Lemma~\ref{lem:uniform_anti_concentration_gaussian_hilbert_squared_norm}, then
\[
\lim_{\rho\downarrow 0} \sup_{P\in\cP}\sup_{u\ge 0} \pr_P\left(\left|\|G_P\|-u\right|\le \rho\right) = 0.
\]
\end{corollary}

\begin{proof}
Set
\[
M \coloneqq \sup_{P\in\cP}\E_P\left[\left\|G_P\right\|^2\right] < \infty,
\qquad
\alpha(\rho)\coloneqq
\sup_{P\in\cP}\sup_{u\ge 0}
\pr_P\left(\left|\left\|G_P\right\|-u\right|\le \rho\right).
\]
Fix \(A>\rho>0\). We split the supremum in \(\alpha(\rho)\) into the regions
\(0\le u\le A\) and \(u>A\).

First consider \(0\le u\le A\). If \(\left|\left\|G_P\right\|-u\right|\le \rho\), then
\[
\left|\left\|G_P\right\|^2-u^2\right|
=
\left|\left\|G_P\right\|-u\right|
\left(\left\|G_P\right\|+u\right)
\le
\rho\left(2A+\rho\right),
\]
because \(\left|\left\|G_P\right\|-u\right|\le \rho\) and \(u\le A\) imply
\(\left\|G_P\right\|\le u+\rho\le A+\rho\). Therefore,
\[
\sup_{P\in\cP}\sup_{0\le u\le A}
\pr_P\left(\left|\left\|G_P\right\|-u\right|\le \rho\right)
\le
\sup_{P\in\cP}\sup_{t\in\bbR}
\pr_P\left(\left|\left\|G_P\right\|^2-t\right|
\le
\rho\left(2A+\rho\right)\right).
\]
By Lemma~\ref{lem:uniform_anti_concentration_gaussian_hilbert_squared_norm}, the right-hand side converges to \(0\) as \(\rho\downarrow 0\), for every fixed \(A>0\).

Next consider \(u>A\). On the event
\(\left\{\left|\left\|G_P\right\|-u\right|\le \rho\right\}\), we have
\(\left\|G_P\right\|\ge u-\rho>A-\rho\). Hence, by Markov's inequality,
\[
\sup_{P\in\cP}\sup_{u>A}
\pr_P\left(\left|\left\|G_P\right\|-u\right|\le \rho\right)
\le
\sup_{P\in\cP}
\pr_P\left(\left\|G_P\right\|>A-\rho\right)
\le
\frac{M}{(A-\rho)^2}.
\]

Combining the two bounds, we obtain
\[
\alpha(\rho)
\le
\sup_{P\in\cP}\sup_{t\in\bbR}
\pr_P\left(\left|\left\|G_P\right\|^2-t\right|
\le
\rho\left(2A+\rho\right)\right)
+
\frac{M}{(A-\rho)^2}.
\]

Now let \(\varepsilon>0\). Choose \(A>0\) so large that \(M/A^2<\varepsilon/4\). Then choose \(\rho_0\in(0,A/2)\) so small that
\(M/(A-\rho)^2<\varepsilon/2\) for all \(0<\rho<\rho_0\). Since
\(\rho(2A+\rho)\to 0\) as \(\rho\downarrow 0\), Lemma~\ref{lem:uniform_anti_concentration_gaussian_hilbert_squared_norm} yields \(\rho_1\in(0,\rho_0)\) such that
\[
\sup_{P\in\cP}\sup_{t\in\bbR}
\pr_P\left(\left|\left\|G_P\right\|^2-t\right|
\le
\rho\left(2A+\rho\right)\right)
<
\frac{\varepsilon}{2}
\qquad
\text{for all } 0<\rho<\rho_1.
\]
Hence \(\alpha(\rho)<\varepsilon\) for all \(0<\rho<\rho_1\). Since \(\varepsilon>0\) was arbitrary, this proves
\[
\lim_{\rho\downarrow 0}\alpha(\rho)=0.
\]
\end{proof}

\begin{lemma}[Uniform convergence of squared norms via bounded Lipschitz metric]
\label{lem:uniform_kolmogorov_distance_squared_norm}
Let $\cH$ be a separable Hilbert space, and let $\cP$ be a family of probability laws. For each $P\in\cP$, let $X_{n,P}$ be a random element in $\cH$, and let $G_P$ be a centered Gaussian random element in $\cH$ with covariance operator $C_P$. Assume that the covariance operators are uniformly nondegenerate, i.e.,
\(
\inf_{P\in\cP}\|C_P\|_{\mathrm{op}}>0.
\)
Let $\Delta_n \coloneqq \sup_{P\in\cP} d_{\mathrm{BL},\cH}(X_{n,P}, G_P)$. If $\lim_{n\to\infty} \Delta_n = 0$, then
\[
\lim_{n\to\infty} \sup_{P\in\cP}\sup_{t\in\bbR}
\left|
\pr_P\left(\|X_{n,P}\|^2\le t\right)
-
\pr_P\left(\|G_P\|^2\le t\right)
\right|
= 0.
\]
\end{lemma}

\begin{proof}
As \(\|X_{n,P}\|^2, \|G_P\|^2 \ge 0\) almost surely, both probabilities vanish for \(t<0\). It thus suffices to consider the supremum over \(t\ge 0\). Fix \(\rho\in(0,1]\) and \(t\ge 0\), and write \(r=\sqrt{t}\). Define
\[
\varphi_{r,\rho}^{+}(x)
\coloneqq
\left(1-\frac{(\|x\|-r)_+}{\rho}\right)_+,
\qquad
\varphi_{r,\rho}^{-}(x)
\coloneqq
\left(1-\frac{\left(\|x\|-(r-\rho)_+\right)_+}{\rho}\right)_+,
\qquad x\in\cH.
\]
Then, for every \(x\in\cH\),
\[
\mathds 1\left\{\|x\|\le (r-\rho)_+\right\}
\le
\varphi_{r,\rho}^{-}(x)
\le
\mathds 1\left\{\|x\|\le r\right\}
\le
\varphi_{r,\rho}^{+}(x)
\le
\mathds 1\left\{\|x\|\le r+\rho\right\}.
\]
Equivalently,
\[
\mathds 1\left\{\|x\|^2\le (r-\rho)_+^2\right\}
\le
\varphi_{r,\rho}^{-}(x)
\le
\mathds 1\left\{\|x\|^2\le t\right\}
\le
\varphi_{r,\rho}^{+}(x)
\le
\mathds 1\left\{\|x\|^2\le (r+\rho)^2\right\}.
\]
Moreover,
\[
\|\varphi_{r,\rho}^{\pm}\|_{\infty}\le 1,
\qquad
\mathrm{Lip}\left(\varphi_{r,\rho}^{\pm}\right)\le \rho^{-1}.
\]

We now invoke the definition of \(d_{\mathrm{BL},\cH}\). Let \(f:\cH\to\bbR\) be measurable with
\(\|f\|_{\infty}\le 1\) and \(\mathrm{Lip}(f)\le \rho^{-1}\). Set \(g\coloneqq \rho f\). Since \(\rho\in(0,1]\), we have
\[
\|g\|_{\infty}\le \rho\|f\|_{\infty}\le 1,
\qquad
\mathrm{Lip}(g)\le \rho\,\mathrm{Lip}(f)\le 1.
\]
Hence \(g\in \mathrm{BL}_1(\cH)\), and therefore, by the definition of \(\Delta_n\),
\[
\left|
\E_P\left[g(X_{n,P})\right]
-
\E_P\left[g(G_P)\right]
\right|
\le
\Delta_n.
\]
Dividing by \(\rho\), we obtain
\[
\left|
\E_P\left[f(X_{n,P})\right]
-
\E_P\left[f(G_P)\right]
\right|
\le
\rho^{-1}\Delta_n.
\]

Applying the bound to \(f=\varphi_{r,\rho}^{\pm}\) yields, respectively,
\begin{align*}
\pr_P\left(\|X_{n,P}\|^2\le t\right) &\le \E_P\left[\varphi_{r,\rho}^{+}(G_P)\right] + \rho^{-1}\Delta_n \le \pr_P\left(\|G_P\|\le r+\rho\right) + \rho^{-1}\Delta_n, \\
\pr_P\left(\|X_{n,P}\|^2\le t\right) &\ge \E_P\left[\varphi_{r,\rho}^{-}(G_P)\right] - \rho^{-1}\Delta_n \ge \pr_P\left(\|G_P\|\le (r-\rho)_+\right) - \rho^{-1}\Delta_n.
\end{align*}

Combining the two bounds yields, for every \(P\in\cP\) and \(t\ge 0\),
\[
\left|
\pr_P\left(\|X_{n,P}\|^2\le t\right)
-
\pr_P\left(\|G_P\|^2\le t\right)
\right|
\le
\rho^{-1}\Delta_n
+
\pr_P\left(\left|\|G_P\|-\sqrt{t}\right|\le \rho\right).
\]
Taking \(\sup_{t\ge 0}\) and then \(\sup_{P\in\cP}\), we obtain
\begin{equation}
\label{eq:lem_uniform_BL_to_squared_norm_basic_bound}
\sup_{P\in\cP}\sup_{t\in\bbR}
\left|
\pr_P\left(\|X_{n,P}\|^2\le t\right)
-
\pr_P\left(\|G_P\|^2\le t\right)
\right|
\le
\rho^{-1}\Delta_n+\alpha(\rho),
\end{equation}
where
\[
\alpha(\rho)
\coloneqq
\sup_{P\in\cP}\sup_{u\ge 0}
\pr_P\left(\left|\|G_P\|-u\right|\le \rho\right).
\]

By Corollary~\ref{cor:uniform_anti_concentration_gaussian_hilbert_norm}, we have
\[
\alpha(\rho) \to 0 \qquad \text{as } \rho \downarrow 0.
\]
For each fixed \(\rho\in(0,1]\), taking \(n\to\infty\) in \eqref{eq:lem_uniform_BL_to_squared_norm_basic_bound} yields
\[
\limsup_{n\to\infty} \sup_{P\in\cP}\sup_{t\in\bbR} \left| \pr_P\left(\|X_{n,P}\|^2\le t\right) - \pr_P\left(\|G_P\|^2\le t\right) \right| \le \alpha(\rho).
\]
The claim follows by letting \(\rho\downarrow 0\).
\end{proof}

\runinhead{Quantile regularity of the Gaussian limit}
\begin{lemma}[Uniform quantile regularity for the Gaussian limit family]
\label{lem:uniform_quantile_regularity_gaussian_limit}
Fix \(u\in(0,1)\). Under Assumptions~\ref{ass:fixed-basis-uniform_tail} and
\ref{ass:non_degenerate_unif}, for every \(\eta>0\),
\[
\inf_{P\in\tilde{\cP}_0}\left\{F_P\left(q_P+\eta\right)-u\right\}>0,
\qquad
\inf_{P\in\tilde{\cP}_0}\left\{u-F_P\left(q_P-\eta\right)\right\}>0.
\]
\end{lemma}
\begin{proof}
Set
\[
\cC_0\coloneqq \left\{C_P^{(\infty)}:P\in\tilde{\cP}_0\right\}\subset \cS_1^+(\cH).
\]
For \(C\in \overline{\cC_0}^{\|\cdot\|_{\mathrm{tr}}}\), define
\[
F_C(t)\coloneqq \pr\left(\left\|\cN(0,C)\right\|^2\le t\right),
\qquad
q_C\coloneqq \inf\left\{t\in\bbR:F_C(t)\ge u\right\}.
\]
Then, for \(P\in\tilde{\cP}_0\), we have \(F_P=F_{C_P^{(\infty)}}\) and
\(q_P=q_{C_P^{(\infty)}}\).

We first verify that \(\overline{\cC_0}^{\|\cdot\|_{\mathrm{tr}}}\) is compact. By the
reproducing property and Jensen's inequality,
\(\|\Psi\|^2\le 16K_X^2K_Y^2K_Z^2\). Hence
\[
\sup_{P\in\tilde{\cP}_0}\left\|C_P^{(\infty)}\right\|_{\mathrm{tr}}
=
\sup_{P\in\tilde{\cP}_0}\E_P\left[\left\|\Pi_\infty\Psi\right\|^2\right]
\le
\sup_{P\in\tilde{\cP}_0}\E_P\left[\left\|\Psi\right\|^2\right]
\le
16K_X^2K_Y^2K_Z^2<\infty.
\]
Let \(\{\tilde e_p\}_{p=1}^{L_\infty^X}\), \(\{\tilde f_q\}_{q=1}^{L_\infty^Y}\), and
\(\{\tilde g_r\}_{r\ge 1}\) be as in
Assumption~\ref{ass:fixed-basis-uniform_tail}, and set
\(u_{pqr}\coloneqq \tilde e_p\otimes \tilde f_q\otimes \tilde g_r\). For
\(P\in\tilde{\cP}_0\), define
\(a_{pqr}^{(P)}\coloneqq \left\langle \Pi_\infty\Psi,u_{pqr}\right\rangle\). Then
\[
\left\langle C_P^{(\infty)}u_{pqr},u_{pqr}\right\rangle
=
\E_P\left[\left(a_{pqr}^{(P)}\right)^2\right],
\qquad
a_{pqr}^{(P)}
=
\left\langle \epsilon_{X\mid Z},\tilde e_p\right\rangle
\left\langle \epsilon_{Y\mid Z},\tilde f_q\right\rangle
\tilde g_r(Z).
\]
Using
\[
\sum_{q=1}^{L_\infty^Y}\left\langle \epsilon_{Y\mid Z},\tilde f_q\right\rangle^2
\le \left\|\epsilon_{Y\mid Z}\right\|^2\le 4K_Y^2,
\qquad
\sum_{r=1}^\infty \tilde g_r(Z)^2
=
\left\|k_{\cZ}(Z,\cdot)\right\|^2
\le K_Z^2,
\]
we obtain
\[
\sum_{p>K}\sum_{q,r}\E_P\left[\left(a_{pqr}^{(P)}\right)^2\right]
\le
4K_Y^2K_Z^2\,
\E_P\left[\sum_{p>K}\left\langle \epsilon_{X\mid Z},\tilde e_p\right\rangle^2\right].
\]
Now write \(m_p(Z)\coloneqq \E_P[\tilde e_p(X)\mid Z]\). Since
\(\left\langle \epsilon_{X\mid Z},\tilde e_p\right\rangle=\tilde e_p(X)-m_p(Z)\),
\[
\sum_{p>K}\left\langle \epsilon_{X\mid Z},\tilde e_p\right\rangle^2
\le
2\sum_{p>K}\tilde e_p(X)^2+2\sum_{p>K}m_p(Z)^2.
\]
Taking expectations and using Jensen's inequality yields
\(\E_P\left[\sum_{p>K}m_p(Z)^2\right]
\le
\E_P\left[\sum_{p>K}\tilde e_p(X)^2\right]\), hence
\[
\sup_{P\in\tilde{\cP}_0}
\sum_{p>K}\sum_{q,r}\E_P\left[\left(a_{pqr}^{(P)}\right)^2\right]
\le
16K_Y^2K_Z^2\,
\sup_{P\in\tilde{\cP}_0}\sum_{p>K}\E_P\left[\tilde e_p(X)^2\right].
\]
By the same argument,
\[
\sup_{P\in\tilde{\cP}_0}
\sum_{p,q>K,r}\E_P\left[\left(a_{pqr}^{(P)}\right)^2\right]
\le
16K_X^2K_Z^2\,
\sup_{P\in\tilde{\cP}_0}\sum_{q>K}\E_P\left[\tilde f_q(Y)^2\right],
\]
and
\[
\sup_{P\in\tilde{\cP}_0}
\sum_{p,q,r>K}\E_P\left[\left(a_{pqr}^{(P)}\right)^2\right]
\le
16K_X^2K_Y^2\,
\sup_{P\in\tilde{\cP}_0}\sum_{r>K}\E_P\left[\tilde g_r(Z)^2\right].
\]
Assumption~\ref{ass:fixed-basis-uniform_tail} therefore implies
\[
\sup_{P\in\tilde{\cP}_0}
\sum_{\max\{p,q,r\}>K}
\left\langle C_P^{(\infty)}u_{pqr},u_{pqr}\right\rangle
\to 0
\qquad
(K\to\infty).
\]
Together with the uniform trace bound, this yields the relative compactness of
\(\cC_0\) in trace norm, so \(\overline{\cC_0}^{\|\cdot\|_{\mathrm{tr}}}\) is compact.

Next, Assumption~\ref{ass:non_degenerate_unif} gives
\[
\inf_{P\in\tilde{\cP}_0}\left\|C_P^{(\infty)}\right\|_{\mathrm{op}}>0.
\]
Since \(\|\cdot\|_{\mathrm{op}}\le \|\cdot\|_{\mathrm{tr}}\), the operator norm is
continuous with respect to the trace norm. Hence
\[
\inf_{C\in \overline{\cC_0}^{\|\cdot\|_{\mathrm{tr}}}}\left\|C\right\|_{\mathrm{op}}>0.
\]

Fix \(C\in \overline{\cC_0}^{\|\cdot\|_{\mathrm{tr}}}\), and let
\(S_C\coloneqq \left\|\cN(0,C)\right\|^2\). Since \(\left\|C\right\|_{\mathrm{op}}>0\),
we may write
\[
S_C=\lambda_1(C)Z_1^2+R_C,
\]
where \(\lambda_1(C)=\left\|C\right\|_{\mathrm{op}}>0\), \(Z_1\sim \cN(0,1)\), and
\(R_C\ge 0\) is independent of \(Z_1\). It follows that the law of \(S_C\) is
continuous. Therefore, for every \(\eta>0\),
\[
F_C\left(q_C+\eta\right)>u,
\qquad
F_C\left(q_C-\eta\right)<u.
\]

Finally, by Lemma 12 in \cite{Lundborgetal2022}, the map \(C\mapsto q_C\) is
continuous with respect to the trace norm. Moreover, if \(C_n\to C\) in trace norm,
then \(\left\|N(0,C_n)\right\|^2\Rightarrow \left\|N(0,C)\right\|^2\), so
\(F_{C_n}(t)\to F_C(t)\) for every \(t\in\bbR\), because \(F_C\) is continuous.
Hence the maps
\[
C\mapsto F_C\left(q_C+\eta\right)-u,
\qquad
C\mapsto u-F_C\left(q_C-\eta\right)
\]
are continuous on the compact set \(\overline{\cC_0}^{\|\cdot\|_{\mathrm{tr}}}\).
Since they are strictly positive everywhere on this set, they attain strictly
positive minima there. Restricting back to \(C=C_P^{(\infty)}\) yields
\[
\inf_{P\in\tilde{\cP}_0}\left\{F_P\left(q_P+\eta\right)-u\right\}>0,
\qquad
\inf_{P\in\tilde{\cP}_0}\left\{u-F_P\left(q_P-\eta\right)\right\}>0.
\]
This completes the proof.
\end{proof}
\subsection{Local alternatives under diverging truncation}
\label{subsec:local_power_diverging}
Let \(\{P_n\}\subset \tilde{\cP}_1\) be a sequence of alternatives, and write
\[
\Sigma_n\coloneqq \Sigma_{XYZ\mid Z,P_n}.
\]

\begin{assumption}[Local-alternative regime under diverging truncation]
\label{ass:local_power_diverging}
There exist \(\gamma>0\) and \(h\in \cH_{XYZ}\) such that
\[
n_1^\gamma \Sigma_n \to h
\qquad\text{in } \cH_{XYZ}.
\]
\end{assumption}

\begin{theorem}[Local power under diverging truncation]
\label{thm:local_power_diverging}
Suppose Assumption~\ref{ass:local_power_diverging} holds, and that the sequence of local alternatives $P_n$ satisfies Assumptions~2--5 pointwise. Assume further that all conditional expectations in $\hat T_{n_1}$ and $\hat T_{n_1}^{\ast}$ are estimated from an auxiliary sample independent of the data used to construct the test and bootstrap statistics.

\begin{enumerate}
\item
If $\gamma<1/2$, then
\begin{equation}
\pr_{P_n}\bigl(\phi_n^{(B_n)}=1\bigr)\to 1.
\end{equation}

\item
If $\gamma=1/2$, then
\begin{equation}
n_1\hat T_{n_1}\overset{d}{\to} \|Z+h\|^2,
\end{equation}
and hence
\begin{equation}
\pr_{P_n}\bigl(\phi_n^{(B_n)}=1\bigr)
\to
\pr\bigl(\|Z+h\|^2>q_F(1-\alpha)\bigr).
\end{equation}

\item
If $\gamma>1/2$, then $n_1^\gamma \Sigma_n\to h$ implies
\begin{equation}
\sqrt{n_1}\Sigma_n\to 0,
\end{equation}
so the signal is asymptotically negligible on the $n_1^{-1/2}$-scale. In particular, no nontrivial local power conclusion can be deduced from this assumption alone.
\end{enumerate}
\end{theorem}
\begin{remark}
    Note that under our pointwise asymptotic framework along the sequence $P_n$, the requirements of Assumption~6 become substantially weaker compared to a uniform setting.
\end{remark}

\begin{proof}
First, we address the estimation of the conditional expectations. By assumption, all conditional expectations in $\hat T_{n_1}$ and $\hat T_{n_1}^{\ast}$ are estimated using an independent auxiliary sample. Conditional on this auxiliary sample, the estimated conditional expectations act as fixed functions. Since $P_n$ satisfies Assumptions~2--5 pointwise, the convergence rates of these estimators are sufficiently fast such that the bias and estimation error introduced by replacing the true conditional expectations with their estimates are of order $o_{P_n}(n_1^{-1/2})$. Consequently, their contribution to the test statistic is asymptotically negligible, and we can analyze the leading terms as if the true conditional expectations were known, up to an $o_{P_n}(1)$ remainder in $n_1\hat T_{n_1}$.

Next, we work under the condition that $\|\hat\Pi_n - \Pi_\infty\|_{\mathrm{op}} = o_{P_n}(1)$. By the definition, $\Pi_\infty \Sigma_n = \Sigma_n$ for all $n$, meaning the signal lies entirely within the image of $\Pi_\infty$. From the local alternative assumption
\begin{equation}
n_1^\gamma \Sigma_n\to h, \quad \text{in $\cH_{XYZ}$.}
\end{equation}

By the asymptotic negligibility of the conditional expectation estimation errors, the test statistic satisfies
\begin{equation}
n_1\hat T_{n_1}
=
\left\|
\sqrt{n_1}\hat\Pi_n\Sigma_n
+
\hat\Pi_n\left(
n_1^{-1/2}\sum_{i\in I_1}(\Psi_i-\Sigma_n)
\right)
\right\|^2 + o_{P_n}(1).
\end{equation}
Since $\Pi_\infty \Sigma_n = \Sigma_n$, the projected signal decomposes as $\sqrt{n_1}\hat\Pi_n\Sigma_n = \sqrt{n_1}\Sigma_n + \sqrt{n_1}(\hat\Pi_n-\Pi_\infty)\Sigma_n$. By Assumption~2, $\|\hat\Pi_n-\Pi_\infty\|_{\mathrm{op}} = o_{P_n}(1)$, which implies that the remainder $\sqrt{n_1}(\hat\Pi_n-\Pi_\infty)\Sigma_n$ is $o_{P_n}(1)$ provided $\sqrt{n_1}\|\Sigma_n\|=O(1)$. We now analyze the three regimes for $\gamma$.

If $\gamma=1/2$, then $\sqrt{n_1}\Sigma_n = n_1^{1/2}\Sigma_n \to h$. Thus, the remainder vanishes, yielding $\sqrt{n_1}\hat\Pi_n\Sigma_n \to h$ in $P_n$-probability. Since the centered empirical process $\hat\Pi_n( n_1^{-1/2}\sum_{i\in I_1}(\Psi_i-\Sigma_n) ) \overset{d}{\to} Z$ under $P_n$, Slutsky's theorem and the continuous mapping theorem imply $n_1\hat T_{n_1} \overset{d}{\to} \|Z+h\|^2$. Combined with the bootstrap critical value convergence $Q_{n,1-\alpha}^{\ast(B_n)} \to q_{F_{P_n}}(1-\alpha)$ in probability (Theorem~\ref{thm:abstract-transfer-squared-norm-bootstrap}), we obtain
\begin{equation}
\pr_{P_n}\bigl(\phi_n^{(B_n)}=1\bigr) \to \pr\bigl(\|Z+h\|^2>q_{F_{P_n}}(1-\alpha)\bigr).
\end{equation}

If $\gamma<1/2$ and $h \neq 0$, then $\sqrt{n_1}\|\Sigma_n\| = n_1^{1/2-\gamma}\|n_1^\gamma\Sigma_n\| \to \infty$. By the reverse triangle inequality, $\sqrt{n_1}\|\hat\Pi_n\Sigma_n\| \ge \sqrt{n_1}\|\Sigma_n\| ( 1 - \|\hat\Pi_n-\Pi_\infty\|_{\mathrm{op}} ) \to \infty$ in $P_n$-probability. Since the signal strictly dominates the $O_{P_n}(1)$ centered process and the $O_{P_n}(1)$ bootstrap critical value, the test is consistent:
\begin{equation}
\pr_{P_n}\bigl(\phi_n^{(B_n)}=1\bigr) \to 1.
\end{equation}

Finally, if $\gamma>1/2$, then $\sqrt{n_1}\Sigma_n = n_1^{1/2-\gamma}(n_1^\gamma\Sigma_n) \to 0$. The signal vanishes on the testing scale, meaning no nontrivial local-power conclusion can be deduced from this assumption alone.
\end{proof}

\begin{remark}
The case \(\gamma=1/2\) is the genuine boundary local-alternative regime. The cases \(\gamma<1/2\) and \(\gamma>1/2\) correspond, respectively, to supercritical signals leading to consistency and subcritical signals that are asymptotically negligible on the \(n_1^{-1/2}\)-scale.
\end{remark}

\section{A conditional central limit theorem for the Hilbert-space-valued bootstrap}

In this section, we establish the conditional weak convergence of the bootstrap distribution in a Hilbert space \(\cH\). To make the logical dependence of the argument transparent, we organize our analysis into two modular components.

First, we formulate an abstract transfer theorem. This result demonstrates that the conditional weak convergence of an \(\cH\)-valued bootstrap law to a Gaussian limit implies the uniform validity of the bootstrap distribution for squared norms, applicable to both distribution functions and quantiles.

Second, we provide an abstract verification theorem outlining how to establish this required conditional weak convergence in \(\cH\). We show that it can be systematically deduced from scalar conditional central limit theorems, covariance matching, conditional uniform tightness, and Gaussian covariance replacement.

This modular organization cleanly isolates the abstract transfer mechanism from the model-specific verification steps, clarifying the overarching structure of the bootstrap argument.

\subsection{Common setup and abstract transfer argument}

Throughout this subsection, let \(\cH\) be a separable Hilbert space. For each \(P\in\cP\), let \((\Omega_0,\cF_0,\pr_P)\) be a probability space carrying the data, and let \(\cD_n\subset \cF_0\) be a data \(\sigma\)-field for each \(n\ge 1\). Let \((\Omega_\ast,\cF_\ast,\pr^\ast)\) be an auxiliary probability space, independent of \(P\), carrying the bootstrap randomness. We work on the product extension
\[
(\Omega,\cF,\pr_P)
\coloneqq
(\Omega_0\times\Omega_\ast,\cF_0\otimes\cF_\ast,\pr_P\otimes\pr^\ast),
\]
and, by abuse of notation, identify \(\cD_n\) with \(\cD_n\otimes \{\varnothing,\Omega_\ast\}\subset\cF\). For each \(n\ge 1\) and \(P\in\cP\), let \(X_{n,P}^\ast\) be an \(\cH\)-valued bootstrap random element on \((\Omega,\cF,\pr_P)\), and define
\[
\pi_{n,P}
\coloneqq
\cL_P^\ast(X_{n,P}^\ast)
\coloneqq
\pr_P^\ast\left(X_{n,P}^\ast\in\cdot\mid \cD_n\right),
\]
where \(\cL_P^\ast(X_{n,P}^\ast)\) denotes a regular conditional law of \(X_{n,P}^\ast\) given \(\cD_n\). Also let
\[
\gamma_P\coloneqq \cN(0,C_P)
\]
denote the centered Gaussian probability measure on \(\cH\) with covariance operator \(C_P\).

Let \(\{\rho_{n,P}^\ast:n\ge 1,\ P\in\cP\}\) be a family of \(\cD_n\)-measurable random Borel probability measures on \(\cH\). We say that \(\{\rho_{n,P}^\ast:n\ge 1,\ P\in\cP\}\) is \emph{conditionally uniformly tight over \(\cP\)} if, for every \(\varepsilon>0\) and every \(\delta>0\), there exists a compact set \(K\subset \cH\) such that
\begin{equation}
\label{eq:def-cond-unif-tight}
\limsup_{n\to\infty}\sup_{P\in\cP}
\pr_P\left(\rho_{n,P}^\ast(K^c)>\delta\right)<\varepsilon.
\end{equation}

For \(t\in\bbR\), let \(B_t \coloneqq \{x\in\cH : \|x\|^2\le t\}\) and define
\[
\hat F_{n,P}^\ast(t) \coloneqq \pi_{n,P}(B_t), \qquad F_P(t) \coloneqq \gamma_P(B_t).
\]

Fix a level \(u\in(0,1)\), and define the target Gaussian quantile
\[
q_P
\coloneqq
\inf\left\{t\in\bbR:F_P(t)\ge u\right\}.
\]

When finite-\(B_n\) bootstrap resampling is used, we work on a further product extension, not relabeled, carrying conditionally i.i.d.\ copies
\[
X_{n,P}^{\ast(1)},\dots,X_{n,P}^{\ast(B_n)}
\]
given \(\cD_n\), each with conditional law \(\pi_{n,P}\). Define
\[
\hat F_{n,P}^{\ast(B_n)}(t)
\coloneqq
\frac{1}{B_n}\sum_{b=1}^{B_n}
\mathds{1}\left\{
\left\|X_{n,P}^{\ast(b)}\right\|^2\le t
\right\},
\qquad t\in\bbR,
\]
together with its generalized inverse
\[
\hat q_{n,P}^{\ast(B_n)}
\coloneqq
\inf\left\{t\in\bbR:\hat F_{n,P}^{\ast(B_n)}(t)\ge u\right\}.
\]

The abstract transfer theorem below will be formulated under the following three conditions.

\begin{assumption}[Conditional Hilbert-space BL convergence]
\label{ass:bootstrap-BL-transfer}
For every \(\eta>0\),
\[
\sup_{P\in\cP}
\pr_P\left(
d_{\mathrm{BL},\cH}\left(
\pi_{n,P},\,
\gamma_P
\right)>\eta
\right)\to 0
\qquad (n\to\infty).
\]
\end{assumption}

\begin{assumption}[Uniform anti-concentration for the Gaussian norm]
\label{ass:bootstrap-AC-transfer}
Assume that
\[
\alpha(\rho) \coloneqq \sup_{P\in\cP}\sup_{u\ge 0} \gamma_P\left(\left\{x\in\cH : \left|\|x\|-u\right|\le \rho\right\}\right) \to 0
\qquad \text{as } \rho\downarrow 0.
\]
\end{assumption}

\begin{assumption}[Uniform quantile regularity at level \(u\)]
\label{ass:bootstrap-QR-transfer}
For every \(\eta>0\),
\[
\inf_{P\in\cP}
\left\{
F_P\left(q_P+\eta\right)-u
\right\}
>0,
\qquad
\inf_{P\in\cP}
\left\{
u-F_P\left(q_P-\eta\right)
\right\}
>0.
\]
\end{assumption}

\begin{theorem}[Abstract transfer for bootstrap validity of squared norms]
\label{thm:abstract-transfer-squared-norm-bootstrap}
If Assumptions~\ref{ass:bootstrap-BL-transfer} and \ref{ass:bootstrap-AC-transfer} hold and \(B_n\to\infty\), then for every \(\varepsilon>0\),
\[
\sup_{P\in\cP}
\pr_P\left(
\sup_{t\in\bbR}
\left|
\hat F_{n,P}^{\ast(B_n)}(t)-F_P(t)
\right|>\varepsilon
\right)\to 0
\qquad (n\to\infty).
\]

If, moreover, Assumption~\ref{ass:bootstrap-QR-transfer} holds, then for every \(\eta>0\),
\[
\sup_{P\in\cP}
\pr_P\left(
\left|
\hat q_{n,P}^{\ast(B_n)}-q_P
\right|>\eta
\right)\to 0
\qquad (n\to\infty).
\]
\end{theorem}

\subsection{Verification of conditional Hilbert-space weak convergence}

We now specialize the bootstrap element to the multiplier-bootstrap form
\[
S_{n,P}^\ast
\coloneqq
\frac{1}{\sqrt n}\sum_{i=1}^n W_i^\ast U_{n,i,P}.
\]
Thus, in Theorem~\ref{thm:abstract-transfer-squared-norm-bootstrap}, we take
\[
X_{n,P}^\ast=S_{n,P}^\ast.
\]
Let
\[
\nu_{n,P}^\ast \coloneqq \cN(0,\widehat C_{n,P}),
\]
where \(\widehat C_{n,P}\) is an \(\cF\)-measurable self-adjoint positive trace-class operator on \(\cH\). The Gaussian limit law remains \(\gamma_P\), with covariance operator \(C_P\) as in the common setup.

The verification theorem below will be formulated under the following conditions.

\begin{assumption}[Uniform scalar conditional CLT in every fixed direction]
\label{ass:bootstrap-verification-V1}
For every \(h\in\cH\) and every \(\eta>0\),
\[
\sup_{P\in\cP}
\pr_P\left(
d_{\mathrm{BL},\bbR}\left(
\cL_P^\ast\left(\left\langle S_{n,P}^\ast,h\right\rangle\right),\,
N\left(0,\frac{1}{n}\sum_{i=1}^n \left\langle U_{n,i,P},h\right\rangle^2\right)
\right)>\eta
\right)\to 0
\qquad (n\to\infty).
\]
\end{assumption}

\begin{remark}
Assumption~\ref{ass:bootstrap-verification-V1} is implied by the scalar Lindeberg condition in
Proposition~\ref{prop:uniform-conditional-scalar-clt-fixed-h-P}.
\end{remark}

\begin{assumption}[Covariance matching for scalar projections]
\label{ass:bootstrap-verification-V2}
For every \(h\in\cH\) and every \(P \in \cP\)
\[
\frac{1}{n}\sum_{i=1}^n \left\langle U_{n,i,P},h\right\rangle^2
=
\left\langle \widehat C_{n,P}h,h\right\rangle
\qquad
\pr_P\text{-a.s.}
\]
\end{assumption}

\begin{assumption}[Conditional uniform tightness via a fixed orthonormal basis]
\label{ass:bootstrap-verification-V3}
Assumptions \emph{(T1)}--\emph{(T3)} of
Proposition~\ref{prop:conditional-uniform-tightness-fixed-onb-P}
hold for the random measures \(\pi_{n,P}\), \(\nu_{n,P}^\ast\), and the operator family \(\widehat C_{n,P}\).
\end{assumption}

\begin{assumption}[Uniform trace-norm covariance replacement]
\label{ass:bootstrap-verification-V4}
For every \(\varepsilon>0\),
\[
\sup_{P\in\cP}
\pr_P\left(
\left\|\widehat C_{n,P}-C_P\right\|_{\mathrm{tr}}>\varepsilon
\right)\to 0
\qquad (n\to\infty).
\]
\end{assumption}

\begin{theorem}[Verification of conditional Hilbert-space BL convergence]
\label{thm:verification-conditional-hilbert-BL}
Adopt the multiplier-bootstrap setup above. If
Assumptions~\ref{ass:bootstrap-verification-V1}--\ref{ass:bootstrap-verification-V4}
hold, then for every \(\eta>0\),
\[
\sup_{P\in\cP}
\pr_P\left(
d_{\mathrm{BL},\cH}\left(
\pi_{n,P},\,
\gamma_P
\right)>\eta
\right)\to 0
\qquad (n\to\infty).
\]
In particular, Assumption~\ref{ass:bootstrap-BL-transfer} holds with \(X_{n,P}^\ast=S_{n,P}^\ast\).
\end{theorem}

\subsection{Auxiliary results}

We now collect the technical ingredients used in the proofs of
Theorems~\ref{thm:abstract-transfer-squared-norm-bootstrap}
and \ref{thm:verification-conditional-hilbert-BL}.

\begin{proposition}[BL-to-CDF transfer for squared norms]
\label{prop:conditional_BL_H_to_squared_norm_cdf_gaussian}
Adopt the common setup of Theorem~\ref{thm:abstract-transfer-squared-norm-bootstrap}. Assume that Assumptions~\ref{ass:bootstrap-BL-transfer} and \ref{ass:bootstrap-AC-transfer} hold. Then, for every \(\varepsilon>0\),
\[
\sup_{P\in\cP}
\pr_P\left(
\sup_{t\in\bbR}
\left|
\hat F_{n,P}^\ast(t)-F_P(t)
\right|
>\varepsilon
\right)\to 0.
\]
\end{proposition}

\begin{proof}
For each \(P\in\cP\), define
\(
\Delta_{n,P}
\coloneqq
d_{\mathrm{BL},\cH}\left(
\pi_{n,P},\,
\gamma_P
\right)
\).

Repeating the argument leading to \eqref{eq:lem_uniform_BL_to_squared_norm_basic_bound}, with
\(\cL_P(X_{n,P})\) replaced by the conditional law \(\pi_{n,P}\), we obtain that, for every
\(\rho\in(0,1]\) and every \(P\in\cP\),
\begin{equation}
\label{eq:conditional_BL_H_to_squared_norm_cdf_basic_bound}
\sup_{t\in\bbR}
\left|
\hat F_{n,P}^\ast(t)-F_P(t)
\right|
\le
\rho^{-1}\Delta_{n,P}
+
\alpha(\rho),
\end{equation}
where \(\alpha(\rho)\) is as in Assumption~\ref{ass:bootstrap-AC-transfer}. Here we use that
\[
\int_{\cH} f\,\mathrm d\pi_{n,P}
=
\E_P\left[f(X_{n,P}^\ast)\mid \cD_n\right]
\]
for bounded measurable \(f\).

By Assumption~\ref{ass:bootstrap-AC-transfer},
\(
\alpha(\rho)\to 0
,\enspace\rho\downarrow 0
\). Fix \(\varepsilon>0\). Choose \(\rho\in(0,1]\) so that \(\alpha(\rho)<\varepsilon/2\). Then
\eqref{eq:conditional_BL_H_to_squared_norm_cdf_basic_bound} implies
\[
\left\{
\sup_{t\in\bbR}
\left|
\hat F_{n,P}^{\ast}(t)-F_P(t)
\right|
>\varepsilon
\right\}
\subseteq
\left\{
\Delta_{n,P}>\frac{\varepsilon\rho}{2}
\right\}.
\]
Hence
\[
\sup_{P\in\cP}
\pr_P\left(
\sup_{t\in\bbR}
\left|
\hat F_{n,P}^{\ast}(t)-F_P(t)
\right|
>\varepsilon
\right)
\le
\sup_{P\in\cP}
\pr_P\left(
\Delta_{n,P}>\frac{\varepsilon\rho}{2}
\right).
\]
The right-hand side converges to \(0\) by the assumed conditional BL convergence, applied with
\(\eta=\varepsilon\rho/2\). This proves the claim.
\end{proof}

\begin{theorem}[Abstract conditional uniform weak convergence]
\label{thm:abstract-assembly-conditional-uniform-hilbert-P}
For each \(n\ge 1\) and \(P\in\cP\), let
\(
\pi_{n,P},\lambda_{n,P}
\)
be \(\cF\)-measurable random Borel probability measures on \(\cH\). For \(h\in \cH\), define \(\ell_h:\cH\to\bbR\) by \(\ell_h(x)\coloneqq \langle x,h\rangle\). Assume the following:
\begin{enumerate}
\item[(i)] For every \(h\in \cH\) and every \(\eta>0\),
\begin{equation}\label{eq:abstract-A-scalar-all-h-P}
\sup_{P\in\cP}
\pr_P\left(
d_{\mathrm{BL},\bbR}\left((\ell_h)_\#\pi_{n,P},\ (\ell_h)_\#\lambda_{n,P}\right)>\eta
\right)\to 0
\qquad (n\to\infty).
\end{equation}

\item[(ii)] Both families \(\{\pi_{n,P}:n\ge 1,\ P\in\cP\}\) and \(\{\lambda_{n,P}:n\ge 1,\ P\in\cP\}\) are conditionally uniformly tight over \(\cP\).
\end{enumerate}

Then, for every \(\eta>0\),
\begin{equation}\label{eq:abstract-conclusion-unif-BL}
\sup_{P\in\cP}
\pr_P\left(
d_{\mathrm{BL},\cH}(\pi_{n,P},\lambda_{n,P})>\eta
\right)\to 0
\qquad (n\to\infty).
\end{equation}
\end{theorem}

\begin{proof}
We argue by contradiction and assume \eqref{eq:abstract-conclusion-unif-BL} fails. Then there exist \(\eta_0>0\), \(\varepsilon_0>0\), and a sequence \(\{P^{(n)}\}_{n\ge 1}\subset\cP\) such that
\[
\pr_{P^{(n)}}\left(d_{\mathrm{BL},\cH}(\pi_{n,P^{(n)}},\lambda_{n,P^{(n)}})>\eta_0\right)\ge \varepsilon_0
\qquad\text{for all }n\ge 1.
\]
To construct a suitable separating class, fix an orthonormal basis \(\{e_j\}_{j\ge 1}\) of \(\cH\) and enumerate the countable set
\[
\cH_0\coloneqq
\left\{
\sum_{j=1}^m t_j e_j:\ m\ge 1,\ t_1,\dots,t_m\in\bbQ
\right\}
\]
as \(\{h_1,h_2,\dots\}\). Setting \(\delta_r\coloneqq 2^{-(r+2)}\) for \(r\ge 1\), conditional uniform tightness ensures the existence of compact sets \(K_r^\pi,K_r^\lambda\subset \cH\) such that
\[
\limsup_{n\to\infty}\sup_{P\in\cP}
\pr_P\left(\pi_{n,P}\left((K_r^\pi)^c\right)>\delta_r\right)<\delta_r,
\]
and similarly for \(\lambda_{n,P}\). Letting \(K_r\coloneqq K_r^\pi\cup K_r^\lambda\), it follows that \(K_r\) is compact and \(\limsup_{n\to\infty}\sup_{P\in\cP} \pr_P(\pi_{n,P}(K_r^c)>\delta_r)<\delta_r\), with an analogous bound for \(\lambda_{n,P}\). Defining \(\Delta_{n,P}^{(r)} \coloneqq d_{\mathrm{BL}}((\ell_{h_r})_\#\pi_{n,P},\ (\ell_{h_r})_\#\lambda_{n,P})\), condition \eqref{eq:abstract-A-scalar-all-h-P} implies \(\sup_{P\in\cP}\pr_P(\Delta_{n,P}^{(r)}>a)\to 0\) for every \(a>0\). We can thus recursively choose a strictly increasing subsequence \(n_k\) such that for all \(1\le r\le k\),
\[
\sup_{P\in\cP}\pr_P\left(\Delta_{n_k,P}^{(r)}>2^{-k}\right)\le 2^{-k},
\qquad
\sup_{P\in\cP}\pr_P\left(\pi_{n_k,P}(K_r^c)>\delta_r\right)\le 2^{-k},
\]
and \(\sup_{P\in\cP}\pr_P(\lambda_{n_k,P}(K_r^c)>\delta_r)\le 2^{-k}\).

Next, we realize these sequences almost surely on a product space. Letting \((\Omega_k,\cF_k,\overline{\pr}_k) \coloneqq (\Omega,\cF,\pr_{P^{(n_k)}})\), we define
\[
(\overline\Omega,\overline{\cF},\overline{\pr})
\coloneqq
\bigotimes_{k=1}^\infty (\Omega_k,\cF_k,\overline{\pr}_k)
\]
with coordinate projections \(\pr_k\). Realizing the row-wise quantities on \(\overline\Omega\) via pullback, the summability of the previous probability bounds allows us to apply the Borel--Cantelli lemma. This guarantees the existence of a set \(\Omega_0\in\overline{\cF}\) with \(\overline{\pr}(\Omega_0)=1\) such that, for every \(\omega\in\Omega_0\) and \(r\ge 1\), \(\Delta_k^{(r)}(\omega)\to 0\) and eventually
\[
\widetilde\pi_k(\omega)(K_r^c)\le \delta_r,
\qquad
\widetilde\lambda_k(\omega)(K_r^c)\le \delta_r,
\]
where
\[
\widetilde\pi_k(\omega)\coloneqq \pi_{n_k,P^{(n_k)}}\bigl(\pr_k(\omega)\bigr),
\qquad
\widetilde\lambda_k(\omega)\coloneqq \lambda_{n_k,P^{(n_k)}}\bigl(\pr_k(\omega)\bigr).
\]
Because \(K_r\) is compact and \(\delta_r\downarrow 0\), these latter bounds imply that both deterministic sequences \(\{\widetilde\pi_k(\omega)\}_{k\ge 1}\) and \(\{\widetilde\lambda_k(\omega)\}_{k\ge 1}\) are tight in \(\cP(\cH)\) for any fixed \(\omega\in\Omega_0\).

Consequently, for any such \(\omega\) and any subsequence of indices, Prokhorov's theorem \citep[Theorem~11.5.4]{Dudley2002} yields a further subsequence (not relabeled) and limit measures \(\Pi,\Lambda\in\cP(\cH)\) such that \(\widetilde\pi_k(\omega)\to_w \Pi\) and \(\widetilde\lambda_k(\omega)\to_w \Lambda\). Since \(\ell_{h_r}\) is continuous, this weak convergence pushes forward to \(\bbR\). Combined with \(\Delta_k^{(r)}(\omega)\to 0\), which metrizes weak convergence on \(\bbR\), we obtain
\[
(\ell_{h_r})_\#\Pi=(\ell_{h_r})_\#\Lambda
\qquad\text{for every }r\ge 1.
\]
By considering the finite-dimensional projections \(X^{(m)}(x)\coloneqq (\langle x,e_1\rangle,\dots,\langle x,e_m\rangle)\), this one-dimensional equality extends via characteristic functions to
\[
(X^{(m)})_\#\Pi=(X^{(m)})_\#\Lambda
\qquad\text{for every }m\ge 1.
\]
As the Borel \(\sigma\)-field on \(\cH\) is generated by these coordinate maps, we conclude that \(\Pi=\Lambda\).

This identification of limits forces
\[
d_{\mathrm{BL},\cH}(\widetilde\pi_k(\omega),\widetilde\lambda_k(\omega))\to 0.
\]
Indeed, if this failed, there would exist \(\gamma>0\) and a subsequence \(k_j\) for which the distance is bounded below by \(\gamma\); however, by the tightness established above, a further subsequence would converge to identical limits \(\widetilde\Pi=\widetilde\Lambda\), driving the \(d_{\mathrm{BL},\cH}\) distance to zero and producing a contradiction. Since this pathwise convergence holds for all \(\omega\in\Omega_0\), it holds \(\overline{\pr}\)-a.s., and in particular in \(\overline{\pr}\)-probability. Finally, because the \(k\)-th coordinate marginal of \(\overline{\pr}\) is precisely \(\pr_{P^{(n_k)}}\), this implies
\[
\pr_{P^{(n_k)}}\left(
d_{\mathrm{BL},\cH}(\pi_{n_k,P^{(n_k)}},\lambda_{n_k,P^{(n_k)}})>\eta_0
\right)\to 0,
\]
which contradicts our initial assumption along the subsequence \(n_k\). This completes the proof.
\end{proof}

\begin{proposition}[Uniform conditional scalar CLT for a fixed direction]
\label{prop:uniform-conditional-scalar-clt-fixed-h-P}
Fix \(h\in\cH\). Define \(a_{n,i,P}(h) \coloneqq \langle U_{n,i,P},h\rangle\), \(s_{n,P}^2(h) \coloneqq \frac{1}{n}\sum_{i=1}^n a_{n,i,P}(h)^2\), and for \(\delta>0\),
\begin{equation}
\label{eq:scalar-UL-P}
L_{n,P}(h,\delta)
\coloneqq
\frac{1}{n}\sum_{i=1}^n a_{n,i,P}(h)^2
\,\E^\ast\left[
(W_1^\ast)^2
\mathds 1\left\{
|a_{n,i,P}(h)W_1^\ast| > \delta\sqrt{n}
\right\}
\right].
\end{equation}
If \(\sup_{P\in\cP}\pr_P(L_{n,P}(h,\delta)>t) \to 0\) as \(n\to\infty\) for all \(\delta, t > 0\), then for every \(\eta>0\),
\[
\sup_{P\in\cP}
\pr_P\left(
d_{\mathrm{BL},\bbR}\left(
\cL_P^\ast\left(\left\langle S_{n,P}^\ast,h\right\rangle\right),\,
N\left(0,s_{n,P}^2(h)\right)
\right)
>\eta
\right)
\to 0
\qquad \text{as } n\to\infty.
\]
\end{proposition}

\begin{remark}[Uniform boundedness implies the scalar Lindeberg condition]
If, for the fixed \(h\in \cH\), there exists \(A_h<\infty\) such that
\[
\sup_{n\ge 1}\sup_{P\in\cP}\max_{1\le i\le n}|a_{n,i,P}(h)|
\le A_h
\qquad \pr_P\text{-a.s. for all }P\in\cP,
\]
then \eqref{eq:scalar-UL-P} follows automatically from the assumption \(\E^\ast[(W_1^\ast)^2]=1\). Indeed, for every \(\delta>0\),
\[
L_{n,P}(h,\delta)
\le
A_h^2
\E^\ast\left[
(W_1^\ast)^2
\mathds 1\left\{
|W_1^\ast|>\delta\sqrt n/A_h
\right\}
\right]
\to 0
\]
uniformly in \(P\), by dominated convergence, since \(\delta\sqrt n/A_h\to\infty\) and
\[
0\le (W_1^\ast)^2\mathds 1\{|W_1^\ast|>\delta\sqrt n/A_h\}\le (W_1^\ast)^2,
\qquad
\E^\ast[(W_1^\ast)^2]<\infty.
\]
\end{remark}

\begin{proof}
At the outset of the proof, let
\[
D_{n,P}(h)
\coloneqq
d_{\mathrm{BL},\bbR}\left(
\cL_P^\ast\left(\left\langle S_{n,P}^\ast,h\right\rangle\right),\,
N\left(0,s_{n,P}^2(h)\right)
\right)
\]
denote the bounded Lipschitz distance between the conditional law and its normal approximation. Assume the conclusion fails. Then there exist \(\eta_0>0\), \(\varepsilon_0>0\), and a sequence \(\{P^{(n)}\}_{n\ge 1}\subset\cP\) such that
\[
\pr_{P^{(n)}}\left(D_{n,P^{(n)}}(h)>\eta_0\right)\ge \varepsilon_0
\qquad\text{for all }n\ge 1.
\]
To construct a diagonal subsequence for rational thresholds, enumerate \(\bbQ_{>0}=\{q_1,q_2,\dots\}\). By \eqref{eq:scalar-UL-P}, for each \((m,j)\in\bbN^2\), there exists \(N(m,j)\) such that \(\sup_{P\in\cP} \pr_P(L_{n,P}(h,q_m)>2^{-j})\le 2^{-j}\) for all \(n\ge N(m,j)\). We can thus choose a strictly increasing sequence \(\{r_k\}_{k\ge 1}\) satisfying \(r_k\ge \max_{1\le m\le k} N(m,k)\), which ensures that for every fixed \(m\) and all \(k\ge m\),
\[
\pr_{P^{(r_k)}}\left(L_{r_k,P^{(r_k)}}(h,q_m)>2^{-k}\right)\le 2^{-k}.
\]

Next, we realize these random variables almost surely on a product space. Letting \((\Omega_k,\cF_k,\overline{\pr}_k) \coloneqq (\Omega,\cF,\pr_{P^{(r_k)}})\), we define the product space \((\overline\Omega,\overline{\cF},\overline{\pr}) \coloneqq \bigotimes_{k=1}^\infty (\Omega_k,\cF_k,\overline{\pr}_k)\) with coordinate maps \(\pr_k\). Since the quantities \(L_{r_k,P^{(r_k)}}(h,q_m)\) are data-measurable, we evaluate them along \(\pr_k\). For each fixed \(m\ge 1\), the summability of our previous bound implies \(\sum_{k=m}^\infty \overline{\pr}(L_{r_k,P^{(r_k)}}(h,q_m)\circ \pr_k > 2^{-k}) < \infty\). The first Borel--Cantelli lemma therefore dictates that \(L_{r_k,P^{(r_k)}}(h,q_m)(\pr_k(\omega))\to 0\) for \(\overline{\pr}\)-a.e.\ \(\omega\). Intersecting these events over all \(m\in\bbN\), we obtain a set \(\Omega_0\in\overline{\cF}\) with \(\overline{\pr}(\Omega_0)=1\) on which this convergence holds for every rational \(q_m\). Furthermore, since the map \(\delta\mapsto L_{r_k,P^{(r_k)}}(h,\delta)(\pr_k(\omega))\) is nonincreasing on \((0,\infty)\), bounding any \(\delta>0\) by a smaller rational \(q_m<\delta\) extends this result to all real thresholds. That is, for every \(\omega\in\Omega_0\) and every \(\delta>0\),
\[
L_{r_k,P^{(r_k)}}(h,\delta)(\pr_k(\omega))\to 0
\qquad (k\to\infty).
\]

This almost sure convergence allows us to establish a pathwise real-valued triangular-array central limit theorem. Fix \(\omega\in\Omega_0\) and define \(b_{k,i}(\omega)\coloneqq a_{r_k,i,P^{(r_k)}}(h)(\pr_k(\omega))/\sqrt{r_k}\) for \(i=1,\dots,r_k\). The sum \(\sum_{i=1}^{r_k} b_{k,i}(\omega)W_i^\ast\) forms a real-valued triangular array under the multiplier law \(\pr^{\ast\otimes\infty}\) with row variance \(\sum_{i=1}^{r_k} b_{k,i}(\omega)^2 = s_{r_k,P^{(r_k)}}^2(h)(\pr_k(\omega))\). The limit established above is exactly the Lindeberg condition for this deterministic array:
\[
\sum_{i=1}^{r_k} b_{k,i}(\omega)^2
\E^\ast\left[
(W_1^\ast)^2
\mathds 1\left\{
|b_{k,i}(\omega)W_1^\ast|>\delta
\right\}
\right]
\to 0
\qquad\text{for every }\delta>0.
\]
Consequently, the classical Lindeberg--Feller theorem yields
\[
d_{\mathrm{BL},\bbR}\left(
\cL_\ast\left(\sum_{i=1}^{r_k} b_{k,i}(\omega)W_i^\ast\right),\,
N\left(0,\sum_{i=1}^{r_k} b_{k,i}(\omega)^2\right)
\right)\to 0,
\]
where \(\cL_\ast(\cdot)\) denotes the law under \(\pr^{\ast\otimes\infty}\). Equivalently, \(D_{r_k,P^{(r_k)}}(h)(\pr_k(\omega))\to 0\). Since this holds for all \(\omega\in\Omega_0\), the convergence occurs \(\overline{\pr}\)-a.s., and thus in \(\overline{\pr}\)-probability. Because \(D_{r_k,P^{(r_k)}}(h)\) is \(\cD_{r_k}\)-measurable, its distribution under \(\overline{\pr}\) exactly matches its law under \(\pr_{P^{(r_k)}}\), implying
\[
\pr_{P^{(r_k)}}\left(D_{r_k,P^{(r_k)}}(h)>\eta_0\right)\to 0.
\]
This contradicts our initial assumption along the subsequence \(r_k\), completing the proof.
\end{proof}

\begin{proposition}[Conditional uniform tightness from trace and tail control]
\label{prop:conditional-uniform-tightness-fixed-onb-P}
Let \(\cH\) be a separable Hilbert space with a fixed orthonormal basis
\(\{e_j\}_{j\ge 1}\). For each \(n\ge 1\) and \(P\in\cP\), let \(\pi_{n,P}\) and \(\lambda_{n,P}\) be \(\cF\)-measurable random Borel probability measures on \(\cH\), and let
\(\widehat C_{n,P}\) be an \(\cF\)-measurable self-adjoint positive trace-class operator on \(\cH\). Define
\[
d_{n,P,j}\coloneqq \left\langle \widehat C_{n,P}e_j,e_j\right\rangle,
\qquad j\ge 1.
\]

Assume:
\begin{enumerate}
\item[(T1)] \emph{Uniform trace tightness:}
\begin{equation}\label{eq:tightness-T1-P}
\lim_{M\to\infty}\limsup_{n\to\infty}\sup_{P\in\cP}
\pr_P\left(\sum_{j=1}^\infty d_{n,P,j}>M\right)=0.
\end{equation}

\item[(T2)] \emph{Uniform fixed-ONB tail vanishing:} for every \(\eta>0\),
\begin{equation}\label{eq:tightness-T2-P}
\lim_{J\to\infty}\limsup_{n\to\infty}\sup_{P\in\cP}
\pr_P\left(\sum_{j>J} d_{n,P,j}>\eta\right)=0.
\end{equation}

\item[(T3)] \emph{Second-moment identities along the fixed ONB:} for every \(j\ge 1\),
\begin{equation}\label{eq:tightness-T3-mu-P}
\int_{\cH}\left\langle x,e_j\right\rangle^2\,\pi_{n,P}(\diff x)=d_{n,P,j}
\qquad \pr_P\text{-a.s.,}
\end{equation}
and
\begin{equation}\label{eq:tightness-T3-nu-P}
\int_{\cH}\left\langle x,e_j\right\rangle^2\,\lambda_{n,P}(\diff x)=d_{n,P,j}
\qquad \pr_P\text{-a.s.}
\end{equation}
\end{enumerate}

Then both families \(\{\pi_{n,P}:n\ge 1,\ P\in\cP\}\) and
\(\{\lambda_{n,P}:n\ge 1,\ P\in\cP\}\) are conditionally uniformly tight over \(\cP\).
\end{proposition}

\begin{proof}
Fix \(\varepsilon>0\) and \(\delta>0\). We begin by choosing positive sequences \(\{\alpha_m\}_{m\ge 0}\), \(\{\beta_m\}_{m\ge 0}\), and \(\{\eta_m\}_{m\ge 1}\) such that
\begin{equation*}
\sum_{m=0}^\infty \alpha_m\le \delta,
\qquad
\sum_{m=0}^\infty \beta_m<\varepsilon,
\qquad
\eta_m\downarrow 0.
\end{equation*}
By the uniform trace tightness condition \eqref{eq:tightness-T1-P}, there exists \(R>0\) such that
\begin{equation}
\label{eq:tightness-proof-radius-P}
\limsup_{n\to\infty}\sup_{P\in\cP}
\pr_P\left(\sum_{j=1}^\infty d_{n,P,j}>\alpha_0R^2\right)<\beta_0.
\end{equation}
Similarly, for each \(m\ge 1\), applying the uniform fixed-ONB tail vanishing condition \eqref{eq:tightness-T2-P} with threshold \(\alpha_m\eta_m^2\) guarantees the existence of an integer \(J_m\in\bbN\) satisfying
\begin{equation}
\label{eq:tightness-proof-tail-thresholds-P}
\limsup_{n\to\infty}\sup_{P\in\cP}
\pr_P\left(\sum_{j>J_m} d_{n,P,j}>\alpha_m\eta_m^2\right)<\beta_m.
\end{equation}
By replacing \(J_m\) with \(\max\{J_1,\dots,J_m\}\), we may assume without loss of generality that \(J_1<J_2<\cdots\) and \(J_m\to\infty\).

With these parameters, we construct a candidate compact set. Letting \(\Pi_{>J}\) denote the orthogonal projection onto \(\operatorname{span}\{e_j:j>J\}\), we define
\begin{equation*}
K
\coloneqq
\left\{
x\in \cH:\ \|x\|\le R,\ \|\Pi_{>J_m}x\|\le \eta_m\ \text{for all }m\ge 1
\right\}.
\end{equation*}
To verify that \(K\) is compact, note first that it is closed because the mappings \(x\mapsto \|x\|\) and \(x\mapsto \|\Pi_{>J_m}x\|\) are continuous. For total boundedness, fix \(\rho>0\) and choose \(m_0\) such that \(\eta_{m_0}<\rho/2\). Defining \(\Pi_{\le J_{m_0}}=I-\Pi_{>J_{m_0}}\), the image \(\Pi_{\le J_{m_0}}H\) is finite-dimensional. For any \(x\in K\), we have \(\|\Pi_{\le J_{m_0}}x\|\le R\) and \(\|\Pi_{>J_{m_0}}x\|<\rho/2\), which implies that the projection \(\Pi_{\le J_{m_0}}K\) is totally bounded. We can therefore select a \(\rho/2\)-net \(y_1,\dots,y_N\in \Pi_{\le J_{m_0}}H\) for \(\Pi_{\le J_{m_0}}K\). For any \(x\in K\), choosing an index \(\ell\) such that \(\|\Pi_{\le J_{m_0}}x-y_\ell\|<\rho/2\) yields
\begin{equation*}
\|x-y_\ell\|
\le
\|\Pi_{\le J_{m_0}}x-y_\ell\|+\|\Pi_{>J_{m_0}}x\|
<
\rho.
\end{equation*}
Thus, \(K\) is totally bounded. Since \(H\) is complete, it follows that \(K\) is compact.

Next, we bound the mass outside this compact set for \(\pi_{n,P}\). Since the complement \(K^c\) satisfies
\begin{equation*}
K^c\subset \{x:\|x\|>R\}\cup \bigcup_{m=1}^\infty \{x:\|\Pi_{>J_m}x\|>\eta_m\},
\end{equation*}
the union bound yields
\begin{equation}
\label{eq:tightness-proof-cover-P}
\pi_{n,P}(K^c)
\le
\pi_{n,P}(\|x\|>R)
+
\sum_{m=1}^\infty \pi_{n,P}(\|\Pi_{>J_m}x\|>\eta_m).
\end{equation}
Applying Markov's inequality, Parseval's identity, Tonelli's theorem, and \eqref{eq:tightness-T3-mu-P}, we can bound the norm term by
\begin{equation}
\label{eq:tightness-proof-markov-norm-P}
\pi_{n,P}(\|x\|>R)
\le
\frac{1}{R^2}\int_{\cH} \|x\|^2\,\pi_{n,P}(\diff x)
=
\frac{1}{R^2}\sum_{j=1}^\infty d_{n,P,j},
\end{equation}
and, for each \(m\ge 1\), the tail term by
\begin{equation}
\label{eq:tightness-proof-markov-tail-P}
\pi_{n,P}(\|\Pi_{>J_m}x\|>\eta_m)
\le
\frac{1}{\eta_m^2}\int_{\cH} \|\Pi_{>J_m}x\|^2\,\pi_{n,P}(\diff x)
=
\frac{1}{\eta_m^2}\sum_{j>J_m} d_{n,P,j}.
\end{equation}
Consequently, on the event
\begin{equation*}
A_{n,P}
\coloneqq
\left\{\sum_{j=1}^\infty d_{n,P,j}\le \alpha_0R^2\right\}
\cap
\bigcap_{m=1}^\infty
\left\{\sum_{j>J_m} d_{n,P,j}\le \alpha_m\eta_m^2\right\},
\end{equation*}
we obtain \(\pi_{n,P}(K^c)\le \alpha_0+\sum_{m=1}^\infty \alpha_m\le \delta\). This implies that the probability of exceeding \(\delta\) is bounded by the probability of the complement of \(A_{n,P}\):
\begin{equation}
\label{eq:tightness-proof-prob-bound-mu-P}
\pr_P\left(\pi_{n,P}(K^c)>\delta\right)
\le
\pr_P\left(\sum_{j=1}^\infty d_{n,P,j}>\alpha_0R^2\right)
+
\sum_{m=1}^\infty
\pr_P\left(\sum_{j>J_m} d_{n,P,j}>\alpha_m\eta_m^2\right).
\end{equation}
Taking \(\limsup_{n\to\infty}\sup_{P\in\cP}\) on both sides and applying our initial bounds \eqref{eq:tightness-proof-radius-P}, \eqref{eq:tightness-proof-tail-thresholds-P}, and the fact that \(\sum_m\beta_m<\varepsilon\), we conclude
\begin{equation}
\label{eq:tightness-proof-final-mu-P}
\limsup_{n\to\infty}\sup_{P\in\cP}
\pr_P\left(\pi_{n,P}(K^c)>\delta\right)
<\varepsilon.
\end{equation}

Finally, by replacing \eqref{eq:tightness-T3-mu-P} with \eqref{eq:tightness-T3-nu-P} and following an identical argument, the same bound holds for \(\lambda_{n,P}\):
\begin{equation}
\label{eq:tightness-proof-final-nu-P}
\limsup_{n\to\infty}\sup_{P\in\cP}
\pr_P\left(\lambda_{n,P}(K^c)>\delta\right)
<\varepsilon.
\end{equation}
Since \(\varepsilon\) and \(\delta\) were arbitrary, both families are conditionally uniformly tight over \(\cP\), completing the proof.
\end{proof}

\begin{proposition}[Uniform Gaussian covariance replacement]
\label{prop:uniform-gaussian-cov-replacement-trace}
For each \(n\ge 1\) and \(P\in\cP\), let \(\widehat C_{n,P}\) be an \(\cF\)-measurable self-adjoint positive trace-class operator on \(\cH\), and let \(C_P\) be a deterministic self-adjoint positive trace-class operator on \(\cH\). Define
\begin{equation*}
\nu_{n,P}^\ast \coloneqq \cN(0,\widehat C_{n,P}).
\end{equation*}
Assume that for every \(\varepsilon>0\),
\begin{equation*}
\sup_{P\in\cP}
\pr_P\left(\left\|\widehat C_{n,P}-C_P\right\|_{\mathrm{tr}}>\varepsilon\right)\to 0
\qquad (n\to\infty).
\end{equation*}
Then, for every \(\eta>0\),
\begin{equation*}
\sup_{P\in\cP}
\pr_P\left(
d_{\mathrm{BL},\cH}\left(\nu_{n,P}^\ast,\cN(0,C_P)\right)>\eta
\right)\to 0
\qquad (n\to\infty).
\end{equation*}
\end{proposition}

\begin{proof}
We first prove the deterministic bound
\begin{equation}
\label{eq:det-gaussian-BL-trace-bound}
d_{\mathrm{BL},\cH}\left(\cN(0,A),\cN(0,B)\right)
\le
\|A-B\|_{\mathrm{tr}}^{1/2}
\end{equation}
for any self-adjoint positive trace-class operators \(A,B\) on \(\cH\). Fix an orthonormal basis \(\{e_j\}_{j\ge 1}\) of \(\cH\), and let \(\{\xi_j\}_{j\ge 1}\) be i.i.d.\ \(N(0,1)\) random variables on an auxiliary probability space. Define
\begin{equation*}
X_A \coloneqq \sum_{j=1}^\infty \xi_j A^{1/2}e_j,
\qquad
X_B \coloneqq \sum_{j=1}^\infty \xi_j B^{1/2}e_j.
\end{equation*}
For any \(f\in \mathrm{BL}_1(\cH)\), the property \(\mathrm{Lip}(f)\le 1\) implies
\begin{equation*}
\left|
\E\left[f(X_A)\right]-\E\left[f(X_B)\right]
\right|
\le
\E\left\|X_A-X_B\right\|
\le
\left(\E\left\|X_A-X_B\right\|^2\right)^{1/2}.
\end{equation*}
Moreover, expanding the expected squared norm yields
\begin{equation*}
\E\left\|X_A-X_B\right\|^2
=
\sum_{j=1}^\infty \left\|(A^{1/2}-B^{1/2})e_j\right\|^2
=
\left\|A^{1/2}-B^{1/2}\right\|_{\mathrm{HS}}^2.
\end{equation*}
Taking the supremum over \(f\in \mathrm{BL}_1(\cH)\) and applying the standard operator inequality \(\|A^{1/2}-B^{1/2}\|_{\mathrm{HS}}^2 \le \|A-B\|_{\mathrm{tr}}\), we establish \eqref{eq:det-gaussian-BL-trace-bound}.

Now, applying \eqref{eq:det-gaussian-BL-trace-bound} pathwise with \(A=\widehat C_{n,P}\) and \(B=C_P\), and recalling that \(\nu_{n,P}^\ast=\cN(0,\widehat C_{n,P})\), we obtain
\begin{equation*}
d_{\mathrm{BL},\cH}\left(\nu_{n,P}^\ast,\cN(0,C_P)\right)
\le
\left\|\widehat C_{n,P}-C_P\right\|_{\mathrm{tr}}^{1/2}
\qquad \pr_P\text{-a.s.}
\end{equation*}
Therefore, for every \(\eta>0\),
\begin{equation*}
\pr_P\left(
d_{\mathrm{BL},\cH}\left(\nu_{n,P}^\ast,\cN(0,C_P)\right)>\eta
\right)
\le
\pr_P\left(
\left\|\widehat C_{n,P}-C_P\right\|_{\mathrm{tr}}>\eta^2
\right).
\end{equation*}
Taking \(\sup_{P\in\cP}\) on both sides and using the assumed uniform trace-norm convergence, the right-hand side vanishes as \(n\to\infty\), which yields the desired conclusion.
\end{proof}

\begin{lemma}[Uniform consistency of the finite-\(B_n\) bootstrap quantile under QR]
\label{lem:finiteB_bootstrap_quantile_uniform_consistency}
Assume the following.
\begin{enumerate}
\item[(i)] For every \(\varepsilon>0\),
\[
\sup_{P\in\tilde{\cP}}
\pr_P\left(
\sup_{t\in\bbR}
\left|\hat F_{n,P}^\ast(t)-F_P(t)\right|>\varepsilon
\right)\to 0.
\]

\item[(ii)] For every \(\eta>0\),
\[
\Delta_\eta^+
\coloneqq
\inf_{P\in\tilde{\cP}}
\left\{F_P\left(q_P+\eta\right)-u\right\}>0,
\qquad
\Delta_\eta^-
\coloneqq
\inf_{P\in\tilde{\cP}}
\left\{u-F_P\left(q_P-\eta\right)\right\}>0.
\]
\end{enumerate}

Then, for every \(\eta>0\),
\[
\sup_{P\in\tilde{\cP}}
\pr_P\left(
\left|\hat q_{n,P}^{\ast(B_n)}-q_P\right|>\eta
\right)\to 0.
\]
\end{lemma}

\begin{proof}
Fix \(\eta>0\), and define
\begin{equation*}
\Delta_\eta\coloneqq \min\left\{\Delta_\eta^+,\Delta_\eta^-\right\}>0
\end{equation*}
by assumption (ii). We first show that the finite-\(B_n\) empirical bootstrap cdf uniformly approximates the ideal conditional bootstrap cdf. Conditionally on \(\cD_n\), the bootstrap replicates \(\left\{\|X_{n,P}^{\ast(b)}\|^2\right\}_{b=1}^{B_n}\) are i.i.d.\ with cdf \(\hat F_{n,P}^{\ast}\). By the Dvoretzky--Kiefer--Wolfowitz inequality \citep{Massart1990} applied conditionally on \(\cD_n\), the exceedance probability is bounded by \(2e^{-2B_n\delta^2}\) almost surely. Since this upper bound is deterministic and independent of both the data \(\cD_n\) and the underlying distribution \(P\), taking the unconditional expectation and the supremum over \(\tilde{\cP}\) directly yields
\begin{equation*}
\sup_{P\in\tilde{\cP}}
\pr_P\left(
\sup_{t\in\bbR}
\left|\hat F_{n,P}^{\ast(B_n)}(t)-\hat F_{n,P}^{\ast}(t)\right|>\delta
\right)
\le
2e^{-2B_n\delta^2}
\to 0
\end{equation*}
as \(B_n\to\infty\).

Next, we establish a uniform cdf approximation to \(F_P\). By the triangle inequality,
\begin{equation*}
\sup_{t\in\bbR}
\left|\hat F_{n,P}^{\ast(B_n)}(t)-F_P(t)\right|
\le
\sup_{t\in\bbR}
\left|\hat F_{n,P}^{\ast(B_n)}(t)-\hat F_{n,P}^{\ast}(t)\right|
+
\sup_{t\in\bbR}
\left|\hat F_{n,P}^{\ast}(t)-F_P(t)\right|.
\end{equation*}
Hence, with \(\delta_\eta\coloneqq \Delta_\eta/4\),
\begin{align*}
&\sup_{P\in\tilde{\cP}}
\pr_P\left(
\sup_{t\in\bbR}
\left|\hat F_{n,P}^{\ast(B_n)}(t)-F_P(t)\right|>\frac{\Delta_\eta}{2}
\right) \\
&\qquad \le
\sup_{P\in\tilde{\cP}}
\pr_P\left(
\sup_{t\in\bbR}
\left|\hat F_{n,P}^{\ast(B_n)}(t)-\hat F_{n,P}^{\ast}(t)\right|>\frac{\Delta_\eta}{4}
\right)
+
\sup_{P\in\tilde{\cP}}
\pr_P\left(
\sup_{t\in\bbR}
\left|\hat F_{n,P}^{\ast}(t)-F_P(t)\right|>\frac{\Delta_\eta}{4}
\right)
\to 0,
\end{align*}
where the first term vanishes by the DKW bound shown above, and the second vanishes by assumption (i).

Finally, we invert these bounds using the uniform quantile regularity. Define the event
\begin{equation*}
A_{n,P}(\eta)
\coloneqq
\left\{
\sup_{t\in\bbR}
\left|\hat F_{n,P}^{\ast(B_n)}(t)-F_P(t)\right|
<
\frac{\Delta_\eta}{2}
\right\}.
\end{equation*}
On \(A_{n,P}(\eta)\), we have
\begin{equation*}
\hat F_{n,P}^{\ast(B_n)}\left(q_P-\eta\right)
\le
F_P\left(q_P-\eta\right)+\frac{\Delta_\eta}{2}
\le
u-\Delta_\eta^-+\frac{\Delta_\eta}{2}
\le
u-\frac{\Delta_\eta}{2}
<u.
\end{equation*}
Since \(\hat q_{n,P}^{\ast(B_n)}\) is the generalized inverse of \(\hat F_{n,P}^{\ast(B_n)}\), it follows that \(\hat q_{n,P}^{\ast(B_n)} > q_P-\eta\). Similarly,
\begin{equation*}
\hat F_{n,P}^{\ast(B_n)}\left(q_P+\eta\right)
\ge
F_P\left(q_P+\eta\right)-\frac{\Delta_\eta}{2}
\ge
u+\Delta_\eta^+-\frac{\Delta_\eta}{2}
\ge
u+\frac{\Delta_\eta}{2}
>u,
\end{equation*}
which implies \(\hat q_{n,P}^{\ast(B_n)}\le q_P+\eta\). Therefore, on \(A_{n,P}(\eta)\), we obtain \(\left|\hat q_{n,P}^{\ast(B_n)}-q_P\right|\le \eta\). Equivalently,
\begin{equation*}
\left\{\left|\hat q_{n,P}^{\ast(B_n)}-q_P\right|>\eta\right\}
\subset
A_{n,P}(\eta)^c.
\end{equation*}
Taking probabilities and then the supremum over \(P\in\tilde{\cP}\), we conclude
\begin{equation*}
\sup_{P\in\tilde{\cP}}
\pr_P\left(
\left|\hat q_{n,P}^{\ast(B_n)}-q_P\right|>\eta
\right)
\le
\sup_{P\in\tilde{\cP}}
\pr_P\left(
\sup_{t\in\bbR}
\left|\hat F_{n,P}^{\ast(B_n)}(t)-F_P(t)\right|
\ge
\frac{\Delta_\eta}{2}
\right)
\to 0.
\end{equation*}
This completes the proof.
\end{proof}

\subsection{Proofs of the main results}

\begin{proof}[of  Theorem~\ref{thm:abstract-transfer-squared-norm-bootstrap}]
We first prove the finite-\(B_n\) cdf consistency. Fix \(\varepsilon>0\). Conditionally on \(\cD_n\), the bootstrap replicates
\(
\left\{\left\|X_{n,P}^{\ast(b)}\right\|^2\right\}_{b=1}^{B_n}
\)
are i.i.d.\ with conditional cdf \(\hat F_{n,P}^\ast\). By the conditional Dvoretzky--Kiefer--Wolfowitz inequality,
\[
\sup_{P\in\cP}
\pr_P\left(
\sup_{t\in\bbR}
\left|
\hat F_{n,P}^{\ast(B_n)}(t)-\hat F_{n,P}^\ast(t)
\right|>\delta
\right)
\le 2e^{-2B_n\delta^2}\to 0.
\]
Therefore, by the triangle inequality,
\begin{align*}
&\sup_{P\in\cP}
\pr_P\left(
\sup_{t\in\bbR}
\left|
\hat F_{n,P}^{\ast(B_n)}(t)-F_P(t)
\right|>\varepsilon
\right) \\
&\le
\sup_{P\in\cP}
\pr_P\left(
\sup_{t\in\bbR}
\left|
\hat F_{n,P}^{\ast(B_n)}(t)-\hat F_{n,P}^\ast(t)
\right|>\frac{\varepsilon}{2}
\right)
+
\sup_{P\in\cP}
\pr_P\left(
\sup_{t\in\bbR}
\left|
\hat F_{n,P}^\ast(t)-F_P(t)
\right|>\frac{\varepsilon}{2}
\right)\to 0,
\end{align*}
where the second term converges to \(0\) by Proposition~\ref{prop:conditional_BL_H_to_squared_norm_cdf_gaussian}.

Finally, assume also Assumption~\ref{ass:bootstrap-QR-transfer}. Then Lemma~\ref{lem:finiteB_bootstrap_quantile_uniform_consistency} applies with \(\tilde{\cP}=\cP\), because its assumption (i) is exactly the ideal cdf consistency already proved in Proposition~\ref{prop:conditional_BL_H_to_squared_norm_cdf_gaussian}, and its assumption (ii) is Assumption~\ref{ass:bootstrap-QR-transfer}. Therefore,
\[
\sup_{P\in\cP}
\pr_P\left(
\left|
\hat q_{n,P}^{\ast(B_n)}-q_P
\right|>\eta
\right)\to 0
\qquad \text{for every }\eta>0.
\]
This proves the theorem.
\end{proof}

\begin{proof}[of  Theorem~\ref{thm:verification-conditional-hilbert-BL}]
We first show that
\[
\sup_{P\in\cP}
\pr_P\left(
d_{\mathrm{BL},\cH}\left(
\pi_{n,P},\nu_{n,P}^\ast
\right)>\eta
\right)\to 0
\qquad \text{for every }\eta>0.
\]

To this end, we verify the assumptions of
Theorem~\ref{thm:abstract-assembly-conditional-uniform-hilbert-P}
with \(\pi_{n,P}=\pi_{n,P}\) and \(\lambda_{n,P}=\nu_{n,P}^\ast\).

By Assumption~\ref{ass:bootstrap-verification-V1} and
Assumption~\ref{ass:bootstrap-verification-V2}, for every \(\eta>0\),
\[
\sup_{P\in\cP}
\pr_P\left(
d_{\mathrm{BL},\bbR}\left(
(\ell_h)_\#\pi_{n,P},\,
(\ell_h)_\#\nu_{n,P}^\ast
\right)>\eta
\right)\to 0.
\]
Thus assumption (i) of
Theorem~\ref{thm:abstract-assembly-conditional-uniform-hilbert-P}
holds.

Next, Assumption~\ref{ass:bootstrap-verification-V3} and
Proposition~\ref{prop:conditional-uniform-tightness-fixed-onb-P}
imply that both \(\{\pi_{n,P}\}\) and \(\{\nu_{n,P}^\ast\}\) are conditionally uniformly tight over \(\cP\). Thus assumption (ii) of
Theorem~\ref{thm:abstract-assembly-conditional-uniform-hilbert-P}
also holds.

Applying Theorem~\ref{thm:abstract-assembly-conditional-uniform-hilbert-P},
we obtain
\[
\sup_{P\in\cP}
\pr_P\left(
d_{\mathrm{BL},\cH}\left(
\pi_{n,P},\nu_{n,P}^\ast
\right)>\eta
\right)\to 0
\qquad \text{for every }\eta>0.
\]

It remains to replace \(\nu_{n,P}^\ast=\cN(0,\widehat C_{n,P})\) by
\(\cN(0,C_P)\). By Assumption~\ref{ass:bootstrap-verification-V4} and
Proposition~\ref{prop:uniform-gaussian-cov-replacement-trace},
\[
\sup_{P\in\cP}
\pr_P\left(
d_{\mathrm{BL},\cH}\left(
\nu_{n,P}^\ast,\cN(0,C_P)
\right)>\eta
\right)\to 0
\qquad \text{for every }\eta>0.
\]
Therefore, by the triangle inequality,
\begin{align*}
&\sup_{P\in\cP}
\pr_P\left(
d_{\mathrm{BL},\cH}\left(
\pi_{n,P},\cN(0,C_P)
\right)>\eta
\right) \\
&\le
\sup_{P\in\cP}
\pr_P\left(
d_{\mathrm{BL},\cH}\left(
\pi_{n,P},\nu_{n,P}^\ast
\right)>\frac{\eta}{2}
\right)
+
\sup_{P\in\cP}
\pr_P\left(
d_{\mathrm{BL},\cH}\left(
\nu_{n,P}^\ast,\cN(0,C_P)
\right)>\frac{\eta}{2}
\right)\to 0.
\end{align*}
Since \(\pi_{n,P}=\cL_P^\ast(S_{n,P}^\ast)\), this is exactly the desired conclusion.
\end{proof}
\section{Characteristic-kernel constructions for random objects}
We present (i) concise proofs for the main results of Section~\ref{sec:kernels} and (ii) a brief discussion of implications for weak Fr\'echet regression in \citet{Bhattacharjeeetal2025}.

\subsection{Proofs of the kernel-construction results}
Throughout, we write the kernel mean embedding for a measure $P$ with respect to a kernel $k$ as $\mu^k(P)$ to make the dependence on both the kernel and the measure explicit. We first establish a key lemma that is foundational for the subsequent results: the property of being characteristic is preserved when a kernel is restricted to a Borel subset.

\begin{proof}[of Theorem~\ref{thm:pullback}]
Let \(P,Q\in \cP_{\phi^\ast k_{\cN}}(\cM)\). For \(R\in\{P,Q\}\), define the push-forward
\(
\phi_{\sharp}R\coloneqq R\circ \phi^{-1},
\)
which is a probability measure on \(\cN\).

Since
\begin{align*}
\int_{\cN} \sqrt{k_{\cN}(z,z)}\,(\phi_{\sharp}R)(\diff z)
&=
\int_{\cM} \sqrt{k_{\cN}(\phi(u),\phi(u))}\,R(\mathrm du)\\
&=
\int_{\cM} \sqrt{(\phi^\ast k_{\cN})(u,u)}\,R(\mathrm du)
<\infty,
\end{align*}
the kernel mean embedding \(\mu^{k_{\cN}}(\phi_{\sharp}R)\) is well defined.

Moreover, for every \(w\in\cM\),
\begin{align*}
\mu^{\phi^\ast k_{\cN}}(R)(w)
&=
\int_{\cM} (\phi^\ast k_{\cN})(u,w)\,R(\mathrm du)\\
&=
\int_{\cM} k_{\cN}(\phi(u),\phi(w))\,R(\mathrm du)\\
&=
\int_{\cN} k_{\cN}(z,\phi(w))\,(\phi_{\sharp}R)(\diff z)\\
&=
\mu^{k_{\cN}}(\phi_{\sharp}R)(\phi(w)).
\end{align*}

We first prove the ``only if'' direction. Suppose that
\[
\mu^{\phi^\ast k_{\cN}}(P)=\mu^{\phi^\ast k_{\cN}}(Q).
\]
Set
\[
h\coloneqq \mu^{k_{\cN}}(\phi_{\sharp}P)-\mu^{k_{\cN}}(\phi_{\sharp}Q)
\in \cH_{k_{\cN}}.
\]
Then the identity above yields
\[
h(\phi(w))=0
\qquad \text{for every } w\in\cM.
\]
Equivalently,
\[
\langle h,k_{\cN}(\,\cdot\,,\phi(w))\rangle_{\cH_{k_{\cN}}}=0
\qquad \text{for every } w\in\cM.
\]

On the other hand, by the definition of kernel mean embeddings,
\[
\mu^{k_{\cN}}(\phi_{\sharp}R)
=
\int_{\cM} k_{\cN}(\,\cdot\,,\phi(u))\,R(\mathrm du),
\qquad R\in\{P,Q\},
\]
and therefore \(h\) belongs to the closed linear span of
\[
\left\{k_{\cN}(\,\cdot\,,\phi(w)):w\in\cM\right\}.
\]
Since \(h\) is orthogonal to every generator of this subspace, it follows that \(h=0\). Hence
\[
\mu^{k_{\cN}}(\phi_{\sharp}P)=\mu^{k_{\cN}}(\phi_{\sharp}Q).
\]
Because \(k_{\cN}\) is characteristic on \(\cN\) (more generally, it is enough that its restriction to \(\phi(\cM)\) be characteristic), we conclude that
\[
\phi_{\sharp}P=\phi_{\sharp}Q.
\]

Conversely, suppose that $\phi_{\sharp}P=\phi_{\sharp}Q$. Then $\mu^{k_{\cN}}(\phi_{\sharp}P)=\mu^{k_{\cN}}(\phi_{\sharp}Q)$, and so, for every $w\in\cM$,
\begin{equation}\mu^{\phi^\ast k_{\cN}}(P)(w)
=
\mu^{k_{\cN}}(\phi_{\sharp}P)(\phi(w))
=
\mu^{k_{\cN}}(\phi_{\sharp}Q)(\phi(w))
=
\mu^{\phi^\ast k_{\cN}}(Q)(w).
\end{equation}
Thus $\mu^{\phi^\ast k_{\cN}}(P)=\mu^{\phi^\ast k_{\cN}}(Q)$.

This proves that, for any \(P,Q\in \cP_{\phi^\ast k_{\cN}}(\cM)\),
\[
\mu^{\phi^\ast k_{\cN}}(P)=\mu^{\phi^\ast k_{\cN}}(Q)
\quad \text{if and only if} \quad
\phi_{\sharp}P=\phi_{\sharp}Q.
\]

For the final claim, assume that
\begin{equation}
\phi:(\cM,\cB(\cM))
\to
(\phi(\cM),\cB(\cN)|_{\phi(\cM)})
\end{equation}
is an isomorphism of measurable spaces. We note that this condition is automatically satisfied when our spaces $\cM$ and $\cN$ are Polish and $\phi$ is an injective Borel measurable map \citep[Proposition~4.5.1 and Theorem~4.5.4]{Srivastava1998}.
Let \(B\in \cB(\cM)\).
Since \(\phi^{-1}\) is measurable, we have
\[
\phi(B)=(\phi^{-1})^{-1}(B)\in \cB(\cN)|_{\phi(\cM)}.
\]
Hence, if \(\phi_{\sharp}P=\phi_{\sharp}Q\), then
\begin{align*}
P(B)
&= P\bigl(\phi^{-1}(\phi(B))\bigr) \\
&= \phi_{\sharp}P(\phi(B)) \\
&= \phi_{\sharp}Q(\phi(B)) \\
&= Q\bigl(\phi^{-1}(\phi(B))\bigr) \\
&= Q(B).
\end{align*}
Therefore \(P=Q\).
It follows that \(\phi^\ast k_{\cN}\) is characteristic on \(\cM\) with respect to
\(\cP_{\phi^\ast k_{\cN}}(\cM)\).
\end{proof}

\begin{proof}[of Remark~\ref{rem:pullback-ispd-spd}]
Let \(k_{\phi(\cM)}\) denote the restriction of \(k_{\cN}\) to \(\phi(\cM)\times\phi(\cM)\). For every finite signed Borel measure \(\eta\) on \(\cM\) such that the relevant integrals are finite,
\[
\iint_{\cM\times\cM} (\phi^\ast k_{\cN})(u,v)\,\eta(\mathrm du)\eta(\mathrm dv)
=
\iint_{\phi(\cM)\times\phi(\cM)} k_{\phi(\cM)}(x,y)\,(\phi_\sharp\eta)(\diff x)(\phi_\sharp\eta)(\diff y).
\]
Indeed, this is immediate from the definition \((\phi^\ast k_{\cN})(u,v)=k_{\cN}(\phi(u),\phi(v))\) and the change-of-variables identity for push-forward measures.

If \(\phi:\cM\to\phi(\cM)\) is a Borel isomorphism, then \(\phi_\sharp\) is bijective between finite signed Borel measures on \(\cM\) and on \(\phi(\cM)\), with inverse \((\phi^{-1})_\sharp\). Hence the quadratic form associated with \(\phi^\ast k_{\cN}\) vanishes at a nonzero finite signed measure on \(\cM\) if and only if the quadratic form associated with \(k_{\phi(\cM)}\) vanishes at a nonzero finite signed measure on \(\phi(\cM)\). This proves the equivalence for integrally strict positive definiteness.

For strict positive definiteness, let \(u_1,\dots,u_m\in\cM\) be distinct and \(a_1,\dots,a_m\in\bbR\), not all zero. Since \(\phi\) is injective, the points \(\phi(u_1),\dots,\phi(u_m)\) are distinct, and
\[
\sum_{i,j=1}^m a_i a_j (\phi^\ast k_{\cN})(u_i,u_j)
=
\sum_{i,j=1}^m a_i a_j k_{\phi(\cM)}(\phi(u_i),\phi(u_j)).
\]
Therefore \(\phi^\ast k_{\cN}\) is strictly positive definite on \(\cM\) if and only if \(k_{\phi(\cM)}\) is strictly positive definite on \(\phi(\cM)\).

Finally, integrally strict positive definiteness implies characteristicity by applying the defining strict positivity to finite signed measures of the form \(P-Q\), where \(P\) and \(Q\) are distinct probability measures.
\end{proof}

\begin{proof}[of Proposition~\ref{prop:cmonotone_nonconstant_characteristic}]
We first show that, for every \(s>0\), the exponential kernel
\[
k_s(u,v)\coloneqq \exp\left(-s\,\rho(u,v)\right),
\qquad u,v\in\cM,
\]
is integrally strictly positive definite on \(\cM\).

Fix a base point \(x_0\in\cM\) and define
\[
\tilde k_{\rho}(u,v)\coloneqq
\frac{1}{2}\left\{\rho(u,x_0)+\rho(v,x_0)-\rho(u,v)\right\},
\qquad u,v\in\cM.
\]
By Lemma~12 of \cite{Sejdinovicetal2013}, \(\tilde k_{\rho}\) is symmetric and positive definite. Hence there exist an RKHS \(\widetilde{\cH}\) and the canonical feature map
\[
\psi(u)\coloneqq \tilde k_{\rho}(u,\cdot)\in \widetilde{\cH},
\qquad u\in\cM.
\]
Moreover,
\[
\left\|\psi(u)-\psi(v)\right\|_{\widetilde{\cH}}^2
=
\tilde k_{\rho}(u,u)+\tilde k_{\rho}(v,v)-2\tilde k_{\rho}(u,v)
=
\rho(u,v),
\qquad u,v\in\cM.
\]
Since \(\rho\) is continuous, the map \(\psi\) is continuous, hence Borel measurable. Because \(\cM\) is Polish, \(\widetilde{\cH}\) is separable by \cite[Theorem~2.4]{OwhadiandScovel2017}. Furthermore, Proposition~14 of \cite{Sejdinovicetal2013} yields that \(\psi\) is injective and that
\[
\rho(u,v)=\left\|\psi(u)-\psi(v)\right\|_{\widetilde{\cH}}^2,
\qquad u,v\in\cM.
\]

Now fix \(s>0\), and consider the Gaussian kernel on \(\widetilde{\cH}\),
\[
\tilde k_s(x,y)\coloneqq
\exp\left(-s\,\left\|x-y\right\|_{\widetilde{\cH}}^2\right),
\qquad x,y\in\widetilde{\cH}.
\]
Its pull-back along \(\psi\) is
\[
(\psi^\ast \tilde k_s)(u,v)
=
\tilde k_s(\psi(u),\psi(v))
=
\exp\left(-s\,\left\|\psi(u)-\psi(v)\right\|_{\widetilde{\cH}}^2\right)
=
\exp\left(-s\,\rho(u,v)\right)
=
k_s(u,v).
\]
By Theorem~3.1 of \citet{Ziegeletal2024}, Gaussian kernels on separable Hilbert spaces are integrally strictly positive definite with respect to finite signed measures. Since \(\psi\) is injective and Borel measurable, and \(\cM\) is Polish, Remarks~\ref{rem:pullback-ispd-spd} and \ref{rem:pullback-borel-iso-auto} imply that its pull-back \(k_s=\psi^\ast \tilde k_s\) is integrally strictly positive definite on \(\cM\). Thus \(k_s\) is integrally strictly positive definite for every \(s>0\).

We now prove the claim for
\[
k_f(u,v)\coloneqq f\left(\rho(u,v)\right),
\qquad u,v\in\cM.
\]
Since \(f\) is completely monotone on \((0,\infty)\) and admits the finite limit
\[
f(0)=\lim_{t\downarrow 0}f(t)<\infty,
\]
Bernstein's theorem \citep[Theorem~1.4]{Schillingetal2012} yields a unique positive Borel measure \(\mu\) on \([0,\infty)\) such that
\[
f(t)=\int_0^\infty e^{-st}\mu(\diff s),
\qquad t>0.
\]
By monotone convergence,
\[
f(0)
=
\lim_{t\downarrow 0}\int_0^\infty e^{-st}\mu(\diff s)
=
\int_0^\infty 1\,\mu(\diff s)
=
\mu([0,\infty)),
\]
so \(\mu\) is finite. Since \(f\) is nonconstant, we must have \(\mu((0,\infty))>0\); otherwise \(\mu=c\,\delta_0\) and \(f\equiv c\).

For all \(u,v\in\cM\),
\[
k_f(u,v)
=
f\left(\rho(u,v)\right)
=
\int_0^\infty e^{-s\rho(u,v)}\mu(\diff s)
=
\int_0^\infty k_s(u,v)\mu(\diff s).
\]
Because \(0\le k_s(u,v)\le 1\) for all \(s\ge 0\), it follows that
\[
0\le k_f(u,v)\le \mu([0,\infty))=f(0)<\infty,
\]
so \(k_f\) is bounded. Continuity of \(\rho\) and continuity of \(f\) on \([0,\infty)\) imply that \(k_f\) is continuous. Also, for each \(s\ge 0\), the kernel \(k_s\) is positive definite by Schoenberg's theorem \citep[Theorem~2.2]{Bergetal1984}, and therefore the nonnegative mixture \(k_f\) is positive definite.

It remains to prove that \(k_f\) is integrally strictly positive definite. Let \(\eta\) be a nonzero finite signed Borel measure on \(\cM\). Since \(k_f\) is bounded, all the integrals below are well defined. Using Tonelli's theorem, we obtain
\[
\iint_{\cM\times\cM} k_f(u,v)\,\eta(\mathrm du)\eta(\mathrm dv)
=
\int_0^\infty
\left(
\iint_{\cM\times\cM} k_s(u,v)\,\eta(\mathrm du)\eta(\mathrm dv)
\right)\mu(\diff s).
\]
Now fix such an \(\eta\neq 0\). Since \(k_s\) is integrally strictly positive definite for every \(s>0\), we have
\[
\iint_{\cM\times\cM} k_s(u,v)\,\eta(\mathrm du)\eta(\mathrm dv)>0
\qquad \text{for every } s>0.
\]
Because \(\mu((0,\infty))>0\), it follows that
\[
\iint_{\cM\times\cM} k_f(u,v)\,\eta(\mathrm du)\eta(\mathrm dv)
=
\int_0^\infty
\left(
\iint_{\cM\times\cM} k_s(u,v)\,\eta(\mathrm du)\eta(\mathrm dv)
\right)\mu(\diff s)
>0.
\]
Hence \(k_f\) is integrally strictly positive definite on \(\cM\).
\end{proof}
\begin{remark}[Replacing continuity by metric separability]
In Theorem~5.1, the continuity of $\rho$ was used only to guarantee separability of the Hilbert space underlying the Gaussian kernel constructed from the negative-type structure. This continuity assumption can be replaced by a purely metric assumption on $(\cM,\rho)$.

Indeed, suppose $\rho$ is a semimetric of negative type on $\cM$, and define
\[
d_{\rho}(u,v)\coloneqq \rho(u,v)^{1/2},\qquad u,v\in\cM.
\]
By the standard representation theorem for negative-type semimetrics, there exist a real Hilbert space $\cH$ and a map
$\psi\colon\cM\to\cH$ such that
\[
\rho(u,v)=\left\|\psi(u)-\psi(v)\right\|_{\cH}^{2}
\quad\text{for all }u,v\in\cM.
\]
In particular,
\[
d_{\rho}(u,v)=\left\|\psi(u)-\psi(v)\right\|_{\cH},
\]
so $\psi$ is an isometric embedding of $(\cM,d_{\rho})$ into $\cH$. Hence $\psi$ is a homeomorphism from $(\cM,d_{\rho})$ onto its image $\psi(\cM)$, and the subspace topology on $\psi(\cM)$ induced by $\cH$ coincides with the metric topology of $d_{\rho}$.

If, in addition, $(\cM,d_{\rho})$ is separable, then $\psi(\cM)$ is separable, and so is the closed linear span
$\overline{\operatorname{span}(\psi(\cM))}$. Therefore the Gaussian kernel on this separable Hilbert space is characteristic, and its pull-back along $\psi$ coincides with $k_{\rho}(u,v)=\exp(-\gamma\,\rho(u,v))$, which is thus characteristic on $\cM$.
Consequently, in Theorem~5.1 the continuity of $\rho$ can be replaced by the assumption that $(\cM,d_{\rho})$ is a separable metric space.
\end{remark}
\begin{remark}[Bernstein representation and complete monotonicity]The following are equivalent:(A) $f$ is non-constant, completely monotone on $(0,\infty)$, and has the finite limit $f(0)\coloneqq\lim_{t\downarrow 0}f(t)$.(B) There exists a unique finite positive Borel measure $\mu$ on $[0,\infty)$ with $\mu((0,\infty))>0$ such that $f(t)=\int_0^{\infty} e^{-st}\mu(\diff s)$ for all $t\in[0,\infty)$.The direction $(A)\Rightarrow(B)$ is already contained in the proof. For completeness, $(B)\Rightarrow(A)$ follows as:(i) Continuity at $0$: by monotone convergence, $\lim_{t\downarrow 0}f(t)=\int_0^{\infty}1\mu(\diff s)=\mu([0,\infty))<\infty$.(ii) Complete monotonicity: for $t>0$ and $n\in\bbN_0$,$$f^{(n)}(t)=\int_0^{\infty}\frac{\partial^n}{\partial t^n}e^{-st}\mu(\diff s)=(-1)^n\int_0^{\infty}s^ne^{-st}\mu(\diff s),$$where differentiation under the integral is justified by dominated convergence (fix $t_0>0$ and bound by $s^ne^{-st_0/2}\in L^1(\mu)$). Hence $(-1)^nf^{(n)}(t)\ge 0$.(iii) Non-constancy: if $\mu((0,\infty))=0$ then $\mu=c\,\delta_0$ and $f\equiv c$; if $\mu((0,\infty))>0$ then for $0<t_1<t_2$, $f(t_1)-f(t_2)=\int_{(0,\infty)}\left(e^{-st_1}-e^{-st_2}\right)\mu(\diff s)>0$. Uniqueness of $\mu$ is part of Bernstein’s theorem.
\end{remark}

\begin{proof}[of Corollary~\ref{cor:tensor_char_cmon}]
Since each \(\cN_{\ell}\) is Polish, the finite product \(\cN=\prod_{\ell=1}^L\cN_{\ell}\) is Polish. 
By Proposition~\ref{prop:cmonotone_nonconstant_characteristic}, for every \(\ell\) the kernel
\[
k^{(\ell)}(u_{\ell},v_{\ell}) = f_{\ell}(\rho_{\ell}(u_{\ell},v_{\ell})),
\qquad u_{\ell},v_{\ell}\in \cN_{\ell},
\]
is bounded, continuous, positive definite, and integrally strictly positive definite on \(\cN_\ell\).

Boundedness and continuity of
\[
k_{\otimes,L}(u,v) \coloneqq \prod_{\ell=1}^L k^{(\ell)}(u_{\ell},v_{\ell}),
\qquad u=(u_1,\dots,u_L),\ v=(v_1,\dots,v_L),
\]
follow from being a finite product of bounded continuous functions. 
For positive definiteness, Lemma~4.6 in \cite{SteinwartandChristmann2008} yields that the tensor product of kernels is again a positive definite kernel on the product space; hence \(k_{\otimes,L}\) is positive definite.

It remains to prove that \(k_{\otimes,L}\) is integrally strictly positive definite.
For each \(\ell\), by Bernstein's theorem there exists a finite positive Borel measure \(\mu_{\ell}\) on \([0,\infty)\) with \(\mu_{\ell}((0,\infty))>0\) such that
\[
f_{\ell}(t)=\int_0^\infty e^{-s_\ell t}\mu_{\ell}(\diff s_\ell),
\qquad t>0.
\]
Let \(\mu\coloneqq \bigotimes_{\ell=1}^L \mu_{\ell}\) on \([0,\infty)^L\). By Tonelli's theorem,
\[
k_{\otimes,L}(u,v)
=
\prod_{\ell=1}^L \int_0^\infty e^{-s_\ell \rho_\ell(u_\ell,v_\ell)}\mu_{\ell}(\diff s_\ell)
=
\int_{[0,\infty)^L}
\exp\left(-\sum_{\ell=1}^L s_\ell \rho_\ell(u_\ell,v_\ell)\right)
\mu(\diff s),
\]
where \(s=(s_1,\dots,s_L)\in[0,\infty)^L\).

For \(s\in[0,\infty)^L\), define
\[
\rho_s(u,v)\coloneqq \sum_{\ell=1}^L s_\ell\,\rho_\ell(u_\ell,v_\ell),
\qquad
k_s(u,v)\coloneqq \exp\left(-\rho_s(u,v)\right).
\]
Each \(\rho_\ell\) is a continuous semimetric of negative type, so every nonnegative linear combination \(\rho_s\) is again a continuous function of negative type. Moreover, if \(s\in(0,\infty)^L\), then
\[
\rho_s(u,v)=0
\quad\Longrightarrow\quad
\rho_\ell(u_\ell,v_\ell)=0 \text{ for all } \ell
\quad\Longrightarrow\quad
u=v.
\]
Hence, for every \(s\in(0,\infty)^L\), \(\rho_s\) is a continuous semimetric of negative type on \(\cN\). Applying Proposition~\ref{prop:cmonotone_nonconstant_characteristic} with \(f(t)=e^{-t}\), it follows that \(k_s\) is integrally strictly positive definite on \(\cN\) for every \(s\in(0,\infty)^L\).

Now let \(\eta\) be a nonzero finite signed Borel measure on \(\cN\). Since \(k_{\otimes,L}\) is bounded, all the integrals below are well defined, and Tonelli's theorem yields
\[
\iint_{\cN\times\cN} k_{\otimes,L}(u,v)\,\eta(\diff u)\eta(\diff v)
=
\int_{[0,\infty)^L}
\left(
\iint_{\cN\times\cN} k_s(u,v)\,\eta(\diff u)\eta(\diff v)
\right)\mu(\diff s).
\]
Since \(\mu_\ell((0,\infty))>0\) for every \(\ell\), we have
\[
\mu((0,\infty)^L)=\prod_{\ell=1}^L \mu_\ell((0,\infty))>0.
\]
For every \(s\in(0,\infty)^L\), the kernel \(k_s\) is integrally strictly positive definite, and hence
\[
\iint_{\cN\times\cN} k_s(u,v)\,\eta(\diff u)\eta(\diff v)>0
\qquad\text{for every nonzero } \eta.
\]
Therefore,
\[
\iint_{\cN\times\cN} k_{\otimes,L}(u,v)\,\eta(\diff u)\eta(\diff v)
=
\int_{[0,\infty)^L}
\left(
\iint_{\cN\times\cN} k_s(u,v)\,\eta(\diff u)\eta(\diff v)
\right)\mu(\diff s)
>0.
\]
Hence \(k_{\otimes,L}\) is integrally strictly positive definite on \(\cN\).
\end{proof}

\subsection{Examples of characteristic kernels for random objects}
\runinhead{Negative-type semimetrics on the Hilbert sphere and Fisher--Rao distance}\label{negative_definite_Hilbert_sphere}
Let \(\cH\) be a real Hilbert space and define the Hilbert sphere and its geodesic distance as 
\begin{equation*}
\bbS(\cH) \coloneqq \left\{x\in\cH \colon \ \|x\|_{\cH}=1\right\} ,\qquad d_{\bbS}(u,v) = \arccos(\langle u,v\rangle_{\cH}),\qquad u,v\in\bbS(\cH).
\end{equation*}

\begin{proposition}\label{prop:sphere_nd}
For every \(0<q\le 1\), the semimetric \(d_{\bbS}(u,v)^{q}\) on \(\bbS(\cH)\) is of negative type.
\end{proposition}
\begin{proof}
It suffices to prove the case \(q=1\); the case \(0<q<1\) then follows from \citet[Corollary~2.10]{Bergetal1984}. Fix \(x_1,\dots,x_n\in\bbS(\cH)\) and \(\alpha_1,\dots,\alpha_n\in\bbR\) with \(\sum_{i=1}^n\alpha_i=0\). Using
\begin{equation*}
\arccos t=\frac{\pi}{2}-\arcsin t= \frac{\pi}{2}-\sum_{k=0}^{\infty} c_k\,t^{\,2k+1}\enspace(|t|\le 1),\enspace
c_k\coloneqq \frac{(2k)!}{4^{k}(k!)^2(2k+1)}>0,
\end{equation*}
we obtain
\begin{equation*}
\sum_{i,j=1}^n\alpha_i\alpha_jd_{\bbS}(x_i,x_j)
=
\underbrace{\frac{\pi}{2}\left(\sum_{i=1}^n\alpha_i\right)\left(\sum_{j=1}^n\alpha_j\right)}_{=\,0}
-\sum_{k=0}^{\infty}c_k\sum_{i,j=1}^n\alpha_i\alpha_j\,\langle x_i,x_j\rangle^{2k+1}.
\end{equation*}
For each \(m\in\bbN\), the Gram matrix \(G=(\langle x_i,x_j\rangle)_{i,j}\) is positive semidefinite. By the Schur product theorem, all Hadamard powers \(G^{\circ m}\) are again positive semidefinite \cite[Theorem~7.5.3(a)]{HornandJohnson1985}. Since \(\bigl(G^{\circ m}\bigr)_{ij}=\langle x_i,x_j\rangle^m\), it follows that 
\begin{equation*}
\sum_{i,j=1}^n\alpha_i\alpha_j\,\langle x_i,x_j\rangle^{2k+1}\geq 0, \quad \text{for every \(k\geq 0\).}
\end{equation*}
Since all coefficients \(c_k>0\) and \(|\langle x_i,x_j\rangle|\le 1\), the series is absolutely convergent and termwise nonpositive, whence
\(\sum_{i,j}\alpha_i\alpha_jd_{\bbS}(x_i,x_j)\le 0\).
Thus \(d_{\bbS}\) is of negative type.
Finally, if a measurable semimetric \(\rho\) is of negative type, then \(\rho^{\,q}\) is also of negative type for every \(q\in(0,1]\) \cite[Corollary~2.10]{Bergetal1984}. Applying this to \(\rho=d_{\bbS}\) yields the claim for all \(0<q\le 1\).
\end{proof}
\begin{corollary}\label{cor:FR_nd}
Let \((\Omega,\mu)\) be a measure space and set
\begin{equation*}
\mathfrak D \coloneqq \left\{p\geq 0\, \colon  \int_{\Omega} p\,\diff \mu = 1\right\}.
\end{equation*}
Define the square-root embedding \(\Phi\colon \mathfrak D\to \bbS(L^2(\mu))\) by \(\Phi(p)=\sqrt{p}\).
With the Fisher--Rao distance
\begin{equation*}
d_{\mathrm{FR}}(p,q)\coloneqq \arccos\left(\left\langle \sqrt{p},\sqrt{q}\right\rangle_{L^2}\right),
\end{equation*}
the function \(d_{\mathrm{FR}}^{\,q}\) is of negative type on \(\mathfrak D\) for every \(0<q\le 1\).
\end{corollary}
\begin{proof}
For all \(p,q\in\mathfrak D\),
\begin{equation}
\left\|\Phi(p)\right\|_{L^2(\mu)}^2=\int_{\Omega}p\,\diff \mu=1,
\quad
d_{\mathrm{FR}}(p,q)=\arccos(\left\langle \Phi(p),\Phi(q)\right\rangle_{L^2})
= d_{\bbS}(\Phi(p),\Phi(q)),
\end{equation}
so \(\Phi\) is an isometric embedding from \((\mathfrak D,d_{\mathrm{FR}})\) into the unit sphere \((\bbS(L^2(\mu)),d_{\bbS})\).
By Proposition~\ref{prop:sphere_nd}, \(d_{\bbS}^{\,q}\) is of negative type for all \(0<q\le 1\).
Negative type is preserved under isometric pullbacks, hence \(d_{\mathrm{FR}}^{\,q}\) is of negative type on \(\mathfrak D\).
\end{proof}
In view of Proposition~\ref{prop:sphere_nd} and Corollary~\ref{cor:FR_nd}, when the underlying Hilbert space \(\cH\) (in particular, $L^2(\mu)$) is separable, Theorem~\ref{thm:pullback} applies. Consequently, one can naturally construct a characteristic kernel by pulling back a kernel from the Hilbert sphere via the square-root embedding.

\runinhead{The $L^p$ space for time-varying random objects}
Let \((\cT,\rho_{\cT})\) be a separable metric space equipped with a finite Radon measure \(\nu\), and let \((\cM,d_{\cM})\) be a metric space. For a fixed \(1\le p<\infty\), we consider the class of \(\nu\)-measurable maps \(f\colon \cT \to \cM\), denoted by \(\mathsf{Meas}_{\nu}(\cT,\cM)\).

\begin{definition}[Metric-valued \(L^p\) space]
We define the metric-valued \(L^p\) space as the set of equivalence classes of measurable maps,
\begin{equation*}
L^p(\cT,\cM)
\coloneqq
\left\{f\in \mathsf{Meas}_{\nu}(\cT,\cM): 
\int_{\cT} d_{\cM}(f(t),o)^p\,\nu(\diff t)<\infty\ \text{for some }o\in\cM\right\}\big/\sim,
\end{equation*}
where the equivalence relation is given by \(f\sim g\) if and only if \(d_{\cM}(f(t),g(t))=0\) for \(\nu\)-almost every \(t\in\cT\). The space is equipped with the distance
\begin{equation*}
D_p(f,g)\coloneqq
\left(\int_{\cT} d_{\cM}(f(t),g(t))^p\,\nu(\diff t)\right)^{1/p}.
\end{equation*}
\end{definition}

We establish the following properties for the space $(L^p(\cT,\cM),D_p)$:
\begin{enumerate}
    \item It is a metric space.
    \item The space is well-defined regardless of the choice of \(o\in\cM\).
    \item It is complete if \(\cM\) is complete.
    \item It is separable if \(\cM\) is separable.
    \item $D_p^{\,q}$ is of negative type for any $q \in (0,p]$ if $d_{\cM}^{\,p}$ is of negative type.
\end{enumerate}
Consequently, for a Polish space \(\cM\) where $d_{\cM}^{\,p}$ is of negative type, the distance-induced exponential kernel $\exp(-\gamma D_p(f,g)^q)$ and its generalization for any $q \in (0,p]$ are characteristic on this space. While the topological properties of \(L^p\) spaces taking values in a general metric space follow from arguments analogous to those for Banach-valued Bochner spaces \citep[see, e.g.,][]{Heinonenetal2015}, which rely solely on the triangle inequality rather than linear structure, finding a self-contained reference is surprisingly difficult. We provide the full proofs here for completeness.

\begin{proposition}\label{prop:metric}
\(D_p\) is a metric on \(L^p(\cT,\cM)\).
\end{proposition}
\begin{proof}
Non-negativity and symmetry are immediate.
For the triangle inequality, for all \(t\) we have
\(d_{\cM}(f(t),h(t))\le d_{\cM}(f(t),g(t))+d_{\cM}(g(t),h(t))\).
Taking the \(L^p\)-norm in \(t\) (Minkowski’s inequality) yields \(D_p(f,h)\le D_p(f,g)+D_p(g,h)\).
Finally, if \(D_p(f,g)=0\), then \(d_{\cM}(f(t),g(t))=0\) \(\nu\)-a.e.; by the definition of the quotient, this means \(f=g\) in \(L^p(\cT,\cM)\).
\end{proof}

\begin{proposition}[Independence of the base point]\label{prop:basepoint}
Assume \(\nu(\cT)<\infty\).
If \begin{equation}
    \int d_{\cM}(f(t),o)^p\,\nu(\diff t)<\infty, 
\end{equation} 
for some \(o\in\cM\), then for every \(o'\in\cM\),
\(\int d_{\cM}(f(t), o')^p\,\nu(\diff t)<\infty\).
Thus the definition of \(L^p(\cT,\cM)\) does not depend on the base point \(o\).
\end{proposition}
\begin{proof}
For all \(t\),
\(d_{\cM}(f(t), o')\le d_{\cM}(f(t),o)+d_{\cM}(o,o')\).
Hence, using \((a+b)^p\le 2^{p-1}(a^p+b^p)\),
\begin{equation*}
\int_{\cT} d_{\cM}(f(t), o')^p\,\nu(\diff t)
\leq  
2^{p-1}\int_{\cT} d_{\cM}(f(t),o)^p\,\nu(\diff t)
 + 
2^{p-1}d_{\cM}(o,o')^p\,\nu(\cT)
 <  \infty.
\end{equation*}
\end{proof}

\begin{lemma}\label{lem:meas-limit}
Let \(f_n\colon \cT\to\cM\) be \(\nu\)-measurable and suppose \(f_n(t)\to f(t)\) for \(\nu\)-a.e.\ \(t\).
Then \(f\) is \(\nu\)-measurable.
\end{lemma}

\begin{proof}
Define
\begin{equation*}
A\coloneqq\bigl\{t\in\cT: \{f_n(t)\}_{n\geq1} \text{converges in } \cM\bigr\}.
\end{equation*}
Since \(t\mapsto d_{\cM}(f_n(t),f_m(t))\) is measurable for all \(n,m\) and
\begin{equation*}
A=\bigcap_{k=1}^{\infty}\ \bigcup_{N=1}^{\infty}\ \bigcap_{n,m\geq N}
\Bigl\{t:\ d_{\cM}(f_n(t),f_m(t))<\tfrac1k\Bigr\},
\end{equation*}
the set \(A\) is measurable.

Fix \(x_0\in\cM\) and define \(g\,\colon \cT\to\cM\) by
\begin{equation*}
g(t)=\begin{cases}
\lim_{n\to\infty} f_n(t),& t\in A,\\
o,& t\notin A.
\end{cases}
\end{equation*}
We claim that \(g\) is measurable. Let \(F\subset\cM\) be closed and set
\(h_n(t)\coloneqq d_{\cM}(f_n(t),F)\) and \(h(t)\coloneqq \liminf_{n\to\infty} h_n(t)\).
Each \(h_n\) is measurable (distance to a closed set is continuous), hence \(h\) is measurable.
For \(t\in A\) we have \(g(t)=\lim_n f_n(t)\), so \(d_{\cM}(g(t),F)=\lim_n h_n(t)=h(t)\).
Therefore
\begin{equation*}
\{t:g(t)\in F\}
=(A\cap\{h=0\})\cup (A^{c}\cap\{o\in F\}),
\end{equation*}
which is measurable. As closed sets generate the Borel \(\sigma\)-algebra on \(\cM\),
\(g\) is Borel measurable.

By assumption there exists a \(\nu\)-null set \(N\) such that \(f_n(t)\to f(t)\) for all \(t\in \cT\setminus N\).
In particular \(\cT\setminus N\subset A\) and \(g(t)=f(t)\) for \(t\in \cT\setminus N\).
Modifying \(f\) on the null set \(N\) (which does not affect \(\nu\)-measurability in the completion),
we may assume \(f=g\). Hence \(f\) is \(\nu\)-measurable.
\end{proof}

\begin{proposition}[Completeness]\label{thm:completeness}
Assume \(\nu(\cT)<\infty\) and that \((\cM,d_{\cM})\) is complete.
Then \(\left(L^p(\cT,\cM),D_p\right)\) is complete.
\end{proposition}

\begin{proof}
Let \(\{f_n\}\) be a \(D_p\)-Cauchy sequence.
Choose a subsequence \(\{f_{n_k}\}\) with \(D_p(f_{n_{k+1}},f_{n_k})<2^{-k}\).
Define \(h_k(t)\coloneqq d_{\cM}(f_{n_{k+1}}(t),f_{n_k}(t))\geq 0\).
By Hölder and \(\nu(\cT)<\infty\),
\begin{equation*}
\int_{\cT} h_k\,\diff \nu
\leq  
\nu(\cT)^{1-1/p}\left(\int_{\cT} h_k^p\,\diff \nu\right)^{1/p}
 = 
\nu(\cT)^{1-1/p} D_p(f_{n_{k+1}},f_{n_k})
\leq  
\nu(\cT)^{1-1/p}2^{-k}.
\end{equation*}
Thus \(\sum_k\int h_k\,\diff \nu<\infty\), and by Tonelli, \(\sum_k h_k(t)<\infty\) for \(\nu\)-a.e.\ \(t\).
Hence \(\{f_{n_k}(t)\bigr\}\) is Cauchy in \(\cM\) for \(\nu\)-a.e.\ \(t\), and completeness of \(\cM\) yields a pointwise limit \(f(t)\).
By Lemma~\ref{lem:meas-limit}, \(f\) is \(\nu\)-measurable.

We claim \(D_p(f_{n_k},f)\to 0\).
For a.e.\ \(t\), \(d_{\cM}\left(f_{n_k}(t),f(t)\right)\le \sum_{j\geq k}h_j(t)\).
By Minkowski’s inequality for finite sums and then by monotone convergence,
\begin{equation*}
D_p(f_{n_k},f)
=
\left\|d_{\cM}(f_{n_k},f)\right\|_{L^p}
\le
\Bigl\|\sum_{j\geq k} h_j\Bigr\|_{L^p}
\le
\sum_{j\geq k} \left\|h_j\right\|_{L^p}
=
\sum_{j\geq k} D_p(f_{n_{j+1}},f_{n_j})
 \xrightarrow[k\to\infty]{} 0.
\end{equation*}
Finally, since \(\{f_n\}\) is Cauchy, for any \(\varepsilon>0\) choose \(K\) such that \(D_p(f_n,f_{n_k})<\varepsilon/2\) for all \(n\geq N\), and pick \(K\) large so that \(D_p(f_{n_k},f)<\varepsilon/2\).
Then \(D_p(f_n,f)\le D_p(f_n,f_{n_k})+D_p(f_{n_k},f)<\varepsilon\) for \(n\geq N\).
\end{proof}
\begin{proposition}[Separability]\label{thm:separability}
Assume \((\cT,\rho_{\cT})\) is a separable metric space and \(\nu\) is a finite Radon measure on \(\cT\).
If \((\cM,d_{\cM})\) is separable, then \(L^p(\cT,\cM)\) equipped with \(D_p\) is separable for every \(1\le p<\infty\).
\end{proposition}

\begin{proof}
Since \(\cM\) is separable, fix a countable dense subset \(D=\{y_m\}_{m\geq1}\subset\cM\).
Because \(\cT\) is separable metric, it is second countable; let \(\cB=\{B_n\}_{n\geq1}\) be a countable base of open sets in \(\cT\).
Let \(\cA=\mathrm{alg}(\cB)\) be the algebra generated by \(\cB\) (finite unions and relative complements in \(\cT\)); then \(\cA\) is countable.
Define
\begin{equation*}
\mathfrak S\coloneqq\left\{\sum_{j=1}^{k} y_j\mathds 1_{A_j} : k\in\bbN, y_j\in D, A_j\in\cA~~\text{pairwise disjoint,  value }o\in D\text{ on }\cT\setminus\textstyle\bigcup_jA_j\right\}.
\end{equation*}
Clearly \(\mathfrak S\) is countable. It remains to show that \(\mathfrak S\) is dense in \(L^p(\cT,\cM)\). Here, since \((\cM,d_{\cM})\) is merely a metric space (not a vector space), the notation \(\sum_{j=1}^{k} y_j\,\mathds 1_{A_j}\) denotes the piecewise-constant map \(t\mapsto y_j\) for \(t\in A_j\) (and \(o\) otherwise); the pairwise disjointness of \(\{A_j\}\) ensures that this is well-defined, even without the linear structure of a vector space.

\emph{Density.}
Fix \(f\in L^p(\cT,\cM)\) and \(o\in\cM\) with \(\int_{\cT} d_{\cM}(f(t),o)^p\,\nu(\diff t)<\infty\).
Let \(\varepsilon>0\) be arbitrary.

By monotone convergence, choose \(R>0\) so that
\begin{equation}\label{eq:trancation}
\int_{\cT\setminus E_{R}} d_{\cM}(f(t),o)^p\,\nu(\diff t) <  \left(\frac{\varepsilon}{4}\right)^p,
\qquad
E_{R}\coloneqq\left\{t\in\cT:\ d_{\cM}(f(t),o)\le R\right\}.
\end{equation}

Fix \(\delta>0\) to be specified later.
Because \(D\) is dense, \(\{B_{\cM}(y_m,\delta)\}_{m\geq1}\) covers \(\cM\).
Define \(A'_m\coloneqq f^{-1}\left(B_{\cM}\left(y_m,\delta\right)\right)\cap E_{R}\) (measurable), and define recursively
\begin{equation*}
  A_1\coloneqq A'_1,\qquad
  A_m\coloneqq A'_m\setminus \bigcup_{j=1}^{m-1} A_j\quad(m\geq 2).
\end{equation*}
This makes the family \(\{A_m\}_{m\geq1}\) pairwise disjoint, and
\begin{equation*}
  E_{R} = \bigsqcup_{m\geq 1} A_m,
\end{equation*}
where the union is disjoint.

Pick \(N\) so that 
\begin{equation}\label{eq:eta_E_R_A_n}
\nu\left(E_{R}\setminus \bigcup_{m=1}^{N} A_m\right)<\eta,
\end{equation}
where \(\eta>0\) will be chosen later.
Define the finite-valued map
\begin{equation*}
s_{R,\delta,N}(t)\coloneqq\begin{cases}
y_m,& t\in A_m,\ m=1,\dots,N,\\
o,& t\in \cT\setminus \bigcup_{m=1}^{N} A_m.
\end{cases}
\end{equation*}
Then, by \eqref{eq:trancation}, \eqref{eq:eta_E_R_A_n} and the construction of $\{A_m\}$,
\begin{equation*}
\begin{aligned}
D_p\left(f,s_{R,\delta,N}\right)^p
&\le \int_{\bigcup_{m=1}^{N} A_m}
        d_{\cM}\left(f(t),y_m\right)^p\,\nu(\diff t)\\
&\quad + \int_{E_{R}\setminus \bigcup_{m=1}^{N} A_m}
        d_{\cM}(f(t),o)^p\,\nu(\diff t)\\
&\quad + \int_{\cT\setminus E_{R}}
        d_{\cM}(f(t),o)^p\,\nu(\diff t)\\
&\le \delta^p\,\nu(\cT) + R^p\eta + \left(\frac{\varepsilon}{4}\right)^p.
\end{aligned}
\end{equation*}

As \(\nu(\cT)<\infty\), choosing \(\delta,\eta\) sufficiently small gives \(D_p\left(f,s_{R,\delta,N}\right)<\varepsilon/2\).

Fix a countable base \(\{B_n\}_{n\geq 1}\) of open sets in \(\cT\) and let \(\cA\coloneqq \mathrm{alg}\left(\{B_n\}\right)\) be the algebra they generate (finite unions and relative complements in \(\cT\)); then \(\cA\) is countable.
Given a measurable \(A\subset \cT\) and \(\theta>0\), inner and outer regularity of \(\nu\) provide a compact \(K\subset A\) and an open \(O\supset A\) with
\begin{equation*}
  \nu\left(A\setminus K\right)<\frac{\theta}{2},
  \qquad
  \nu\left(O\setminus A\right)<\frac{\theta}{2}.
\end{equation*}
Because \(\{B_n\}\) is a base and \(K\subset O\), there are indices \(n_1,\dots,n_{J}\) with
\begin{equation*}
  K\subset  U \coloneqq  \bigcup_{j=1}^{J} B_{n_j}\subset  O,
\end{equation*}
hence $U\in\cA$ and
\begin{equation*}
\nu(A \triangle U)\leq   \nu(A\setminus K)+\nu(O\setminus A) <  \theta,
\end{equation*}
where $\triangle$ denotes the symmetric difference.

Apply this with \(A=A_m\) to obtain \(S_m\in\cA\) such that \(\nu(A_m\triangle S_m)<\theta\) for \(m=1,\dots,N\).
Define
\begin{equation*}
  \widetilde S_1\coloneqq S_1,
  \qquad
  \widetilde S_m\coloneqq S_m\setminus \bigcup_{j<m} S_j\quad(m\geq 2),
\end{equation*}
so that \(\widetilde S_1,\dots,\widetilde S_n\in\cA\) are pairwise disjoint.
Define \(E_m\coloneqq A_m\triangle S_m\).
Then, for each \(m\),
\begin{equation*}
  A_m\triangle \widetilde S_m
  \subset  E_m \cup \left(S_m\cap \bigcup_{j<m} S_j\right),
\end{equation*}
and therefore
\begin{equation*}
  \nu\left(A_m\triangle \widetilde S_m\right)
  \leq   \nu(E_m)\ +\ \sum_{j<m}\nu\left(S_m\cap S_j\right).
\end{equation*}
Moreover, if \(x\in S_m\cap S_j\) and \(x\notin E_m\cup E_j\), then \(x\in A_m\cap A_j=\varnothing\) (since the \(A_{r}\) are disjoint), which is impossible; hence
\begin{equation*}
 S_m\cap S_j\subset  E_m\cup E_j
 \implies
 \nu(S_m\cap S_j)\leq  \nu(E_m)+\nu(E_j).
\end{equation*}
Summing in \(m\) yields
\begin{equation*}
  \sum_{m=1}^{N}\nu\left(A_m\triangle \widetilde S_m\right)
  \le \sum_{m=1}^{N}\nu(E_m)
     + \sum_{m=1}^{N}\sum_{j<m}\bigl(\nu(E_m)+\nu\left(E_j\right)\bigr)
  = N\,\sum_{m=1}^{N}\nu(E_m).
\end{equation*}
In particular, if \(\nu(E_m)\le \theta\) for all \(m\), then
\begin{equation*}
  \sum_{m=1}^{N}\nu\left(A_m\triangle \widetilde S_m\right)\leq   N^2\theta.
\end{equation*}

Define \(s\in \mathfrak S\) by \(s(t)=y_m\) on \(\widetilde S_m\) \((m=1,\dots,N)\) and \(s(t)=o\) elsewhere.
Note that whenever \(\widetilde S_m\neq\emptyset\), necessarily \(A_m\neq\emptyset\), hence there exists \(t\in A_m\) with
\(d_{\cM}(y_m,o)\le d_{\cM}(y_m,f(t))+d_{\cM}(f(t),o)<\delta+R\).
Therefore \(d_{\cM}(s_{R,\delta,N}(t),o)\le R+\delta\) and \(d_{\cM}(s(t),o)\le R+\delta\) for all \(t\).

Let \(U\coloneqq\bigcup_{m=1}^{N}\left(A_m\triangle \widetilde S_m\right)\).
Then \(s_{R,\delta,N}=s\) on \(\cT\setminus U\), and on \(U\),
\(
d_{\cM}\left(s_{R,\delta,N}(t),s(t)\right)\le 2(R+\delta).
\)
Hence
\begin{equation*}
\begin{aligned}
D_p\left(s_{R,\delta,N},s\right)^p
&=\int_U d_{\cM}\left(s_{R,\delta,N}(t),s(t)\right)^p\,\nu(\diff t)
\leq   \left(2(R+\delta)\right)^p\,\nu(U)\\
&\le \left(2(R+\delta)\right)^p\,\sum_{m=1}^{N}\nu\left(A_m\triangle \widetilde S_m\right)
\leq   \left(2(R+\delta)\right)^p\,N^2\theta.
\end{aligned}
\end{equation*}
Choose
\(
\theta\coloneqq \dfrac{\left(\varepsilon/2\right)^p}{\left(2(R+\delta)\right)^pN^2}
\)
to get \(D_p\left(s_{R,\delta,N},s\right)<\varepsilon/2\).

By the triangle inequality,
\begin{equation*}
D_p(f,s)\leq   D_p(f,s_{R,\delta,N})+D_p(s_{R,\delta,N},s) <  \frac{\varepsilon}{2}+\frac{\varepsilon}{2} = \varepsilon.
\end{equation*}
Since \(s\in\mathfrak S\) and \(\varepsilon>0\) was arbitrary, \(\mathfrak S\) is dense in \(L^p(\cT,\cM)\).
As \(\mathfrak S\) is countable, \(L^p(\cT,\cM)\) is separable.
\end{proof}

\begin{corollary}[The Polish property and kernels]\label{cor:polish}
If \(\cM\) is Polish, then \(L^p(\cT,\cM)\) is Polish.
Moreover, if \(d_{\cM}^{\,p}\) is of negative type, then for any \(q\in(0,p]\), \(D_p^{\,q}\) is of negative type on \(L^p(\cT,\cM)\). Thus, \(h\colon[0,\infty)\to\bbR\) is non-constant, completely monotone on \((0,\infty)\), and satisfies \(h(0)=\lim_{t\downarrow 0}h(t)\), then
\begin{equation*}
k_{h}(f,g)\coloneqq h(D_p(f,g)^q)
\end{equation*}
is a bounded continuous characteristic kernel.
\end{corollary}
\begin{proof}
Since Polishness follows from Theorem~\ref{thm:completeness} and \ref{thm:separability}, to show that the generalized exponential kernel is positive definite and characteristic it suffices, by Theorem~5.4, to prove that the distance $D_p^p$ is of negative type.

Fix $n\in\bbN$, $f_1,\dots,f_n\in L^p(\cT,\cM)$, and $c_1,\dots,c_n\in\bbR$ with $\sum_{i=1}^nc_i=0$.
Define
\begin{equation*}
S(t)\coloneqq\sum_{i=1}^n\sum_{j=1}^n c_i c_j d_{\cM}(f_i(t),f_j(t))^p\qquad(t\in\cT).
\end{equation*}
Since $d_{\cM}^{\,p}$ is of negative type on $\cM$, we have $S(t)\le 0$ for each fixed $t$.

To justify integrability and interchange of the (finite) sum with the integral, fix $o\in\cM$ and note that
\begin{equation*}
d_{\cM}(f_i(t),f_j(t))^p
 \leq  2^{p-1}\left\{d_{\cM}(f_i(t),o)^p+d_{\cM}\left(f_j(t),o\right)^p\right\},
\end{equation*}
so each map $t\mapsto d_{\cM}(f_i(t),f_j(t))^p$ lies in $L^1(\nu)$ (because $f_i\in L^p(\cT, \cM)$).
Hence $S\in L^1(\nu)$, and by linearity of the Lebesgue integral for finite sums,
\begin{equation*}
\sum_{i=1}^n\sum_{j=1}^n c_i c_j\, D_p\left(f_i,f_j\right)^p
=\sum_{i,j} c_i c_j \int_{\cT} d_{\cM}(f_i(t),f_j(t))^p\,\nu(\diff t)
=\int_{\cT} S(t)\,\nu(\diff t)\leq   0.
\end{equation*}
Therefore, $D_p^{\,p}$ is of negative type on $L^p(\cT,\cM)$.
\end{proof}

\runinhead{Manifold-valued data}

In what follows, we show that characteristic kernels can be constructed on a general manifold by patching local pullback kernels through a partition of unity.

Let \(\cM\) be a Hausdorff, second-countable \(C^0\)-manifold modelled on a separable Hilbert space \(\cH\). Since \(\cH\) is metrizable and \(\cM\) is locally homeomorphic to \(\cH\), the manifold \(\cM\) is metrizable; in particular, it is paracompact. Hence every open cover of \(\cM\) admits a locally finite open refinement and a continuous partition of unity subordinate to it. In particular, any countable atlas \(\left\{(U_n,\varphi_n)\right\}_{n\ge 1}\) admits a continuous partition of unity \(\left\{\eta_n\right\}_{n\ge 1}\) such that
\[
0\le \eta_n \le 1,\qquad \operatorname{supp}(\eta_n)\subset U_n,\qquad \sum_{n=1}^\infty \eta_n(x)=1 \quad \text{for all } x\in\cM,
\]
and the family \(\left\{\operatorname{supp}(\eta_n)\right\}_{n\ge 1}\) is locally finite.

The following proposition gives the desired construction.
\begin{proposition}\label{prop:patched_manifold_kernel}
Let \(\cM\) be a Hausdorff, second-countable \(C^0\)-manifold modelled on a separable Hilbert space \(\cH\). Let \(\left\{(U_n,\varphi_n)\right\}_{n\ge 1}\) be a countable atlas on \(\cM\), where each
\[
\varphi_n:U_n\to V_n\subset \cH
\]
is a homeomorphism onto an open subset \(V_n\). Let \(\left\{\eta_n\right\}_{n\ge 1}\) be a continuous partition of unity subordinate to \(\left\{U_n\right\}_{n\ge 1}\). Let \(k:\cH\times \cH\to\bbR\) be a bounded continuous positive definite kernel.

For each \(n\ge 1\), define
\[
K_n(x,y)\coloneqq 
\begin{cases}
\eta_n(x)\eta_n(y)\,k\left(\varphi_n(x),\varphi_n(y)\right), & x,y\in U_n,\\
0, & \text{otherwise},
\end{cases}
\qquad x,y\in\cM,
\]
and set
\[
K(x,y)\coloneqq \sum_{n=1}^\infty K_n(x,y), \qquad x,y\in\cM.
\]

Then the following statements hold.

\begin{enumerate}
\item The kernel \(K\) is well defined, bounded, continuous, and positive definite on \(\cM\).

\item If \(k\) is ISPD on \(\cH\) with respect to finite signed Borel measures, then \(K\) is ISPD on \(\cM\). Consequently, \(K\) is characteristic on \(\cP(\cM)\).
\end{enumerate}
\end{proposition}
\begin{remark}\label{rem:patched_kernel_ispd_needed}
In general, the assumption that \(k\) be ISPD cannot be replaced by the weaker assumption that \(k\) is merely characteristic. The reason is that, in the proof above, the localized measures
\[
\lambda_n=\left(\eta_n\lambda\right)\circ \varphi_n^{-1}
\]
are arbitrary finite signed measures on \(V_n\), rather than signed measures of total mass zero. Characteristicity of \(k\) only controls differences of probability measures (equivalently, signed measures with total mass zero), whereas the patching argument requires injectivity on all finite signed measures. This is precisely why ISPD is the natural assumption in Proposition~\ref{prop:patched_manifold_kernel}.

However, this causes no essential difficulty in practice: by Proposition~\ref{prop:characteristic_plus_constant_ispd}, if \(k\) is bounded, continuous, positive definite, and characteristic, then \(k+c\) is ISPD for every \(c>0\). Thus one may replace \(k\) by \(k+c\) and then apply Proposition~\ref{prop:patched_manifold_kernel}.
\end{remark}
\begin{proof}
Let $M > 0$ be an upper bound for the kernel $k$, so that $\sup_{x,y} |k(x,y)| \leq M$.

We first show that the kernel \(K\) is pointwise well defined. Fix \(x,y\in\cM\). Since \(\left\{\operatorname{supp}(\eta_n)\right\}_{n\ge 1}\) is locally finite, the sets
\[
I_x\coloneqq \left\{n\ge 1:\eta_n(x)\neq 0\right\},
\qquad
I_y\coloneqq \left\{n\ge 1:\eta_n(y)\neq 0\right\}
\]
are finite. If \(K_n(x,y)\neq 0\), then necessarily \(\eta_n(x)\neq 0\) and \(\eta_n(y)\neq 0\), hence
\[
n\in I_x\cap I_y.
\]
Therefore only finitely many terms in the series defining \(K(x,y)\) are nonzero. Thus \(K\) is pointwise well defined.

Next, we verify the continuity of each local kernel \(K_n\). Fix \(n\ge 1\). On \(U_n\times U_n\), the map
\[
(x,y)\mapsto \eta_n(x)\eta_n(y)\,k\left(\varphi_n(x),\varphi_n(y)\right)
\]
is continuous, because \(\eta_n\), \(\varphi_n\), and \(k\) are continuous. It remains to check continuity across the boundary of \(U_n\times U_n\).

For all \(x,y\in\cM\), one has
\[
\left|K_n(x,y)\right|
\le
M\,\eta_n(x)\eta_n(y).
\]
Indeed, if \(x,y\in U_n\), this is immediate from the definition; if \(x\notin U_n\) or \(y\notin U_n\), then \(K_n(x,y)=0\), while \(\eta_n(x)=0\) or \(\eta_n(y)=0\) because \(\operatorname{supp}(\eta_n)\subset U_n\).

Now let \((x_m,y_m)\to(x,y)\) in \(\cM\times\cM\), and suppose that \(x\notin U_n\) or \(y\notin U_n\). Then \(\eta_n(x)=0\) or \(\eta_n(y)=0\). By continuity of \(\eta_n\),
\[
\eta_n(x_m)\eta_n(y_m)\to 0,
\]
and therefore
\[
\left|K_n(x_m,y_m)\right|
\le
M\,\eta_n(x_m)\eta_n(y_m)\to 0 = K_n(x,y).
\]
Hence \(K_n\) is continuous on \(\cM\times\cM\).

To see that \(K\) is bounded, observe that for every \(x,y\in\cM\),
\[
\left|K(x,y)\right|
\le
\sum_{n=1}^\infty \left|K_n(x,y)\right|
\le
M\sum_{n=1}^\infty \eta_n(x)\eta_n(y)
\le
M\left(\sum_{n=1}^\infty \eta_n(x)\right)\left(\sum_{n=1}^\infty \eta_n(y)\right)
=
M.
\]
Thus \(K\) is bounded.

We now prove the continuity of \(K\). To do so, fix \((x,y)\in\cM\times\cM\). By local finiteness, there exist neighborhoods \(O_x\ni x\) and \(O_y\ni y\) such that only finitely many supports \(\operatorname{supp}(\eta_n)\) meet \(O_x\cup O_y\). Hence, on \(O_x\times O_y\), only finitely many kernels \(K_n\) are nonzero. Therefore \(K\) is locally a finite sum of continuous functions, and is thus continuous on \(\cM\times\cM\).

To establish that \(K\) is positive definite, let \(m\ge 1\), let \(x_1,\dots,x_m\in\cM\), and let \(c_1,\dots,c_m\in\bbR\). Since the family \(\left\{\operatorname{supp}(\eta_n)\right\}_{n\ge 1}\) is locally finite, only finitely many indices \(n\) satisfy \(\eta_n(x_i)\neq 0\) for at least one \(i\). Hence
\[
\sum_{i,j=1}^m c_i c_j K(x_i,x_j)
=
\sum_{n=1}^\infty \sum_{i,j=1}^m c_i c_j K_n(x_i,x_j),
\]
and the outer sum is actually finite.

Fix \(n\). Then
\[
\sum_{i,j=1}^m c_i c_j K_n(x_i,x_j)
=
\sum_{\substack{1\le i,j\le m\\ x_i,x_j\in U_n}}
\left(c_i\eta_n(x_i)\right)\left(c_j\eta_n(x_j)\right)
k\left(\varphi_n(x_i),\varphi_n(x_j)\right).
\]
Since \(k\) is positive definite on \(\cH\), the right-hand side is nonnegative. Summing over \(n\) yields
\[
\sum_{i,j=1}^m c_i c_j K(x_i,x_j)\ge 0.
\]
Thus \(K\) is positive definite.

Turning to the second statement, we first establish an energy decomposition for finite signed measures. Let \(\lambda\) be a finite signed Borel measure on \(\cM\). For each \(n\), define a finite signed Borel measure on \(V_n\) by
\[
\lambda_n\coloneqq \left(\eta_n\lambda\right)\circ \varphi_n^{-1},
\]
where \(\eta_n\lambda\) is the signed measure on \(\cM\) given by
\[
\left(\eta_n\lambda\right)(A)\coloneqq \int_A \eta_n(x)\,\lambda(\diff x).
\]

Since \(K\) is bounded, the energy
\[
\iint_{\cM\times\cM} K(x,y)\,\lambda(\diff x)\lambda(\diff y)
\]
is well defined. We claim that
\[
\iint_{\cM\times\cM} K(x,y)\,\lambda(\diff x)\lambda(\diff y)
=
\sum_{n=1}^\infty
\iint_{V_n\times V_n} k(u,v)\,\lambda_n(\mathrm du)\lambda_n(\mathrm dv).
\]
To see this, first note that
\[
\sum_{n=1}^\infty \left|K_n(x,y)\right|
\le
M\sum_{n=1}^\infty \eta_n(x)\eta_n(y)
\le M
\]
for all \(x,y\in\cM\). Hence the partial sums \(\sum_{n=1}^N K_n(x,y)\) are dominated by the integrable function \(M\) with respect to \(\left|\lambda\right|\otimes \left|\lambda\right|\). Dominated convergence therefore gives
\[
\iint K(x,y)\,\lambda(\diff x)\lambda(\diff y)
=
\sum_{n=1}^\infty \iint K_n(x,y)\,\lambda(\diff x)\lambda(\diff y).
\]
For each \(n\),
\[
\iint_{\cM\times\cM} K_n(x,y)\,\lambda(\diff x)\lambda(\diff y)
=
\iint_{V_n\times V_n} k(u,v)\,\lambda_n(\mathrm du)\lambda_n(\mathrm dv)
\]
by the definition of push-forward. This proves the claim.

Using this decomposition, we now show that \(K\) is ISPD whenever \(k\) is ISPD. Assume that \(k\) is ISPD on \(\cH\). Let \(\lambda\) be a finite signed Borel measure on \(\cM\) such that
\[
\iint_{\cM\times\cM} K(x,y)\,\lambda(\diff x)\lambda(\diff y)=0.
\]
By the decomposition above,
\[
0=
\sum_{n=1}^\infty
\iint_{V_n\times V_n} k(u,v)\,\lambda_n(\mathrm du)\lambda_n(\mathrm dv).
\]
Each term is nonnegative because \(k\) is positive definite. Hence every term must be zero:
\[
\iint_{V_n\times V_n} k(u,v)\,\lambda_n(\mathrm du)\lambda_n(\mathrm dv)=0
\qquad \text{for all } n\ge 1.
\]

We now use the ISPD property. Since \(V_n\subset \cH\), each \(\lambda_n\) may be regarded as a finite signed Borel measure on \(\cH\) by extension by zero outside \(V_n\). The corresponding energy with respect to \(k\) is unchanged. Therefore the ISPD property of \(k\) implies
\[
\lambda_n=0
\qquad \text{for all } n\ge 1.
\]
Equivalently,
\[
\eta_n\lambda=0
\qquad \text{for all } n\ge 1.
\]

It remains to show that \(\lambda=0\). Let \(f:\cM\to\bbR\) be any bounded Borel function. Since
\[
0\le \sum_{n=1}^N \eta_n(x)\le 1
\qquad \text{and} \qquad
\sum_{n=1}^N \eta_n(x)\to 1
\]
for every \(x\in\cM\), dominated convergence with respect to \(\left|\lambda\right|\) yields
\[
\int_{\cM} f(x)\,\lambda(\diff x)
=
\lim_{N\to\infty}
\int_{\cM} f(x)\left(\sum_{n=1}^N \eta_n(x)\right)\lambda(\diff x)
=
\sum_{n=1}^\infty \int_{\cM} f(x)\eta_n(x)\,\lambda(\diff x).
\]
Since \(\eta_n\lambda=0\) for every \(n\), each term on the right-hand side vanishes. Thus
\[
\int_{\cM} f(x)\,\lambda(\diff x)=0
\qquad \text{for all bounded Borel } f,
\]
which implies \(\lambda=0\). Hence \(K\) is ISPD.

Finally, we deduce the characteristic property of \(K\) on \(\cP(\cM)\). Since \(K\) is bounded and ISPD, if \(P,Q\in\cP(\cM)\) satisfy \(\mu^K(P)=\mu^K(Q)\), then for \(\lambda\coloneqq P-Q\),
\[
0
=
\left\|\mu^K(P)-\mu^K(Q)\right\|_{\cH_K}^2
=
\iint_{\cM\times\cM} K(x,y)\,\lambda(\diff x)\lambda(\diff y),
\]
and the ISPD property gives \(P=Q\).
\end{proof}
\begin{proposition}[Adding a positive constant upgrades characteristicity to ISPD]
\label{prop:characteristic_plus_constant_ispd}
Let \((\cX,\cB)\) be a measurable space, and let
\(k:\cX\times\cX\to\bbR\) be a bounded measurable positive definite kernel
that is characteristic on \(\cP(\cX)\).
For \(c>0\), define
\[
\tilde k(x,y)\coloneqq k(x,y)+c,
\qquad x,y\in\cX.
\]
Then \(\tilde k\) is bounded, measurable, positive definite, and ISPD with
respect to finite signed measures. In particular, \(\tilde k\) is
characteristic on \(\cP(\cX)\).
\end{proposition}

\begin{proof}
Boundedness and measurability are immediate. To see that \(\tilde k\) is positive definite, fix \(m\ge 1\), \(x_1,\dots,x_m\in\cX\), and \(a_1,\dots,a_m\in\bbR\). Then
\[
\sum_{i,j=1}^m a_i a_j \tilde k(x_i,x_j)
=
\sum_{i,j=1}^m a_i a_j k(x_i,x_j)
+
c\left(\sum_{i=1}^m a_i\right)^2
\ge 0,
\]
since \(k\) is positive definite.

We next prove that \(\tilde k\) is ISPD. Let \(\lambda\) be a finite signed measure on \((\cX,\cB)\), and suppose that
\[
\iint_{\cX\times\cX} \tilde k(x,y)\,\lambda(\diff x)\lambda(\diff y)=0.
\]
Since \(\tilde k=k+c\), this becomes
\[
0
=
\iint_{\cX\times\cX} k(x,y)\,\lambda(\diff x)\lambda(\diff y)
+
c\,\lambda(\cX)^2.
\]
Now the first term is nonnegative because \(k\) is positive definite, and the second term is also nonnegative because \(c>0\). Hence both terms must vanish. In particular,
\[
\lambda(\cX)=0
\qquad\text{and}\qquad
\iint_{\cX\times\cX} k(x,y)\,\lambda(\diff x)\lambda(\diff y)=0.
\]

Write the Jordan decomposition of \(\lambda\) as \(\lambda=\lambda^+-\lambda^-\). Since \(\lambda(\cX)=0\), the measures \(\lambda^+\) and \(\lambda^-\) have the same total mass; denote it by \(m\ge 0\). If \(m=0\), then \(\lambda=0\), so there is nothing to prove. Assume \(m>0\), and define probability measures \(P\coloneqq \lambda^+/m\) and \(Q\coloneqq \lambda^-/m\). Then \(\lambda=m(P-Q)\), and therefore
\[
0
=
\iint_{\cX\times\cX} k(x,y)\,\lambda(\diff x)\lambda(\diff y)
=
m^2
\iint_{\cX\times\cX} k(x,y)\,(P-Q)(\diff x)(P-Q)(\diff y).
\]
Since \(m>0\), it follows that
\[
\iint_{\cX\times\cX} k(x,y)\,(P-Q)(\diff x)(P-Q)(\diff y)=0.
\]

Because \(k\) is bounded, kernel mean embeddings are well defined for all probability measures, and the last display equals \(\|\mu_k(P)-\mu_k(Q)\|_{\cH_k}^2\). Hence \(\mu_k(P)=\mu_k(Q)\), and the characteristic property of \(k\) yields \(P=Q\). Thus \(\lambda=m(P-Q)=0\). This proves that \(\tilde k\) is ISPD.

Finally, every bounded ISPD kernel is characteristic on \(\cP(\cX)\): indeed, if \(P,Q\in\cP(\cX)\) satisfy \(\mu_{\tilde k}(P)=\mu_{\tilde k}(Q)\), then for \(\lambda\coloneqq P-Q\),
\[
0
=
\|\mu_{\tilde k}(P)-\mu_{\tilde k}(Q)\|_{\cH_{\tilde k}}^2
=
\iint_{\cX\times\cX} \tilde k(x,y)\,\lambda(\diff x)\lambda(\diff y),
\]
and the ISPD property implies \(\lambda=0\), that is, \(P=Q\).
\end{proof}

\subsection{Implications for weak Fréchet regression}\label{subsubsec:discussion_weak_frechet}

\cite{Bhattacharjeeetal2025} recently introduced the \emph{weak conditional Fr\'echet mean}, which replaces the full conditional expectation with its orthogonal projection onto an RKHS \(\cH^{(k)}\). They established that this weak mean coincides with the standard conditional Fr\'echet mean if \(\cH^{(k)}+\bbR\) is dense in \(L^2(P_X)\). 

To satisfy this density requirement on metric spaces, \cite{Bhattacharjeeetal2025} suggested using \emph{cc-universal} kernels \citep{Zhangetal2024}. However, on a non-compact predictor domain, cc-universality only guarantees uniform density on compact subsets and does \emph{not} imply \(L^2(P_X)\)-density unless \(P_X\) has compact support \citep{Sriperumbuduretal2011}.

A more direct and mathematically robust approach is to require the kernel \(k\) to be \emph{characteristic}. By \citet[Proposition~5]{Fukumizuetal2009}, a bounded measurable positive-definite kernel is characteristic if and only if \(\cH^{(k)}+\bbR\) is dense in \(L^2(P)\) for \emph{every} probability measure \(P\). Therefore, the characteristic kernels we construct on metric spaces in this paper provide exactly the \emph{support-free} \(L^2\)-density required. This directly validates the theoretical framework of \cite{Bhattacharjeeetal2025} without imposing any restrictive compact-support assumptions on the DGP.
\section{Spectral-gap conditions for empirical truncation}\label{sec:spectral_gap_growing_gap}
\subsection{General spectral-gap results}
We consider a parametric family for the law of \(X\). If the parameter space is compact and the densities are Hellinger continuous in \(\theta\), then the eigenvalues of the associated covariance operator depend continuously on \(\theta\). Therefore, whenever the relevant spectral gap is positive throughout the parameter space, its infimum is attained and remains strictly positive. In that case, one can select truncation levels \(L_{X,n}\) so that Assumption~\ref{ass:simplicity_gap_unif} is satisfied.

Let \(\Theta_0\) be a compact metric space. For each \(\theta\in\Theta_0\), let \(\rho_{\theta}\) be a Borel probability measure on \((\cX,\cB)\). Define the integral operator
\begin{equation*}
  T_{\theta}\colon L^2(\rho_{\theta})\to L^2(\rho_{\theta}),
  \quad
  (T_{\theta}f)(x)
  \coloneqq \int_{\cX} k(x,y)\,f(y)\diff \rho_{\theta}(y).
\end{equation*}
Assume that each \(T_{\theta}\) is self-adjoint and compact on \(L^2(\rho_{\theta})\), so that its eigenvalues
\begin{equation*}
  \lambda_1(\theta)\ge\lambda_2(\theta)\ge\cdots\downarrow 0
\end{equation*}
are well-defined, listed with algebraic multiplicity.
For a fixed \(P\in\bbN\), define the spectral gap
\begin{equation*}
  \gamma_P(\theta)\coloneqq\lambda_P(\theta)-\lambda_{P+1}(\theta).
\end{equation*}
We impose the following assumptions.
\begin{assumption}\label{ass:spectral_unif_sufficient}
\begin{itemize}
  \item[(A1)]
  There exists a $\sigma$-finite measure $\mu$ on $(\cX,\cB)$ such that
  \begin{equation*}
    \rho_{\theta}\ll\mu
    \quad\text{for all }\theta\in\Theta_0.
  \end{equation*}
  Hence, for each $\theta\in\Theta_0$, there exists a density
  \begin{equation*}
    p_{\theta}\coloneqq\frac{\diff\rho_{\theta}}{\diff\mu}\ge 0
    \quad\text{with}\quad
    \int_{\cX} p_{\theta}\diff \mu = 1.
  \end{equation*}

  \item[(A2)]
  The map
  \begin{equation*}
    \Theta_0\ni\theta\ \longmapsto\ \sqrt{p_{\theta}}\in L^2(\mu)
  \end{equation*}
  is continuous, i.e.
  \begin{equation*}
    \lim_{\vartheta\to\theta}
    \int_{\cX}
      \left(
        \sqrt{p_{\theta}(x)}-\sqrt{p_{\vartheta}(x)}
      \right)^2
    \diff \mu(x)
    = 0
    \quad\text{for all }\theta\in\Theta_0.
  \end{equation*}

  \item[(A3)]
  The kernel $k$ is bounded.

  \item[(A4)]
  For any $P\in\bbN$, we assume
  \begin{equation*}
    \lambda_P(\theta)>\lambda_{P+1}(\theta)
    \quad\text{for all }\theta\in\Theta_0.
  \end{equation*}
\end{itemize}
\end{assumption}

\begin{theorem}[Uniform spectral gap at level $P$]\label{thm:uniform_spectral_gap}
Suppose Assumption~\ref{ass:spectral_unif_sufficient} holds. Then, we have
\begin{equation}
  \inf_{\theta\in\Theta_0} \bigl(\lambda_P(\theta)-\lambda_{P+1}(\theta)\bigr) > 0.
\end{equation}
\end{theorem}
This theorem implies that there exists an increasing sequence of integers $L_{X,n} \to \infty$ such that 
\begin{equation*}
    \sqrt{n}g_{L_{X,n}}^X \to \infty.
\end{equation*}

If we further restrict our attention to translation invariant kernels (i.e., $k(x,y) = \kappa(x-y)$), the assumptions can be extended to accommodate location shifts. This is because, for such kernels, the eigenvalues of the integral operator do not depend on the mean parameter (a location shift), as formalized in the following proposition.
\begin{proposition}[Mean invariance for translation invariant kernels]\label{prop:mean_invariance_translation_kernel}
Let $\rho$ be a Borel probability measure on $\left(\bbR^d,\cB(\bbR^d)\right)$.
For each $\mu\in\bbR^d$, define the translated measure $\rho_{\mu}$ by
\begin{equation*}
  \rho_{\mu}(A)\coloneqq \rho(A-\mu),
  \qquad A\in\cB(\bbR^d).
\end{equation*}
Let $\kappa\colon\bbR^d\to\bbR$ be measurable and define the translation invariant kernel
\begin{equation*}
  k(x,y)\coloneqq \kappa(x-y),
  \qquad x,y\in\bbR^d.
\end{equation*}
For each $\mu\in\bbR^d$, define the integral operator
\begin{equation*}
  T_{\mu}:L^2(\rho_\mu)\to L^2(\rho_\mu),
  \qquad
  (T_{\mu}f)(x)
  \coloneqq \int_{\bbR^d} k(x,y) f(y)\rho_{\mu}(\diff y),
\end{equation*}
and assume $T_{\mu}$ is well-defined and bounded. Define also
\begin{equation*}
  T_0:L^2(\rho)\to L^2(\rho),
  \qquad
  \left(T_0g\right)(x)
  \coloneqq \int_{\bbR^d} k(x,y) g(y)\rho(\diff y).
\end{equation*}

Then for each $\mu\in\bbR^d$ there exists a unitary operator
\begin{equation*}
  U_{\mu}:L^2(\rho_\mu)\to L^2(\rho),
  \qquad
  (U_{\mu}f)(x)\coloneqq f(x+\mu),
\end{equation*}
such that
\begin{equation*}
  U_{\mu}T_{\mu}U_{\mu}^{-1} = T_0.
\end{equation*}
In particular, $T_{\mu}$ and $T_0$ are unitarily equivalent for all $\mu$, and hence
have identical spectra, including eigenvalues with their algebraic multiplicities.
Thus the eigenvalues of $T_{\mu}$ do not depend on the mean parameter $\mu$.
\end{proposition}
\begin{proof}
Let $f\in L^2(\rho_\mu)$. Then
\begin{equation*}
  \left\lVert U_{\mu}f \right\rVert_{L^2(\rho)}^2
  =
  \int_{\bbR^d} \left\lvert f(x+\mu)\right\rvert^2\rho(\diff x).
\end{equation*}
By the definition of $\rho_{\mu}$,
\begin{equation*}
  \int_{\bbR^d} g(y)\rho_{\mu}(\diff y)
  =
  \int_{\bbR^d} g(x+\mu)\rho(\diff x)
  \quad\text{for all bounded measurable } g.
\end{equation*}
Taking $g(y)=\left\lvert f(y)\right\rvert^2$ gives
\begin{equation*}
  \int_{\bbR^d} \left\lvert f(x+\mu)\right\rvert^2\rho(\diff x)
  =
  \int_{\bbR^d} \left\lvert f(y)\right\rvert^2\rho_{\mu}(\diff y)
  =
  \left\lVert f \right\rVert_{L^2(\rho_\mu)}^2.
\end{equation*}
Hence $U_{\mu}$ is an isometry.

For surjectivity, let $g\in L^2(\rho)$ and define $f\colon\bbR^d\to\bbC$ by
$f(y)\coloneqq g(y-\mu)$. Then
\begin{equation*}
  \int_{\bbR^d} \left\lvert f(y)\right\rvert^2 d\rho_{\mu}(y)
  =
  \int_{\bbR^d} \left\lvert g(x)\right\rvert^2 \rho(\diff x)
  <\infty,
\end{equation*}
so $f\in L^2(\rho_\mu)$ and
\begin{equation*}
  (U_{\mu}f)(x)
  =
  f(x+\mu)
  =
  g(x).
\end{equation*}
Thus $U_{\mu}$ is onto. Being an onto isometry, $U_{\mu}$ is unitary.

Let $g\in L^2(\rho)$ and put $f\coloneqq U_{\mu}^{-1} g$, so that
$f(y)=g(y-\mu)$. Then
\begin{equation*}
  (T_{\mu} f)(x)
  =
  \int_{\bbR^d} k(x,y) f(y)\rho_{\mu}(\diff y)
  =
  \int_{\bbR^d} \kappa(x-y) g(y-\mu)\rho_{\mu}(\diff y).
\end{equation*}
Using $\rho_{\mu}(A)=\rho(A-\mu)$ and the change of variables
$y=z+\mu$, we obtain
\begin{equation*}
  \left(T_{\mu} f\right)(x)
  =
  \int_{\bbR^d} \kappa\left(x-\left(z+\mu\right)\right) g(z)\rho(\diff z).
\end{equation*}
Now
\begin{equation*}
  \left(U_{\mu} T_{\mu} U_{\mu}^{-1} g\right)(x)
  =
  \left(U_{\mu} T_{\mu} f\right)(x)
  =
  \left(T_{\mu} f\right)(x+\mu)
\end{equation*}
and hence
\begin{equation*}
  \left(U_{\mu} T_{\mu} U_{\mu}^{-1} g\right)(x)
  =
  \int_{\bbR^d}
    \kappa\left((x+\mu)-\left(z+\mu\right)\right)
    g(z)\rho(\diff z).
\end{equation*}
By translation invariance of the kernel,
$\kappa\left((x+\mu)-\left(z+\mu\right)\right)=\kappa(x-z)$, so
\begin{equation*}
  \left(U_{\mu} T_{\mu} U_{\mu}^{-1} g\right)(x)
  =
  \int_{\bbR^d} \kappa(x-z) g(z)\rho(\diff z)
  =
  \left(T_0 g\right)(x).
\end{equation*}
Thus $U_{\mu} T_{\mu} U_{\mu}^{-1} = T_0$.

Since $U_{\mu}$ is unitary, $T_{\mu}$ and $T_0$ are unitarily equivalent.
Therefore they have identical spectra, including algebraic multiplicities of eigenvalues.
In particular, the eigenvalues of $T_{\mu}$ do not depend on $\mu$.
\end{proof}


The proof of~\ref{thm:uniform_spectral_gap} relies on mapping the operator \(T_{\theta}\), which acts on a \(\theta\)-dependent space \(L^2(\rho_{\theta})\), to an equivalent operator \(B_{\theta}\) on a fixed Hilbert space \(L^2(\mu)\).

Define the isometry \(J_{\theta}:L^2(\rho_{\theta})\to L^2(\mu)\) by \((J_{\theta}f)(x) \coloneqq \sqrt{p_{\theta}(x)}\,f(x)\), and its adjoint \(J_{\theta}^{\ast}:L^2(\mu)\to L^2(\rho_{\theta})\).
Let \(B_{\theta} \coloneqq J_{\theta}T_{\theta}J_{\theta}^{\ast}\). This is a self-adjoint Hilbert--Schmidt operator on \(L^2(\mu)\).

Two key lemmas, (Lemma \ref{lem:eig_coincide} and \ref{lem:hs_cont} below), establish the following:
\begin{enumerate}
    \item The nonzero eigenvalues of \(T_{\theta}\) and \(B_{\theta}\) coincide, including multiplicities (Lemma \ref{lem:eig_coincide}).
    \item Under assumptions (A1)--(A3), the map \(\Theta_0\ni\theta\mapsto B_{\theta}\) is continuous in the Hilbert--Schmidt norm, i.e., \(\lim_{\vartheta\to\theta} \lVert B_{\theta}-B_{\vartheta}\rVert_{\mathrm{HS}} = 0\) (Lemma \ref{lem:hs_cont}).
\end{enumerate}

Since the Hilbert--Schmidt norm dominates the operator norm (\(\lVert \cdot \rVert_{\mathrm{op}} \le \lVert \cdot \rVert_{\mathrm{HS}}\)), the map \(\theta\mapsto B_{\theta}\) is also continuous in the operator norm.
By standard perturbation results for compact self-adjoint operators, this implies that, for each fixed \(j\in\bbN\), the eigenvalue map
\begin{equation*}
  \Theta_0\ni\theta\ \longmapsto\ \lambda_j(\theta)
\end{equation*}
is continuous. Consequently, the spectral gap function
\begin{equation*}
  \Theta_0\ni\theta\ \longmapsto\ \gamma_P(\theta)
  =\lambda_P(\theta)-\lambda_{P+1}(\theta)
\end{equation*}
is continuous.

By assumption (A4), we have \(\gamma_P(\theta)>0\) for every \(\theta\in\Theta_0\).
Since \(\Theta_0\) is compact and \(\gamma_P\) is continuous, \(\gamma_P\) attains its minimum at some \(\theta^{\ast}\in\Theta_0\). By (A4), \(\gamma_P(\theta) > 0\) for all \(\theta\in\Theta_0\), so in particular \(\gamma_P(\theta^{\ast}) > 0\). Therefore,
\begin{equation*}
  c_P \coloneqq \inf_{\theta\in\Theta_0} \gamma_P(\theta) = \gamma_P(\theta^{\ast}) > 0,
\end{equation*}
which completes the proof.
\qed

The proof above relies on the following two lemmas.

\begin{lemma} \label{lem:eig_coincide}
For each fixed \(\theta\in\Theta_0\), the nonzero eigenvalues of \(T_{\theta}\) and \(B_{\theta}\)
coincide, including algebraic multiplicities.
\end{lemma}

\begin{proof}
Let \(\lambda\neq 0\).

If \(T_{\theta}f=\lambda f\) with \(f\neq 0\), then \(g\coloneqq J_{\theta}f\neq 0\) (since \(J_\theta\) is an isometry) and
\begin{equation*}
  B_{\theta}g
  =J_{\theta}T_{\theta}J_{\theta}^{\ast}J_{\theta}f
  =J_{\theta}T_{\theta}f
  =\lambda J_{\theta}f
  =\lambda g.
\end{equation*}
Conversely, if \(B_{\theta}g=\lambda g\) with \(g\neq 0\), put \(f\coloneqq J_{\theta}^{\ast}g\).
Then
\begin{equation*}
  T_{\theta}f
  =T_{\theta}J_{\theta}^{\ast}g
  =J_{\theta}^{\ast}J_{\theta}T_{\theta}J_{\theta}^{\ast}g
  =J_{\theta}^{\ast}B_{\theta}g
  =\lambda J_{\theta}^{\ast}g
  =\lambda f.
\end{equation*}
We must ensure \(f \neq 0\). If \(f=0\), then \(J_{\theta}^{\ast}g=0\). This implies
\(\langle B_{\theta}g,g\rangle_{\mu}
  =\langle J_{\theta}T_{\theta}J_{\theta}^{\ast}g,g\rangle_{\mu}
  =\langle T_{\theta}J_{\theta}^{\ast}g,J_{\theta}^{\ast}g\rangle_{\rho_{\theta}}=0\).
From \(B_{\theta}g=\lambda g\), we get \(\lambda\|g\|_{\mu}^2=0\), which contradicts \(\lambda\neq 0\) and \(g\neq 0\).
Thus \(f\neq 0\), and \(\lambda\) is an eigenvalue of \(T_{\theta}\).
The correspondences \(f\mapsto J_{\theta}f\) and \(g\mapsto J_{\theta}^{\ast}g\) restrict to inverse
linear isomorphisms between the nonzero eigenspaces, so multiplicities coincide.
\end{proof}

\begin{lemma}\label{lem:hs_cont}
Under \textnormal{(A1)}--\textnormal{(A3)}, the map
\(\Theta_0\ni\theta\mapsto B_{\theta}\) is continuous in the Hilbert--Schmidt norm:
\begin{equation*}
  \lim_{\vartheta\to\theta}
  \lVert B_{\theta}-B_{\vartheta}\rVert_{\mathrm{HS}}
  =
  \lim_{\vartheta\to\theta}
  \lVert K_{\theta}-K_{\vartheta}\rVert_{L^2(\mu\times\mu)}
  = 0.
\end{equation*}
where \(K_{\theta}(x,y) \coloneqq \sqrt{p_{\theta}(x)}\,k(x,y)\,\sqrt{p_{\theta}(y)}\) is the kernel of \(B_\theta\).
\end{lemma}

\begin{proof}
We have
\begin{equation*}
  K_{\theta}(x,y)
  =
  \sqrt{p_{\theta}(x)}\,k(x,y)\,\sqrt{p_{\theta}(y)}.
\end{equation*}
Since \(\lvert k(x,y)\rvert\le K\) (by (A3)), we obtain
\begin{equation*}
  \lVert K_{\theta}-K_{\vartheta}\rVert_{L^2(\mu \times \mu)}
  \le
  M\,
  \left\lVert
    \sqrt{p_{\theta}}\otimes\sqrt{p_{\theta}}
    -
    \sqrt{p_{\vartheta}}\otimes\sqrt{p_{\vartheta}}
  \right\rVert_{L^2(\mu \times \mu)},
\end{equation*}
where \(\otimes\) denotes the tensor product.
We use the triangle inequality:
\begin{equation*}
  \sqrt{p_{\theta}}\otimes\sqrt{p_{\theta}}
  -
  \sqrt{p_{\vartheta}}\otimes\sqrt{p_{\vartheta}}
  =
  \left(\sqrt{p_{\theta}}-\sqrt{p_{\vartheta}}\right)\otimes\sqrt{p_{\theta}}
  +
  \sqrt{p_{\vartheta}}\otimes
  \left(\sqrt{p_{\theta}}-\sqrt{p_{\vartheta}}\right).
\end{equation*}
Thus, by the properties of the \(L^2\) norm of tensor products,
\begin{align*}
  \left\lVert
    \sqrt{p_{\theta}}\otimes\sqrt{p_{\theta}}
    -
    \sqrt{p_{\vartheta}}\otimes\sqrt{p_{\vartheta}}
  \right\rVert_2
  &\le
  \left\lVert (\sqrt{p_{\theta}}-\sqrt{p_{\vartheta}})\otimes\sqrt{p_{\theta}} \right\rVert_2
  +
  \left\lVert \sqrt{p_{\vartheta}}\otimes (\sqrt{p_{\theta}}-\sqrt{p_{\vartheta}}) \right\rVert_2
  \\
  &=
  \lVert\sqrt{p_{\theta}}-\sqrt{p_{\vartheta}}\rVert_2 \lVert\sqrt{p_{\theta}}\rVert_2
  +
  \lVert\sqrt{p_{\vartheta}}\rVert_2 \lVert\sqrt{p_{\theta}}-\sqrt{p_{\vartheta}}\rVert_2
  \\
  &=
  \left(
    \lVert\sqrt{p_{\theta}}\rVert_2
    +
    \lVert\sqrt{p_{\vartheta}}\rVert_2
  \right)
  \lVert\sqrt{p_{\theta}}-\sqrt{p_{\vartheta}}\rVert_2.
\end{align*}
By (A1), \(\lVert\sqrt{p_{\theta}}\rVert_2^2=\int p_{\theta}\diff \mu=1\) (and similarly for \(\vartheta\)). Thus, the term in parentheses is \(1+1=2\).
\begin{equation*}
  \left\lVert
    \sqrt{p_{\theta}}\otimes\sqrt{p_{\theta}}
    -
    \sqrt{p_{\vartheta}}\otimes\sqrt{p_{\vartheta}}
  \right\rVert_2
  \le 2\,
  \lVert\sqrt{p_{\theta}}-\sqrt{p_{\vartheta}}\rVert_2.
\end{equation*}
By assumption (A2) (Hellinger continuity), the right-hand side converges to \(0\) as \(\vartheta\to\theta\).
Hence \(\lVert K_{\theta}-K_{\vartheta}\rVert_2\to 0\), which yields
\(\lVert B_{\theta}-B_{\vartheta}\rVert_{\mathrm{HS}}\to 0\).
\end{proof}

\subsection{Gaussian-kernel eigenvalues under Gaussian measures}\label{sec:gaussian_case_eigengap}
We now examine the divergence rate of $L_{X,n}$ necessary to fulfill the uniform growing gap condition for the Gaussian kernel under the family of Gaussian measures. As established in Proposition~\ref{prop:mean_invariance_translation_kernel}, the eigenvalues remain invariant with respect to mean shifts. Consequently, we fix the mean $\mu$ and define our parameter set solely by the variance.

For \(\gamma>0\), define the Gaussian kernel as
\begin{equation*}
  k_{\gamma}(x,y)\coloneqq \exp(-\gamma^{-2}(x-y)^2),\qquad x,y\in\bbR.
\end{equation*}
For mean \(\mu\in\bbR\) and variance \(\sigma^2\in[\sigma_{\min}^2,\sigma_{\max}^2]\subset\left(0,\infty\right)\), let \(\nu_{\mu,\sigma}=N(\mu,\sigma^2)\) and define the integral operator
\begin{equation*}
  \left(T_{\mu,\sigma}f\right)(x)
  \coloneqq \int_{\bbR} k_\gamma(x,y) f(y)\nu_{\mu,\sigma}(\diff y),
  \qquad T_{\mu,\sigma}:L^2\left(\nu_{\mu,\sigma}\right)\to L^2(\nu_{\mu,\sigma}).
\end{equation*}
Since \(k_\gamma\) is bounded and continuous, \(T_{\mu,\sigma}\) is compact and self-adjoint; thus its nonzero spectrum is a decreasing sequence \(\lambda_1(\sigma)\ge\lambda_2(\sigma)\ge\cdots\downarrow 0\) (independent of \(\mu\); see Proposition~\ref{prop:mean_invariance_translation_kernel}). For \(P\in\bbN\), write the spectral gap
\begin{equation*}
  \gamma_P(\sigma)\coloneqq \lambda_P(\sigma)-\lambda_{P+1}(\sigma),
  \qquad
  c_P\coloneqq \inf_{\sigma\in\left[\sigma_{\min},\sigma_{\max}\right]}\gamma_P(\sigma).
\end{equation*}

It is classical (see, e.g., \citealp[Section~3]{FasshauerandMcCourt2012}) that, after the identification \(L^2\left(\nu_{0,\sigma}\right)\cong L^2\left(\rho_{\alpha}\right)\) with
\begin{equation*}
  \rho_{\alpha}\left(\diff x\right)=\frac{\alpha}{\sqrt{\pi}}\,e^{-\alpha^2x^2}\,\diff x,
  \qquad \alpha=\frac{1}{\sqrt{2}\,\sigma},
\end{equation*}
the eigenfunctions are Hermite functions and the eigenvalues form a geometric sequence:
\begin{equation}\label{eq:eigs}
  \lambda_n(\sigma)
  = \lambda_1(\sigma)\, r(\sigma)^{\,n-1},
  \qquad n\ge 1,
\end{equation}
with ratio \(r(\sigma)\in(0,1)\) given explicitly by
\begin{equation}\label{eq:ratio}
  r(\sigma)
  = \frac{\varepsilon^2}{\varepsilon^2+\alpha^2+\delta^2},
  \qquad
  \delta^2
  = \frac{\alpha^2}{2}\Bigl(\sqrt{1+\left(\tfrac{2\varepsilon}{\alpha}\right)^2}-1\Bigr),
  \quad \alpha=\frac{1}{\sqrt{2}\,\sigma},
\end{equation}
and
\begin{equation}\label{eq:lambda1}
  \lambda_1(\sigma)
  = \sqrt{\frac{\alpha^2}{\varepsilon^2+\alpha^2+\delta^2}}.
\end{equation}
By compactness of the variance interval and continuity of \eqref{eq:ratio}--\eqref{eq:lambda1} in \(\sigma\), there exist
\begin{equation}\label{eq:rstar-cstar}
  r_{\ast}\coloneqq \sup_{\sigma\in\left[\sigma_{\min},\sigma_{\max}\right]} r(\sigma) \in (0,1),
  \qquad
  c_{\ast}\coloneqq \inf_{\sigma\in\left[\sigma_{\min},\sigma_{\max}\right]}
    \lambda_1(\sigma)\left(1-r(\sigma)\right) > 0 .
\end{equation}

From \eqref{eq:eigs},
\begin{equation}\label{eq:gap-sigma}
  \gamma_P(\sigma)
  = \lambda_P(\sigma)-\lambda_{P+1}(\sigma)
  = \lambda_1(\sigma)\left(1-r(\sigma)\right)\, r(\sigma)^{\,P-1}.
\end{equation}
Taking the infimum over \(\sigma\in\left[\sigma_{\min},\sigma_{\max}\right]\) and using \eqref{eq:rstar-cstar} yields
\begin{equation}\label{eq:uniform-gap}
  \gamma_P
  \ge  c_{\ast}\, r_{\ast}^{\,P-1}.
\end{equation}

\begin{theorem}
\label{thm:choice}
Let \(L_{X,n}\to\infty\) be integers. Then \(\sqrt{n}\,c_{L_{X,n}}\to\infty\) holds provided
\begin{equation*}
  r_{\ast}^{\,L_{X,n}}\ \gg\ n^{-1/2}
  \quad\text{equivalently}\quad
  L_{X,n}\ \ll\ \frac{\tfrac12 \log n}{\log\left(1/r_{\ast}\right)} .
\end{equation*}
In particular, for any fixed \(\delta\in\left(0,\tfrac12\right)\),
\begin{equation}\label{eq:Pn}
  L_{X,n}\ \coloneqq\
  \left\lfloor \frac{\left(\tfrac12-\delta\right)\log n}{\log\left(1/r_{\ast}\right)} \right\rfloor
  \quad\Longrightarrow\quad
  \sqrt{n}\,c_{L_{X,n}}\ge  c_{\ast}\, r_{\ast}^{-1}\, n^{\delta}\ \xrightarrow[n\to\infty]{}\ \infty .
\end{equation}
Conversely, if \(L_{X,n}\sim \frac{\tfrac12 \log n}{\log\left(1/r_{\ast}\right)}\), then \(\sqrt{n}\,\gamma_{L_{X,n}}\) stays of constant order; if \(L_{X,n}\) grows faster, then \(\sqrt{n}\,c_{L_{X,n}}\to 0\).
\end{theorem}

\begin{proof}
Inequality \eqref{eq:uniform-gap} gives
\(
  \sqrt{n}\,\gamma_{L_{X,n}}
  \ge c_{\ast}\, r_{\ast}^{\,L_{X,n}-1}\sqrt{n}
  = c_{\ast}\, r_{\ast}^{-1}\, \sqrt{n}\, r_{\ast}^{\,L_{X,n}}.
\)
The stated conditions are precisely those that force the right-hand side to diverge. The borderline and super-logarithmic cases follow similarly.
\end{proof}
\subsection{Projected-signal preservation from projector approximation}
\label{subsec:sufficient_projected_signal_preservation}

For each fixed \(c>0\), let
\[
g^X_{L_{X,n}}(c)
\coloneqq 
\inf_{P\in\tilde{\cP}_{1,n}(c)}
\left\{
\lambda_{L_{X,n}}(P)-\lambda_{L_{X,n}+1}(P)
\right\},
\qquad
g^Y_{L_{Y,n}}(c)
\coloneqq 
\inf_{P\in\tilde{\cP}_{1,n}(c)}
\left\{
\mu_{L_{Y,n}}(P)-\mu_{L_{Y,n}+1}(P)
\right\}.
\]

A convenient sufficient condition for Assumption~\ref{ass:uniform_power_projected_signal_preservation_Pi_n}
is that the empirical projector \(\hat\Pi_n\) approximates \(\Pi_n\) at the rate
\[
\sqrt{n_1}\left\|\hat\Pi_n-\Pi_n\right\|_{\mathrm{op}}
=
O_{\tilde{\cP}_{1,n}(c)}(1).
\]
Indeed, for each \(P\in\tilde{\cP}_{1,n}(c)\),
\[
\left(
\left\|\Pi_n\Sigma_{XYZ\mid Z}(P)\right\|
-
\left\|\hat\Pi_n\Sigma_{XYZ\mid Z}(P)\right\|
\right)_+
\le
\left\|(\Pi_n-\hat\Pi_n)\Sigma_{XYZ\mid Z}(P)\right\|
\le
\left\|\hat\Pi_n-\Pi_n\right\|_{\mathrm{op}}
\left\|\Sigma_{XYZ\mid Z}(P)\right\|,
\]
and \(\Sigma_{XYZ\mid Z}(P)\) is uniformly bounded over \(P\in\tilde{\cP}_1\) under the standing assumptions. Therefore,
\[
\sqrt{n_1}
\left(
\left\|\Pi_n\Sigma_{XYZ\mid Z}\right\|
-
\left\|\hat\Pi_n\Sigma_{XYZ\mid Z}\right\|
\right)_+
=
O_{\tilde{\cP}_{1,n}(c)}(1),
\]
so Assumption~\ref{ass:uniform_power_projected_signal_preservation_Pi_n} follows. Accordingly, whenever one has the projector perturbation bound
\[
\left\|\hat\Pi_n-\Pi_n\right\|_{\mathrm{op}}
=
O_{\tilde{\cP}_{1,n}(c)}
\left(
\frac{1}{
\sqrt{n_2}\,
\min\left\{
g^X_{L_{X,n}}(c),\,g^Y_{L_{Y,n}}(c)
\right\}}
\right),
\]
a sufficient condition for Assumption~\ref{ass:uniform_power_projected_signal_preservation_Pi_n} is
\[
\frac{\sqrt{n_1}}{
\sqrt{n_2}\,
\min\left\{
g^X_{L_{X,n}}(c),\,g^Y_{L_{Y,n}}(c)
\right\}}
=
O(1).
\]

This condition becomes particularly transparent in the Gaussian-kernel/Gaussian-measure setting considered in Supplementary Material, Section~\ref{sec:gaussian_case_eigengap}. In that case, the eigengaps are of geometric order: suppose that there exist constants
\(
a_X,a_Y>0\), \(r_X,r_Y\in(0,1)\), such that, for all sufficiently large \(n\),
\[
g^X_{L_{X,n}}(c)\ge a_X r_X^{L_{X,n}},
\qquad
g^Y_{L_{Y,n}}(c)\ge a_Y r_Y^{L_{Y,n}}.
\]
Then the above display shows that a sufficient condition is
\[
\frac{\sqrt{n_1}}{\sqrt{n_2}}\,r_X^{-L_{X,n}}=O(1),
\qquad
\frac{\sqrt{n_1}}{\sqrt{n_2}}\,r_Y^{-L_{Y,n}}=O(1).
\]
Equivalently, it suffices that
\[
L_{X,n}
\le
\frac{\frac12\log(n_2/n_1)+O(1)}{\log(1/r_X)},
\qquad
L_{Y,n}
\le
\frac{\frac12\log(n_2/n_1)+O(1)}{\log(1/r_Y)}.
\]
In particular, if \(n_1/n_2=O(1)\), then it is sufficient that
\(L_{X,n}=O(1)\), \(L_{Y,n}=O(1)\).

This requirement is stronger than Assumption~\ref{ass:simplicity_gap_unif}, which only asks that
\[
\sqrt{n_2}\,g^X_{L_{X,n}}(c)\to\infty,
\qquad
\sqrt{n_2}\,g^Y_{L_{Y,n}}(c)\to\infty.
\]
Under geometric eigengaps, the latter still allows logarithmically diverging truncation levels, whereas the present sufficient condition for projected-signal preservation does not.
\section{KRR-type CME conditions for Assumption 3}
\label{sec:uniform-krr-cme}

We return to the split-sample notation of Section~4 and write \(n_1 \coloneqq |I_1|\). This section gives uniform sufficient conditions under which Assumption~3 follows from empirical \(L^2\)-control of KRR-type CME estimators. Once the scalar nuisance regressions in Section~4 are implemented through \(\cH_X\)- and \(\cH_Y\)-valued CME estimators, the product-bias requirement reduces to uniform bounds for the corresponding empirical \(L^2\)-errors.

This reduction is useful because it removes any direct dependence on the truncation dimensions
\(L_{X,n}\) and \(L_{Y,n}\). It is therefore enough to control the CME estimators themselves, and the
resulting rate conditions can be stated in terms of smoothness and a \(Z\)-side trace bound. The
main conclusion of the section is that these conditions imply the product-bias requirement in
Assumption~3 whenever \(\rho_X+\rho_Y>1/2\).

We begin with the reduction step. Work conditionally on the \(\sigma\)-field with respect to which
the empirical basis \(\{\hat e_p\}_{p\ge 1}\) is measurable. Recall that \(\hat m_{X,p}\) denotes the
scalar KRR-type estimator associated with the response \(\hat e_p(X)\). Pathwise, the regression function induced by the KRR-type CME estimator \(\hat\mu_{X\mid Z,\lambda_X,P}\) and its corresponding target are respectively given by
\[
\hat m_{X,p}(z) = \left\langle \hat\mu_{X\mid Z,\lambda_X,P}(z), \hat e_p \right\rangle_{\cH_{\cX}}, \quad m_{X,p}(z) = \E_P\left[\hat e_p(X) \mid Z=z, \hat e_p\right] = \left\langle \mu_{X\mid Z,P}(z), \hat e_p \right\rangle_{\cH_{\cX}},
\]
where \(\mu_{X\mid Z,P}\) is the target CME. Therefore, for every \(i\in I_1\),
\[
\sum_{p=1}^{L_{X,n}}
\left\{
\hat m_{X,p}(Z_i)-m_{X,p}(Z_i)
\right\}^2
=
\sum_{p=1}^{L_{X,n}}
\left\langle
\hat\mu_{X\mid Z,\lambda_X,P}(Z_i)-\mu_{X\mid Z,P}(Z_i),
\hat e_p
\right\rangle_{\cH_{\cX}}^2.
\]
Since \(\{\hat e_p\}_{p\ge 1}\) is an orthonormal system in \(\cH_{\cX}\), applying Bessel's inequality pointwise and averaging over \(i\in I_1\) yields
\[
B_{X,n_1}^2
=
\frac{1}{n_1}\sum_{i\in I_1}\sum_{p=1}^{L_{X,n}} \left\{ \hat m_{X,p}(Z_i)-m_{X,p}(Z_i) \right\}^2
\le
\frac{1}{n_1}\sum_{i\in I_1} \left\| \hat\mu_{X\mid Z,\lambda_X,P}(Z_i)-\mu_{X\mid Z,P}(Z_i) \right\|_{\cH_{\cX}}^2
\eqqcolon
R_{X,n_1,P,\lambda_X}.
\]
The same argument gives
\[
B_{Y,n_1}^2
\le
\frac{1}{n_1}\sum_{i\in I_1}
\left\|
\hat\mu_{Y\mid Z,\lambda_Y,P}(Z_i)-\mu_{Y\mid Z,P}(Z_i)
\right\|_{\cH_{\cY}}^2
\eqqcolon
R_{Y,n_1,P,\lambda_Y}.
\]
Hence it is enough to derive uniform convergence rates for these empirical \(L^2\)-errors.

For each \(P\in\tilde{\cP}\) with \(Z\)-marginal \(\pi_{Z,P}\), define the covariance operator
\[
C_{ZZ,P} \coloneqq \E_P\left[\phi_{\cZ}(Z)\otimes \phi_{\cZ}(Z)\right],
\]
and let \(\{(\mu_{P,j},e_{P,j})\}_{j\ge 1}\) be its eigensystem, ordered such that \(\mu_{P,1}\ge \mu_{P,2}\ge \cdots \ge 0\).
For \(V\in\{X,Y\}\), let
\[
\mu_{V\mid Z,P}\in L^2(\pi_{Z,P};\cH_{\cV})
\]
denote the target CME. We write
\[
I_P\coloneqq I_{\pi_{Z,P}}:\cH_{\cZ}\to L^2(\pi_{Z,P})
\]
for the canonical embedding, and let \([H_{Z,P}]^\alpha\) denote the interpolation space in the
sense of \cite{Lietal2022b}. For \(\alpha\ge 0\), write
\[
[G_{\cV,P}]^\alpha
\]
for the corresponding vector-valued interpolation space in
\(L^2(\pi_{Z,P};\cH_{\cV})\).

Whenever \(\mu_{V\mid Z,P}\in [G_{\cV,P}]^\beta\) with \(\beta\ge 1\), we let
\[
C_{V\mid Z,P}\in S_2(\cH_{\cZ},\cH_{\cV})
\]
denote the canonical minimum-Hilbert--Schmidt-norm representative of \(\mu_{V\mid Z,P}\).
Thus
\[
\mu_{V\mid Z,P}(z)
=
C_{V\mid Z,P}\phi_{\cZ}(z)
\qquad
\text{for }\pi_{Z,P}\text{-almost every }z.
\]

Based on the subsample \(I_1\), we define the empirical covariance operators and, for \(\lambda>0\), the regularized conditional mean embedding estimator as
\begin{align*}
\hat C_{ZZ,P} &\coloneqq \frac{1}{n_1}\sum_{i\in I_1} \phi_{\cZ}(Z_i)\otimes \phi_{\cZ}(Z_i), & \hat C_{VZ,P} &\coloneqq \frac{1}{n_1}\sum_{i\in I_1} \phi_{\cV}(V_i)\otimes \phi_{\cZ}(Z_i), \\
\hat C_{V\mid Z,\lambda,P} &\coloneqq \hat C_{VZ,P}\left(\hat C_{ZZ,P}+\lambda I\right)^{-1}, & \hat\mu_{V\mid Z,\lambda,P}(z) &\coloneqq \hat C_{V\mid Z,\lambda,P}\phi_{\cZ}(z).
\end{align*}
We also define the effective dimension
\[
N_{Z,P}(\lambda)
\coloneqq
\mathrm{tr}\left(
C_{ZZ,P}\left(C_{ZZ,P}+\lambda I\right)^{-1}
\right).
\]

For \(V\in\{X,Y\}\), \(P\in\tilde{\cP}\), and \(\lambda>0\), we define
\begin{align*}
B_{V,P,\lambda}
&\coloneqq
\hat C_{V\mid Z,\lambda,P}-C_{V\mid Z,P}, \\
R_{V,P,\lambda}
&\coloneqq
\left\|
B_{V,P,\lambda}C_{ZZ,P}^{1/2}
\right\|_{\mathrm{HS}}^2
=
\left\|
[\hat\mu_{V\mid Z,\lambda,P}] - \mu_{V\mid Z,P}
\right\|_0^2, \\
R_{V,n_1,P,\lambda}
&\coloneqq
\frac{1}{n_1}\sum_{i\in I_1}
\left\|
\hat\mu_{V\mid Z,\lambda,P}(Z_i)-\mu_{V\mid Z,P}(Z_i)
\right\|_{\cH_{\cV}}^2.
\end{align*}

\begin{assumption}[Uniform KRR--CME conditions]
\label{ass:uniform-krr-cme}
There exist \(\beta_V\in[1,2]\), \(p\in(0,1]\), constants
\(Q>0\), \(M_V>0\), \(L_V>0\), and a deterministic sequence
\[
\lambda_{V,n_1}\asymp n_1^{-1/(\beta_V+p)}
\]
such that the following hold.

\begin{enumerate}
\item[(i)] \textbf{Uniform effective-dimension bound.}
\[
N_{Z,P}(\lambda)\le Q\,\lambda^{-p}
\qquad
\text{for all }P\in\tilde{\cP}\text{ and all }\lambda\in(0,1].
\]

\item[(ii)] \textbf{Uniform source condition.}
For every \(P\in\tilde{\cP}\),
\[
\mu_{V\mid Z,P}\in [G_{\cV,P}]^{\beta_V},
\]
and
\[
\sup_{P\in\tilde{\cP}}
\left\|
\mu_{V\mid Z,P}
\right\|_{[G_{\cV,P}]^{\beta_V}}
\le M_V.
\]

\item[(iii)] \textbf{Uniform Li-type finite-sample bound at \(\gamma=0\).}
For every \(\tau\ge 1\), there exists an integer \(n_{0,V}(\tau)\ge 1\),
independent of \(P\in\tilde{\cP}\), such that, for all \(n_1\ge n_{0,V}(\tau)\),
\begin{equation}
\label{eq:uniform-population-L2-bound}
\sup_{P\in\tilde{\cP}}
\pr_P\left(
\left\|
[\hat\mu_{V\mid Z,\lambda_{V,n_1},P}]
-
\mu_{V\mid Z,P}
\right\|_0^2
>
\tau^2 L_V n_1^{-2\rho_V}
\right)
\le
4e^{-\tau},
\qquad
\rho_V\coloneqq \frac{\beta_V}{2(\beta_V+p)}.
\end{equation}
\end{enumerate}
\end{assumption}

\begin{remark}
\label{rem:effective-dimension-evd-equivalence}
The effective-dimension condition in Assumption~\ref{ass:uniform-krr-cme}(i) is equivalent,
up to constants, to the polynomial eigenvalue-decay condition used by \cite{Lietal2022b}. More precisely,
if \(\{\mu_{P,j}\}_{j\ge 1}\) denotes the nonincreasing sequence of eigenvalues of \(C_{ZZ,P}\), then
\[
N_{Z,P}(\lambda)\le Q\lambda^{-p}
\qquad
\text{for all }\lambda\in(0,1]
\]
holds uniformly in \(P\in\tilde{\cP}\) if and only if
\[
\mu_{P,j}\le Q' j^{-1/p},
\qquad
j\ge 1,
\]
for some constant \(Q'>0\), uniformly in \(P\). Thus, for the upper-bound argument developed
here, the effective-dimension formulation is merely a convenient restatement of the Li-type
spectral regularity condition.
\end{remark}

\begin{remark}
\label{rem:uniform-krr-cme-li}
Boundedness of \(k_{\cZ}\) implies the embedding condition at \(\alpha=1\) in \cite{Lietal2022b}.
\end{remark}
Under the uniform source and effective-dimension conditions in
Assumption~\ref{ass:uniform-krr-cme}, together with boundedness of \(k_{\cZ}\), the
\(\gamma=0\) Li-type finite-sample bound in
Assumption~\ref{ass:uniform-krr-cme}(iii) follows from a routine uniformization of the
upper-bound argument of \cite{Lietal2022b}, provided that one also assumes a uniform lower
bound on \(\|C_{ZZ,P}\|_{\mathrm{op}}\). Under these conditions, the embedding requirement is
automatically satisfied, and the constants appearing in the pointwise argument can be chosen
independently of \(P\in\tilde{\cP}\). Consequently, for each fixed \(\tau\ge 1\), the associated
threshold index can also be chosen independently of \(P\), so that the resulting finite-sample
bound holds uniformly over \(\tilde{\cP}\). Since the proof is a straightforward repetition of the
corresponding pointwise argument, with all constants tracked uniformly over \(P\), we omit the
details. Moreover, Assumption~\ref{ass:non_degenerate_unif}, which is already imposed in our asymptotic theory, together with Lemma~\ref{lem:cpinf-controls-czz}, yields the required uniform lower bound on \(\|C_{ZZ,P}\|_{\mathrm{op}}\).

\begin{lemma}[Uniform empirical--population comparison]
\label{lem:uniform-cme-empirical-population-gap}
Define the regularized covariance and the normalized error operator by
\(C_{ZZ,P,\lambda} \coloneqq C_{ZZ,P}+\lambda I\) and
\(E_{P,\lambda} \coloneqq C_{ZZ,P,\lambda}^{-1/2}(\hat C_{ZZ,P}-C_{ZZ,P})C_{ZZ,P,\lambda}^{-1/2}\),
respectively. Then
\begin{equation}
\label{eq:uniform-risk-gap-identity-revised}
R_{V,n_1,P,\lambda}-R_{V,P,\lambda}
=
\mathrm{tr}\left(
B_{V,P,\lambda}
\left(
\hat C_{ZZ,P}-C_{ZZ,P}
\right)
B_{V,P,\lambda}^{\ast}
\right),
\end{equation}
which implies
\begin{equation}
\label{eq:uniform-risk-gap-master-revised}
\left|
R_{V,n_1,P,\lambda}-R_{V,P,\lambda}
\right|
\le
\left\|
E_{P,\lambda}
\right\|_{\mathrm{op}}
\left(
R_{V,P,\lambda}
+
\lambda
\left\|
B_{V,P,\lambda}
\right\|_{\mathrm{HS}}^2
\right).
\end{equation}

Moreover, if a deterministic sequence \(\{\lambda_{n_1}\}_{n_1\ge 1}\subset(0,1]\) satisfies
\begin{equation}
\label{eq:lambda-growth-condition-revised}
\frac{
n_1\lambda_{n_1}
}{
1+\log\left(1+\lambda_{n_1}^{-p}\right)
}
\to\infty,
\end{equation}
then \(\left\|E_{P,\lambda_{n_1}}\right\|_{\mathrm{op}}=o_{\tilde{\cP}}(1)\).
\end{lemma}
\begin{proof}
Fix \(V\in\{X,Y\}\), \(P\in\tilde{\cP}\), and \(\lambda\in(0,1]\). By definition,
\(R_{V,P,\lambda} = \mathrm{tr}(B_{V,P,\lambda} C_{ZZ,P} B_{V,P,\lambda}^{\ast})\) and
\(R_{V,n_1,P,\lambda} = \mathrm{tr}(B_{V,P,\lambda} \hat C_{ZZ,P} B_{V,P,\lambda}^{\ast})\), so their difference yields \eqref{eq:uniform-risk-gap-identity-revised}. Inserting \(C_{ZZ,P,\lambda}^{1/2}C_{ZZ,P,\lambda}^{-1/2}\) around \(\hat C_{ZZ,P}-C_{ZZ,P}\) and defining \(A_{V,P,\lambda}\coloneqq B_{V,P,\lambda}C_{ZZ,P,\lambda}^{1/2}\), we obtain
\(R_{V,n_1,P,\lambda}-R_{V,P,\lambda} = \mathrm{tr}(A_{V,P,\lambda} E_{P,\lambda} A_{V,P,\lambda}^{\ast})\). Since \(E_{P,\lambda}\) is self-adjoint, \(|R_{V,n_1,P,\lambda}-R_{V,P,\lambda}| \le \|E_{P,\lambda}\|_{\mathrm{op}} \|A_{V,P,\lambda}\|_{\mathrm{HS}}^2\). Moreover,
\(\|A_{V,P,\lambda}\|_{\mathrm{HS}}^2 = \mathrm{tr}(B_{V,P,\lambda} C_{ZZ,P,\lambda} B_{V,P,\lambda}^{\ast}) = R_{V,P,\lambda} + \lambda \|B_{V,P,\lambda}\|_{\mathrm{HS}}^2\), which proves \eqref{eq:uniform-risk-gap-master-revised}.

To prove \(\|E_{P,\lambda_{n_1}}\|_{\mathrm{op}}=o_{\tilde{\cP}}(1)\), define \(h_{P,\lambda}(z)\coloneqq C_{ZZ,P,\lambda}^{-1/2}\phi_{\cZ}(z)\) and \(\Sigma_{P,\lambda}\coloneqq \E_P[h_{P,\lambda}(Z)\otimes h_{P,\lambda}(Z)]\). Then
\(E_{P,\lambda} = n_1^{-1}\sum_{i\in I_1}\Xi_{i,P,\lambda}\), where \(\Xi_{i,P,\lambda} \coloneqq h_{P,\lambda}(Z_i)\otimes h_{P,\lambda}(Z_i)-\Sigma_{P,\lambda}\). The operators \(\Xi_{i,P,\lambda}\) are independent, centered, self-adjoint Hilbert--Schmidt operators. By boundedness of \(k_{\cZ}\) and \(C_{ZZ,P,\lambda}\succeq \lambda I\), we have \(\|h_{P,\lambda}(z)\|_{\cH_{\cZ}}^2 \le K_Z^2/\lambda\). Also, \(\Sigma_{P,\lambda} = C_{ZZ,P,\lambda}^{-1/2} C_{ZZ,P} C_{ZZ,P,\lambda}^{-1/2}\), hence \(\|\Sigma_{P,\lambda}\|_{\mathrm{op}}\le 1\). Therefore, since \(\lambda\le 1\),
\[
\|\Xi_{i,P,\lambda}\|_{\mathrm{op}} \le \frac{K_Z^2+1}{\lambda} \eqqcolon U_\lambda \qquad \text{a.s.}
\]

Let \(A_{i,P,\lambda}\coloneqq h_{P,\lambda}(Z_i)\otimes h_{P,\lambda}(Z_i)\). Then \(A_{i,P,\lambda}^2 = \|h_{P,\lambda}(Z_i)\|_{\cH_{\cZ}}^2 A_{i,P,\lambda} \preceq (K_Z^2/\lambda) A_{i,P,\lambda}\), so \(\E_P A_{i,P,\lambda}^2 \preceq (K_Z^2/\lambda)\Sigma_{P,\lambda}\). Since \(\E_P[\Xi_{i,P,\lambda}^2] = \E_P(A_{i,P,\lambda}-\Sigma_{P,\lambda})^2 = \E_P A_{i,P,\lambda}^2-\Sigma_{P,\lambda}^2 \preceq \E_P A_{i,P,\lambda}^2\), it follows that \(\|\E_P[\Xi_{i,P,\lambda}^2]\|_{\mathrm{op}} \le K_Z^2/\lambda\). Hence
\[
\left\|\sum_{i\in I_1}\E_P[\Xi_{i,P,\lambda}^2]\right\|_{\mathrm{op}} \le n_1\frac{K_Z^2}{\lambda} \eqqcolon \sigma_\lambda^2,
\qquad
\mathrm{tr}\left(\sum_{i\in I_1}\E_P[\Xi_{i,P,\lambda}^2]\right)
\le
n_1\frac{K_Z^2}{\lambda}\,\mathrm{tr}(\Sigma_{P,\lambda})
\le
n_1 K_Z^2 Q \lambda^{-(1+p)}.
\]

Set \(S_{P,\lambda}\coloneqq \sum_{i\in I_1}\Xi_{i,P,\lambda}\). Applying Theorem~2.1 and Section~4 of \cite{Minsker2017} to the rescaled operators \(U_\lambda^{-1}\Xi_{i,P,\lambda}\), we obtain for every \(t>0\),
\[
\sup_{P\in\tilde{\cP}} \pr_P\left(\|S_{P,\lambda}\|_{\mathrm{op}}>t\right)
\le
2\,\frac{\mathrm{tr}\left(\sum_{i\in I_1}\E_P[\Xi_{i,P,\lambda}^2]\right)}{\sigma_\lambda^2}
\exp\left(-\frac{t^2}{2\left(\sigma_\lambda^2+U_\lambda t/3\right)}\right)
r_\lambda(t),
\]
where \(r_\lambda(t) \coloneqq 1+ \frac{6U_\lambda^2}{t^2}\log^2\left(1+\frac{U_\lambda t}{\sigma_\lambda^2}\right)\). Since
\[
\frac{\mathrm{tr}\left(\sum_{i\in I_1}\E_P[\Xi_{i,P,\lambda}^2]\right)}{\sigma_\lambda^2}
\le
Q\lambda^{-p},
\]
it follows that
\[
\sup_{P\in\tilde{\cP}} \pr_P\left(\|S_{P,\lambda}\|_{\mathrm{op}}>t\right)
\le
2Q\lambda^{-p}
\exp\left(-\frac{t^2}{2\left(\sigma_\lambda^2+U_\lambda t/3\right)}\right)
r_\lambda(t).
\]

Fix \(\varepsilon>0\) and let \(t=\varepsilon n_1\). Since \(E_{P,\lambda}=n_1^{-1}S_{P,\lambda}\), the event \(\{\|E_{P,\lambda}\|_{\mathrm{op}}>\varepsilon\}\) is identical to \(\{\|S_{P,\lambda}\|_{\mathrm{op}}>\varepsilon n_1\}\). The exponent becomes \(-c_\varepsilon n_1\lambda\), where \(c_\varepsilon \coloneqq \varepsilon^2/\left(2\left(K_Z^2+\varepsilon(K_Z^2+1)/3\right)\right)\). Moreover,
\((U_\lambda \varepsilon n_1)/\sigma_\lambda^2 = \varepsilon (K_Z^2+1)/K_Z^2\), so \(r_{\lambda_{n_1}}(\varepsilon n_1)\) is bounded, and in fact \(r_{\lambda_{n_1}}(\varepsilon n_1)=1+O\left((n_1\lambda_{n_1})^{-2}\right)\). Therefore there exists \(C_\varepsilon<\infty\) such that
\[
\sup_{P\in\tilde{\cP}} \pr_P\left(\|E_{P,\lambda_{n_1}}\|_{\mathrm{op}}>\varepsilon\right)
\le
2Q C_\varepsilon \lambda_{n_1}^{-p} \exp\left(-c_\varepsilon n_1\lambda_{n_1}\right).
\]
Finally, \eqref{eq:lambda-growth-condition-revised} implies \(\log(1+\lambda_{n_1}^{-p}) = o(n_1\lambda_{n_1})\), hence \(\lambda_{n_1}^{-p}\exp\left(-c_\varepsilon n_1\lambda_{n_1}\right)\to 0\). Thus
\[
\sup_{P\in\tilde{\cP}} \pr_P\left(\|E_{P,\lambda_{n_1}}\|_{\mathrm{op}}>\varepsilon\right)\to 0
\qquad\text{for every }\varepsilon>0,
\]
which proves \(\|E_{P,\lambda_{n_1}}\|_{\mathrm{op}}=o_{\tilde{\cP}}(1)\).
\end{proof}

\begin{remark}
\label{rem:lambda-growth-condition-revised}
Under Assumption~\ref{ass:uniform-krr-cme}, the rate \(\lambda_{V,n_1}\asymp n_1^{-1/(\beta_V+p)}\) automatically satisfies \eqref{eq:lambda-growth-condition-revised}. Indeed, since \(\beta_V\ge 1\) and \(p>0\), the polynomial growth \(n_1\lambda_{V,n_1}\asymp n_1^{1-1/(\beta_V+p)}\) strictly dominates the logarithmic term \(\log\left(1+\lambda_{V,n_1}^{-p}\right)=O(\log n_1)\).
\end{remark}

\begin{lemma}[Uniform control of the penalty term]
\label{lem:uniform-krr-cme-operator-control}
Suppose that Assumption~\ref{ass:uniform-krr-cme} holds. Then, for every deterministic sequence \(\{\lambda_{n_1}\}_{n_1\ge 1}\subset(0,1]\) satisfying
\eqref{eq:lambda-growth-condition-revised},
\[
R_{V,n_1,P,\lambda_{n_1}}
+
\lambda_{n_1}
\left\|
B_{V,P,\lambda_{n_1}}
\right\|_{\mathrm{HS}}^2
=
O_{\tilde{\cP}}
\left(
n_1^{-1}\lambda_{n_1}^{-p}
+
\lambda_{n_1}^{\beta_V}
\right).
\]
\end{lemma}

\begin{proof}
Fix \(V\in\{X,Y\}\), \(P\in\tilde{\cP}\), \(n_1\ge 1\), and \(\lambda\in(0,1]\). Define \(\varepsilon_{V,i,P} \coloneqq \phi_{\cV}(V_i)-C_{V\mid Z,P}\phi_{\cZ}(Z_i)\) and \(\Delta_{V,n_1,P} \coloneqq \frac{1}{n_1}\sum_{i\in I_1} \varepsilon_{V,i,P}\otimes \phi_{\cZ}(Z_i)\). Since \(\mu_{V\mid Z,P}(z) = C_{V\mid Z,P}\phi_{\cZ}(z)\) for \(\pi_{Z,P}\)-almost every \(z\), we have \(\E_P[\varepsilon_{V,i,P}\mid Z_i] = 0\) almost surely. Moreover, \(\|\phi_{\cV}(V_i)\|_{\cH_{\cV}} \le K_V\) almost surely, and \(\|C_{V\mid Z,P}\phi_{\cZ}(z)\|_{\cH_{\cV}} = \|\E_P[\phi_{\cV}(V)\mid Z=z]\|_{\cH_{\cV}} \le K_V\). Therefore,
\begin{equation}
\label{eq:uniform-krr-epsilon-bound}
\left\| \varepsilon_{V,i,P} \right\|_{\cH_{\cV}} \le 2K_V \qquad\text{a.s.}
\end{equation}

Next,
\begin{align*}
\hat C_{VZ,P}
= \frac{1}{n_1}\sum_{i\in I_1} \phi_{\cV}(V_i)\otimes \phi_{\cZ}(Z_i)
&= \frac{1}{n_1}\sum_{i\in I_1} \left\{ C_{V\mid Z,P}\phi_{\cZ}(Z_i)+\varepsilon_{V,i,P} \right\}\otimes \phi_{\cZ}(Z_i)\\
&= C_{V\mid Z,P}\hat C_{ZZ,P} + \Delta_{V,n_1,P}.
\end{align*}
Since \(\hat C_{V\mid Z,\lambda,P} ( \hat C_{ZZ,P}+\lambda I ) = \hat C_{VZ,P}\), it follows that
\[
B_{V,P,\lambda} \left( \hat C_{ZZ,P}+\lambda I \right) = \Delta_{V,n_1,P} - \lambda C_{V\mid Z,P}.
\]
Taking the Hilbert--Schmidt inner product with \(B_{V,P,\lambda}\), we obtain
\begin{equation}
\label{eq:uniform-krr-master-identity}
R_{V,n_1,P,\lambda} + \lambda \left\| B_{V,P,\lambda} \right\|_{\mathrm{HS}}^2
= \left\langle \Delta_{V,n_1,P}, B_{V,P,\lambda} \right\rangle_{\mathrm{HS}} - \lambda \left\langle C_{V\mid Z,P}, B_{V,P,\lambda} \right\rangle_{\mathrm{HS}}.
\end{equation}

We evaluate the right-hand side of \eqref{eq:uniform-krr-master-identity}. By inserting \((\hat C_{ZZ,P}+\lambda I)^{-1/2}(\hat C_{ZZ,P}+\lambda I)^{1/2}\) and applying the Cauchy--Schwarz inequality, both terms share the common factor
\[
\left\| B_{V,P,\lambda} \left( \hat C_{ZZ,P}+\lambda I \right)^{1/2} \right\|_{\mathrm{HS}}
= \left( R_{V,n_1,P,\lambda} + \lambda \left\|B_{V,P,\lambda}\right\|_{\mathrm{HS}}^2 \right)^{1/2}.
\]
For the stochastic term, this yields
\[
\left| \left\langle \Delta_{V,n_1,P}, B_{V,P,\lambda} \right\rangle_{\mathrm{HS}} \right|
\le \left\| \Delta_{V,n_1,P} \left( \hat C_{ZZ,P}+\lambda I \right)^{-1/2} \right\|_{\mathrm{HS}} \left( R_{V,n_1,P,\lambda} + \lambda \left\|B_{V,P,\lambda}\right\|_{\mathrm{HS}}^2 \right)^{1/2}.
\]
For the deterministic term, invoking the source condition \(C_{V\mid Z,P} = A_{V\mid Z,P}C_{ZZ,P}^{(\beta_V-1)/2}\) similarly yields
\[
\lambda \left| \left\langle C_{V\mid Z,P}, B_{V,P,\lambda} \right\rangle_{\mathrm{HS}} \right|
\le \lambda \left\| A_{V\mid Z,P}C_{ZZ,P}^{(\beta_V-1)/2} \left( \hat C_{ZZ,P}+\lambda I \right)^{-1/2} \right\|_{\mathrm{HS}} \left( R_{V,n_1,P,\lambda} + \lambda \left\|B_{V,P,\lambda}\right\|_{\mathrm{HS}}^2 \right)^{1/2}.
\]

Define the event \(\cE_{n_1,P,\lambda} \coloneqq \left\{ \|E_{P,\lambda}\|_{\mathrm{op}} \le 1/2 \right\}\). On \(\cE_{n_1,P,\lambda}\),
\[
\left( \hat C_{ZZ,P}+\lambda I \right)^{-1} \preceq 2 \left( C_{ZZ,P}+\lambda I \right)^{-1}.
\]
Therefore,
\[
\left\| A_{V\mid Z,P}C_{ZZ,P}^{(\beta_V-1)/2} \left( \hat C_{ZZ,P}+\lambda I \right)^{-1/2} \right\|_{\mathrm{HS}}^2
\le 2 \left\| A_{V\mid Z,P}C_{ZZ,P}^{(\beta_V-1)/2} \left( C_{ZZ,P}+\lambda I \right)^{-1/2} \right\|_{\mathrm{HS}}^2.
\]
Now let \(c_{\beta_V} \coloneqq \sup_{u\ge 0} \frac{u^{\beta_V-1}}{1+u}\), which is finite because \(\beta_V\in[1,2]\). Using the eigensystem of \(C_{ZZ,P}\) and the definition of \(c_{\beta_V}\), we get
\begin{align*}
\left\| A_{V\mid Z,P}C_{ZZ,P}^{(\beta_V-1)/2} \left( C_{ZZ,P}+\lambda I \right)^{-1/2} \right\|_{\mathrm{HS}}^2
&= \sum_{j:\mu_{P,j}>0} \frac{\mu_{P,j}^{\beta_V-1}}{\mu_{P,j}+\lambda} \left\| A_{V\mid Z,P}e_{P,j} \right\|_{\cH_{\cV}}^2 \\
&\le c_{\beta_V}\lambda^{\beta_V-2} \sum_{j:\mu_{P,j}>0} \left\| A_{V\mid Z,P}e_{P,j} \right\|_{\cH_{\cV}}^2 \\
&= c_{\beta_V}\lambda^{\beta_V-2} \left\| A_{V\mid Z,P} \right\|_{\mathrm{HS}}^2.
\end{align*}
Hence, on \(\cE_{n_1,P,\lambda}\),
\[
\lambda \left| \left\langle C_{V\mid Z,P}, B_{V,P,\lambda} \right\rangle_{\mathrm{HS}} \right|
\le \sqrt{2c_{\beta_V}}\, \left\| A_{V\mid Z,P} \right\|_{\mathrm{HS}} \lambda^{\beta_V/2} \left( R_{V,n_1,P,\lambda} + \lambda \left\|B_{V,P,\lambda}\right\|_{\mathrm{HS}}^2 \right)^{1/2}.
\]

Similarly, on \(\cE_{n_1,P,\lambda}\),
\[
\left\| \Delta_{V,n_1,P} \left( \hat C_{ZZ,P}+\lambda I \right)^{-1/2} \right\|_{\mathrm{HS}}^2
\le 2 \left\| \Delta_{V,n_1,P} \left( C_{ZZ,P}+\lambda I \right)^{-1/2} \right\|_{\mathrm{HS}}^2.
\]
Combining the preceding bounds with \eqref{eq:uniform-krr-master-identity}, we obtain on \(\cE_{n_1,P,\lambda}\),
\[
\left( R_{V,n_1,P,\lambda} + \lambda \left\|B_{V,P,\lambda}\right\|_{\mathrm{HS}}^2 \right)^{1/2}
\le \sqrt{2} \left\| \Delta_{V,n_1,P} \left( C_{ZZ,P}+\lambda I \right)^{-1/2} \right\|_{\mathrm{HS}} + \sqrt{2c_{\beta_V}}\, \left\| A_{V\mid Z,P} \right\|_{\mathrm{HS}} \lambda^{\beta_V/2}.
\]
Therefore, on \(\cE_{n_1,P,\lambda}\),
\begin{equation}
\label{eq:uniform-krr-control-on-good-event}
R_{V,n_1,P,\lambda} + \lambda \left\| B_{V,P,\lambda} \right\|_{\mathrm{HS}}^2
\le 4 \left\| \Delta_{V,n_1,P} \left( C_{ZZ,P}+\lambda I \right)^{-1/2} \right\|_{\mathrm{HS}}^2 + 4c_{\beta_V} \left\| A_{V\mid Z,P} \right\|_{\mathrm{HS}}^2 \lambda^{\beta_V}.
\end{equation}

It remains to bound the first term on the right-hand side. Define \(h_{P,\lambda}(z) \coloneqq ( C_{ZZ,P}+\lambda I )^{-1/2} \phi_{\cZ}(z)\). Then
\[
\Delta_{V,n_1,P} \left( C_{ZZ,P}+\lambda I \right)^{-1/2} = \frac{1}{n_1}\sum_{i\in I_1} \varepsilon_{V,i,P}\otimes h_{P,\lambda}(Z_i).
\]
Conditional on \(\{Z_i\}_{i\in I_1}\), the cross-terms vanish because \(\E_P[\varepsilon_{V,i,P}\mid Z_i] = 0\) and the observations are independent. Using \(\|u\otimes v\|_{\mathrm{HS}}^2 = \|u\|^2\|v\|^2\) and \eqref{eq:uniform-krr-epsilon-bound}, we obtain
\begin{align*}
\E_P\left[ \left\| \Delta_{V,n_1,P} \left( C_{ZZ,P}+\lambda I \right)^{-1/2} \right\|_{\mathrm{HS}}^2 \,\middle|\, \{Z_i\}_{i\in I_1} \right]
&= \frac{1}{n_1^2}\sum_{i\in I_1} \E_P\left[ \left\| \varepsilon_{V,i,P}\otimes h_{P,\lambda}(Z_i) \right\|_{\mathrm{HS}}^2 \,\middle|\, Z_i \right] \\
&\le \frac{4K_V^2}{n_1^2} \sum_{i\in I_1} \left\| h_{P,\lambda}(Z_i) \right\|_{\cH_{\cZ}}^2.
\end{align*}
Taking expectations and using \(\E_P[ \| h_{P,\lambda}(Z) \|_{\cH_{\cZ}}^2 ] = \mathrm{tr}( C_{ZZ,P} ( C_{ZZ,P}+\lambda I )^{-1} ) = N_{Z,P}(\lambda)\), we conclude that
\[
\E_P\left[ \left\| \Delta_{V,n_1,P} \left( C_{ZZ,P}+\lambda I \right)^{-1/2} \right\|_{\mathrm{HS}}^2 \right]
\le \frac{4K_V^2}{n_1} N_{Z,P}(\lambda) \le \frac{4K_V^2Q}{n_1\lambda^p}.
\]
Hence
\begin{equation}
\label{eq:uniform-krr-delta-rate}
\left\| \Delta_{V,n_1,P} \left( C_{ZZ,P}+\lambda I \right)^{-1/2} \right\|_{\mathrm{HS}}^2 = O_{\tilde{\cP}} \left( n_1^{-1}\lambda^{-p} \right).
\end{equation}

Now take \(\lambda=\lambda_{n_1}\). By Lemma~\ref{lem:uniform-cme-empirical-population-gap}, \(\sup_{P\in\tilde{\cP}} \pr_P( \cE_{n_1,P,\lambda_{n_1}}^c ) \to 0\). Also, \(\sup_{P\in\tilde{\cP}} \| A_{V\mid Z,P} \|_{\mathrm{HS}} \le M_V\) by Assumption~\ref{ass:uniform-krr-cme}(ii). Thus \eqref{eq:uniform-krr-control-on-good-event} and \eqref{eq:uniform-krr-delta-rate} imply
\[
R_{V,n_1,P,\lambda_{n_1}} + \lambda_{n_1} \left\| B_{V,P,\lambda_{n_1}} \right\|_{\mathrm{HS}}^2 = O_{\tilde{\cP}} \left( n_1^{-1}\lambda_{n_1}^{-p} + \lambda_{n_1}^{\beta_V} \right),
\]
as claimed.
\end{proof}

\begin{proposition}[Uniform empirical \(L^2\)-rates]
\label{prop:uniform-krr-cme-risk-rates}
Suppose that Assumption~\ref{ass:uniform-krr-cme} holds. Then,
\[
R_{V,n_1,P,\lambda_{V,n_1}}
=
O_{\tilde{\cP}}\left(
n_1^{-2\rho_V}
\right),
\qquad
2\rho_V=\frac{\beta_V}{\beta_V+p}.
\]
More precisely,
\[
\left|
R_{V,n_1,P,\lambda_{V,n_1}}
-
R_{V,P,\lambda_{V,n_1}}
\right|
=
o_{\tilde{\cP}}\left(
n_1^{-2\rho_V}
\right).
\]
\end{proposition}

\begin{proof}
Fix \(V\in\{X,Y\}\). By Assumption~\ref{ass:uniform-krr-cme}(iii),
\[
R_{V,P,\lambda_{V,n_1}}
=
O_{\tilde{\cP}}\left(
n_1^{-2\rho_V}
\right).
\]
Moreover,
\[
\lambda_{V,n_1}\asymp n_1^{-1/(\beta_V+p)}
\]
implies
\[
n_1^{-1}\lambda_{V,n_1}^{-p}
\asymp
n_1^{-\beta_V/(\beta_V+p)}
=
n_1^{-2\rho_V},
\qquad
\lambda_{V,n_1}^{\beta_V}
\asymp
n_1^{-\beta_V/(\beta_V+p)}
=
n_1^{-2\rho_V}.
\]
Hence Lemma~\ref{lem:uniform-krr-cme-operator-control} yields
\[
\lambda_{V,n_1}
\left\|
B_{V,P,\lambda_{V,n_1}}
\right\|_{\mathrm{HS}}^2
=
O_{\tilde{\cP}}\left(
n_1^{-2\rho_V}
\right).
\]

By Remark~\ref{rem:lambda-growth-condition-revised}, \(\lambda_{V,n_1}\) satisfies \eqref{eq:lambda-growth-condition-revised}, so Lemma~\ref{lem:uniform-cme-empirical-population-gap} yields \(\|E_{P,\lambda_{V,n_1}}\|_{\mathrm{op}} = o_{\tilde{\cP}}(1)\). Applying \eqref{eq:uniform-risk-gap-master-revised} and noting that the factor in parentheses is \(O_{\tilde{\cP}}(n_1^{-2\rho_V})\), we obtain
\[
\left| R_{V,n_1,P,\lambda_{V,n_1}} - R_{V,P,\lambda_{V,n_1}} \right|
\le
o_{\tilde{\cP}}(1) \cdot O_{\tilde{\cP}}\left(n_1^{-2\rho_V}\right)
=
o_{\tilde{\cP}}\left(n_1^{-2\rho_V}\right).
\]
Since the population risk satisfies \(R_{V,P,\lambda_{V,n_1}} = O_{\tilde{\cP}}(n_1^{-2\rho_V})\), it immediately follows that \(R_{V,n_1,P,\lambda_{V,n_1}} = O_{\tilde{\cP}}(n_1^{-2\rho_V})\), as required.
\end{proof}

\begin{corollary}[Uniform product-bias bound]
\label{cor:uniform-krr-product-bias}
Under Assumption~\ref{ass:uniform-krr-cme},
\[
n_1
R_{X,n_1,P,\lambda_{X,n_1}}
R_{Y,n_1,P,\lambda_{Y,n_1}}
=
O_{\tilde{\cP}}\left(
n_1^{1-2\rho_X-2\rho_Y}
\right).
\]
Consequently, if \(
\rho_X+\rho_Y>1/2\), then
\[
n_1
R_{X,n_1,P,\lambda_{X,n_1}}
R_{Y,n_1,P,\lambda_{Y,n_1}}
=
o_{\tilde{\cP}}(1).
\]
Moreover, under \(\beta_X,\beta_Y\ge 1\) and \(p\in(0,1]\), the condition
\(\rho_X+\rho_Y>1/2\) fails only in the boundary case
\[
\beta_X=\beta_Y=1
\qquad\text{and}\qquad
p=1.
\]
\end{corollary}

\begin{proof}
By Proposition~\ref{prop:uniform-krr-cme-risk-rates}, we have $R_{X,n_1,P,\lambda_{X,n_1}} = O_{\tilde{\cP}}(n_1^{-2\rho_X})$ and $R_{Y,n_1,P,\lambda_{Y,n_1}} = O_{\tilde{\cP}}(n_1^{-2\rho_Y})$. Multiplying these bounds by $n_1$ yields
\begin{equation}
n_1 R_{X,n_1,P,\lambda_{X,n_1}} R_{Y,n_1,P,\lambda_{Y,n_1}} = O_{\tilde{\cP}}\left( n_1^{1-2\rho_X-2\rho_Y} \right).
\end{equation}
Hence, the left-hand side is $o_{\tilde{\cP}}(1)$ whenever $\rho_X+\rho_Y>1/2$. Now suppose that $\beta_X,\beta_Y\ge 1$ and $p\in(0,1]$. Since $2\rho_V = \beta_V / (\beta_V+p) \ge 1/(1+p)$ for $V\in\{X,Y\}$, we have $\rho_X+\rho_Y \ge 1/(1+p) \ge 1/2$. Equality holds if and only if $\beta_X=\beta_Y=1$ and $p=1$, which completes the proof.
\end{proof}

\begin{remark}
\label{rem:uniform-krr-ass3}
By the reduction at the beginning of this section, \(B_{V,n_1}^2 \le R_{V,n_1,P,\lambda_{V,n_1}}\) for \(V\in\{X,Y\}\). Hence, Corollary~\ref{cor:uniform-krr-product-bias} ensures that Assumption~3 (which requires \(B_{X,n_1}^2, B_{Y,n_1}^2 = o_{\tilde{\cP}}(1)\) and \(n_1 B_{X,n_1}^2 B_{Y,n_1}^2 = o_{\tilde{\cP}}(1)\)) is satisfied whenever \(\rho_X+\rho_Y>1/2\). In the boundary case \(\beta_X=\beta_Y=1\) and \(p=1\), this argument only yields \(B_{X,n_1}B_{Y,n_1} = O_{\tilde{\cP}}(n_1^{-1/2})\), so additional information is needed to recover Assumption~3.
\end{remark}

\begin{lemma}
\label{lem:cpinf-controls-czz}
For every
\(P\in\tilde{\cP}\),
\[
\left\|C_P^{(\infty)}\right\|_{\mathrm{op}}
\le
16 K_X^2 K_Y^2 \left\|C_{ZZ,P}\right\|_{\mathrm{op}}.
\]
Consequently, if Assumption~5 holds, then
\[
\inf_{P\in\tilde{\cP}} \left\|C_{ZZ,P}\right\|_{\mathrm{op}} > 0.
\]
\end{lemma}

\begin{proof}
Recall that \(\Psi=\epsilon_{X\mid Z}\otimes \epsilon_{Y\mid Z}\otimes k_{\cZ}(Z,\cdot)\) and
\(C_P^{(\infty)}=\E_P[(\Pi_\infty\Psi)\otimes(\Pi_\infty\Psi)]\). Since
\(\|\mu_{X\mid Z}(z)\|_{\cH_X}\le K_X\) and \(\|\mu_{Y\mid Z}(z)\|_{\cH_Y}\le K_Y\), we have
\(\|\epsilon_{X\mid Z}\|_{\cH_X}\le 2K_X\) and \(\|\epsilon_{Y\mid Z}\|_{\cH_Y}\le 2K_Y\) almost surely.

Fix \(h\in \cH_X\otimes \cH_Y\otimes \cH_Z\) with \(\|h\|=1\), and identify
\(\cH_X\otimes \cH_Y\otimes \cH_Z\) with \(\mathrm{HS}(\cH_Z,\cH_X\otimes \cH_Y)\). Let
\(T_{\Pi_\infty h}:\cH_Z\to \cH_X\otimes \cH_Y\) be the Hilbert--Schmidt operator corresponding to
\(\Pi_\infty h\). Then
\(\langle \Pi_\infty\Psi,h\rangle
=
\langle \epsilon_{X\mid Z}\otimes \epsilon_{Y\mid Z},\,T_{\Pi_\infty h}k_{\cZ}(Z,\cdot)\rangle\),
so by Cauchy--Schwarz,
\[
\langle C_P^{(\infty)}h,h\rangle
=
\E_P\left[\langle \Pi_\infty\Psi,h\rangle^2\right]
\le
16K_X^2K_Y^2\,\E_P\left[\left\|T_{\Pi_\infty h}k_{\cZ}(Z,\cdot)\right\|^2\right].
\]
For any Hilbert--Schmidt operator \(T:\cH_Z\to \cH_X\otimes \cH_Y\),
\(\E_P[\|Tk_{\cZ}(Z,\cdot)\|^2]
=
\mathrm{tr}(TC_{ZZ,P}T^\ast)
\le
\|C_{ZZ,P}\|_{\mathrm{op}}\|T\|_{\mathrm{HS}}^2\).
Applying this with \(T=T_{\Pi_\infty h}\), and using
\(\|T_{\Pi_\infty h}\|_{\mathrm{HS}}=\|\Pi_\infty h\|\le \|h\|=1\), we obtain
\(\langle C_P^{(\infty)}h,h\rangle \le 16K_X^2K_Y^2\|C_{ZZ,P}\|_{\mathrm{op}}\).
Taking the supremum over \(\|h\|=1\) gives the first claim.

If Assumption~5 holds, then
\(\inf_{P\in\tilde{\cP}}\|C_P^{(\infty)}\|_{\mathrm{op}}>0\). Since
\(\|C_{ZZ,P}\|_{\mathrm{op}}
\ge
\|C_P^{(\infty)}\|_{\mathrm{op}}/(16K_X^2K_Y^2)\) for every \(P\in\tilde{\cP}\),
it follows that \(\inf_{P\in\tilde{\cP}} \|C_{ZZ,P}\|_{\mathrm{op}} > 0\).
\end{proof}

\section{Uniform-tightness condition for Assumption 4}\label{sec:suff_cond_uniform_tightness}

In this section we give a convenient sufficient condition for Assumption~\ref{ass:fixed-basis-uniform_tail} based on uniform tightness of the marginal laws. Let \(\cQ\subset \cP(\cS)\) be a family of Borel probability measures on a Polish space \(\cS\). We say that \(\cQ\) is uniformly tight if, for every \(\varepsilon>0\), there exists a compact set \(C_\varepsilon\subset \cS\) such that \(\sup_{Q\in\cQ}Q(C_\varepsilon^c)\le \varepsilon\). In our setting, when \(L_{X,n},L_{Y,n}\to\infty\) and the \(Z\)-basis is infinite, Assumption~\ref{ass:fixed-basis-uniform_tail} follows once the kernels \(k_{\cX},k_{\cY},k_{\cZ}\) are bounded and continuous and the marginal families \(\{P\circ X^{-1}:P\in\tilde{\cP}\}\), \(\{P\circ Y^{-1}:P\in\tilde{\cP}\}\), and \(\{P\circ Z^{-1}:P\in\tilde{\cP}\}\) are uniformly tight. The point is that these conditions force the tail energy of any fixed orthonormal basis in the corresponding RKHS to vanish uniformly over \(P\in\tilde{\cP}\).

\begin{proposition}[Uniform tail control from uniform tightness]
Let \(\cS\) be a Polish space, let \(k:\cS\times\cS\to\bbR\) be a bounded continuous positive-definite kernel, and let \(\cH_k\) be its RKHS. Let \(\{e_p\}_{p=1}^\infty\) be an orthonormal basis of \(\cH_k\). Let \(\cQ\subset \cP(\cS)\) be uniformly tight. Then
\[
\lim_{K\to\infty}\sup_{Q\in\cQ}\sum_{p=K+1}^\infty \E_Q\left[e_p(S)^2\right]=0.
\]
\end{proposition}

\begin{proof}
Define \(R_K(s)\coloneqq \sum_{p=K+1}^\infty e_p(s)^2\) for \(s\in\cS\). Since \(k\) is continuous, the feature map \(\phi(s)\coloneqq k(s,\cdot)\) is norm-continuous, making each \(e_p(s)=\langle e_p,\phi(s)\rangle_{\cH_k}\) continuous. Parseval's identity yields \(\sum_{p=1}^\infty e_p(s)^2 = k(s,s)\). Thus, \(R_K(s)=k(s,s)-\sum_{p=1}^K e_p(s)^2\) is continuous, decreases monotonically to \(0\) for each \(s\), and is uniformly bounded by \(M\coloneqq \sup_{s\in\cS} k(s,s)<\infty\).

Fix \(\varepsilon>0\). By uniform tightness, there exists a compact set \(C_\varepsilon\subset\cS\) such that \(\sup_{Q\in\cQ}Q(C_\varepsilon^c)\le \varepsilon/(2M)\). By Dini's theorem \citep[Theorem~2.4.10]{Dudley2002}, \(R_K \downarrow 0\) uniformly on \(C_\varepsilon\). Hence, for all sufficiently large \(K\), \(\sup_{s\in C_\varepsilon}R_K(s)\le \varepsilon/2\). Splitting the expectation over \(C_\varepsilon\) and its complement yields
\[
\sup_{Q\in\cQ}\sum_{p=K+1}^\infty \E_Q\left[e_p(S)^2\right]
=
\sup_{Q\in\cQ}\E_Q\left[R_K(S)\right]
\le \sup_{s\in C_\varepsilon}R_K(s)+M \sup_{Q\in\cQ}Q(C_\varepsilon^c) \le \varepsilon.
\]
Since \(\varepsilon>0\) is arbitrary, the result follows.
\end{proof}

\begin{corollary}[Uniform tightness via functions with compact sublevel sets]
\label{cor:compact_sublevel}
Let $\cS$ be a Polish space. Suppose there exists a measurable function $\psi: \cS \to [0, \infty]$ such that for every $c > 0$, the sublevel set $K_c \coloneqq \{s \in \cS : \psi(s) \le c\}$ is compact in $\cS$. If a family of probability measures $\cQ \subset \cP(\cS)$ satisfies
\[
    \sup_{Q \in \cQ} \E_Q[\psi(S)] < \infty,
\]
then $\cQ$ is uniformly tight.
\end{corollary}

\begin{proof}
By Markov's inequality, \(\sup_{Q\in\cQ}Q(\psi(S) > c) \le c^{-1} \sup_{Q\in\cQ}\E_Q[\psi(S)] \to 0\) as \(c \to \infty\). Since the sublevel sets \(K_c\) are compact by assumption, uniform tightness immediately follows.
\end{proof}

\begin{remark}[Sufficient conditions for uniform tightness in various spaces]
Corollary~\ref{cor:compact_sublevel} is a versatile tool for establishing uniform tightness via an appropriate penalty function $\psi$. Notable examples include:
\begin{enumerate}
    \item \textbf{Euclidean space ($\bbR^d$):} Let $\psi(x) = \|x\|^p$ for some $p > 0$. The closed sublevel sets are compact by the Heine--Borel theorem. Thus, a uniform moment bound $\sup_{Q \in \cQ} \E_Q[\|X\|^p] < \infty$ implies uniform tightness.
    
    \item \textbf{Functional data space (e.g., $L^2([0,1])$):} Since closed balls are not compact in infinite dimensions, one may use the Sobolev norm $\psi(f) = \|f\|_{H^1}^2$. By the Rellich--Kondrachov theorem \citep[e.g.,][Theorem~9.16]{Brezis2011}, the embedding \(H^1([0,1]) \hookrightarrow L^2([0,1])\) is compact, making the sublevel sets relatively compact.
    
    \item \textbf{Wasserstein space ($\cW_p(\bbR^d)$):} Relative compactness in the $p$-Wasserstein metric requires both uniform tightness and uniform $p$-integrability. Choosing $\psi(\mu) = \int \|x\|^q \mu(\diff x)$ for any $q > p$ simultaneously guarantees both conditions.
\end{enumerate}
\end{remark}

\section{Projected-signal conditions for the power analysis}\label{sec:suffic_cond_PSS}
In this section we give sufficient conditions for the projected signal strength condition used in the uniform power analysis of Section~\ref{subsec:uniform_power}. The argument is based on a basis expansion of the projected CCCO, which makes explicit how the projected signal is distributed across the retained spectral coordinates. Throughout, for a subclass \(\cA\subset \tilde{\cP}_1\), we write \(\Sigma_P \coloneqq \Sigma_{XYZ\mid Z}\) for \(P\in\cA\).

For each \(P\in\cA\), let \(\{e_{p,P}\}_{p\ge1}\) and \(\{f_{q,P}\}_{q\ge1}\) denote orthonormal eigensystems of \(C_{XX}\) and \(C_{YY}\), respectively. As in the main text, let \(\zeta_{X,p,P}\) and \(\zeta_{Y,q,P}\) be the corresponding residual scores, and define \(c_{pq,P}(z)\coloneqq \E_P[\zeta_{X,p,P}\zeta_{Y,q,P}\mid Z=z]\). Then \(\Sigma_P\) admits the expansion
\[
\Sigma_P
=
\sum_{p,q\ge1} e_{p,P}\otimes f_{q,P}\otimes \Gamma_{pq,P},
\]
where \(\Gamma_{pq,P}\coloneqq \E_P[c_{pq,P}(Z)\,k_{\cZ}(Z,\cdot)]\). We have $\|\Sigma_P\|^2=\sum_{p,q\ge1}\|\Gamma_{pq,P}\|_{\cH_{\cZ}}^2$. 
Moreover, since $\Pi_n=\Pi_{X,L_{X,n}}\otimes \Pi_{Y,L_{Y,n}}\otimes I$, 
we similarly obtain
\[
\|(I-\Pi_n)\Sigma_P\|^2
=
\sum_{\substack{p>L_{X,n}~ \text{or}~ q>L_{Y,n}}}
\|\Gamma_{pq,P}\|_{\cH_{\cZ}}^2.
\]

\begin{assumption}[Weighted score-covariance decay]\label{ass:weighted_score_cov_decay}
For given $\cA\subset \tilde{\cP}_1$, there exist constants $s_X,s_Y>0$ and $M<\infty$ such that
\[
\sup_{P\in\cA}
\sum_{p,q\ge1}
\left\{(1+p)^{2s_X}+(1+q)^{2s_Y}\right\}
\,
\|c_{pq,P}\|_{L^2(P_Z)}^2
\le M.
\]
\end{assumption}
This condition means that the conditional dependence between $X$ and $Y$ is concentrated in low-order eigendirections of the marginal covariance operators, while higher-order score interactions decay sufficiently fast to make the truncation remainder small. 
Define
\[
r_n^2
\coloneqq 
M\left\{(1+L_{X,n})^{-2s_X}+(1+L_{Y,n})^{-2s_Y}\right\}.
\]

\begin{proposition}\label{prop:projected_signal_close_to_full_signal}
Assume that Assumption~\ref{ass:weighted_score_cov_decay} holds on a class $\cA_n \subset \tilde{\cP}_1$. If $\inf_{P\in\cA_n}\|\Sigma_P\|
\ge c/\sqrt{n_1}+ K_Zr_n$ holds for some \(c>0\), then we have $\cA_n\subset \tilde{\cP}_{1,n}(c)$ for sufficiently large $n$. 
\end{proposition}
\begin{remark}
The proposition is useful when the truncation remainder \(r_n\) is small relative to the full CCCO signal. In particular, for fixed alternatives one typically requires \(r_n\to0\), whereas for local alternatives at the \(n_1^{-1/2}\) scale one needs \(r_n=O(n_1^{-1/2})\).
\end{remark}
\begin{remark}
    Under the interpretation in Section~\ref{subsec:CCCO_MMD}, the lower bound on $\Sigma_P$ may be read as requiring that the observed law be sufficiently separated from its conditionally independent coupling. The role of the present subsection is not to derive such a separation, but to show that, once it holds, a sufficiently large portion of the signal is retained after truncation.
\end{remark}
\begin{proof}[of Proposition \ref{prop:projected_signal_close_to_full_signal}]
For any \(P\in\cA_n\) and any \((p,q)\), the reproducing property and Cauchy--Schwarz inequality yield \(\left\|\Gamma_{pq,P}\right\|_{\cH_{\cZ}} = \left\|\E_P\left[c_{pq,P}(Z)\,k_{\cZ}(Z,\cdot)\right]\right\|_{\cH_{\cZ}} \le K_Z\|c_{pq,P}\|_{L^2(P_Z)}\). Since \(\{p>L_{X,n} \text{ or } q>L_{Y,n}\} \subset \{p>L_{X,n}\} \cup \{q>L_{Y,n}\}\), this implies
\begin{align*}
\left\|(I-\Pi_n)\Sigma_P\right\|^2
&\le
K_Z^2 \sum_{p>L_{X,n} \text{ or } q>L_{Y,n}} \|c_{pq,P}\|_{L^2(P_Z)}^2 \\
&\le
K_Z^2 \left( \sum_{p>L_{X,n},\,q\ge1} \|c_{pq,P}\|_{L^2(P_Z)}^2 + \sum_{p\ge1,\,q>L_{Y,n}} \|c_{pq,P}\|_{L^2(P_Z)}^2 \right).
\end{align*}
Because \(1\le (1+L_{X,n})^{-2s_X}(1+p)^{2s_X}\) for \(p>L_{X,n}\), and similarly for \(q>L_{Y,n}\), Assumption~\ref{ass:weighted_score_cov_decay} gives \(\sup_{P\in\cA_n}\|(I-\Pi_n)\Sigma_P\|^2 \le K_Z^2 M\{(1+L_{X,n})^{-2s_X}+(1+L_{Y,n})^{-2s_Y}\} = K_Z^2 r_n^2\).

By the reverse triangle inequality, \(\bigl|\|\Sigma_P\|-\|\Pi_n\Sigma_P\|\bigr| \le \|(I-\Pi_n)\Sigma_P\| \le K_Zr_n\) uniformly over \(P\in\cA_n\). Finally, if \(\inf_{P\in\cA_n}\|\Sigma_P\| \ge c/\sqrt{n_1}+ K_Zr_n\), then for every \(P\in\cA_n\), we have \(\left\|\Pi_n\Sigma_P\right\| \ge \|\Sigma_P\|-K_Zr_n \ge c/\sqrt{n_1}\). This implies \(P\in\tilde{\cP}_{1,n}(c)\), establishing \(\cA_n\subset \tilde{\cP}_{1,n}(c)\).
\end{proof}

The next lemma shows a further sufficient condition for Assumption \ref{ass:weighted_score_cov_decay}. The condition of this lemma is satisfied when conditional dependence is driven by a small number of $Z$-dependent interaction channels, and the corresponding score loadings are concentrated in low-frequency directions relative to the marginal covariance geometry of $X$ and $Y$. In particular, finite-rank score interactions are included as a special case.
\begin{lemma}[A low-rank channel condition implies Assumption~\ref{ass:weighted_score_cov_decay}]
\label{lem:low_rank_channel_implies_weighted_decay}
Let $\cA\subset \tilde{\cP}_1$. Suppose there exist an integer $R\ge1$ and constants
$C_h,C_u,C_v<\infty$ such that, for every $P\in\cA$,
\[
c_{pq,P}(z)
=
\sum_{m=1}^R u_{m,p,P}\,v_{m,q,P}\,h_{m,P}(z)
\qquad\text{in }L^2(P_Z),
\]
where $\sup_{P\in\cA}\max_{1\le m\le R}\|h_{m,P}\|_{L^2(P_Z)}\le C_h$, and
\begin{align*}
&\sup_{P\in\cA}\sum_{m=1}^R\sum_{p\ge1}(1+p)^{2s_X}u_{m,p,P}^2 \le C_u,
\qquad
\sup_{P\in\cA}\sum_{m=1}^R\sum_{q\ge1}(1+q)^{2s_Y}v_{m,q,P}^2 \le C_v, \\
&\sup_{P\in\cA}\sum_{m=1}^R\sum_{p\ge1}u_{m,p,P}^2 \le C_u,
\qquad
\sup_{P\in\cA}\sum_{m=1}^R\sum_{q\ge1}v_{m,q,P}^2 \le C_v.
\end{align*}
Then Assumption~\ref{ass:weighted_score_cov_decay} holds on $\cA$.
\end{lemma}
\begin{proof}[of Lemma \ref{lem:low_rank_channel_implies_weighted_decay}]
Fix $P\in\cA$. By the Cauchy--Schwarz inequality, we have 
\[
\|c_{pq,P}\|_{L^2(P_Z)}^2
=
\left\|
\sum_{m=1}^R u_{m,p,P}v_{m,q,P}h_{m,P}
\right\|_{L^2(P_Z)}^2
\le
R\sum_{m=1}^R u_{m,p,P}^2 v_{m,q,P}^2 \|h_{m,P}\|_{L^2(P_Z)}^2.
\]
This implies that 
\begin{align*}
&\sum_{p,q\ge1}
\left\{(1+p)^{2s_X}+(1+q)^{2s_Y}\right\}
\|c_{pq,P}\|_{L^2(P_Z)}^2 \\
&\le R C_h^2
\sum_{m=1}^R
\sum_{p,q\ge1}
\left\{(1+p)^{2s_X}+(1+q)^{2s_Y}\right\}
u_{m,p,P}^2 v_{m,q,P}^2 \\
&= R C_h^2
\sum_{m=1}^R\left\{\sum_{p\ge1}(1+p)^{2s_X}u_{m,p,P}^2
\sum_{q\ge1}v_{m,q,P}^2
+
\sum_{p\ge1}u_{m,p,P}^2
\sum_{q\ge1}(1+q)^{2s_Y}v_{m,q,P}^2\right\} \\
&\le 2 R C_h^2 C_u C_v.
\end{align*}
Therefore Assumption~\ref{ass:weighted_score_cov_decay} holds with
$M=2RC_h^2C_uC_v$.
\end{proof}

\section{Power of regression-based CI tests}\label{sec:reg_based_power}

Assume that the random vector \((X,Y,Z)\) takes values in \(\bbR^{d_X}\times\bbR^{d_Y}\times\bbR^{d_Z}\). Let \(\cE_{XYZ}\) be the set of all such distributions.

The general alternative space $\cP_1$, representing conditional dependence, is consequently the set of distributions where this condition is not met. That is, $\cP_1$ consists of distributions for which
\begin{equation*}
\E\left[\left(f(X,Z)-\E[f(X,Z) \mid Z]\right)\left(g(Y)-\E[g(Y) \mid Z]\right)\mid Z\right]\neq 0
\quad \text{a.s.-}P_Z
\end{equation*}
holds for \textit{some} functions $f\in L^2_{XZ}$ and $g\in L^2_Y$. In this work, we study the extent to which three specific regression-based CI tests can reject distributions in $\cP_1$ with non-trivial power.

It is worth noting that this regression-based perspective is quite general. For instance, kernel-based CI tests, which estimate CMEs in a RKHS, can also be formulated as regression problems. Viewed this way, these kernel methods can be understood as generalizations of methods such as the Generalized Covariance Measure (GCM) and the Weighted Generalized Covariance Measure (WGCM). Specifically, the kernel $k_{\cZ}$ associated with the conditioning variable $Z$ in the CME approach serves a role analogous to the weight function in WGCM. A detailed exploration of this connection, however, is omitted here, as our investigation focuses on regression-based CI tests that do not explicitly utilize the kernel framework.

\subsection{Population functionals for three regression-based CI tests}
Here, we define the population versions of the statistics employed by these three tests. We begin by defining $\gamma(P)$:
\begin{equation*}\label{eq:gamma}
  \gamma(P)
  \coloneqq \E_P\left[\cov_P(X,Y\mid Z)\right] = \E[(Y-\E[Y \mid Z])\otimes (X- \E[X \mid Z])],
\end{equation*}
whose sample analogue appears in
\citet{ShahandPeters2020}.

Next, we consider $\Gamma(P)(z)$:
\begin{equation*}\label{eq:Gamma}
  \Gamma(P)(z)
  \coloneqq \cov_{P}(X,Y\mid Z = z),
\end{equation*}
estimated by the \emph{WGCM} of
\citet{Scheideggeretal2022};
replacing the weight function by a
kernel function yields the \emph{Kernelized GCM}, as described in, e.g., \citet{FernandezandRivera2024}.

Finally, we introduce $\Delta(P)$:
\begin{equation*}\label{eq:Delta}
  \Delta(P)
  \coloneqq \sup_{f\in L^2_X, g\in L^2_Y}
       \left|\E_{P}\left[
         \cov_P\bigl(f(X),g(Y)\mid Z\bigr)
       \right]\right|,
\end{equation*}
whose empirical version is implemented in
\cite{Shietal2021}.
\footnote{%
  Following \cite{Shietal2021}, we call the resulting statistic
  the \emph{Maximum GCM} (MGCM) to emphasize the maximization
  over function classes.
}
The relationships among the null and alternative hypotheses for these three tests are summarized in Table~\ref{tab:null_alt_en}. Each test’s null class \(\cP_0^{\mathrm{T}}\) is defined by the condition under which its corresponding population functional vanishes, and the associated alternative class \(\cP_1^{\mathrm{T}}\) records those alternatives for which the population target is nonzero. Accordingly, under whatever additional regularity conditions ensure validity and consistency of the corresponding test, membership in \(\cP_1^{\mathrm{T}}\) can be interpreted as indicating that asymptotic power should be expected, whereas vanishing of the target suggests that no such power conclusion can be drawn from that criterion alone.

\begin{table}[ht]
\centering
\caption{Regression-based tests and their detectable alternatives (all \(\cP_1^{\mathrm{T}}\subset\cP_1 \)). The symbol \(\equiv\) denotes equality almost surely (a.s.), and \(\not\equiv\) indicates that the values differ with positive probability.}
\label{tab:null_alt_en}
\small
\setlength{\tabcolsep}{4pt}
\begin{tabular}{@{}lccc@{}}
\toprule
Test & Functional & Null class \(\cP_0^{\mathrm{T}}\) & Alternative \(\cP_1^{\mathrm{T}}\)\\
\midrule
GCM \citep{ShahandPeters2020} & \(\gamma\) & \(\gamma(P)=0\) & \(\gamma(P)\neq 0\)\\
WGCM \citep{Scheideggeretal2022} & \(\Gamma\) & \(\Gamma(P)\equiv 0\) & \(\Gamma(P)\not\equiv 0\)\\
MGCM \citep{Shietal2021} & \(\Delta\) & \(\Delta(P)=0\) & \(\Delta(P)\neq 0\)\\
\bottomrule
\end{tabular}
\end{table}

Each test's null class, \(\cP_0^{\mathrm{T}}\), is defined by the condition under which its corresponding population functional is zero. The null class for MGCM, characterized by \(\Delta(P)=0\), is known as weak conditional independence \citep{Daudin1980}. The other nulls also represent strictly weaker conditions. Specifically, the null for WGCM requires \(\Gamma(P) \equiv 0\) almost surely, while the null for GCM only requires that the conditional covariance is zero on average, \(\gamma(P) = 0\).

Consequently, the alternative hypothesis classes \(\cP_1\) detectable by WGCM and MGCM are strictly broader than that of GCM. This hierarchy is given by:
\begin{equation*}
    \cP_1^{\text{GCM}} \subsetneq \cP_1^{\text{WGCM}} \enspace \text{and} \enspace
    \cP_1^{\text{GCM}} \subsetneq  \cP_1^{\text{MGCM}}
\end{equation*}
However, the functionals \(\Gamma\) and \(\Delta\) are \emph{incomparable}, which means that neither of their corresponding alternative classes is a subset of the other. The following examples are chosen to illustrate this incomparability.
\subsection{Illustrative examples}
\begin{example}[Even-moment dependence: WGCM powerless, MGCM powerful]
\label{ex:even_moment}
Take \(Z=c\) (constant),
\(X\sim N(0,1)\) and \(Y=X^2-1\).
\begin{equation*}
  \Gamma(P)(c)
    = \cov_P(X,Y\mid Z=c)
    = \E_P [XY] - \E_P[X]\E_P[Y]
    = \E_P[X(X^2-1)] - 0 \times 0
    = 0,
\end{equation*}
so \(\Gamma(P)\equiv0\) and the WGCM is powerless against this alternative. Choose \(f(x)=x^2\) and \(g(y)=y\); then
\begin{equation*}
  \cov(f(X),g(Y)\mid Z=c)
    = \E [(X^2-1)(Y-0)]
    = \E [(X^2-1)^2]
    = \var(X^2)
    = 3-1 = 2,
\end{equation*}
so \(\Delta(P)\geq 2>0\) and MGCM \emph{does} have power.
\end{example}

\begin{example}[Binary XOR dependence: MGCM powerless, WGCM powerful]
\label{ex:binary_xor}
Let \(X,Y,Z\in\{0,1\}\) and the conditional probabilities $\pr(X=x, Y=y \mid Z=z)$ for $x, y, z \in \{0, 1\}$ are given by:
\begin{equation*}
\pr(X=x, Y=y \mid Z=z) = 
\setlength{\arraycolsep}{10pt} 
\begin{array}{c|c|cc}
 & & Y=0 & Y=1 \\ \hline
\multirow{2}{*}{Z=0} & X=0 & 1/6 & 1/3 \\
 & X=1 & 1/3 & 1/6 \\ \hline
\multirow{2}{*}{Z=1} & X=0 & 1/3 & 1/6 \\
 & X=1 & 1/6 & 1/3
\end{array}
\end{equation*}

As discussed later, the binary nature of $X$ and $Y$ in this example allows WGCM to guarantee power. In contrast, MGCM cannot do so for this data-generating process, as shown by \citet{Shietal2021}.
\end{example}

\subsection{The binary-case strength of WGCM}
If \(X\) and \(Y\) are binary (or categorical), the equivalence \(\Gamma(P)\equiv0 \iff \text{CI}\) \citep[Section~1.1.1]{Scheideggeretal2022} implies that \(\cP_1^{\text{WGCM}} = \cP_1\), so the WGCM criterion fully characterizes CI in this case. Accordingly, under the regularity conditions ensuring consistency of the WGCM test, one may expect full asymptotic power in this binary setting.

In summary, GCM is simple but blind to alternatives with
\(\E [\cov(X,Y\mid Z)]=0\).
WGCM and MGCM extend power in distinct and complementary
directions:
WGCM thrives when conditional covariance cancels on average
(and dominates for binary data),
while MGCM uncovers higher-order nonlinear dependence.
Neither test alone exhausts the alternative space
\(\cP_1 \), underscoring the value of combining
weighting and maximization strategies in
regression-based CI testing.

\subsection{Detectability of GCM and WGCM in the low-dimensional designs}
\label{subsec:sim_power_explanation}

To clarify the behaviour of GCM and WGCM in the low-dimensional designs of Section~6.2, define
\[
\Gamma(P)(z) \coloneqq \cov_P(X,Y \mid Z=z),
\qquad
\gamma(P) \coloneqq \E_P\left[\Gamma(P)(Z)\right].
\]
GCM targets \(\gamma(P)\), whereas WGCM targets the conditional covariance function \(\Gamma(P)\).

Under the null design of Section~6.2, \(\Gamma(P)(z)=0\) for all \(z\), and hence \(\gamma(P)=0\). Thus both procedures correctly retain the null.
\begin{description}[ leftmargin=0pt, itemindent=0pt, labelsep=0.5em, itemsep=1.5ex, topsep=1ex, parsep=0pt ]
\item[Case 1] \emph{linear perturbation.}
If \(Y\) is replaced by \(Y^{\mathrm{new}}=Y+0.2X\), then
\[
\Gamma(P)(z)
=
\cov_P\left(X, Y+0.2X \mid Z=z\right)
=
0.2\,\var_P\left(X \mid Z=z\right)
=
0.018\,\var(\varepsilon_X)
>
0
\]
for every \(z\). Therefore \(\gamma(P)>0\), so both WGCM and GCM have population-level detectability.

\item[Case 2] \emph{latent even--odd structure.}
Let
\[
X^{\mathrm{new}}=f_a(Z)+0.3\,\varepsilon_X+0.3\left(U^2-1\right),
\qquad
Y^{\mathrm{new}}=f_a(Z)+0.3\,\varepsilon_Y+0.3\,U,
\]
where \(U \sim N(0,1)\) is independent of \(Z,\varepsilon_X,\varepsilon_Y\). Then
\[
\Gamma(P)(z)
=
0.09\,\cov_P\left(U^2-1, U \mid Z=z\right)
=
0
\]
for all \(z\), because \(U \mid Z=z \sim N(0,1)\) has vanishing odd moments. Hence \(\gamma(P)=0\) as well. Thus neither WGCM nor GCM has a population-level guarantee against this alternative.

\item[Case 3] \emph{\(Z\)-coupled latent structure.}
Retain the construction of Case~2, but let \(U=f_1(Z)+\varepsilon_U\), where \(\varepsilon_U \sim N(0,1)\) is independent of \(Z,\varepsilon_X,\varepsilon_Y\). Writing \(\mu=f_1(z)\), we have \(U \mid Z=z \sim N(\mu,1)\), so
\[
\cov_P\left(U^2-1, U \mid Z=z\right)
=
\left(\mu^3+2\mu\right)-\mu^3
=
2\mu.
\]
Therefore
\[
\Gamma(P)(z)
=
0.09 \cdot 2 f_1(z)
=
0.18\,f_1(z).
\]
Since \(f_1\) is not identically zero, \(\Gamma(P)\) is nonzero on a set of positive probability, and WGCM retains population-level detectability. By contrast,
\[
\gamma(P)
=
0.18\,\E\left[f_1(Z)\right]
=
0,
\]
because \(f_1(z)=\exp\left(-z^2/2\right)\sin(z)\) is odd and \(Z\) is symmetric. Hence GCM does not admit a corresponding guarantee.
\end{description}

In summary, Case~1 is detectable by both GCM and WGCM, Case~2 by neither, and Case~3 by WGCM but not by GCM.
\section{Additional simulation results}\label{sec:additional_simulation}
\subsection{Bootstrap multipliers and separation constants}
\label{sec:supp_multiplier_theory}

We briefly record how the multiplier fourth moment can enter the \emph{sufficient} separation constant in the proof of Theorem~\ref{thm:uniform_power_monotone}. The point is not that the exact bootstrap quantile, or the exact power function, is monotone in the multiplier law. Rather, for bootstrap statistics of the form
\[
S_n^\ast
\coloneqq
nT_n^\ast
=
\frac{1}{n}\sum_{i=1}^n\sum_{j=1}^n
W_i^\ast W_j^\ast \langle u_i,u_j\rangle_{\cH}
=
\left\|
\frac{1}{\sqrt n}\sum_{i=1}^n W_i^\ast u_i
\right\|_{\cH}^2,
\]
one can bound the conditional bootstrap critical value by a quantity that is monotone in the fourth moment \(\mu_4\coloneqq \E[(W^\ast)^4]\).

Let \(W_1^\ast,\dots,W_n^\ast\) be i.i.d.\ multipliers, independent of the data, with \(\E[W^\ast]=0\), \(\E[(W^\ast)^2]=1\), and \(\mu_4<\infty\). This includes the Gaussian, Rademacher, and Mammen choices, for which \(\mu_4=3,1,2\), respectively. Fix \(u_1,\dots,u_n\in\cH\), and let \(q_{n,1-\alpha}^\ast(u)\) be the exact conditional \((1-\alpha)\)-quantile of \(S_n^\ast\) given \(u_{1:n}\). A direct fourth-moment calculation gives
\[
\E[S_n^\ast\mid u_{1:n}]
=
\frac{1}{n}\sum_{i=1}^n \|u_i\|_{\cH}^2, \quad 
\var(S_n^\ast\mid u_{1:n})
=
\frac{2}{n^2}\sum_{i\neq j}\langle u_i,u_j\rangle_{\cH}^2
+
\frac{\mu_4-1}{n^2}\sum_{i=1}^n \|u_i\|_{\cH}^4.
\]
Hence the conditional variance is affine, and therefore monotone nondecreasing, in \(\mu_4\). By the one-sided Chebyshev bound (cf.\ Exercise 1.6.4 in \cite{Durrett2019}),
\[
q_{n,1-\alpha}^\ast(u)
\le
\overline q_{n,1-\alpha}^\ast(u;\mu_4)
\coloneqq
\frac{1}{n}\sum_{i=1}^n \|u_i\|_{\cH}^2
+
\sqrt{\frac{1-\alpha}{\alpha}}
\left\{
\frac{2}{n^2}\sum_{i\neq j}\langle u_i,u_j\rangle_{\cH}^2
+
\frac{\mu_4-1}{n^2}\sum_{i=1}^n \|u_i\|_{\cH}^4
\right\}^{1/2}.
\]
Thus \(\overline q_{n,1-\alpha}^\ast(u;\mu_4)\) is an explicit upper bound on the conditional bootstrap critical value, and this upper bound is monotone nondecreasing in \(\mu_4\).

This can be inserted into the proof of Theorem~\ref{thm:uniform_power_monotone}. There, non-rejection implies
\[
\{\phi_n^{(B_n)}=0\}
\subset
\left\{
\Delta_{n,P}+R_{n,P}+\left(Q_{n,1-\alpha}^{\ast(B_n)}\right)^{1/2}\ge c
\right\},
\]
where \(\Delta_{n,P}\) and \(R_{n,P}\) are the projection-error and empirical-fluctuation terms appearing in that proof. If one further bounds \(Q_{n,1-\alpha}^{\ast(B_n)}\) by \(\overline q_{n,1-\alpha}^\ast(u;\mu_4)\), then any constant \(c\) exceeding a sufficiently high-probability envelope of
\[
\Delta_{n,P}+R_{n,P}+\overline q_{n,1-\alpha}^\ast(u;\mu_4)^{1/2}
\]
is sufficient for Part~(i). In particular, for each \(\beta\in(\alpha,1)\), the proof yields a sufficient constant \(c_\beta(\mu_4)\) that may be chosen monotone nondecreasing in \(\mu_4\).

Under the additional bound \(\|u_i\|_{\cH}\le M\) for all \(i\), one obtains
\[
\overline q_{n,1-\alpha}^\ast(u;\mu_4)
\le
M^2
\left[
1+
\sqrt{\frac{1-\alpha}{\alpha}}
\left\{
2+\frac{\mu_4-1}{n}
\right\}^{1/2}
\right],
\]
so that \(c_\beta(\mu_4)\) may be chosen to dominate a corresponding high-probability envelope of
\[
\Delta_{n,P}+R_{n,P}
+
M
\left[
1+
\sqrt{\frac{1-\alpha}{\alpha}}
\left\{
2+\frac{\mu_4-1}{n}
\right\}^{1/2}
\right]^{1/2}.
\]
Hence the sufficient signal threshold delivered by this argument becomes larger as \(\mu_4\) increases. This does not show that the exact power is monotone in \(\mu_4\), but it does show that the proof-based sufficient constant \(c\) can be chosen monotonically in \(\mu_4\). The empirical results in Table~\ref{tab:sgcm_multipliers} are consistent with this ordering, with Gaussian multipliers appearing more conservative than Mammen and Rademacher multipliers.
\subsection{Empirical comparison of multiplier choices}
\label{sec:supp_multiplier_empirical}
Table~\ref{tab:sgcm_multipliers} is broadly consistent with the proof-based sufficient-constant comparison in Section~\ref{sec:supp_multiplier_theory}. Under the null, Gaussian yields the smallest Type~I error across DGPs and \(n\), whereas Rademacher can exceed the nominal level in small samples; under alternatives, Rademacher attains the highest power. We therefore adopt \emph{Gaussian} multipliers to prioritize small-sample level control.

\subsection{Alternative bootstrap schemes}
As also noted by \citet{Leeetal2020}, alternative bootstrap designs exist (e.g., response-perturbation multiplier bootstraps such as \citet{Patileaetal2016}), but in our setting they require recomputing the full statistic at each replicate, so regressions must be repeatedly refit and the computational cost is high.

\begin{table}[htbp]
\centering
\caption{Empirical rejection rates for SGCM with different bootstrap multipliers \(\left(\alpha=0.05\right)\). Across both \texttt{(a2)} and \texttt{(a4)}, the Gaussian multiplier is the best calibrated, whereas the Rademacher multiplier is the most powerful but also the most liberal.}
\label{tab:sgcm_multipliers}
\scriptsize
\renewcommand{\arraystretch}{1.12}
\setlength{\tabcolsep}{5pt}
\sisetup{
  table-format = 1.3,
  detect-weight = true,
  detect-inline-weight = math
}
\begin{tabular}{ll
                S S S S
                S S S S}
\toprule
\multirow{2}{*}{Scenario} & \multirow{2}{*}{Multiplier}
& \multicolumn{4}{c}{\texttt{(a2)}}
& \multicolumn{4}{c}{\texttt{(a4)}} \\
\cmidrule(lr){3-6}\cmidrule(l){7-10}
&
& {100} & {200} & {300} & {400}
& {100} & {200} & {300} & {400} \\
\midrule
\multirow{3}{*}{Size (Null)}
& Gaussian    & 0.059 & 0.053 & 0.069 & 0.073 & 0.073 & 0.085 & 0.095 & 0.071 \\
& Mammen      & 0.075 & 0.068 & 0.074 & 0.077 & 0.095 & 0.091 & 0.104 & 0.072 \\
& Rademacher  & 0.097 & 0.087 & 0.086 & 0.079 & 0.128 & 0.100 & 0.119 & 0.081 \\
\cmidrule(lr){1-10}
\multirow{3}{*}{Power (DGP1-1)}
& Gaussian    & 0.242 & 0.534 & 0.723 & 0.828 & 0.339 & 0.616 & 0.777 & 0.873 \\
& Mammen      & 0.292 & 0.562 & 0.740 & 0.834 & 0.387 & 0.635 & 0.798 & 0.879 \\
& Rademacher  & 0.351 & 0.600 & 0.756 & 0.852 & 0.451 & 0.671 & 0.811 & 0.886 \\
\cmidrule(lr){1-10}
\multirow{3}{*}{Power (DGP1-2)}
& Gaussian    & 0.155 & 0.335 & 0.486 & 0.655 & 0.153 & 0.316 & 0.475 & 0.624 \\
& Mammen      & 0.189 & 0.370 & 0.519 & 0.681 & 0.199 & 0.341 & 0.511 & 0.648 \\
& Rademacher  & 0.255 & 0.423 & 0.549 & 0.697 & 0.251 & 0.395 & 0.549 & 0.674 \\
\cmidrule(lr){1-10}
\multirow{3}{*}{Power (DGP1-3)}
& Gaussian    & 0.326 & 0.694 & 0.921 & 0.979 & 0.255 & 0.544 & 0.805 & 0.907 \\
& Mammen      & 0.383 & 0.731 & 0.927 & 0.981 & 0.298 & 0.582 & 0.824 & 0.917 \\
& Rademacher  & 0.458 & 0.770 & 0.941 & 0.981 & 0.369 & 0.627 & 0.843 & 0.927 \\
\bottomrule
\end{tabular}
\end{table}

\subsection{Simulation results for DGP \texttt{(a6)}}\label{subsec:a6-results}
In addition to the low-dimensional settings \texttt{(a2)} and \texttt{(a4)} in Section~6.2, we also consider the more challenging case \texttt{(a6)}, obtained by the same construction with \(a=6\), which further increases the oscillation frequency of \(f_a\).

Table~\ref{tab:a6_benchmarks} reports empirical size and power for DGP~\texttt{(a6)}. As in the other difficult settings, no method is exactly calibrated at the nominal level \(\alpha=0.05\). Size distortion remains severe for KCI, CDCOV, and MMDCI, whereas GCM, WGCM, and SGCM are comparatively closer to the nominal level. For SGCM, the choice of multiplier matters: the Gaussian version is the best calibrated, Mammen is slightly more liberal, and Rademacher is the most liberal among the three

For power, the same ordering persists. The more liberal SGCM variants achieve slightly larger rejection rates, but this comes with weaker calibration under the null. Thus, within the SGCM family, the Gaussian multiplier gives the most conservative and best-calibrated benchmark, while Mammen provides a slightly more aggressive alternative. Overall, SGCM remains competitive with GCM and WGCM while offering a clearer size--power trade-off than the heavily oversized procedures.

\begin{table}[htbp]
\centering
\caption{Empirical size and power for DGP \texttt{(a6)} at nominal level \(\alpha=0.05\). For SGCM, results are reported separately for Gaussian, Mammen, and Rademacher multipliers.}
\label{tab:a6_benchmarks}
\scriptsize
\renewcommand{\arraystretch}{1.12}
\setlength{\tabcolsep}{6pt}
\sisetup{
  table-format = 1.3,
  detect-weight = true,
  detect-inline-weight = math
}
\begin{tabular}{l
                S S S S
                S S S S}
\toprule
\multirow{2}{*}{Method}
& \multicolumn{4}{c}{Size}
& \multicolumn{4}{c}{Power (Alternative)} \\
\cmidrule(lr){2-5}\cmidrule(l){6-9}
& {100} & {200} & {300} & {400}
& {100} & {200} & {300} & {400} \\
\midrule
\textsc{gcm}               & 0.232 & 0.158 & 0.134 & 0.123 & 0.788 & 0.911 & 0.973 & 0.990 \\
\textsc{wgcm}              & 0.388 & 0.250 & 0.170 & 0.140 & 0.832 & 0.852 & 0.896 & 0.929 \\
\textsc{cdcov}             & 1.000 & 1.000 & 1.000 & 1.000 & \multicolumn{1}{c}{---} & \multicolumn{1}{c}{---} & \multicolumn{1}{c}{---} & \multicolumn{1}{c}{---} \\
\textsc{kci}               & 1.000 & 1.000 & 1.000 & 1.000 & \multicolumn{1}{c}{---} & \multicolumn{1}{c}{---} & \multicolumn{1}{c}{---} & \multicolumn{1}{c}{---} \\
\textsc{mmdci}             & 1.000 & 1.000 & 1.000 & 1.000 & \multicolumn{1}{c}{---} & \multicolumn{1}{c}{---} & \multicolumn{1}{c}{---} & \multicolumn{1}{c}{---} \\
\textsc{ccit}              & 0.454 & 0.444 & 0.430 & 0.401 & \multicolumn{1}{c}{---} & \multicolumn{1}{c}{---} & \multicolumn{1}{c}{---} & \multicolumn{1}{c}{---} \\
\textsc{sgcm} (Gaussian)   & 0.089 & 0.119 & 0.121 & 0.089 & 0.419 & 0.696 & 0.846 & 0.923 \\
\textsc{sgcm} (Mammen)     & 0.113 & 0.138 & 0.134 & 0.105 & 0.467 & 0.723 & 0.858 & 0.925 \\
\textsc{sgcm} (Rademacher) & 0.155 & 0.148 & 0.144 & 0.116 & 0.541 & 0.747 & 0.866 & 0.937 \\
\bottomrule
\end{tabular}
\end{table}

\subsection{Sensitivity to the FVE cutoff}\label{sec:supp_fve}

We examine the sensitivity of \textsc{SGCM} to the FVE cutoff under the Gaussian multiplier. Since the Gaussian multiplier is the best-calibrated choice within the \textsc{SGCM} family, it provides the clearest benchmark for isolating the effect of the truncation rule.

\begin{figure*}[htbp]
  \centering
  \includegraphics[width=\textwidth]{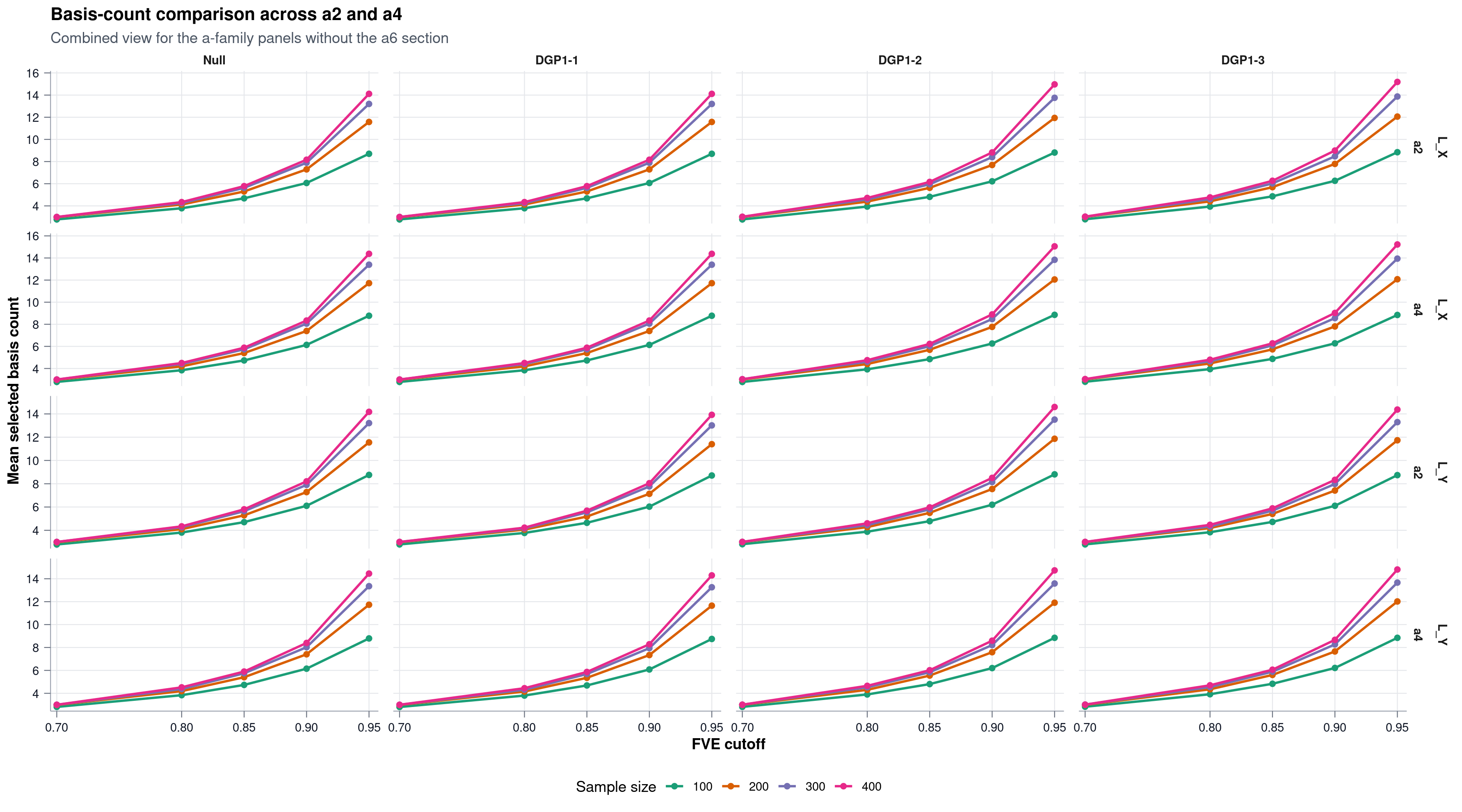}
  \caption{Mean selected basis counts across FVE cutoffs for DGPs \texttt{(a2)} and \texttt{(a4)} under the Gaussian multiplier. Rows correspond to \(L_X\) and \(L_Y\), and columns to the null and three alternatives. Basis counts increase with the FVE cutoff.}
  \label{fig:fve-sensitivity-gaussian-basis-a2a4}
\end{figure*}

Figure~\ref{fig:fve-sensitivity-gaussian-basis-a2a4} shows that, for both DGPs \texttt{(a2)} and \texttt{(a4)}, the mean number of selected basis functions increases steadily with the FVE cutoff, both for \(L_X\) and \(L_Y\). This is the expected mechanical effect of retaining a larger fraction of the empirical spectral variation. Table~\ref{tab:fve-sensitivity-gaussian-a2a4}, however, shows that the corresponding empirical rejection rates change only modestly over the same range of cutoffs. In particular, size remains broadly stable, and the power differences are small relative to the effects of the sample size and the strength of the alternative. Taken together, these results indicate that, once the Gaussian multiplier is fixed, raising the FVE threshold mainly increases the number of retained basis functions, but does not materially affect the inferential conclusions for \textsc{SGCM} in DGPs \texttt{(a2)} and \texttt{(a4)}.

\begin{table*}[htbp]
  \centering
  \caption{Sensitivity of \textsc{SGCM} to the FVE cutoff under the Gaussian multiplier. Results for DGPs \texttt{(a2)} and \texttt{(a4)} are shown side by side. Entries are empirical rejection rates.}
  \label{tab:fve-sensitivity-gaussian-a2a4}
  \scriptsize
  \renewcommand{\arraystretch}{1.08}
  \setlength{\tabcolsep}{5pt}
  \resizebox{\textwidth}{!}{%
  \begin{tabular}{@{}llccccc@{\hspace{1.5em}}ccccc@{}}
    \toprule
    \multirow{2}{*}{Scenario} & \multirow{2}{*}{\(n\)}
      & \multicolumn{5}{c}{\texttt{(a2)}}
      & \multicolumn{5}{c}{\texttt{(a4)}} \\
    \cmidrule(r{0.75em}){3-7}\cmidrule(l{0.75em}){8-12}
      &
      & 0.70 & 0.80 & 0.85 & 0.90 & 0.95
      & 0.70 & 0.80 & 0.85 & 0.90 & 0.95 \\
    \midrule
    \multirow{4}{*}{\shortstack[l]{Size}}
      & 100 & 0.060 & 0.060 & 0.058 & 0.059 & 0.059 & 0.075 & 0.075 & 0.073 & 0.073 & 0.073 \\
      & 200 & 0.056 & 0.053 & 0.053 & 0.053 & 0.053 & 0.083 & 0.086 & 0.085 & 0.085 & 0.084 \\
      & 300 & 0.069 & 0.070 & 0.069 & 0.069 & 0.069 & 0.095 & 0.094 & 0.094 & 0.095 & 0.095 \\
      & 400 & 0.071 & 0.073 & 0.073 & 0.073 & 0.073 & 0.072 & 0.071 & 0.071 & 0.071 & 0.071 \\
    \midrule
    \multirow{4}{*}{\shortstack[l]{Power \\ (DGP1-1)}}
      & 100 & 0.248 & 0.243 & 0.243 & 0.242 & 0.242 & 0.341 & 0.338 & 0.338 & 0.339 & 0.338 \\
      & 200 & 0.540 & 0.534 & 0.533 & 0.534 & 0.534 & 0.624 & 0.616 & 0.617 & 0.616 & 0.616 \\
      & 300 & 0.730 & 0.725 & 0.723 & 0.723 & 0.722 & 0.783 & 0.776 & 0.777 & 0.777 & 0.777 \\
      & 400 & 0.832 & 0.826 & 0.828 & 0.828 & 0.828 & 0.875 & 0.872 & 0.873 & 0.873 & 0.873 \\
    \midrule
    \multirow{4}{*}{\shortstack[l]{Power \\ (DGP1-2)}}
      & 100 & 0.158 & 0.154 & 0.154 & 0.155 & 0.155 & 0.150 & 0.152 & 0.152 & 0.153 & 0.153 \\
      & 200 & 0.333 & 0.336 & 0.334 & 0.335 & 0.335 & 0.314 & 0.314 & 0.316 & 0.316 & 0.317 \\
      & 300 & 0.480 & 0.484 & 0.485 & 0.486 & 0.486 & 0.475 & 0.475 & 0.475 & 0.475 & 0.476 \\
      & 400 & 0.652 & 0.654 & 0.655 & 0.655 & 0.655 & 0.623 & 0.623 & 0.624 & 0.624 & 0.624 \\
    \midrule
    \multirow{4}{*}{\shortstack[l]{Power \\ (DGP1-3)}}
      & 100 & 0.318 & 0.323 & 0.325 & 0.326 & 0.326 & 0.251 & 0.254 & 0.255 & 0.255 & 0.255 \\
      & 200 & 0.681 & 0.694 & 0.693 & 0.694 & 0.694 & 0.547 & 0.545 & 0.546 & 0.544 & 0.544 \\
      & 300 & 0.916 & 0.922 & 0.921 & 0.921 & 0.920 & 0.801 & 0.805 & 0.804 & 0.805 & 0.805 \\
      & 400 & 0.974 & 0.978 & 0.979 & 0.979 & 0.979 & 0.908 & 0.907 & 0.907 & 0.907 & 0.907 \\
    \bottomrule
  \end{tabular}%
  }
\end{table*}

\begin{figure*}[htbp]
  \centering
  \includegraphics[width=\textwidth]{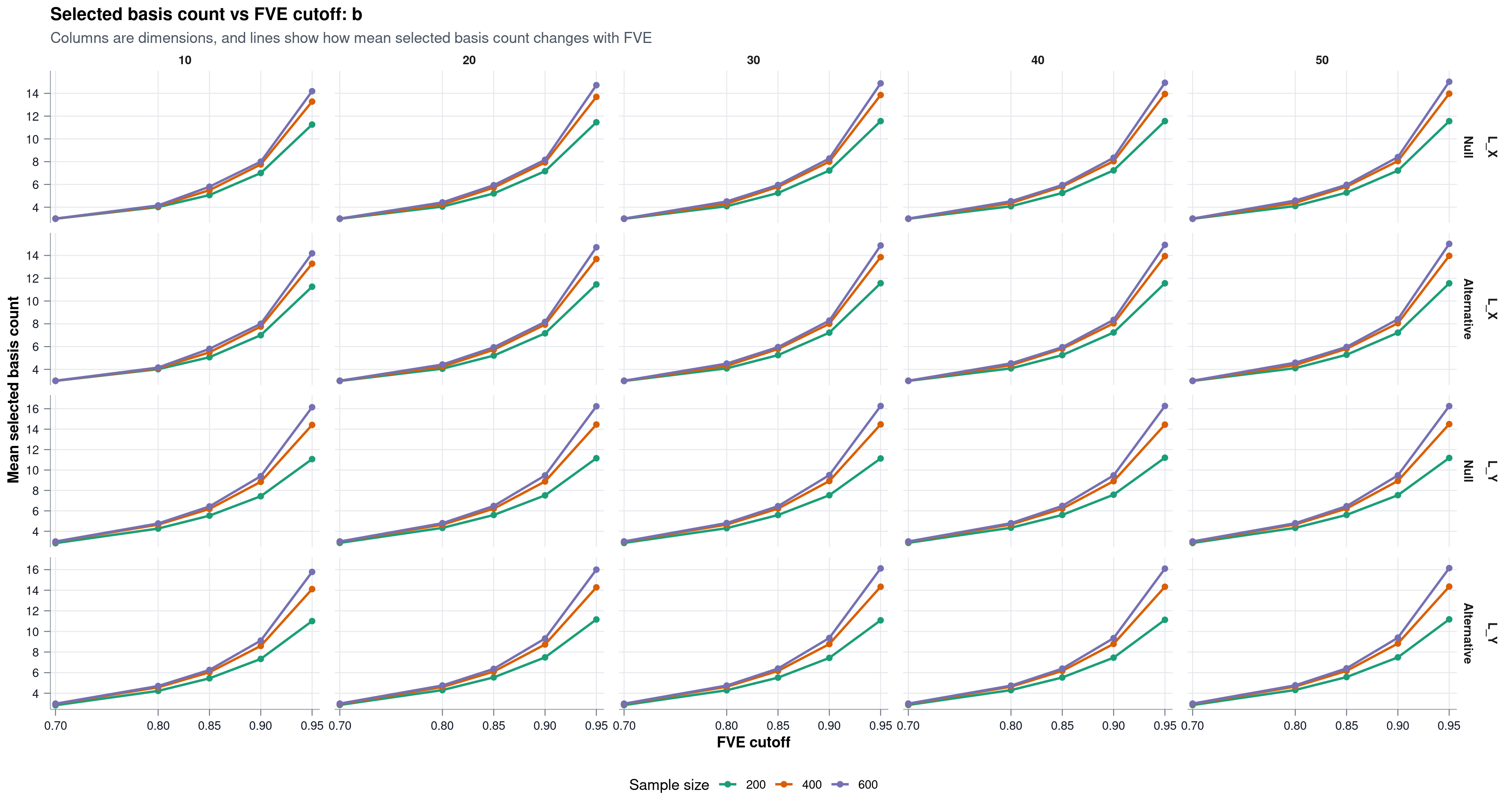}
  \caption{Mean selected basis counts across FVE cutoffs for DGP \texttt{(b)} under the Gaussian multiplier. Columns correspond to the ambient dimension, and rows to \(L_X\) and \(L_Y\) under the null and alternative. Basis counts increase with the FVE cutoff.}
  \label{fig:fve-sensitivity-gaussian-basis-b}
\end{figure*}

Figure~\ref{fig:fve-sensitivity-gaussian-basis-b} shows the same qualitative pattern for DGP \texttt{(b)}: for every ambient dimension, the mean number of selected basis functions increases with the FVE cutoff, both for \(L_X\) and \(L_Y\), under the null and under the alternative. Table~\ref{tab:fve-sensitivity-gaussian-b} nevertheless shows that the empirical rejection rates remain nearly unchanged across the same range of cutoffs. In particular, size is broadly stable over the FVE thresholds considered, while the main changes in power are driven instead by the ambient dimension and the sample size: power decreases with the dimension and increases with the sample size. Thus, as in DGPs \texttt{(a2)} and \texttt{(a4)}, increasing the FVE cutoff mainly enlarges the retained basis set, but has little practical effect on the resulting inference.

\begin{table*}[htbp]
  \centering
  \caption{Sensitivity of \textsc{SGCM} to the FVE cutoff under the Gaussian multiplier for DGP \texttt{(b)}. Entries are empirical rejection rates.}
  \label{tab:fve-sensitivity-gaussian-b}
  \tiny
  \renewcommand{\arraystretch}{1.04}
  \setlength{\tabcolsep}{3.2pt}
  \resizebox{\textwidth}{!}{%
  \begin{tabular}{@{}llccccc@{\hspace{0.9em}}ccccc@{\hspace{0.9em}}ccccc@{}}
    \toprule
    \multirow{2}{*}{Scenario} & \multirow{2}{*}{Dim}
      & \multicolumn{5}{c}{\(n=200\)}
      & \multicolumn{5}{c}{\(n=400\)}
      & \multicolumn{5}{c}{\(n=600\)} \\
    \cmidrule(r{0.45em}){3-7}\cmidrule(lr{0.45em}){8-12}\cmidrule(l{0.45em}){13-17}
      &
      & 0.70 & 0.80 & 0.85 & 0.90 & 0.95
      & 0.70 & 0.80 & 0.85 & 0.90 & 0.95
      & 0.70 & 0.80 & 0.85 & 0.90 & 0.95 \\
    \midrule
    \multirow{5}{*}{\shortstack[l]{Size}}
      & 10 & 0.060 & 0.060 & 0.060 & 0.060 & 0.060 & 0.085 & 0.085 & 0.086 & 0.086 & 0.086 & 0.081 & 0.081 & 0.081 & 0.081 & 0.081 \\
      & 20 & 0.058 & 0.058 & 0.058 & 0.058 & 0.058 & 0.065 & 0.065 & 0.065 & 0.065 & 0.065 & 0.055 & 0.055 & 0.055 & 0.055 & 0.055 \\
      & 30 & 0.033 & 0.033 & 0.033 & 0.033 & 0.033 & 0.054 & 0.055 & 0.054 & 0.054 & 0.054 & 0.052 & 0.052 & 0.052 & 0.052 & 0.052 \\
      & 40 & 0.048 & 0.047 & 0.047 & 0.047 & 0.047 & 0.053 & 0.052 & 0.052 & 0.052 & 0.052 & 0.049 & 0.047 & 0.047 & 0.047 & 0.047 \\
      & 50 & 0.063 & 0.061 & 0.061 & 0.061 & 0.061 & 0.044 & 0.045 & 0.045 & 0.045 & 0.045 & 0.056 & 0.056 & 0.056 & 0.056 & 0.056 \\
    \midrule
    \multirow{5}{*}{\shortstack[l]{Power}}
      & 10 & 0.562 & 0.560 & 0.561 & 0.561 & 0.561 & 0.894 & 0.897 & 0.896 & 0.896 & 0.896 & 0.988 & 0.988 & 0.987 & 0.987 & 0.987 \\
      & 20 & 0.401 & 0.400 & 0.401 & 0.400 & 0.400 & 0.763 & 0.766 & 0.766 & 0.766 & 0.766 & 0.928 & 0.928 & 0.929 & 0.929 & 0.929 \\
      & 30 & 0.341 & 0.342 & 0.342 & 0.342 & 0.342 & 0.672 & 0.672 & 0.673 & 0.673 & 0.673 & 0.861 & 0.863 & 0.863 & 0.863 & 0.863 \\
      & 40 & 0.308 & 0.309 & 0.309 & 0.309 & 0.309 & 0.604 & 0.605 & 0.604 & 0.604 & 0.604 & 0.812 & 0.813 & 0.813 & 0.813 & 0.813 \\
      & 50 & 0.273 & 0.269 & 0.270 & 0.271 & 0.271 & 0.554 & 0.550 & 0.550 & 0.551 & 0.550 & 0.748 & 0.750 & 0.750 & 0.751 & 0.750 \\
    \bottomrule
  \end{tabular}%
  }
\end{table*}

\subsection{Choice of the sample split ratio}
\label{sec:supp_split_ratio}

We also examine the sensitivity of \textsc{SGCM} to the sample split ratio. Write \(n_2\) for the auxiliary sample size used to construct the empirical eigenspaces and \(n_1=n-n_2\) for the sample size used to form the test statistic, and let \(\rho=n_2/n\). A smaller value of \(\rho\) leaves more observations for the testing stage and can therefore improve power. This effect is visible in Table~\ref{tab:split-sensitivity-gaussian-a2a4}: in several alternatives, especially for moderate and large sample sizes, smaller values of \(\rho\) tend to yield larger empirical rejection rates.

This gain in power, however, comes at the cost of reduced eigenspace stability. To assess this trade-off, we carried out a separate fixed-rank stability experiment. For each Monte Carlo replicate \(r=1,\dots,1000\), we split the sample into two disjoint parts \(I_1^{(r)}\) and \(I_2^{(r)}\), and for each \(V\in\{X,Y\}\) constructed the empirical eigenspaces
\[
\widehat{\mathcal E}_{V,I_1}^{(r)}
\qquad\text{and}\qquad
\widehat{\mathcal E}_{V,I_2}^{(r)}.
\]
We then compared these two eigenspaces at fixed ranks \(L\in\{3,5,10\}\) through the normalized Hilbert--Schmidt projector distance
\[
d_{\mathrm{nHS}}^{(r)}
=
\frac{1}{\sqrt{2L}}
\left\|P_{I_1}^{(r)}-P_{I_2}^{(r)}\right\|_{\mathrm{HS}}.
\]
This quantity measures the average discrepancy between the two estimated eigenspaces across the retained directions.

\begin{figure*}[t]
\centering
\begin{minipage}[t]{1.0\textwidth}
\centering
\includegraphics[width=\linewidth]{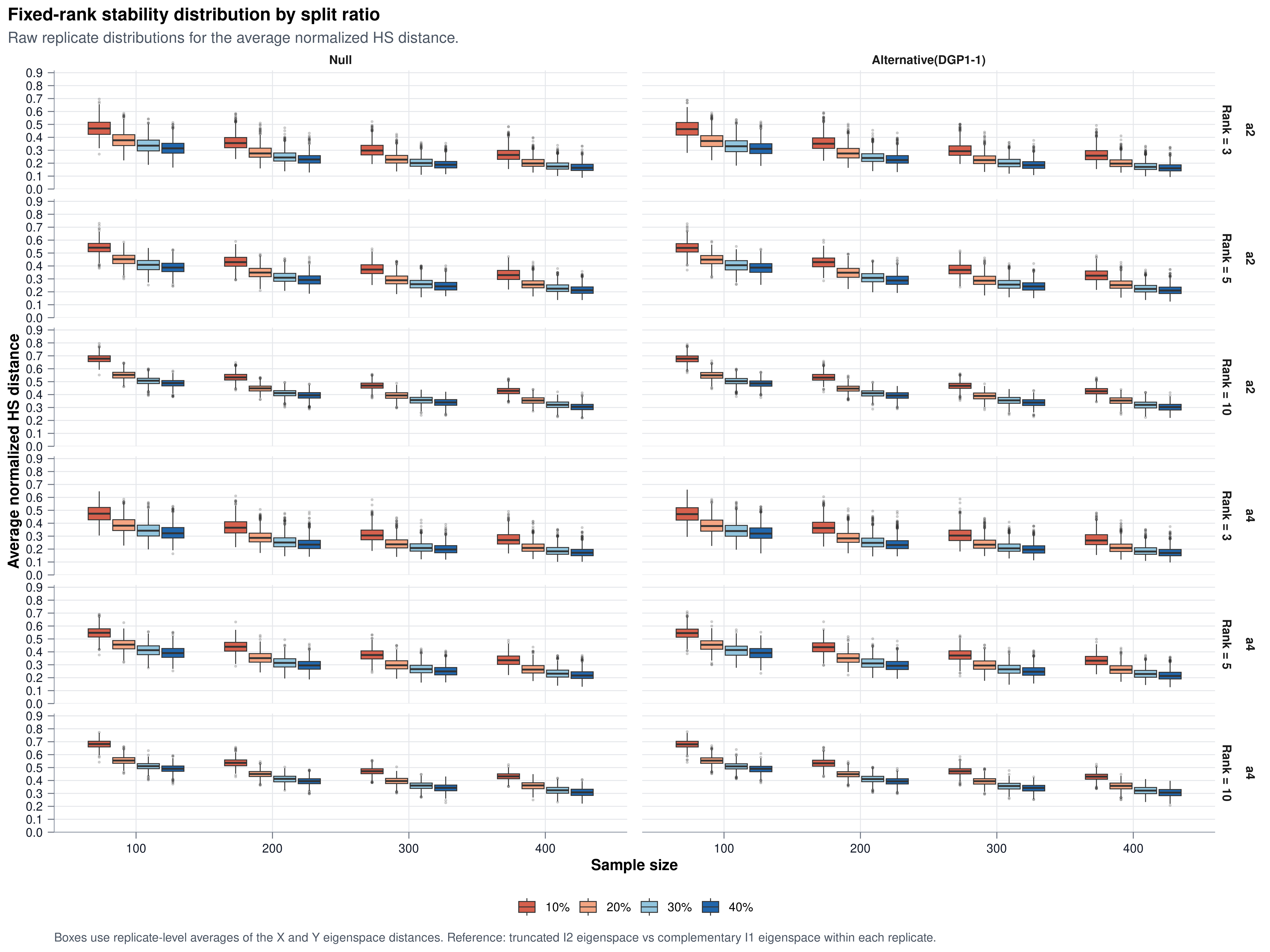}
\end{minipage}
\caption{Fixed-rank eigenspace discrepancy across sample split ratios, based on the normalized Hilbert--Schmidt distance between the empirical eigenspaces estimated on \(I_1\) and \(I_2\) at ranks \(L\in\{3,5,10\}\). Smaller split ratios yield larger discrepancies and hence lower stability.}
\label{fig:split_ratio_stability}
\end{figure*}

Figure~\ref{fig:split_ratio_stability} summarizes the distribution of \(d_{\mathrm{nHS}}^{(r)}\) over Monte Carlo replicates by boxplots. Smaller values of \(\rho\) tend to produce both larger medians and a wider spread, indicating that the eigenspaces estimated from \(I_1\) and \(I_2\) disagree more and exhibit greater variability, when the auxiliary sample is too small. Hence, although a smaller \(I_2\) can be favourable for power, it also makes the estimated eigenspaces less stable and therefore reduces reproducibility.

Taken together, Table~\ref{tab:split-sensitivity-gaussian-a2a4} and Figure~\ref{fig:split_ratio_stability} indicate a clear trade-off: smaller auxiliary splits may modestly improve power, but they also lead to systematically less stable eigenspace estimation. We therefore use a split ratio of \(20\%\) throughout as a practical compromise between power and stability.

\begin{table*}[htbp] 
  \centering
  \caption{Sensitivity of \textsc{SGCM} to the sample split ratio at FVE \(=0.90\) under the Gaussian multiplier. Results for DGPs \texttt{(a2)} and \texttt{(a4)} are shown side by side. Entries are empirical rejection rates.}
  \label{tab:split-sensitivity-gaussian-a2a4}
  \scriptsize
  \renewcommand{\arraystretch}{1.08}
  \setlength{\tabcolsep}{5pt}
  \resizebox{\textwidth}{!}{%
  \begin{tabular}{@{}llcccc@{\hspace{1.5em}}cccc@{}}
    \toprule
    \multirow{2}{*}{Scenario} & \multirow{2}{*}{\(n\)}
      & \multicolumn{4}{c}{\texttt{(a2)}}
      & \multicolumn{4}{c}{\texttt{(a4)}} \\
    \cmidrule(r{0.75em}){3-6}\cmidrule(l{0.75em}){7-10}
      &
      & 0.10 & 0.20 & 0.30 & 0.40
      & 0.10 & 0.20 & 0.30 & 0.40 \\
    \midrule
    \multirow{4}{*}{\shortstack[l]{Size}}
      & 100 & 0.051 & 0.059 & 0.056 & 0.045 & 0.069 & 0.073 & 0.060 & 0.063 \\
      & 200 & 0.053 & 0.053 & 0.058 & 0.069 & 0.072 & 0.085 & 0.079 & 0.077 \\
      & 300 & 0.066 & 0.069 & 0.083 & 0.076 & 0.085 & 0.095 & 0.099 & 0.091 \\
      & 400 & 0.062 & 0.073 & 0.075 & 0.068 & 0.075 & 0.071 & 0.090 & 0.080 \\
    \midrule
    \multirow{4}{*}{\shortstack[l]{Power \\ (DGP1-1)}}
      & 100 & 0.244 & 0.242 & 0.244 & 0.177 & 0.305 & 0.339 & 0.300 & 0.253 \\
      & 200 & 0.516 & 0.534 & 0.487 & 0.414 & 0.614 & 0.616 & 0.568 & 0.496 \\
      & 300 & 0.716 & 0.723 & 0.662 & 0.609 & 0.805 & 0.777 & 0.732 & 0.686 \\
      & 400 & 0.843 & 0.828 & 0.792 & 0.728 & 0.896 & 0.873 & 0.833 & 0.772 \\
    \midrule
    \multirow{4}{*}{\shortstack[l]{Power \\ (DGP1-2)}}
      & 100 & 0.171 & 0.155 & 0.137 & 0.113 & 0.173 & 0.153 & 0.147 & 0.115 \\
      & 200 & 0.355 & 0.335 & 0.288 & 0.267 & 0.324 & 0.316 & 0.298 & 0.248 \\
      & 300 & 0.529 & 0.486 & 0.448 & 0.370 & 0.517 & 0.475 & 0.436 & 0.374 \\
      & 400 & 0.719 & 0.655 & 0.594 & 0.518 & 0.671 & 0.624 & 0.577 & 0.504 \\
    \midrule
    \multirow{4}{*}{\shortstack[l]{Power \\ (DGP1-3)}}
      & 100 & 0.333 & 0.326 & 0.282 & 0.236 & 0.263 & 0.255 & 0.228 & 0.195 \\
      & 200 & 0.706 & 0.694 & 0.659 & 0.557 & 0.579 & 0.544 & 0.489 & 0.425 \\
      & 300 & 0.927 & 0.921 & 0.868 & 0.824 & 0.836 & 0.805 & 0.757 & 0.681 \\
      & 400 & 0.983 & 0.979 & 0.947 & 0.934 & 0.941 & 0.907 & 0.881 & 0.809 \\
    \bottomrule
  \end{tabular}%
  }
\end{table*}

